\documentclass[aps,prd,onecolumn,groupedaddress]{revtex4}

\usepackage{epsfig}
\usepackage{subfigure}
\usepackage{graphicx}
\usepackage{amsmath}

\def\d{{\rm d}}

\def\beq{\begin{equation}}
\def\eeq{\end{equation}}
\def\bea{\begin{eqnarray}}
\def\eea{\end{eqnarray}}

\def\nn{\nonumber}

\def\mG{m_{\tilde{G}}}

\begin{document}
\preprint{FREIBURG-PHENO 08-01, LPSC 07-116}

\title{Flavour Violation in Gauge-Mediated Supersymmetry Breaking Models:\\
       Experimental Constraints and Phenomenology at the LHC}

\author{Benjamin Fuks}
\affiliation{Physikalisches Institut, Albert-Ludwigs-Universit\"at Freiburg, Hermann-Herder-Stra{\ss}e 3,
D-79106 Freiburg im Breisgau, Germany}

\author{Bj\"orn Herrmann}
\affiliation{Laboratoire de Physique Subatomique et de Cosmologie,
 Universit\'e Joseph Fourier/CNRS-IN2P3/INPG,
 53 Avenue des Martyrs, F-38026 Grenoble, France}

\author{Michael Klasen}
\email[]{klasen@lpsc.in2p3.fr}
\affiliation{Laboratoire de Physique Subatomique et de Cosmologie,
 Universit\'e Joseph Fourier/CNRS-IN2P3/INPG,
 53 Avenue des Martyrs, F-38026 Grenoble, France}

\date{\today}

\begin{abstract}
We present an extensive analysis of gauge-mediated supersymmetry breaking models
with minimal and non-minimal flavour violation.  We first demonstrate that
low-energy, precision electroweak, and cosmological constraints exclude large
``collider-friendly'' regions of the minimal parameter space. We then discuss
various possibilities how flavour violation, although naturally suppressed, may
still occur in gauge-mediation models. The introduction of non-minimal flavour
violation at the electroweak scale is shown to relax the stringent experimental
constraints, so that benchmark points, that are also cosmologically viable, can
be defined and their phenomenology, i.e.\ squark and gaugino production cross
sections with flavour violation, at the LHC can be studied.
\end{abstract}

\pacs{12.60.Jv,13.85.Ni,14.80.Ly}

\maketitle

\section{Introduction \label{sec1}}

Weak scale supersymmetry (SUSY) remains a both theoretically and
phenomenologically attractive extension of the Standard Model (SM) of
particle physics \cite{Nilles:1983ge,Haber:1984rc}. Apart from linking
bosons with fermions and unifying internal and external (space-time)
symmetries, SUSY allows for a stabilization of the gap between the Planck
and the electroweak scale and for gauge coupling unification at high
energies. It appears naturally in string theories, includes gravity, and
contains a stable lightest SUSY particle (LSP) as a dark matter candidate.
Spin partners of the SM particles have not yet been observed, and in order
to remain a viable solution to the hierarchy problem, SUSY must be broken at
low energy via soft mass terms in the Lagrangian. As a consequence, the SUSY
particles must be massive in comparison to their SM counterparts, and the
Tevatron and the LHC will perform a conclusive search covering a wide range
of masses up to the TeV scale. After the discovery of SUSY particles, the
revelation of the underlying SUSY-breaking mechanism will be one of the key
challenges in the experimental high-energy physics program.

In gauge-mediated supersymmetry-breaking (GMSB) models, SUSY is broken in a
secluded sector at a scale $\langle F \rangle$ related to the gravitino mass by
$\mG = \langle F \rangle/(\sqrt{3} M_{\rm P})$, where $M_{\rm P}$ is the reduced
Planck mass. The breaking is mediated to the visible sector of squarks, sleptons,
gauginos and gluinos through a gauge-singlet chiral superfield $S$ and $n_q$
quark-like and $n_l$ lepton-like messenger fields \cite{Dine:1993yw,
Dine:1994vc, Dine:1995ag, Giudice:1998bp}. The superfield $S$ is characterized
by its scalar and auxiliary components, which overlap with the gravitino and
acquire vacuum expectation values $\langle S \rangle$ and $\langle F_S \rangle$,
respectively. Yukawa couplings of the messengers to the superfield $S$ then
induce masses of order $M_{\rm mes} \simeq\langle S\rangle$ for the messengers.
Gauginos and sfermions acquire masses through ordinary gauge interactions with
messengers through one- and two-loop self-energy diagrams, respectively. In
these scenarios, the lightest SUSY particle is always the gravitino, which is
thus a natural candidate for the dark matter in our Universe.
Besides $M_{\rm mes}$, $n_q$, and $n_l$, minimal GMSB scenarios are determined
by the ratio of the two Higgs vacuum expectation values, $\tan\beta$, the sign
of the off-diagonal Higgs mass-parameter $\mu$, and by the auxiliary vacuum
expectation value $\langle F_S \rangle$. The latter is related to the mass
splitting of the messenger fields and is considerably smaller than
both the squared mass scale of the messenger fields, $\langle S \rangle^2$, and
the fundamental SUSY-breaking scale, $\langle F\rangle$. It is usually
re-expressed in terms of an effective SUSY-breaking scale, $\Lambda=\langle
F_S\rangle/\langle S\rangle$. An additional free parameter is the gravitino mass,
$\mG$, which is, however, constrained by the fact that the gravitino relic
density $\Omega_{\tilde{G}}h^2$ has to agree with the current WMAP limits and
that the abundances of the light elements should be correctly described, i.e.\
the next-to-lightest SUSY particle (NLSP) must not decay too quickly.

GMSB is an attractive scenario regarding the so-called SUSY flavour problem. SUSY
is usually broken within a few orders of magnitude of the weak scale, whereas the
unrelated flavour-breaking scale can be chosen much higher. This avoids important
flavour-violating terms in the SUSY-breaking Lagrangian and leads to approximately
flavour-conserving mass matrices at the low-energy scale and good agreement with
measurements of flavour-changing neutral current observables. 
However, several possibilities reintroducing flavour-violating terms in the
Minimal Supersymmetric Standard Model (MSSM) with GMSB have been pointed out
\cite{Giudice:1998bp, Tobe:2003nx, Dubovsky:1998nr}. For example, mixing between
messenger and matter fields may lead to important flavour violations in the squark
and slepton sectors.

In SUSY models with non-minimal flavour violation (NMFV), the flavour-violating
off-diagonal terms $\Delta_{ij}$ of the squared sfermion mass matrices, where
$i,j={\rm L,R}$ refer to the helicities of the (SM partners of the) sfermions, are
conveniently considered as arbitrary parameters. Stringent experimental
constraints are then imposed by precise measurements of $K^0-\bar{K}^0$ and
$B^0-\bar{B}^0$ mixing, the first evidence of $D^0-\bar{D}^0$ mixing, and rare
decays \cite{Hagelin:1992tc, Gabbiani:1996hi, Ciuchini:2007ha}. The minimal GMSB
model obviously relies on constrained minimal flavour violation (cMFV), where all
the flavour-violating elements $\Delta_{ij}$ are neglected.
Recently, possible effects of non-minimal flavour violation on the
experimentally allowed minimal supergravity (mSUGRA) parameter
space have been investigated, and all squark and gaugino production cross
sections and decay widths have been recalculated including both helicity and
flavour mixing in the squark sector \cite{Bozzi:2007me}.
The aim of this work is to extend this study to GMSB scenarios,
to evaluate the experimental constraints, discuss the role
of flavour violation, and make numerical predictions for squark- and
gaugino-production cross sections at the LHC.

This paper is organized as follows: In Sec.\ \ref{sec2}, we impose the current
experimental constraints on the minimal GMSB models. We show that these
scenarios are strongly disfavoured due to the very stringent constraint coming
from the rare $b\to s\gamma$ decay. However, the latter can be relaxed by
introducing NMFV in the squark sector, as shown in Sec.\ \ref{sec3}, allowing us
to define benchmark points for NMFV GMSB scenarios. Sec.\ \ref{sec4} is devoted
to the discussion of cosmological implications on the gravitino mass in our
scenarios. In Sec.\ \ref{sec5}, we present numerical predictions for squark and
gaugino hadroproduction cross sections at the LHC. Our conclusions are given in
Sec.\ \ref{sec6}.

\section{Experimental Constraints on GMSB Models with Minimal Flavour Violation \label{sec2}}

In the absence of experimental evidence for Supersymmetry, a large variety of
data can be used to constrain the parameter space of the MSSM. Sparticle mass
limits can be obtained from searches of charginos ($m_{\tilde{\chi}^\pm_1} \ge
150$ GeV from D0), neutralinos ($m_{\tilde{\chi}^0_1} \ge 93$ GeV in GMSB from
the combination of LEP2 results), gluinos ($m_{\tilde g} \ge 195$ GeV from CDF),
stops ($m_{\tilde{t}_1} \ge 95 \dots 96$ GeV for neutral- or charged-current
decays from the combination of LEP2 results), other squarks ($m_{\tilde q}\ge
300$ GeV for gluinos of equal mass from CDF), and gravitinos ($\mG \ge 1.3\cdot
10^{-5}$ eV for $m_{\tilde{q}} = m_{\tilde{g}}  = 200$ GeV) at colliders
\cite{Yao:2006px, Klasen:2006kb}.

Cosmological, electroweak precision, and low energy observables can be used to
put additional constraints on the SUSY parameter space. The theoretically robust
inclusive branching ratio 
\beq
    {\rm BR}(b\to s\gamma) ~=~ (3.55 \pm 0.26) \cdot 10^{-4},
\label{eq.bsg}
\eeq
obtained from the combined measurements of BaBar, Belle, and CLEO
\cite{Barberio:2006bi}, can be confronted to theoretical predictions
including two-loop QCD and one-loop SUSY contributions \cite{Hahn:2005qi,%
Kagan:1998bh}. Squarks contribute here already at the one-loop level, as do the
SM contributions. A second observable, sensitive to the squark-mass splitting
within isospin doublets, is the electroweak $\rho$-parameter with
\beq
    \Delta\rho ~=~ \frac{\Sigma_Z(0)}{m_Z^2} - \frac{\Sigma_W(0)}{m_W^2},
\eeq
where $m_{Z,W}$ and $\Sigma_{Z,W}(0)$ denote the $Z$- and $W$-boson masses and
self-energies at zero momentum, respectively.
New physics contributions are constrained by the latest combined
electroweak precision measurements to $T = -0.13 \pm 0.11$ or
\beq
    \Delta\rho ~=~ -\alpha T ~=~ (1.02 \pm 0.86) \cdot 10^{-3}
\label{eq.drho}    
\eeq
for $\alpha(m_Z) = 1/127.918$ \cite{Yao:2006px}. This value is compared to
theoretical calculations including SUSY two-loop corrections
\cite{Heinemeyer:2004by}. A third variable sensitive to new physics loop
contributions is the anomalous magnetic moment of the muon, for which 
we require the SUSY contribution $a_{\mu}^{\rm SUSY}$, known up to the two-loop
level \cite{Heinemeyer:2003dq, Heinemeyer:2004yq}, to close the gap between
recent BNL experimental data and the SM prediction \cite{Yao:2006px},
\beq
    \Delta a_{\mu} ~=~ (29.2 \pm 8.6)\cdot 10^{-10} .
\label{eq.amu}
\eeq
Note that the one-loop SUSY contributions are approximatively given by
\cite{Moroi:1995yh} 
\beq
    a_{\mu}^{\rm SUSY, 1-loop} ~\simeq~ 13\cdot 10^{-10} \biggr(
    \frac{100{\rm ~GeV}}{M_{\rm SUSY}} \biggr)^2 \tan\beta ~{\rm sgn}(\mu), 
\eeq
if the relevant SUSY particles have masses of the order of $M_{\rm SUSY}$. As a
consequence, negative values of $\mu$ then increase, not decrease, the
disagreement between the experimental measurements and the theoretical value
of $a_{\mu}$, so that the region $\mu<0$ is strongly disfavoured in all SUSY
models. In addition, this region is also virtually excluded by the $b\to
s\gamma$ constraint at the $2\sigma$ confidence level. We therefore restrict
ourselves to positive values of $\mu$ throughout this analysis. 

The above experimental limits are imposed at the 2$\sigma$ confidence level on
the minimal GMSB model with $\mu>0$ and four free parameters $\Lambda$,
$M_{\rm mes}$,
$N_{\rm mes} \equiv n_q = n_l$, and $\tan\beta$. The renormalization
group equations (RGEs) are solved  numerically to two-loop order using 
the computer programme {\tt SPheno 2.2.3} \cite{Porod:2003um}, which computes the
soft SUSY-breaking masses at the electroweak scale with the complete one-loop formulas,
supplemented by two-loop contributions for the neutral Higgs bosons and the
$\mu$-parameter. We then diagonalize the mass matrices and compute the
electroweak precision and low-energy observables with the computer programme
{\tt FeynHiggs 2.6.4} \cite{Heinemeyer:1998yj}. For the SM input
parameters, i.e.\ the masses and widths of the electroweak gauge bosons and
quarks, the angles of the CKM-matrix and its $CP$-violating phase, and the
Fermi coupling constant, we refer the reader to Ref.\ \cite{Yao:2006px}. 

\begin{figure}
\begin{center}
	\includegraphics[scale=0.27]{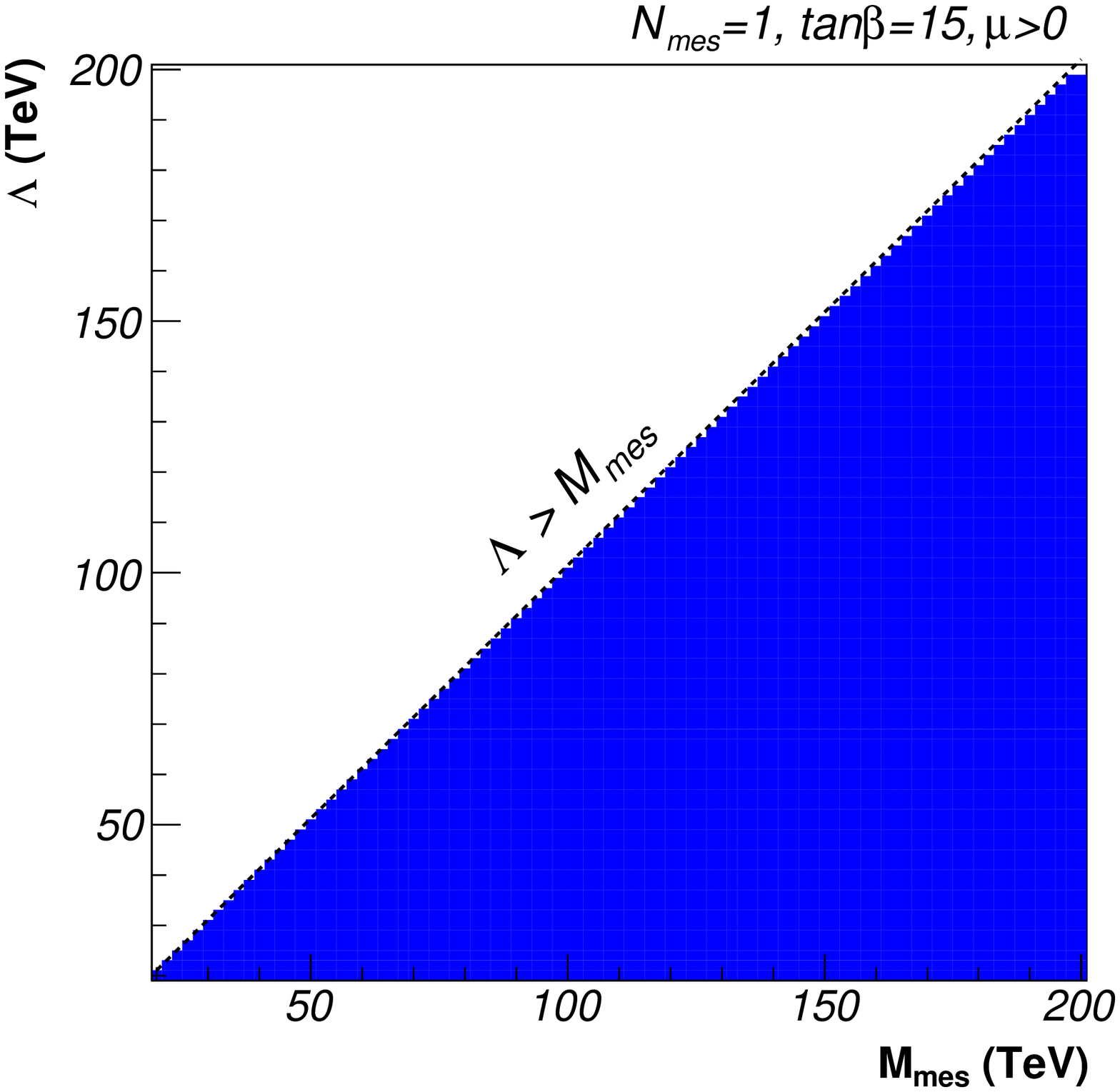}
	\includegraphics[scale=0.27]{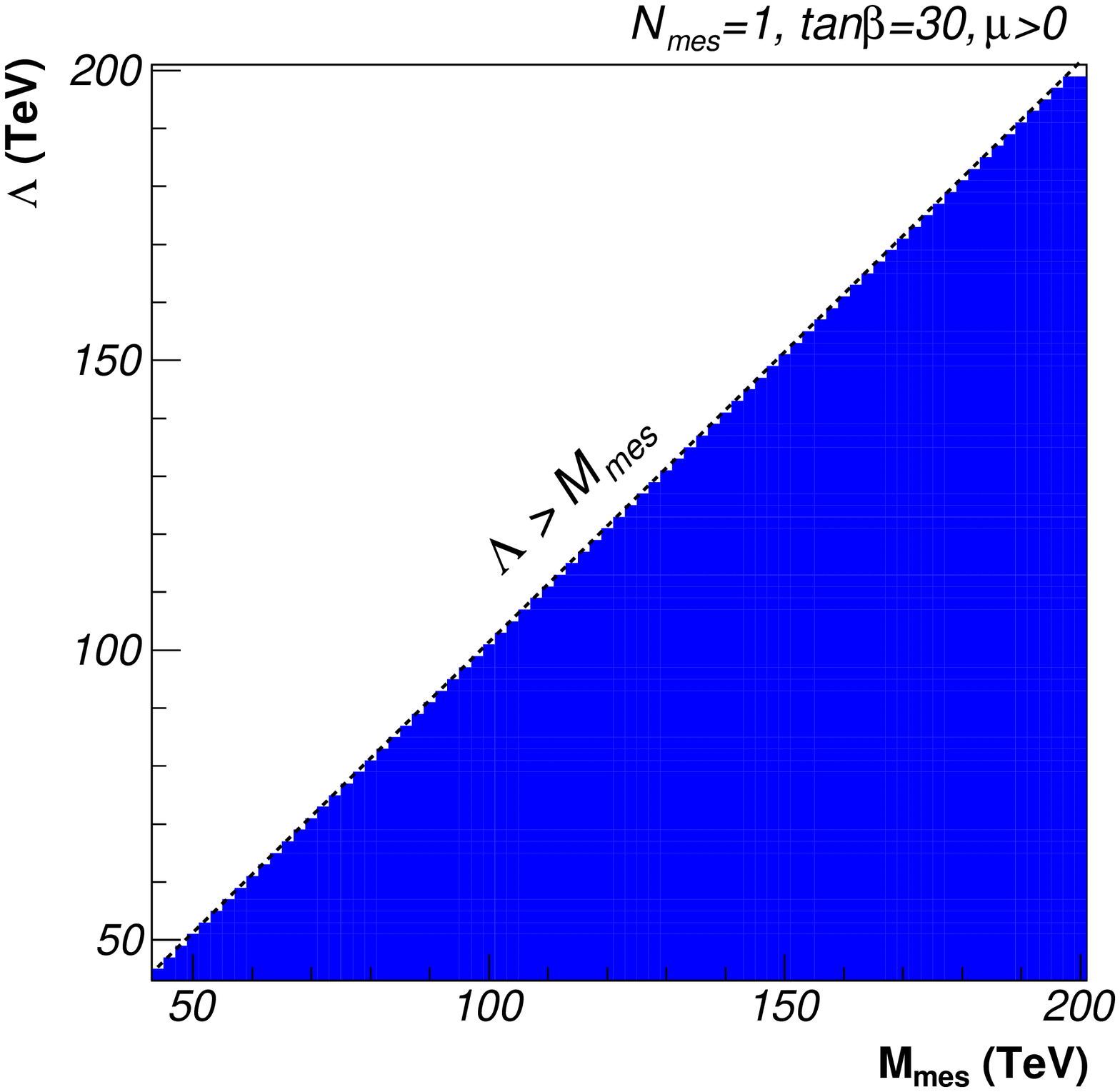}
	\includegraphics[scale=0.27]{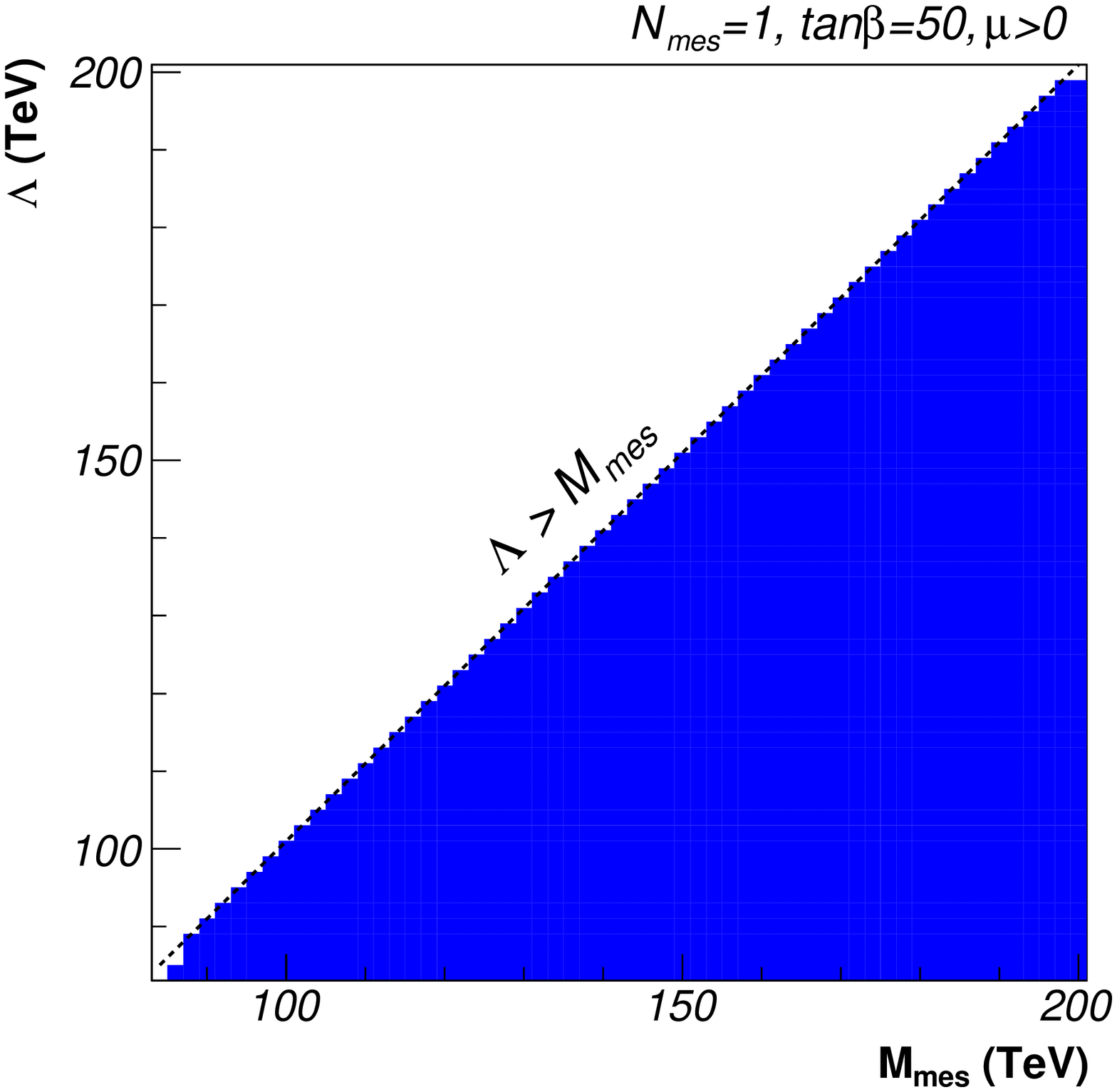}
	\includegraphics[scale=0.27]{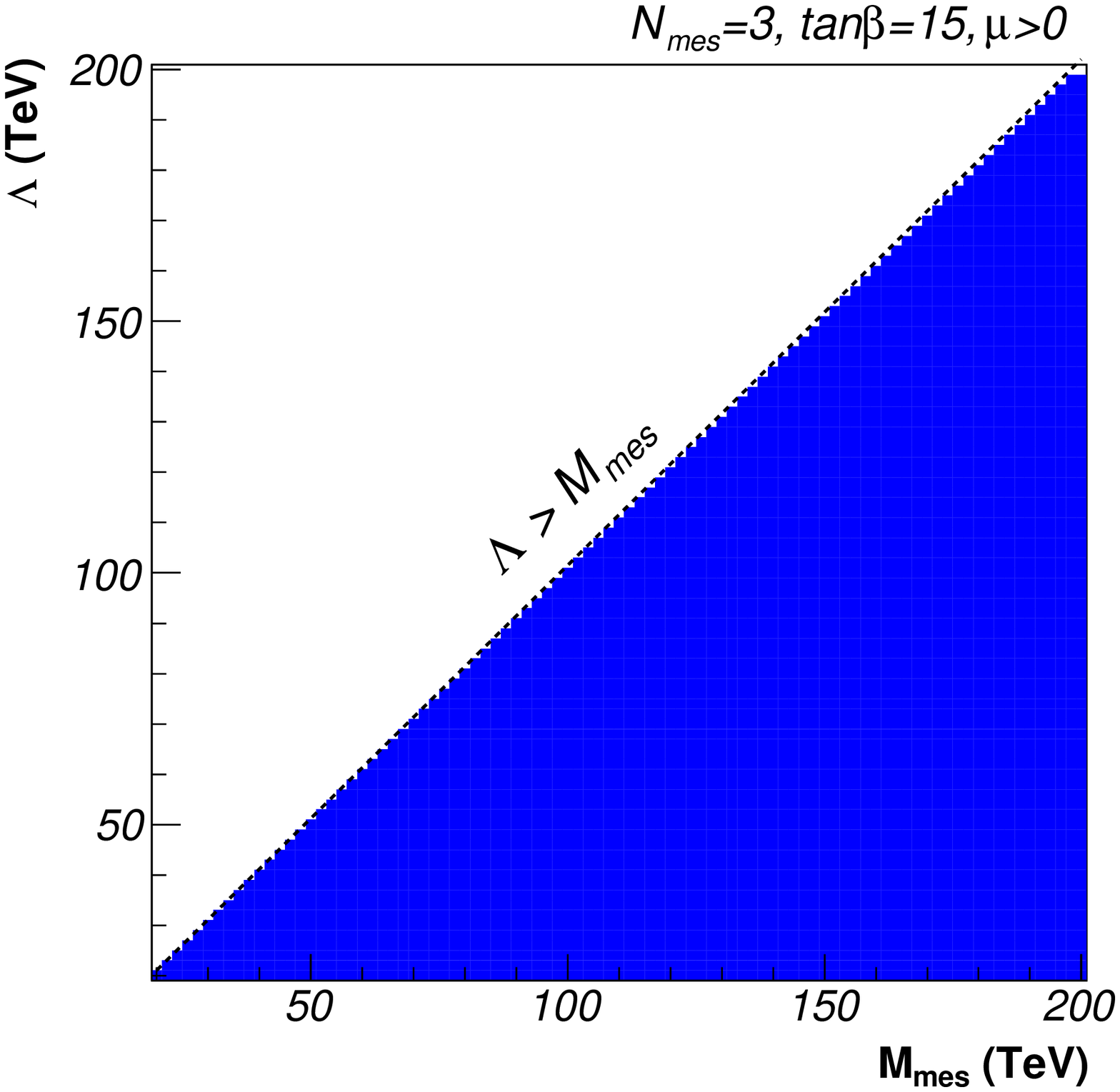}
	\includegraphics[scale=0.27]{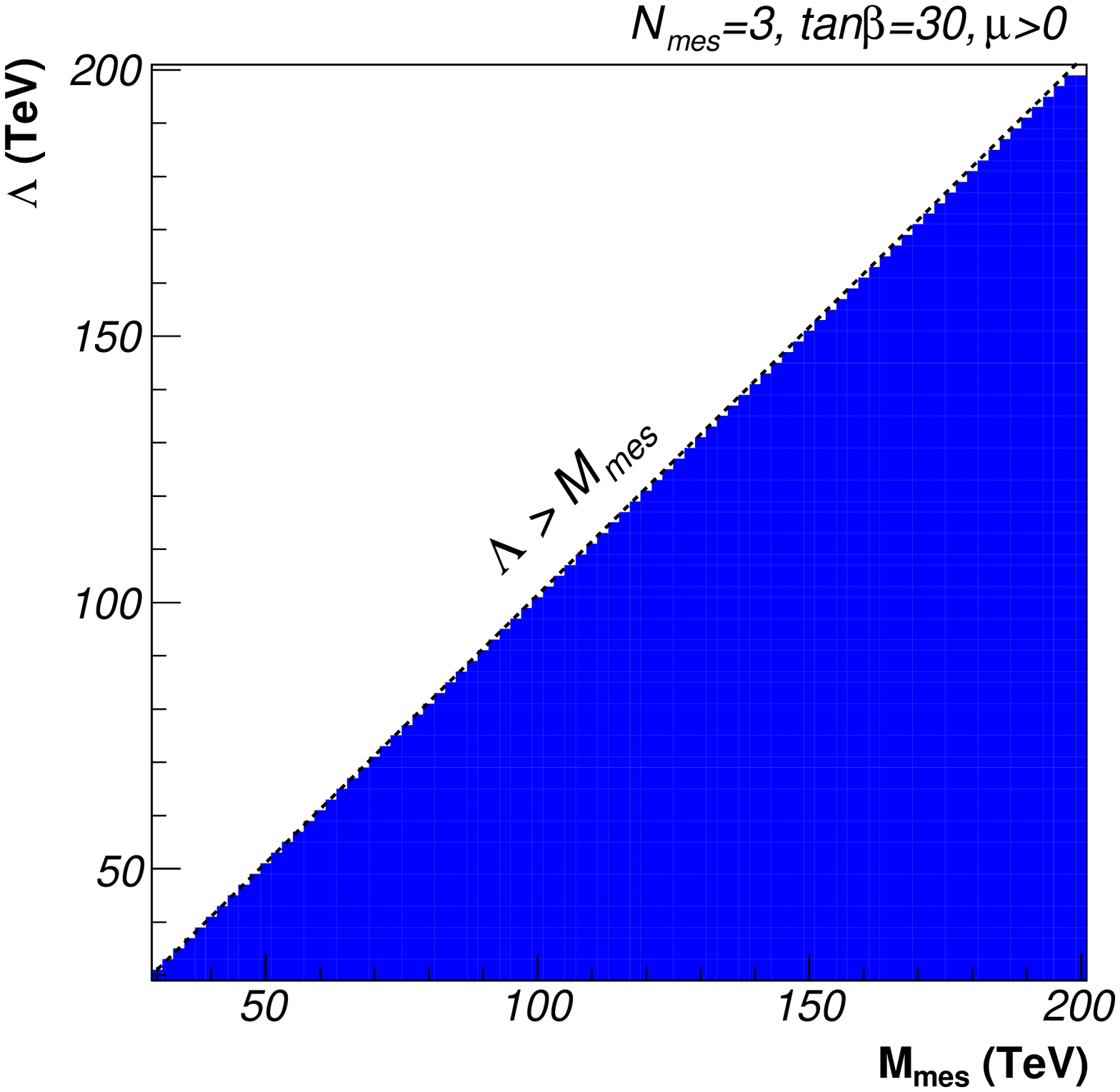}
	\includegraphics[scale=0.27]{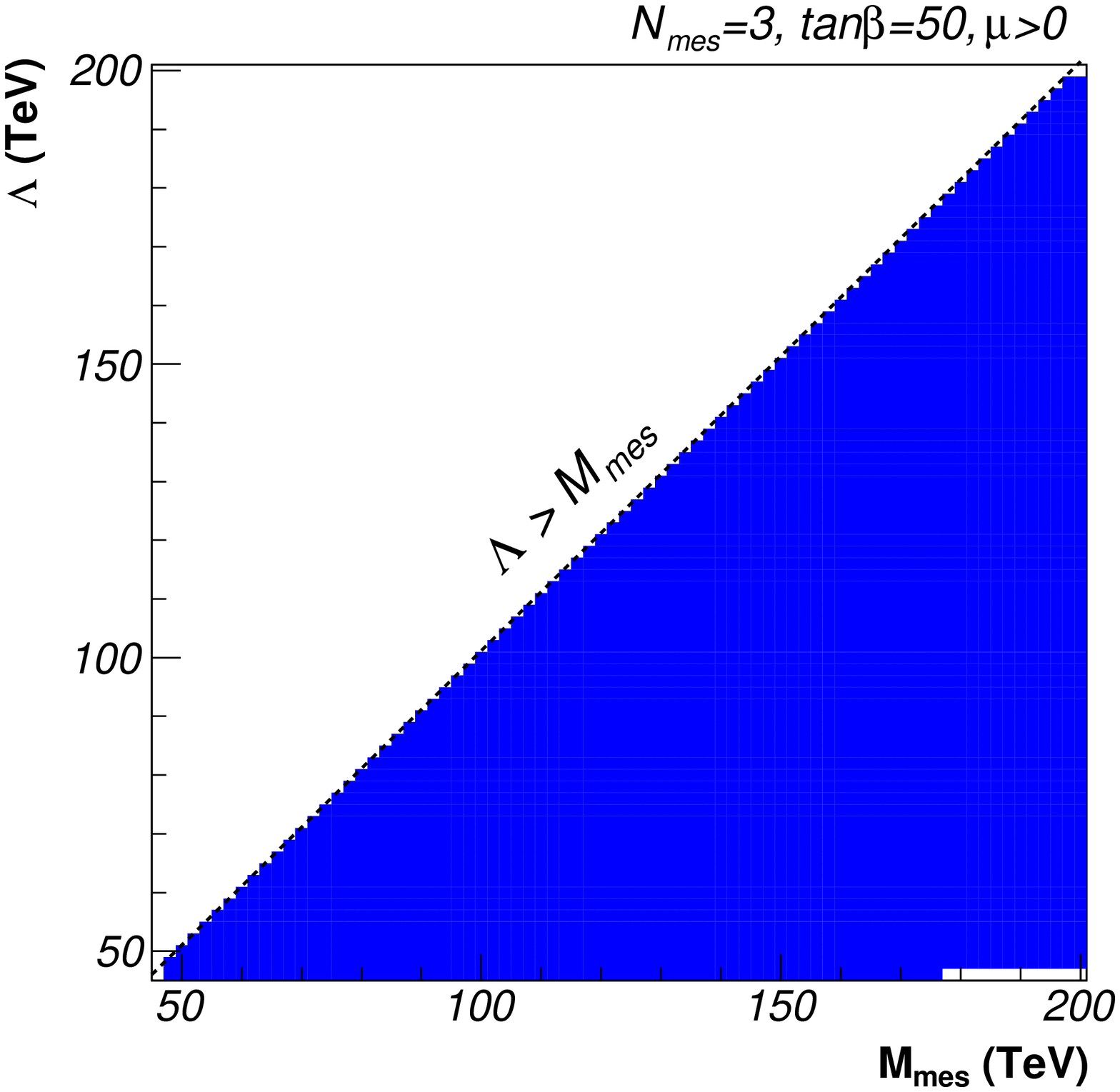}
\caption{The $\Lambda$--$M_{\rm mes}$ planes for $\mu>0$ and different values of
 $\tan\beta$ and $N_{\rm mes}$, assuming constrained minimal flavour violation.
 Dark (blue) regions are excluded by the constraint coming from the $b\to s
 \gamma$ branching ratio. The regions where $\Lambda > M_{\rm mes}$ does not
 allow for physical solutions of the RGEs.}
\label{fig1}
\end{center}
\end{figure}


In Fig.\ \ref{fig1}, we show typical scans of the minimal GMSB parameter space in
$\Lambda$ and $M_{\rm mes}$ for different values of $\tan\beta$ (15, 30, and 50)
and $N_{\rm mes}$ (1 and 3). The six panels reveal that these scenarios
are strongly disfavoured by the measurements of the $b\to s\gamma$ branching
ratio. 
In particular, the Snowmass benchmark points \cite{Allanach:2002nj} SPS 7 ($\Lambda=40$
TeV, $M_{\rm mes}=80$ TeV, $\tan\beta=15$, $\mu>0$, and $N_{\rm mes}=3$) and SPS
8 ($\Lambda=100$ TeV, $M_{\rm mes}=200$ TeV, $\tan\beta=15$, $\mu>0$, and $N_{\rm
mes}=1$) lead to values of BR$(b\to s\gamma) = 6.97 \cdot 10^{-4}$ and $6.77
\cdot 10^{-4}$, which are both excluded beyond the 5$\sigma$ level,
even if both of these points lie well within 2$\sigma$ of the
experimentally allowed range for the anomalous magnetic moment of the muon with
$a_{\mu}^{\rm SUSY} = 22.8\cdot 10^{-10}$ and $a_{\mu}^{\rm SUSY} = 16.31\cdot
10^{-10}$. Note that the regions with $\Lambda > M_{\rm mes}$ are theoretically
excluded, since they do not allow for physical solutions of the RGEs.

Recently, a detailed study of electroweak precision observables, including
scenarios with minimal GMSB, has been performed \cite{Heinemeyer:2008fb}.
Scanning also over $\tan\beta$ and allowing for higher values of $N_{\rm mes}
\le 8$, the authors show that experimentally favoured scenarios can be achieved
at low messenger scales, which, however, implies a certain amount of fine
tuning at the weak scale, e.g.\ in the Higgs sector. Note that for $N_{\rm mes}
\gtrsim 8$ problems with perturbativity of the gauge interactions arise at very
high scales \cite{Giudice:1998bp}.

\section{GMSB Models with Non-Minimal Flavour Violation\label{sec3}}

The minimal GMSB is known to suppress flavour-changing neutral currents as
suggested by measurements and thus to avoid the SUSY ``flavour problem'', which
arises naturally in models where SUSY-breaking is mediated by gravity. Models
beyond the minimal GMSB can, however, reintroduce flavour-breaking terms at the
electroweak scale. In this Section, we first review flavour violation in the MSSM,
present its implementation at the electroweak scale, and elaborate on different
non-minimal GMSB models including flavour violation. We then re-analyze the
parameter space and show how NMFV can provide a way to relax the stringent
constraints challenging the minimal GMSB models.

\subsection{Theoretical Framework \label{sec3a}}

In constrained minimal flavour violation (cMFV) SUSY models, the
only source of flavour violation arises through the rotation of
the quark interaction eigenstates into the basis of physical mass
eigenstates, where the Yukawa matrices are diagonal, as in the SM,
and the flavour-violating entries of the squark mass matrices are
neglected both at the SUSY-breaking and the weak scale. In SUSY
with NMFV, these flavour-violating entries $\Delta_{ij}^{qq'}$ are
considered as free parameters. The squared squark mass matrices
are then given by
\renewcommand{\arraystretch}{1.2}
\beq
    \hspace*{-2.2mm} M_{\tilde{q}}^2 ~=~ \left( \begin{array}{ccc|ccc}
        M^2_{L_1} & \Delta^{12}_{LL} & \Delta^{13}_{LL} & m_1 X_1 & \Delta^{12}_{LR} & \Delta^{13}_{LR} \\
        \Delta^{12*}_{LL} & M^2_{L_2} & \Delta^{23}_{LL} & \Delta^{12*}_{RL} & m_2 X_2 & \Delta^{23}_{LR} \\
            \Delta^{13*}_{LL} &  \Delta^{23*}_{LL} & M^2_{L_3} & \Delta^{13*}_{RL} & \Delta^{23*}_{RL} & m_3 X_3 \\
        \hline
        m_1 X_1^* & \Delta^{12}_{RL} & \Delta^{13}_{RL} & M^2_{R_1} & \Delta^{12}_{RR} & \Delta^{13}_{RR} \\
        \Delta^{12*}_{LR}~& m_2 X_2^* &  \Delta^{23}_{RL} & \Delta^{12*}_{RR} & M^2_{R_2} & \Delta^{23}_{RR} \\
        \Delta^{13*}_{LR}~& \Delta^{23*}_{LR} & m_3 X_3^* & \Delta^{13*}_{RR} & \Delta^{23*}_{RR} & M^2_{R_3}
        \end{array} \right) ,
\eeq
where $M_{L_k}^2$ and $M_{R_k}^2$ denote their usual diagonal entries, 
\bea
	M^2_{L_k} &=& M^2_{Q_k} + m_k^2 + \cos 2\beta \ m^2_Z \left( T^3_k - e_k
	\sin^2\theta_W \right), \\
	M^2_{R_k} &=& M^2_{U_k,D_k} + m_k^2 + \cos 2\beta \ m^2_Z e_k
	\sin^2\theta_W,
\eea
and helicity mixing is generated by the elements
\beq
	X_q ~=~ A_q^* - \mu \left\{ \begin{array}{lcl} \cot\beta & \quad & {\rm
	for\ up-type\ squarks,} \\ \tan\beta & & {\rm for\ down-type\ squarks.}
	\end{array} \right. 
\eeq
Here, $\theta_W$ is the electroweak mixing angle,
$M_{Q,U,D}$ are the usual SUSY-breaking squark masses, $A_q$ is the trilinear
coupling, and
$m_k$, $T^3_k$, and $e_k$ denote the mass, weak isospin, and electric
charge of the quark $q_k$, the index $k$ referring to the (s)quark generation. The
flavour-violating elements $\Delta_{ij}^{qq'}$ are usually normalized to the
diagonal entries \cite{Gabbiani:1996hi},  
\beq
    \Delta^{qq'}_{ij} ~=~ \lambda^{qq'}_{ij} M_{i_q} M_{j_{q'}} ,
\eeq
so that NMFV is governed by 24 arbitrary complex dimensionless parameters
$\lambda^{qq'}_{ij}$.

The diagonalization of the mass matrices $M_{\tilde{u}}^2$ and
$M_{\tilde{d}}^2$ requires the introduction of two additional
$6\times 6$ matrices $R^u$ and $R^d$, relating the helicity and flavour
eigenstates to the physical mass eigenstates through
\bea
    \hspace*{-2.2mm}
    (\tilde{u}_1,\tilde{u}_2,\tilde{u}_3,\tilde{u}_4,\tilde{u}_5,\tilde{u}_6)^T & = &
    R^u (\tilde{u}_L, \tilde{c}_L, \tilde{t}_L, \tilde{u}_R, \tilde{c}_R, \tilde{t}_R)^T , \\
    \hspace*{-2.2mm}
    (\tilde{d}_1,\tilde{d}_2,\tilde{d}_3,\tilde{d}_4,\tilde{d}_5,\tilde{d}_6)^T & = &
    R^d (\tilde{d}_L, \tilde{s}_L, \tilde{b}_L, \tilde{d}_R, \tilde{s}_R, \tilde{b}_R)^T .
\eea
By convention, the squark mass eigenstates are labeled according
to $m_{\tilde{q}_1} < ... < m_{\tilde{q}_6}$ for $q=u,d$. For a detailed review
of flavour violation in the MSSM see e.g.\ Ref.\ \cite{delAguila:2008iz}.
We stress that we do not employ the mass-insertion approximation with its
perturbative expansion in the parameters $\lambda_{ij}^{qq'}$, but rather
perform the diagonalization of the squark-mass matrices numerically.
Relatively strong constraints on NMFV SUSY models can be obtained from low-energy
and electroweak precision observables, e.g.\ upper limits from the neutral kaon
sector, on $B$- and $D$-meson oscillations, various rare decays, and electric
dipole moments. In several publications \cite{Gabbiani:1996hi,Ciuchini:2007ha,%
Foster:2006ze}, rather complete analyses have been presented, pointing out that
the down-squark sector is particularly constrained from $K$- and $B$-physics
processes with external down-type quarks and that within the mass-insertion
approximation the only substantial mixing in the squark sector occurs between the
second and third generations in the left-left and right-right chiral sectors.
Note that the latter is suppressed in gravity mediation models by the scaling of
the corresponding entries $\Delta_{ij}^{qq'}$ with the SUSY breaking scale,
while in gauge mediation models the mixing in the left-right chiral sector,
induced by $A$-terms, is small. The up-squark sector is in general
less experimentally constrained. This situation may change once
additional information from neutral and charged Higgs production and decay
becomes available \cite{Hahn:2005qi,Dittmaier:2007uw}.
In our analysis, we apply $SU(2)$ gauge invariance to the left-chiral sector and
take implicitly into account the above mentioned constraints by restricting
ourselves, also for the sake of simplicity, to the case of two real NMFV
parameters,
\beq
	\lambda_{\rm LL} \equiv \lambda_{\rm LL}^{sb} \simeq \lambda_{\rm LL}^{ct}
	\lesssim 0.2
	\qquad {\rm and} \qquad 
	\lambda_{\rm RR} \equiv \lambda_{\rm RR}^{sb} \simeq \lambda_{\rm RR}^{ct}
	\lesssim 0.2,
\eeq
while all other $\lambda_{ij}^{qq'}$ are zero.

Although the gauge interactions are flavour-blind, it has been shown that there
are several possibilities for flavour violation in both the squark and slepton
sectors to arise within GMSB models. For
example, for very high messenger scales $M_{\rm mes} \gtrsim 10^{15}$ GeV gravity
is no longer negligible with respect to gauge interactions
\cite{Giudice:1998bp}. As a consequence, flavour-violating terms are reintroduced 
through gravity mediation as in mSUGRA models. However, scenarios with such high
messenger scales are rather unattractive from a phenomenological point of view
due to the resulting very high SUSY masses.
Second, flavour violation can also be induced from heavy right-handed neutrinos
participating in leptogenesis \cite{Tobe:2003nx}. If these are lighter than the
messenger scale, flavour off-diagonal mass terms are introduced into the slepton
mass matrices. A third possibility might be to consider broken messenger number
invariance, that implies that the lightest messenger is not stable and 
introduces flavour-violating terms in the Lagrangian at the weak scale
\cite{Giudice:1998bp}. A disadvantage of this model is that the now
unstable lightest messenger may not be a viable candidate for cold dark matter
in the case of a very light gravitino, which cannot account for the observed
relic abundance.

For our study, we focus on the model proposed in Ref.\ \cite{Dubovsky:1998nr},
based on the introduction of a mixing between messenger and matter fields. In
the case of fundamental messenger multiplets belonging to $\bf 5$ and
$\bf\overline{5}$ representations of $SU(5)$, the messengers carry quantum
numbers of left-handed leptons and right-handed down-type quarks. As a
consequence, flavour
violation is introduced into the chiral sectors of right-handed sleptons and
left-handed (up- and down-type) squarks. In our analysis of squark flavour violation, this
corresponds to including a variation of the parameter $\lambda_{\rm LL}$, while
$\lambda_{\rm RR}$ is set to zero. An alternative model with antisymmetric
messenger multiplets belonging to $\bf 10$ and $\bf\overline{10}$
representations can also be considered. In this case, the messengers share
quantum numbers with right-handed leptons, left-handed up- and down-type quarks,
and right-handed up-type quarks, leading to flavour mixing for left-handed
sleptons as well as for both left- and right-handed up- and down-type squarks.
Note that, in this antisymmetric scenario, flavour
mixing in the sector of right-handed down-type squarks may be parameterized
independently of the other squarks.  In our analysis, however, we use for
simplicity the same flavour violation parameter $\lambda_{\rm LL} = \lambda_{\rm
RR}$ for both chiral sectors. We should stress that
in both scenarios flavour violation is completely governed by the parameter
$\lambda_{\rm LL}$.  

\subsection{Scans of the Parameter Space and Benchmark Points \label{sec3b}}

We now re-investigate the constraints discussed in Sec.\ \ref{sec2} and
include non-minimal flavour violation as discussed above. 
Allowed regions for the parameters $\Lambda$, $M_{\rm mes}$, $N_{\rm mes}$,
$\tan\beta$, ${\rm sgn}(\mu)$, $\lambda_{\rm LL}$, and $\lambda_{\rm RR}$ are 
obtained by explicitly imposing the constraints from $b\to s\gamma$,
$\Delta\rho$, and $a_{\mu}$, that are sensitive to flavour-violating terms.
In particular, the branching ratio ${\rm BR}(b\to s\gamma)$ is directly
affected by the allowed squark mixing between the second 
and third generation. Squarks enter the calculation at the one-loop level, as
do the SM contributions, so that the dependence on the NMFV-parameters is rather
important. A second important consequence of NMFV in the MSSM is the generation
of a large splitting between squark-mass eigenvalues, which directly influences
the electroweak precision variable $\Delta\rho$. Concerning $a_{\mu}^{\rm SUSY}$,
the squark contributions are suppressed with respect to the slepton
contributions, so that its dependence on flavour violation is less important. 
The renormalization
group running is again performed with {\tt SPheno 2.2.3}. The flavour 
violating terms $\lambda_{\rm LL}$ and $\lambda_{\rm RR}$ are included in the
squark sector at the weak scale as discussed in Sec.\ \ref{sec3a} before
diagonalizing the mass matrices and computing the low-energy and electroweak
precision observables with {\tt FeynHiggs 2.6.4}. 

\begin{figure}\begin{center}
	\includegraphics[scale=0.27]{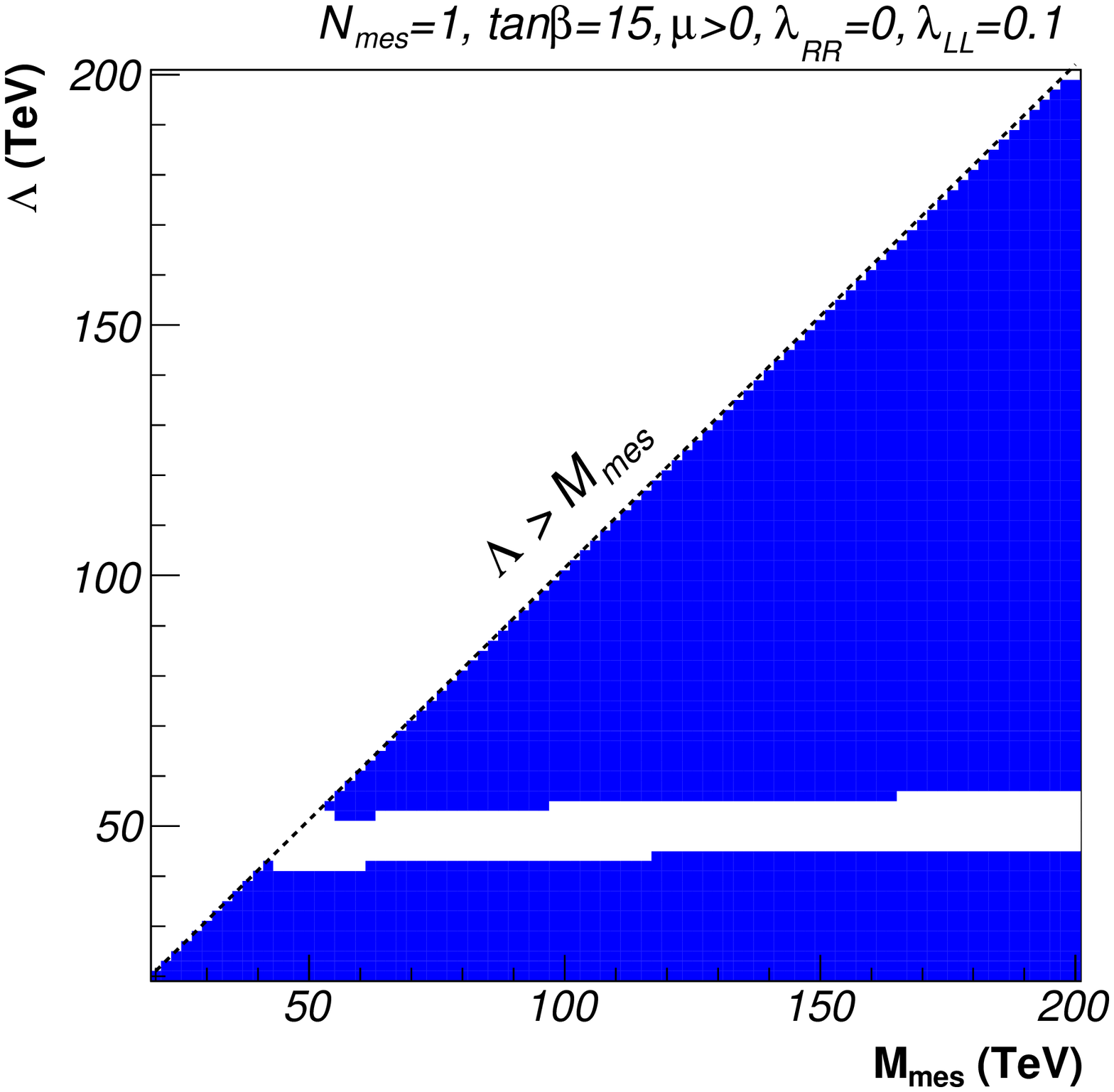}
	\includegraphics[scale=0.27]{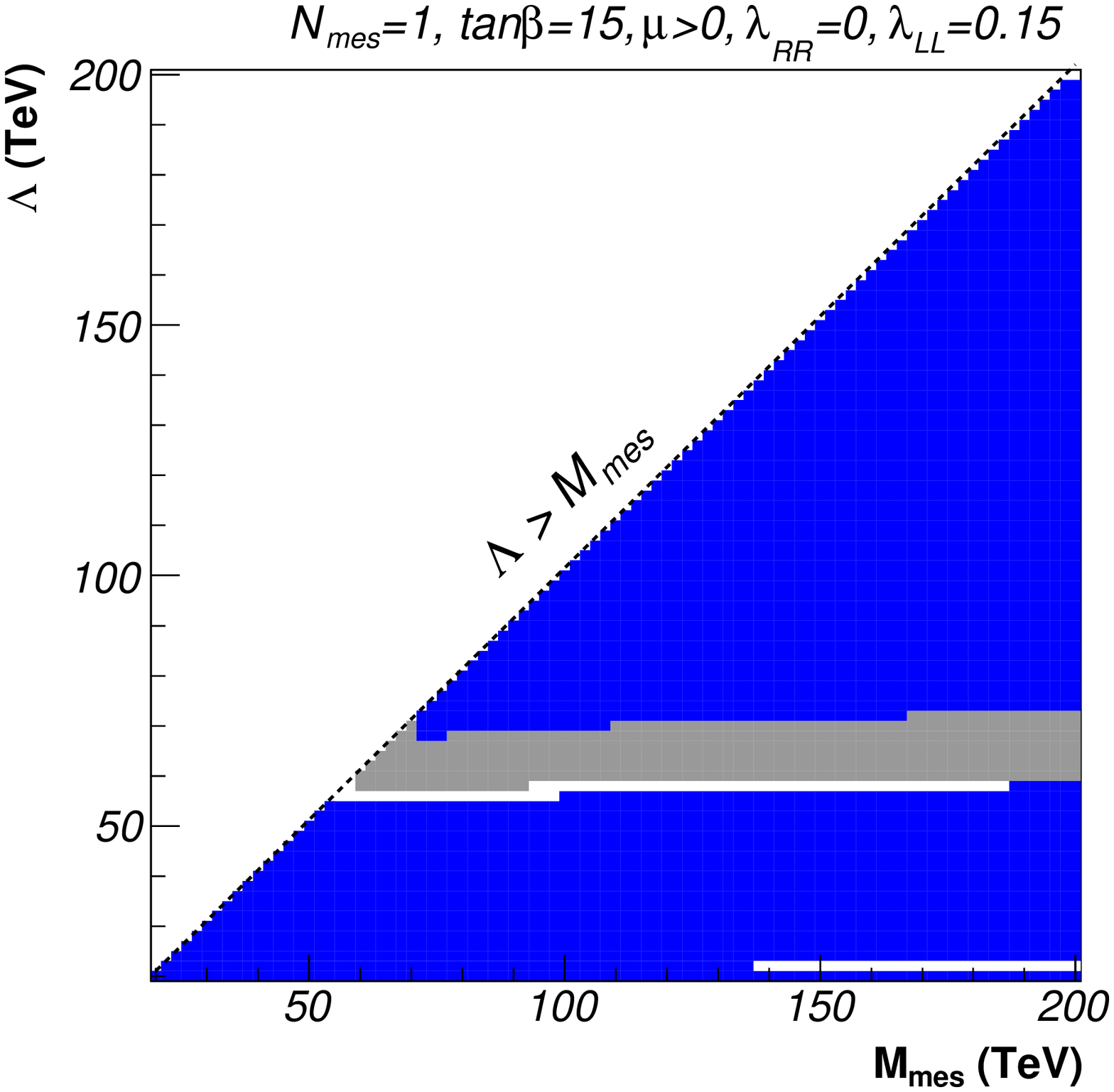}
	\includegraphics[scale=0.27]{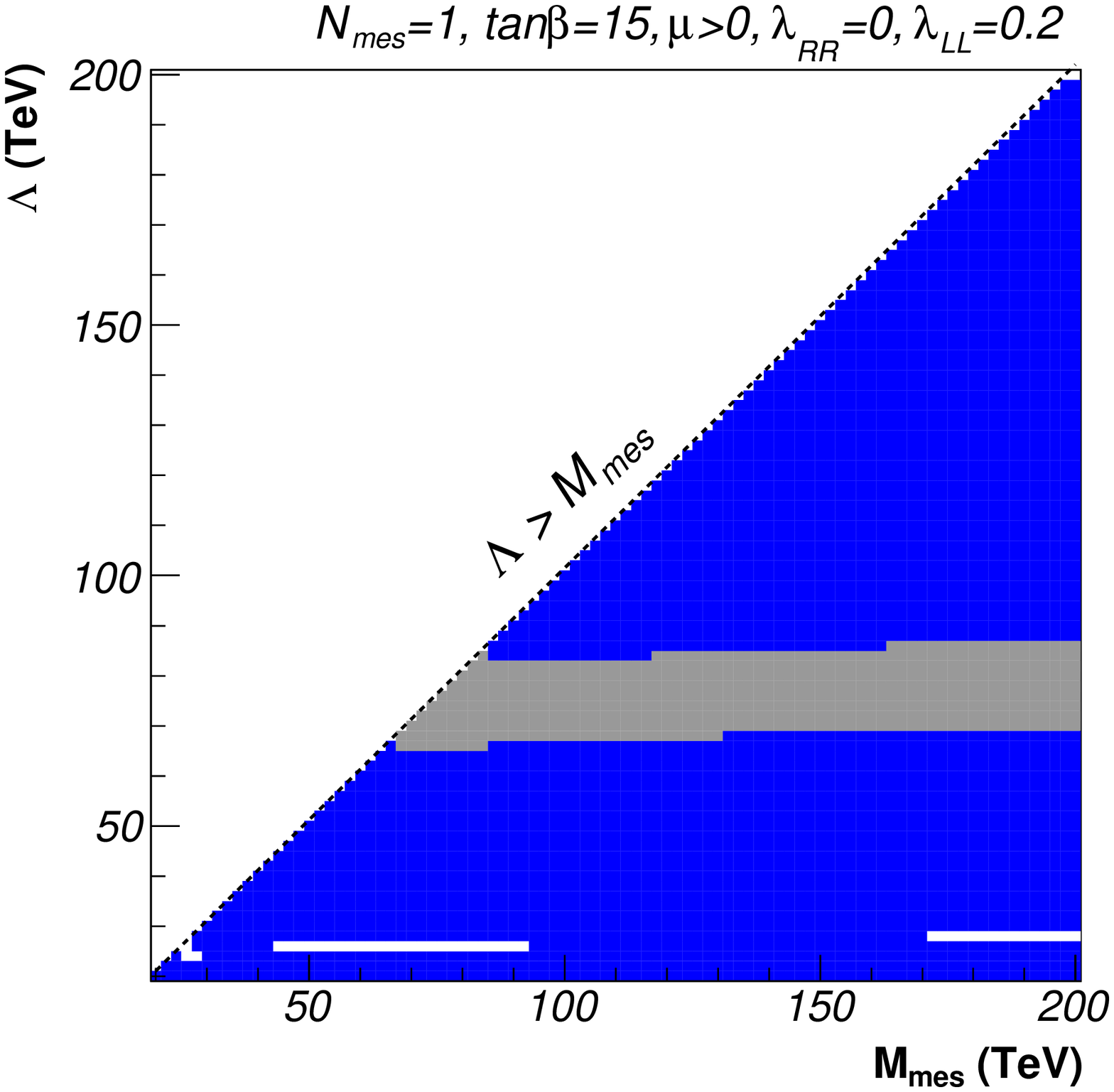}
	\includegraphics[scale=0.27]{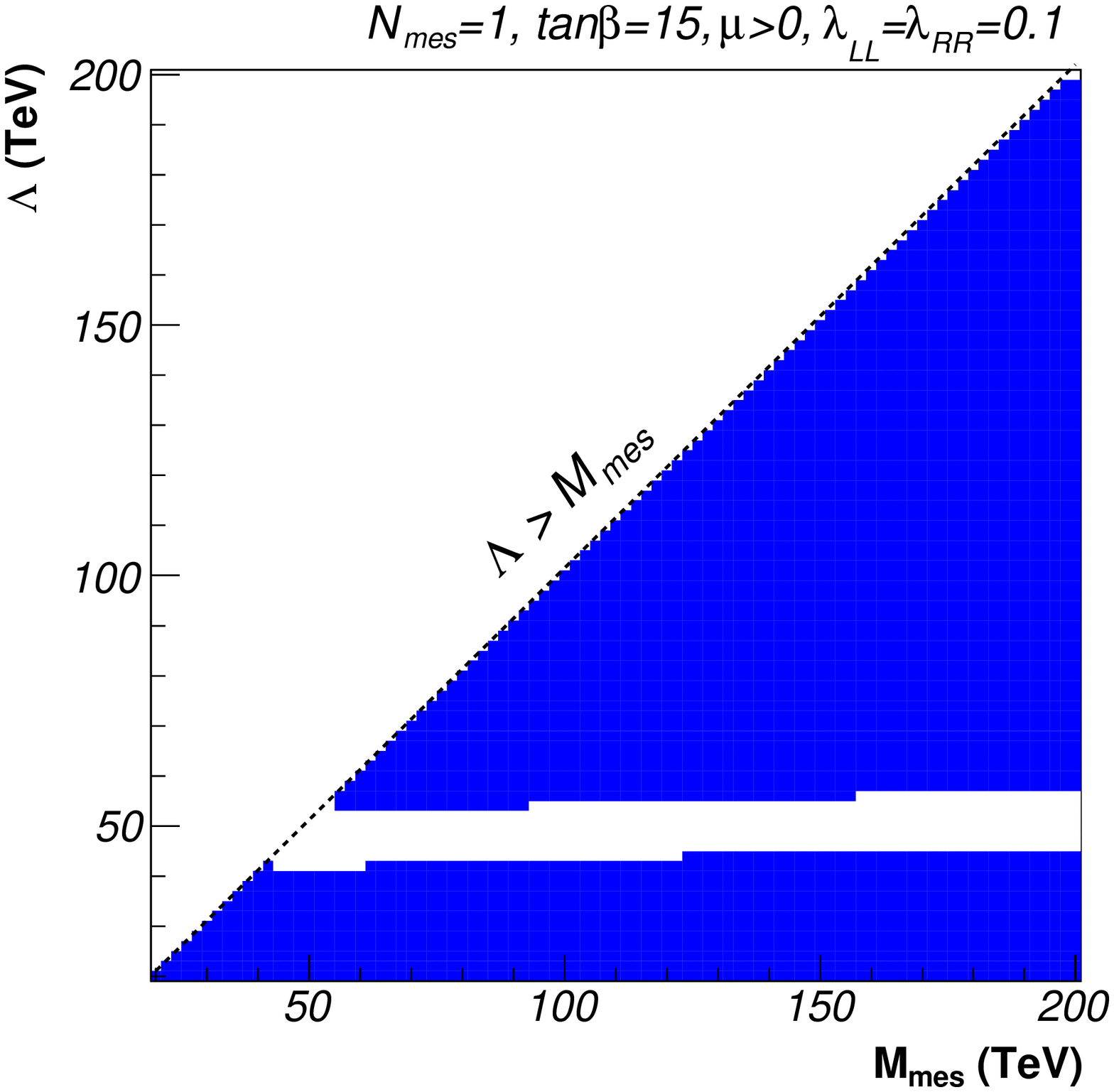}
	\includegraphics[scale=0.27]{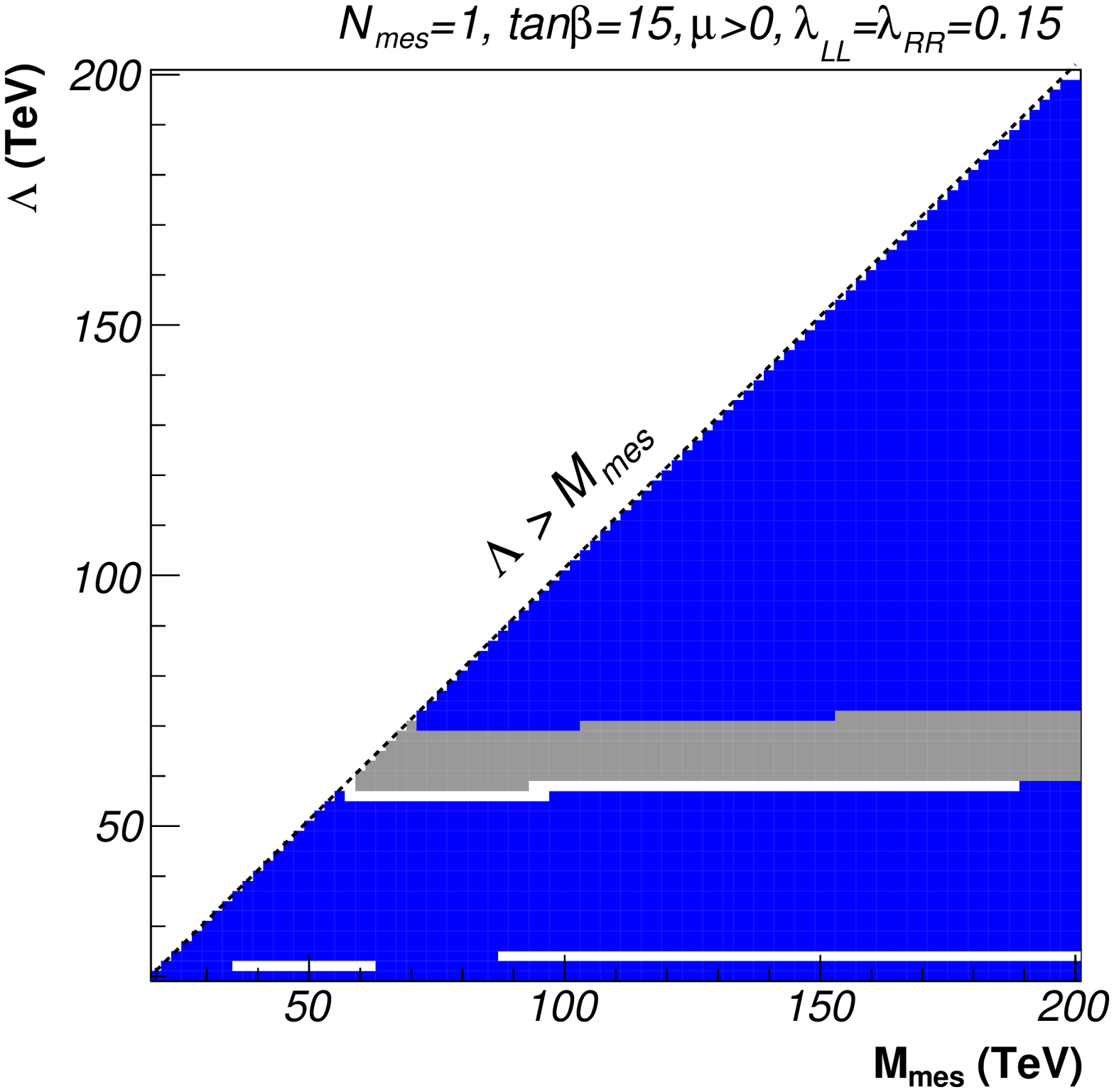}
	\includegraphics[scale=0.27]{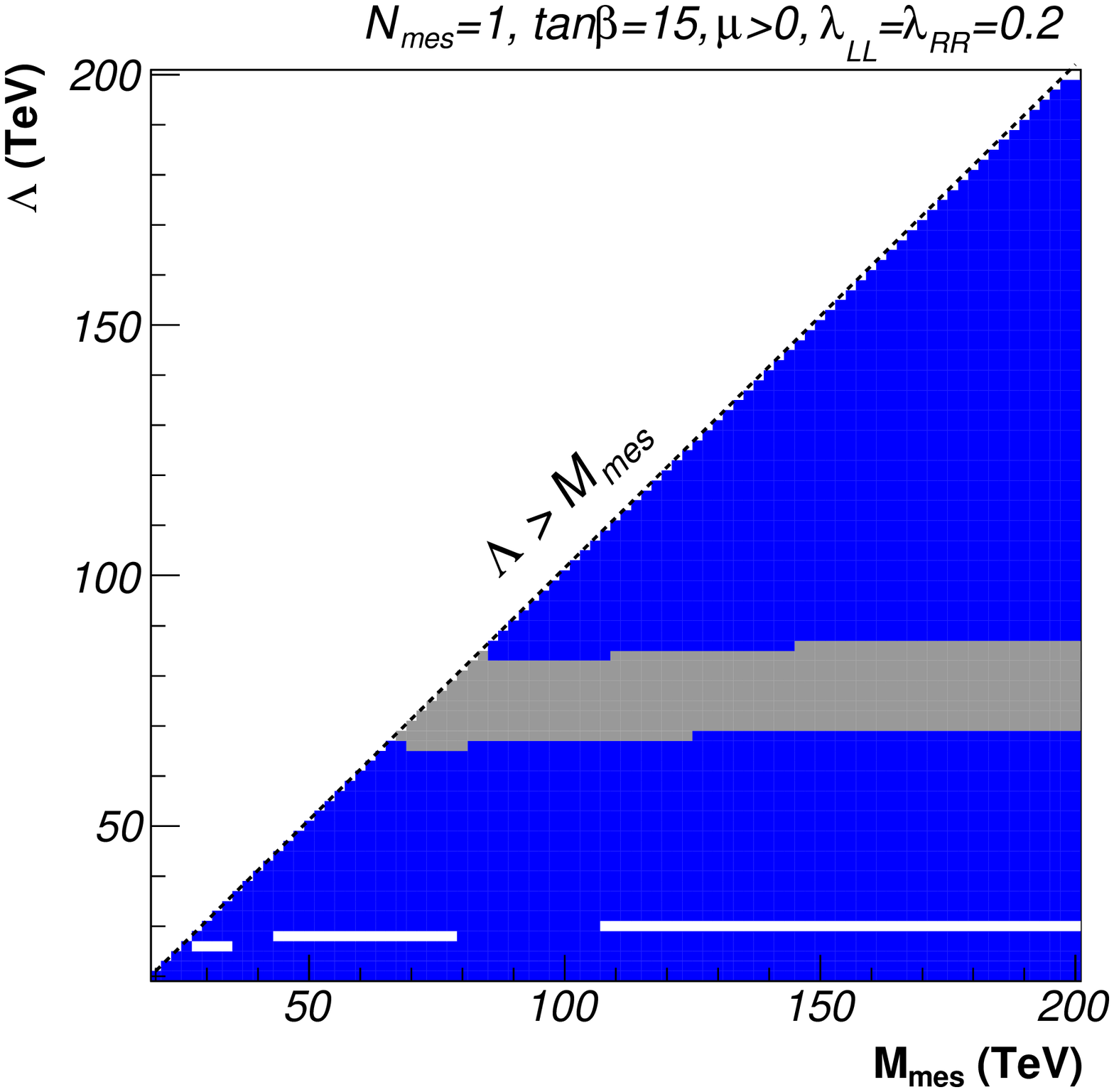}
\caption{The $\Lambda$--$M_{\rm mes}$ planes for $N_{\rm mes}=1$,
$\tan\beta=15$, $\mu>0$, and $\lambda_{\rm LL} = 0.1, 0.15$ and $0.2$. We show
$a_{\mu}$ favoured (light/grey) and $b\to s\gamma$ excluded (dark/blue) regions of the
GMSB parameter space with non-minimal flavour violation in either the
left-left chiral ($\lambda_{\rm RR}=0$, top) or both the left-left
and right-right chiral ($\lambda_{\rm RR}=\lambda_{\rm LL}$, bottom) squark
sectors. The region where $\Lambda > M_{\rm mes}$ does not allow for physical
solutions of the RGEs.}
\label{fig2}
\end{center}\end{figure}

\begin{figure}\begin{center}
	\includegraphics[scale=0.27]{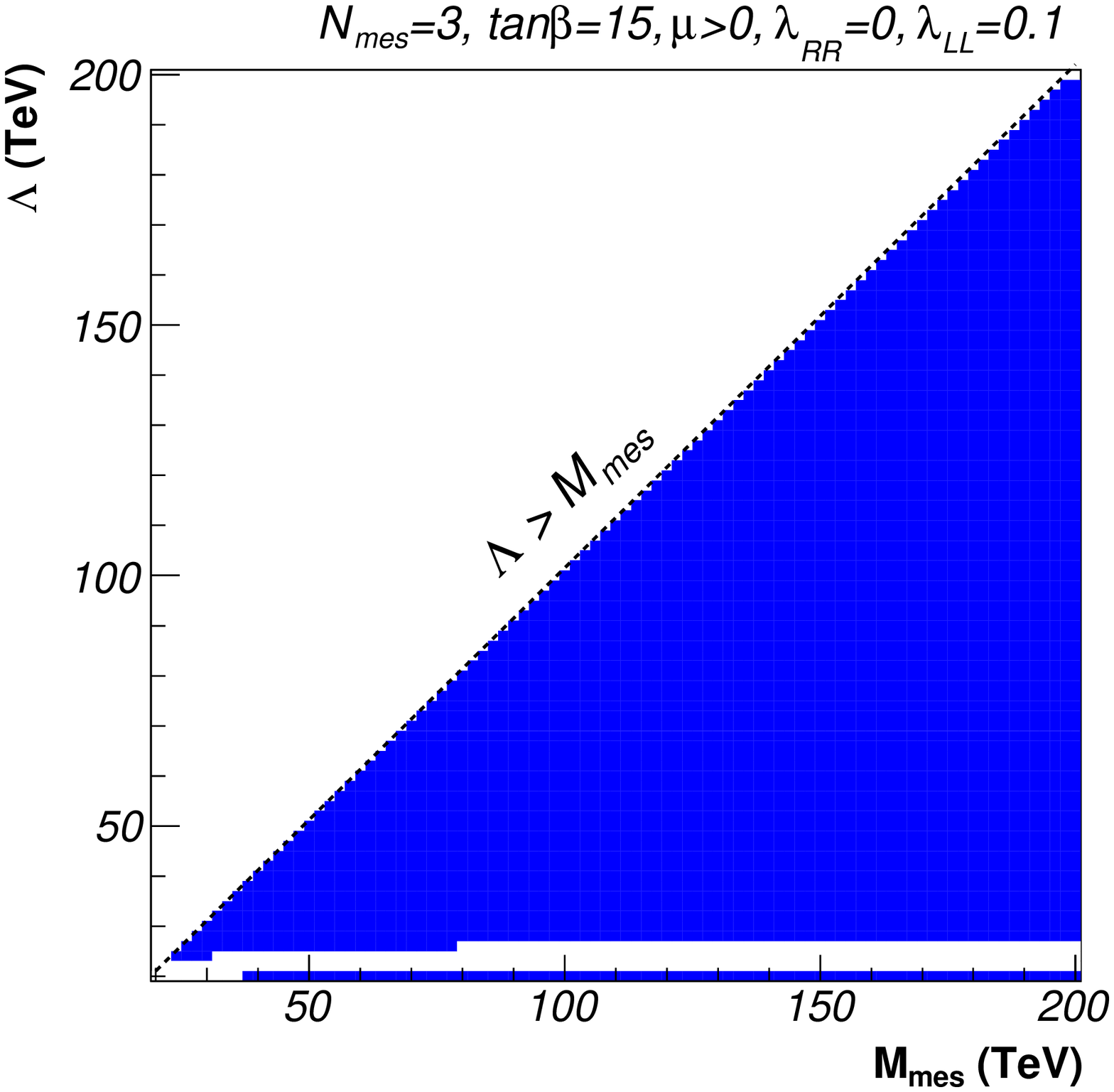}
	\includegraphics[scale=0.27]{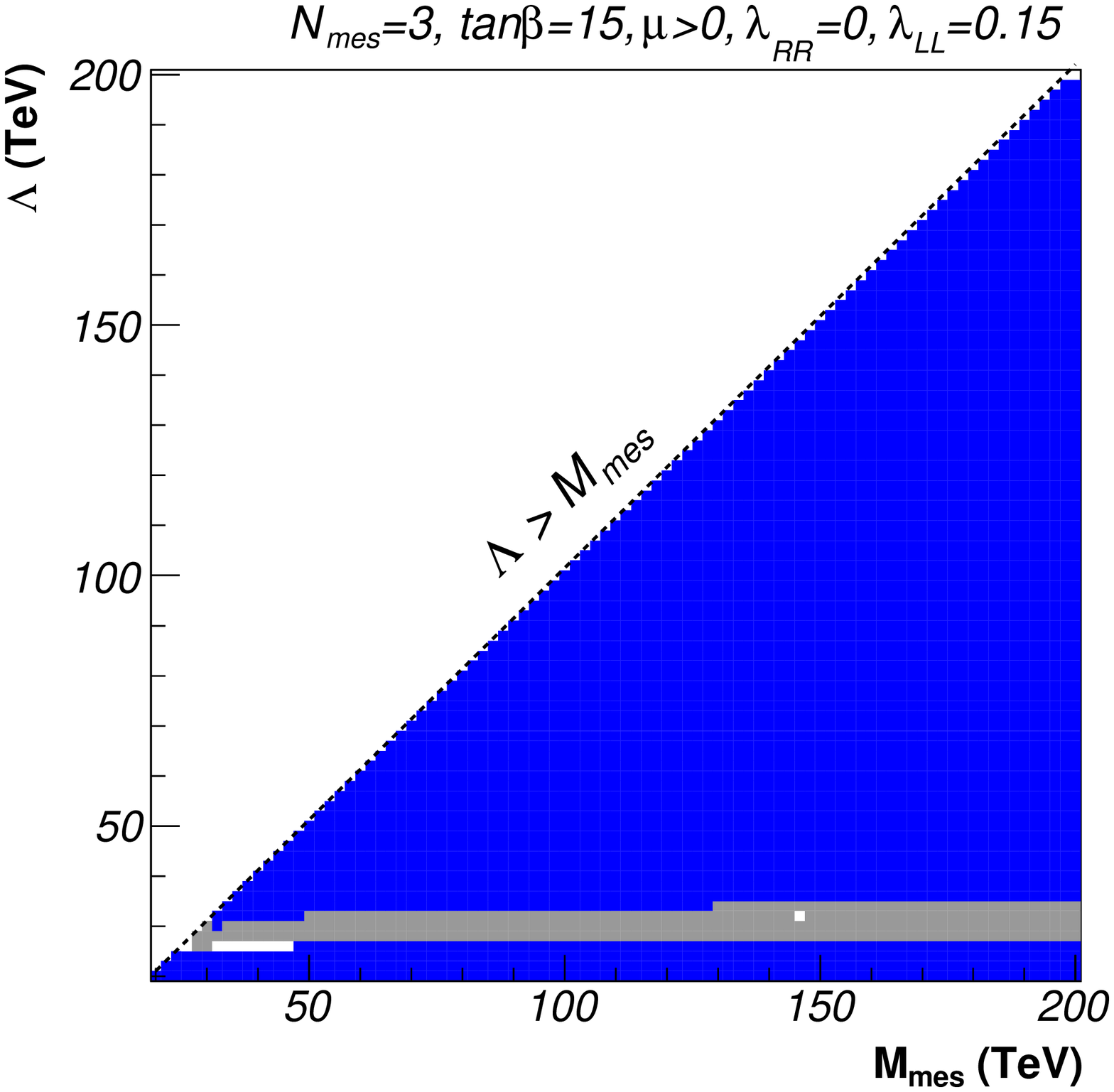}
	\includegraphics[scale=0.27]{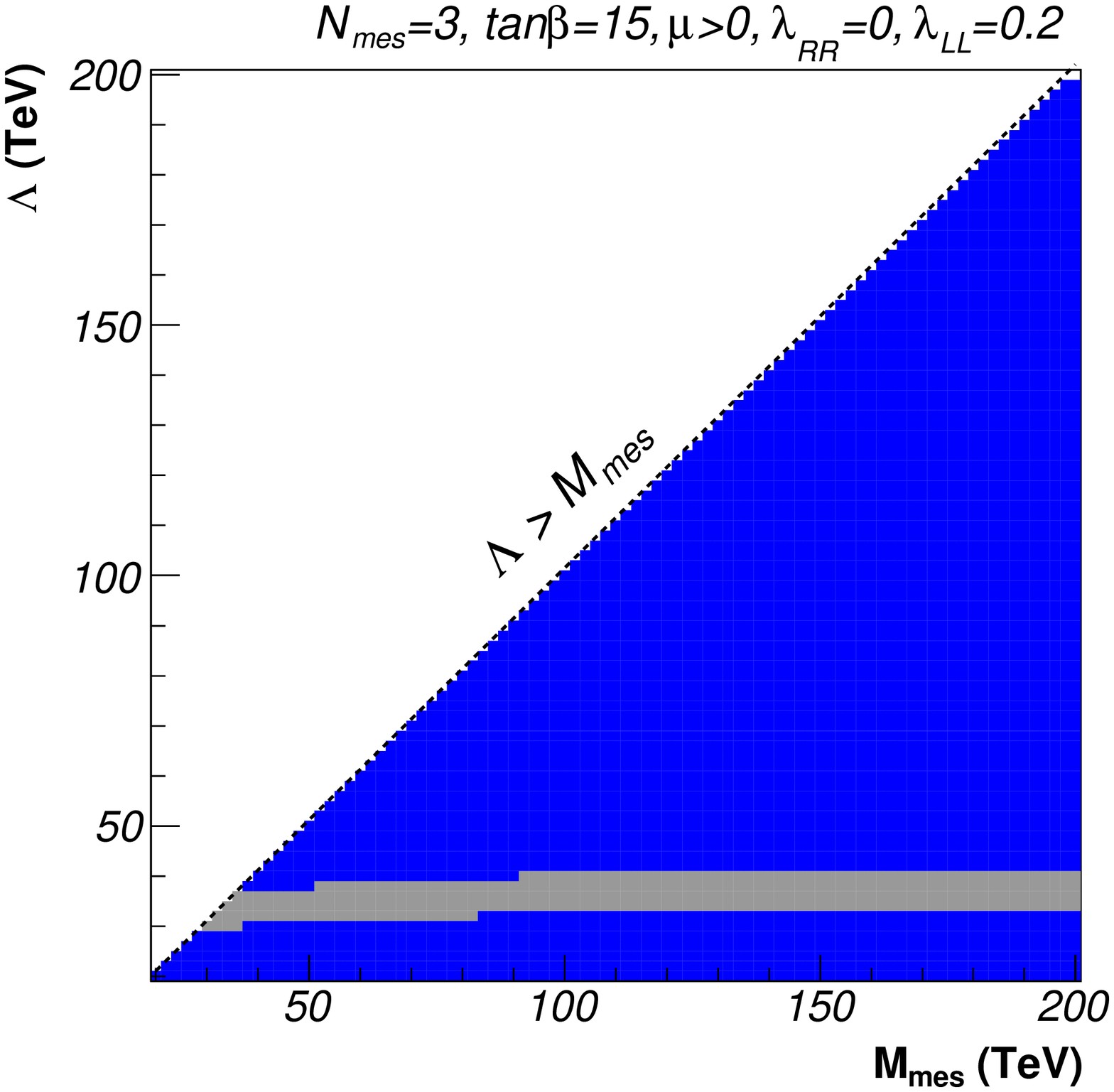}
	\includegraphics[scale=0.27]{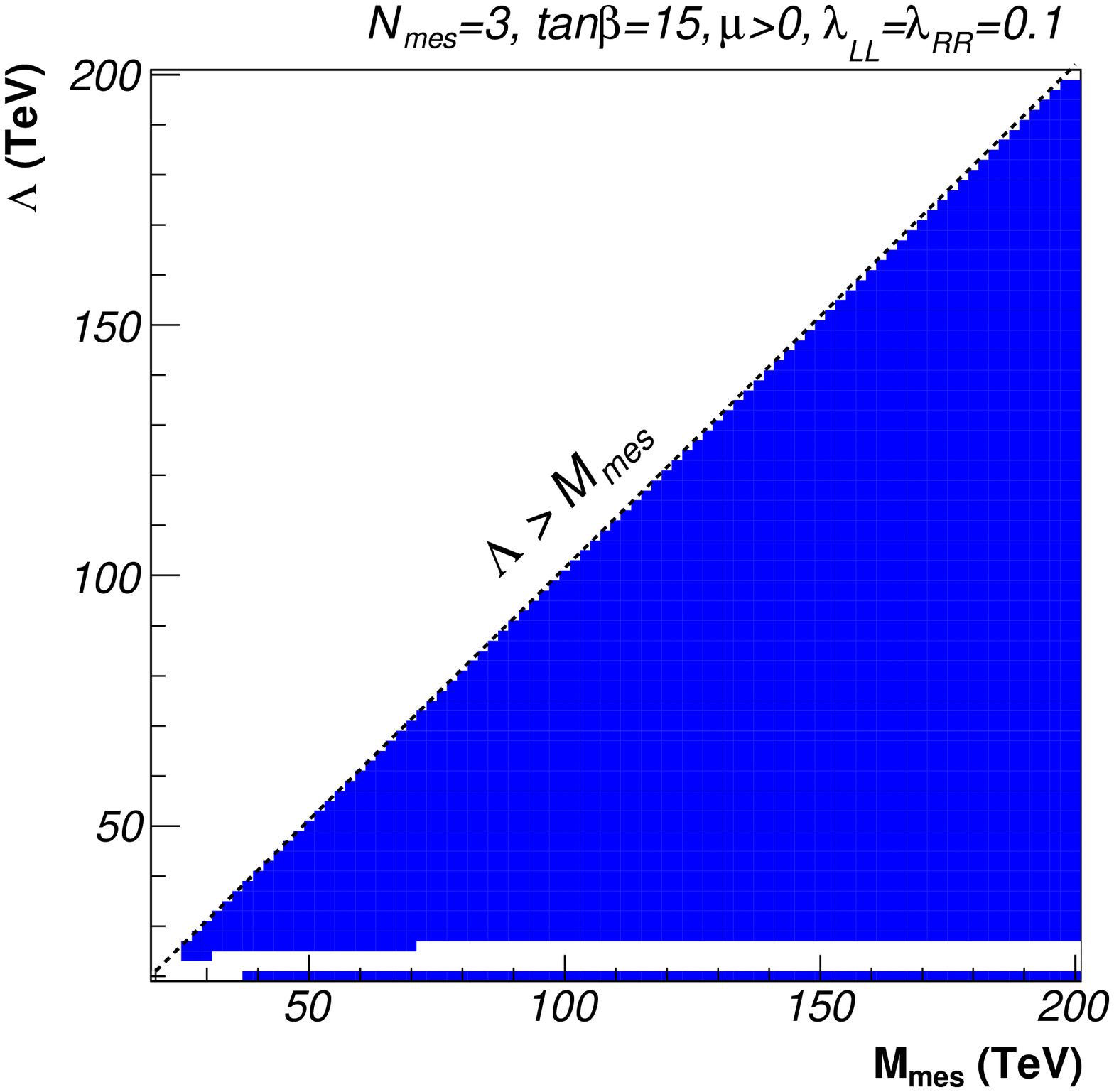}
	\includegraphics[scale=0.27]{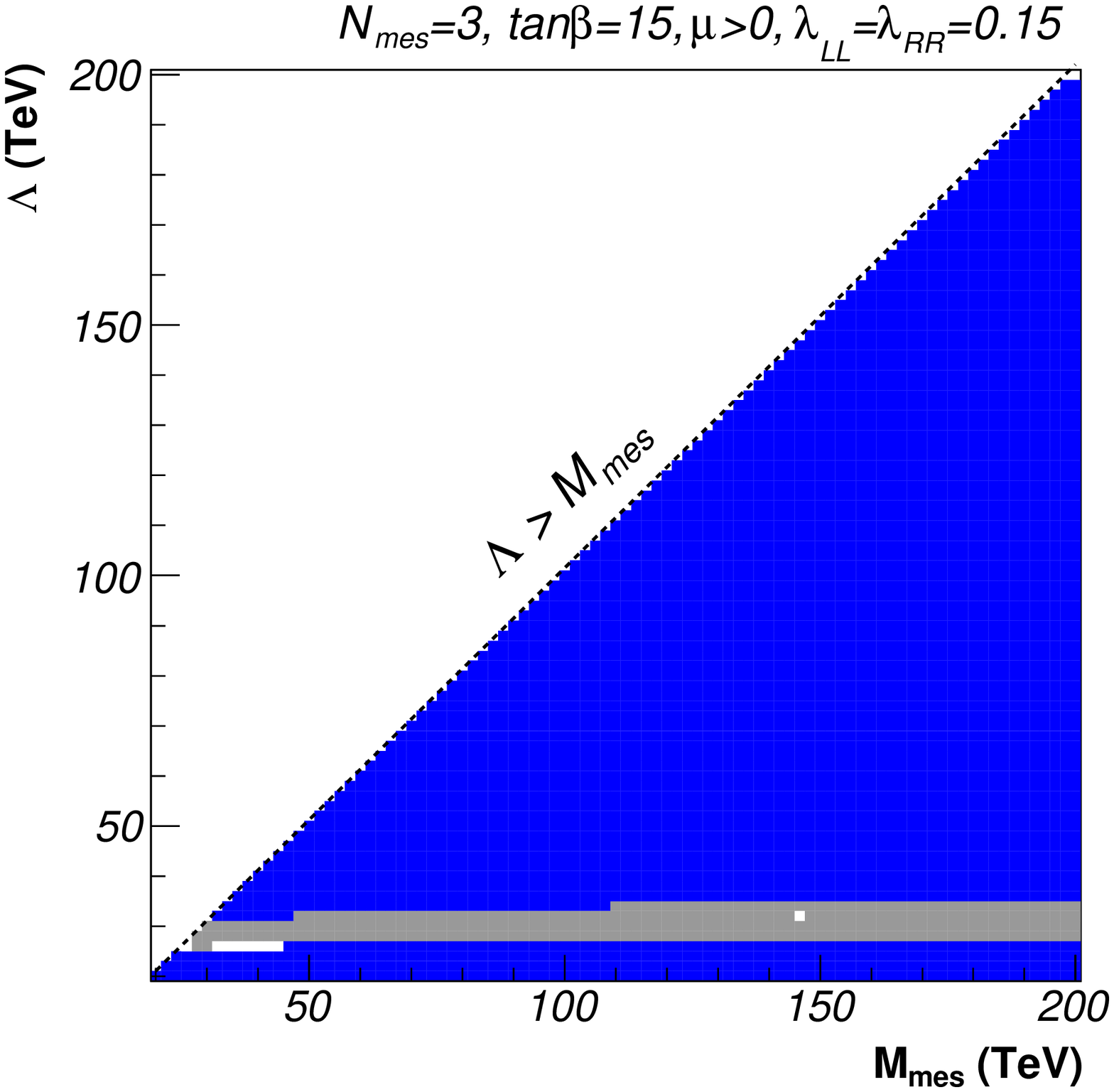}
	\includegraphics[scale=0.27]{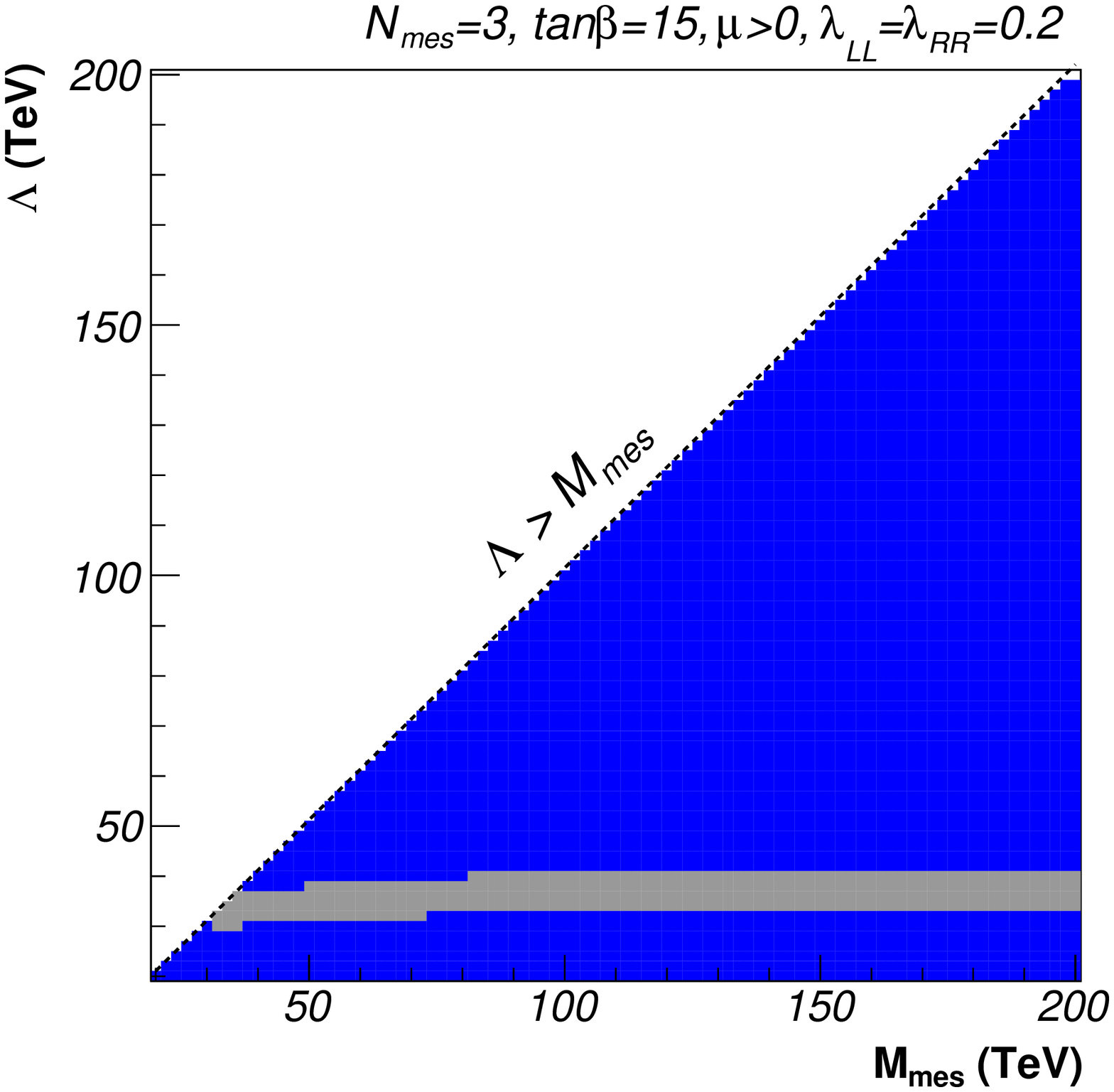}
\caption{Same as Fig.\ \ref{fig2} for $N_{\rm mes}=3$ and $\tan\beta=15$.}
\label{fig3}
\end{center}\end{figure}

\begin{figure}\begin{center}
	\includegraphics[scale=0.27]{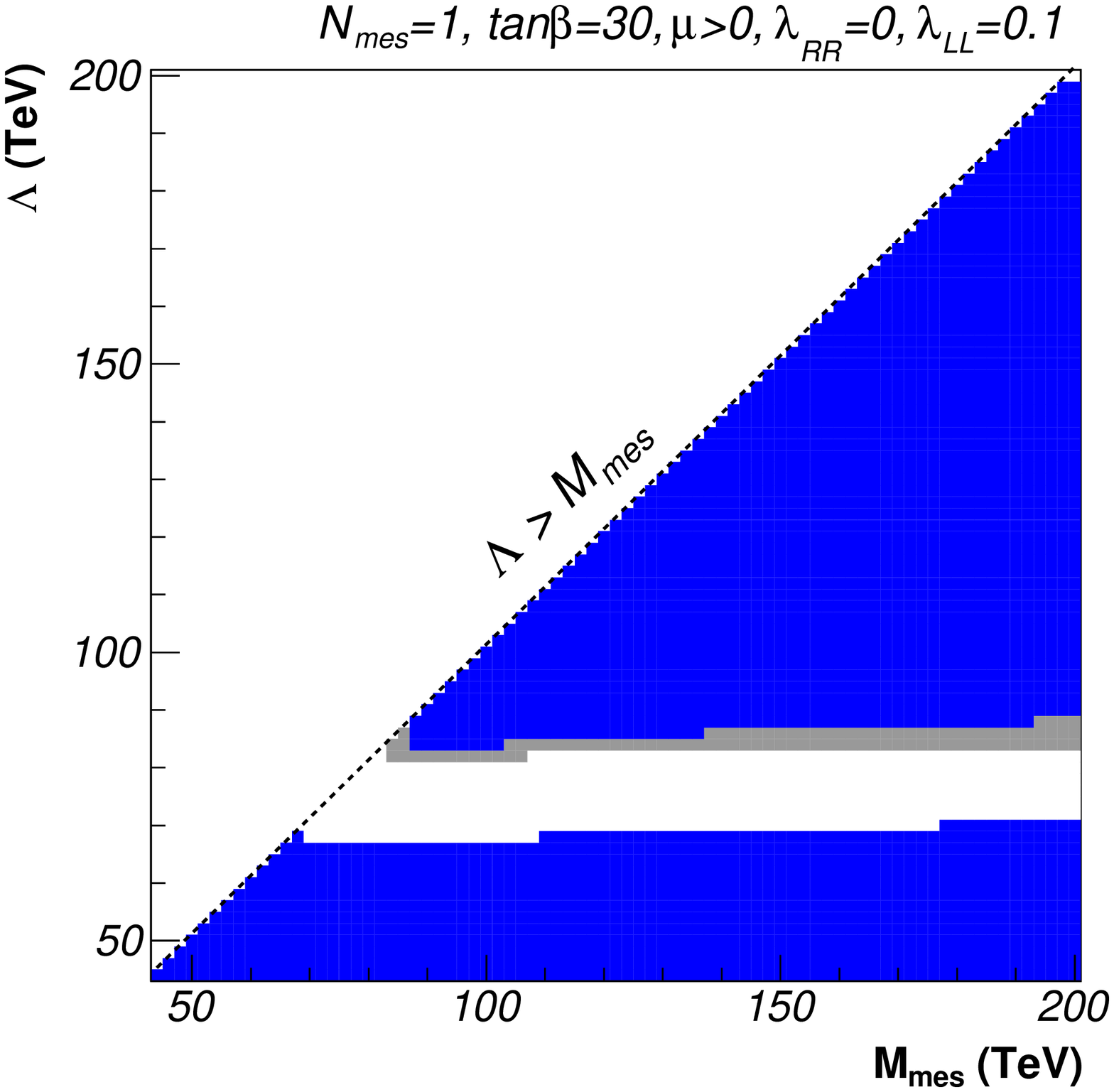}
	\includegraphics[scale=0.27]{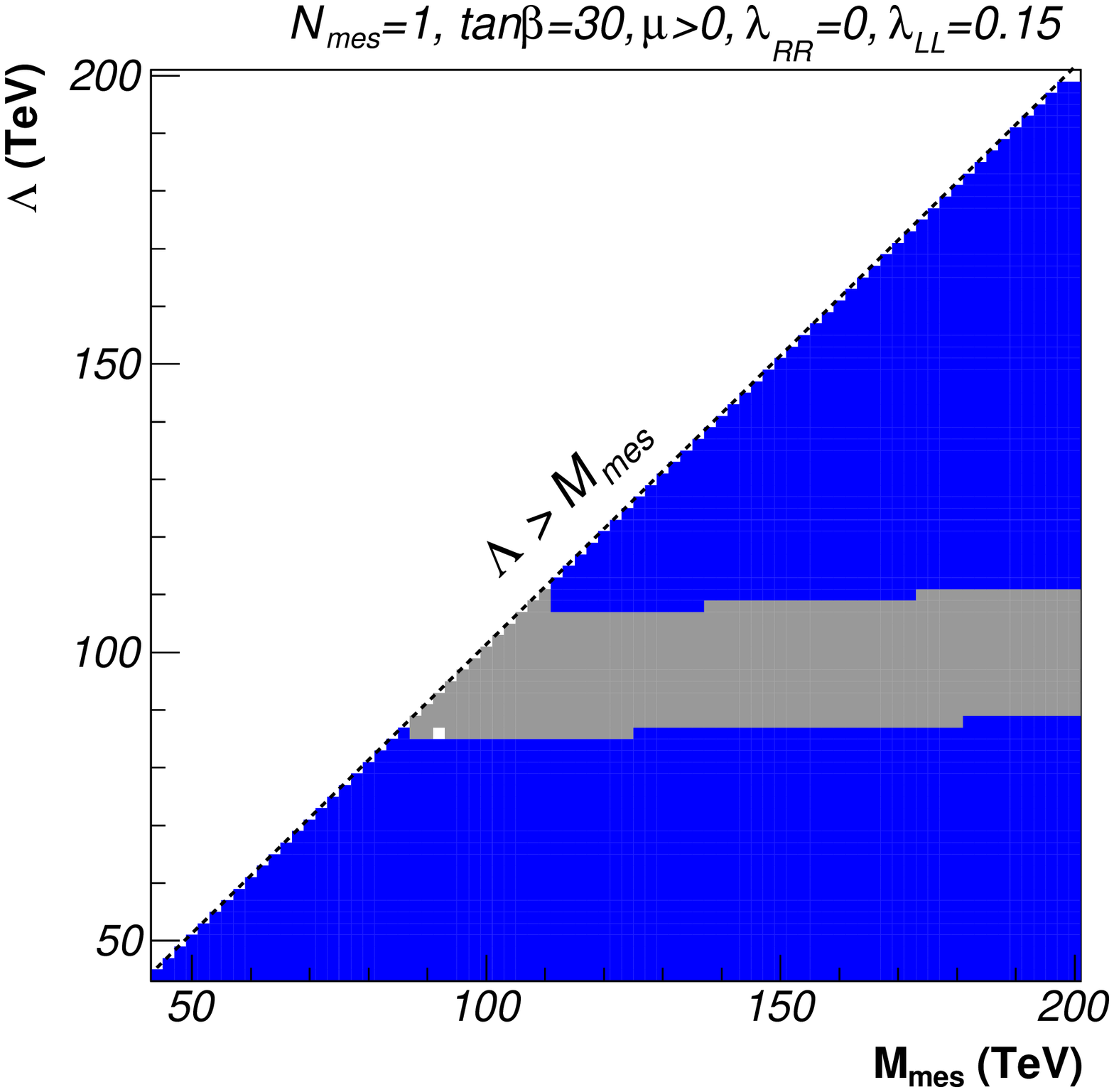}
	\includegraphics[scale=0.27]{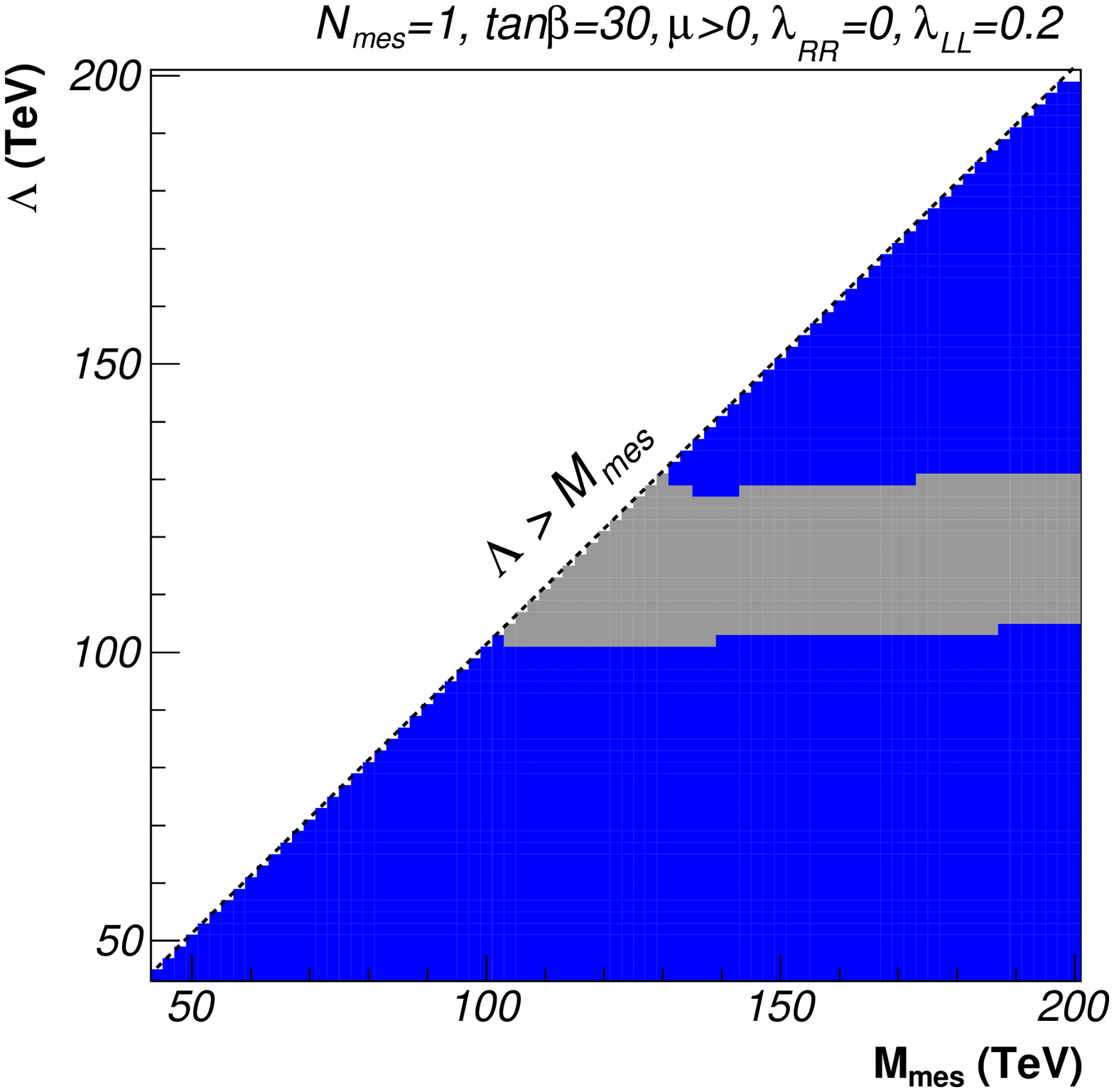}
	\includegraphics[scale=0.27]{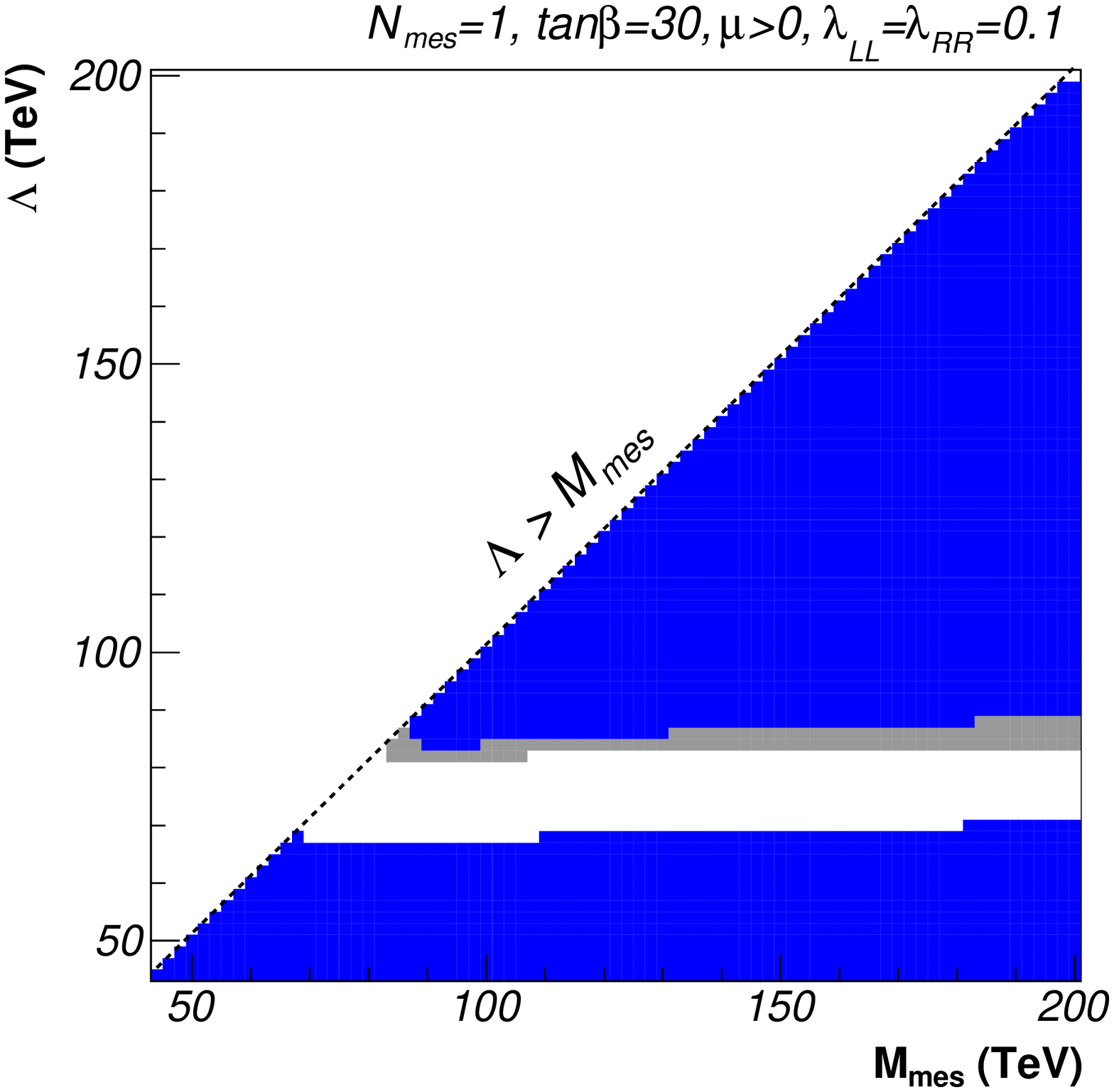}
	\includegraphics[scale=0.27]{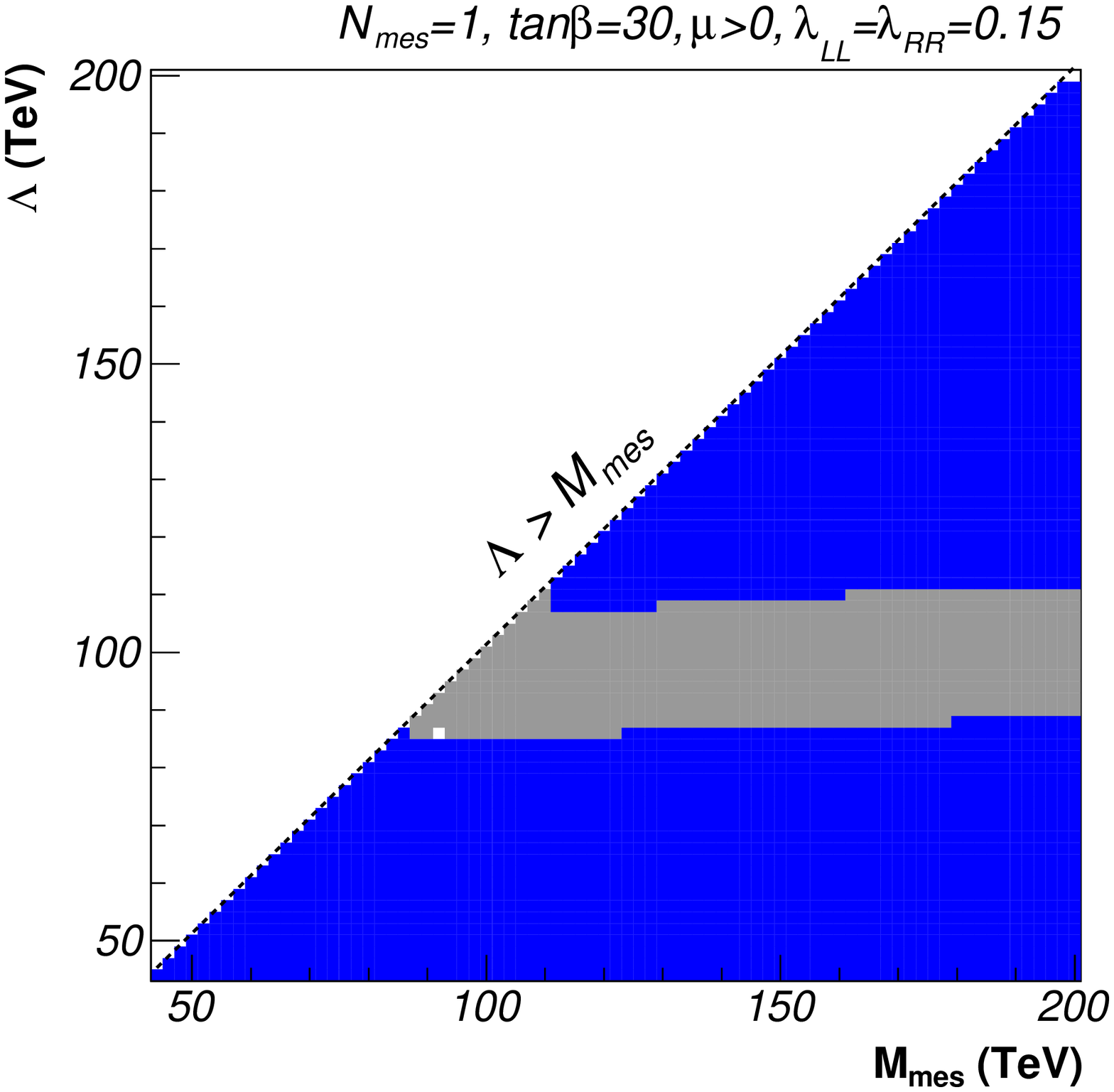}
	\includegraphics[scale=0.27]{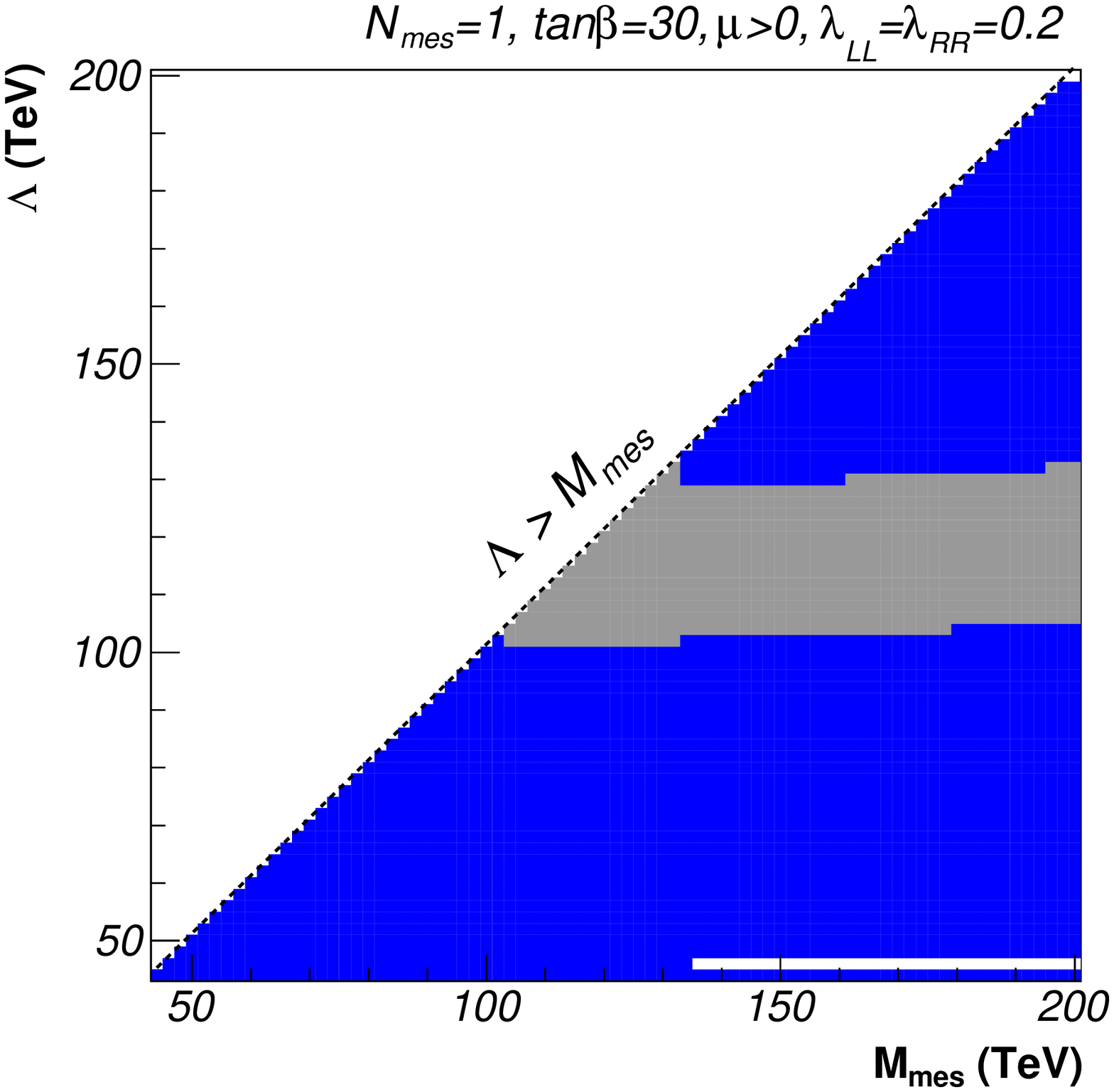}
\caption{Same as Fig.\ \ref{fig2} for $N_{\rm mes}=1$ and $\tan\beta=30$.}
\label{fig4}
\end{center}\end{figure}

\begin{figure}\begin{center}
	\includegraphics[scale=0.27]{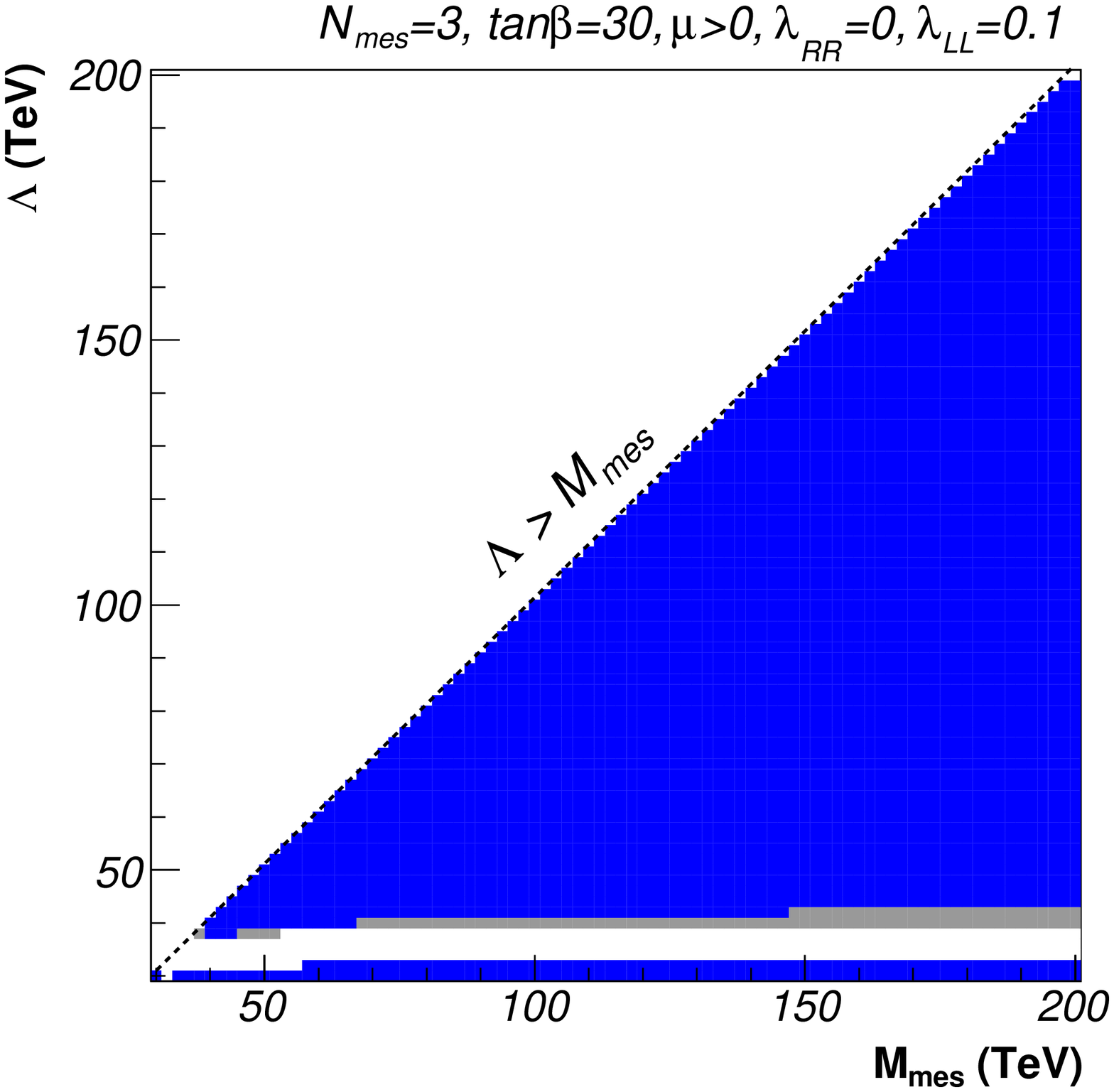}
	\includegraphics[scale=0.27]{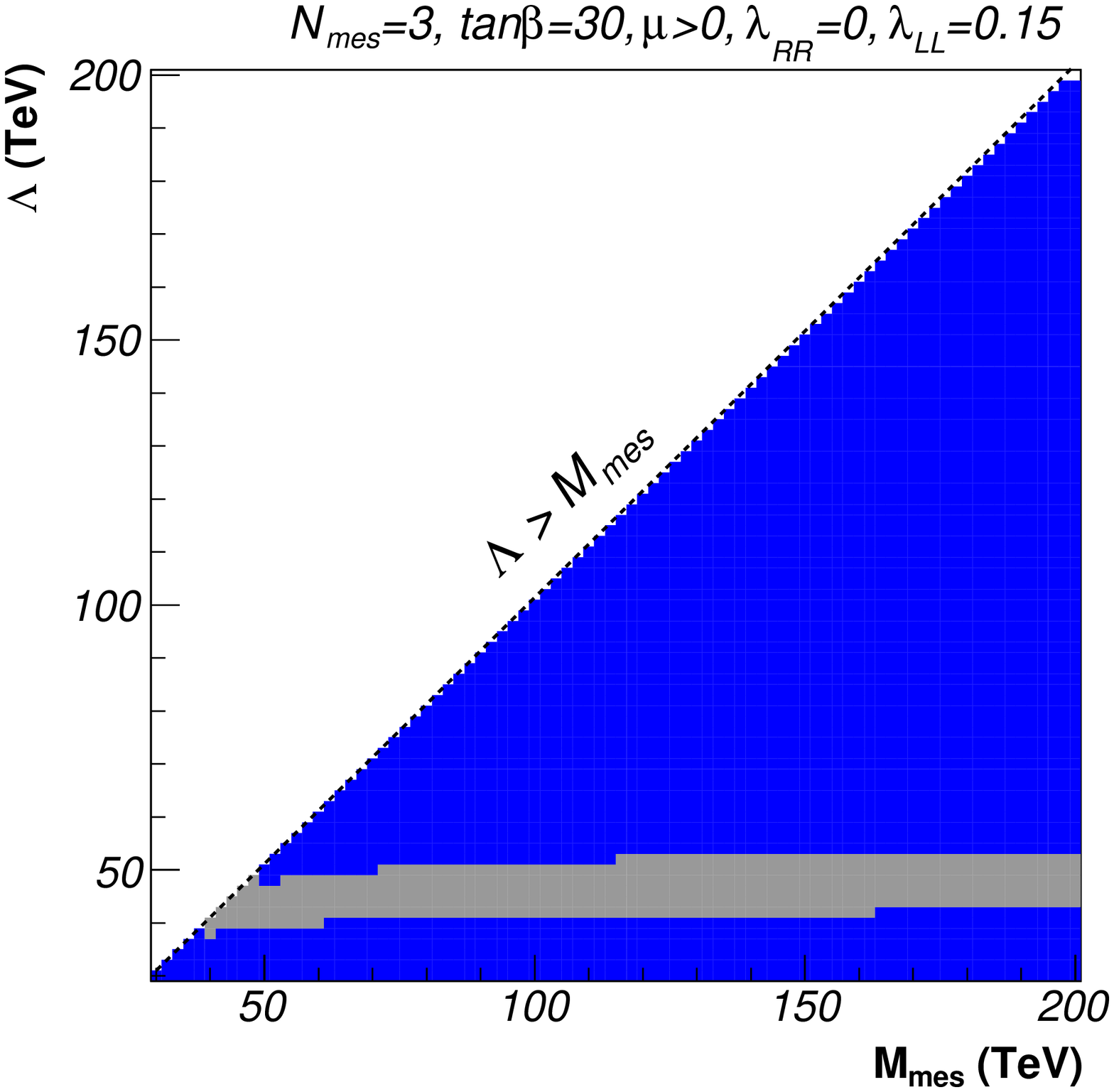}
	\includegraphics[scale=0.27]{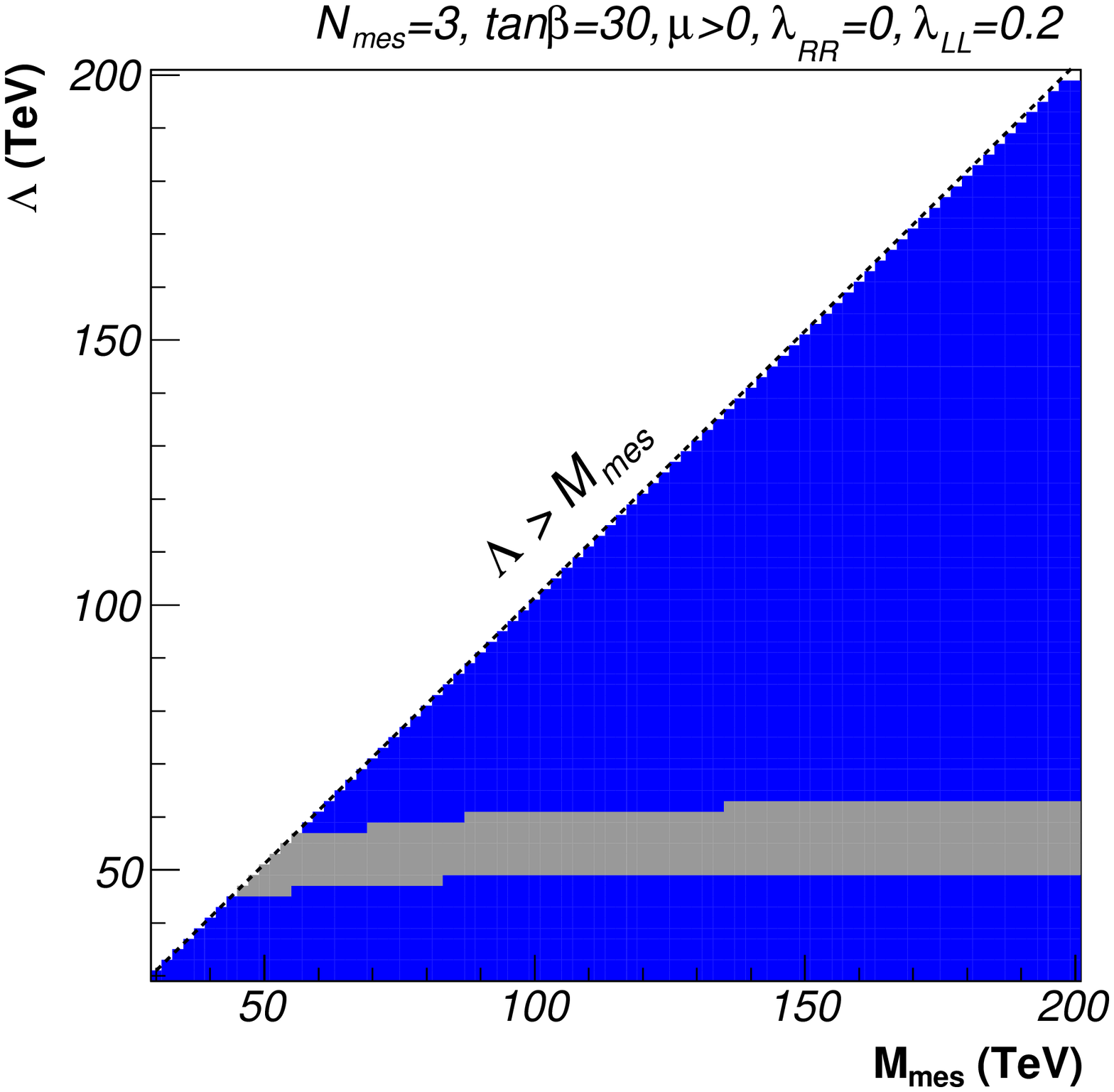}
	\includegraphics[scale=0.27]{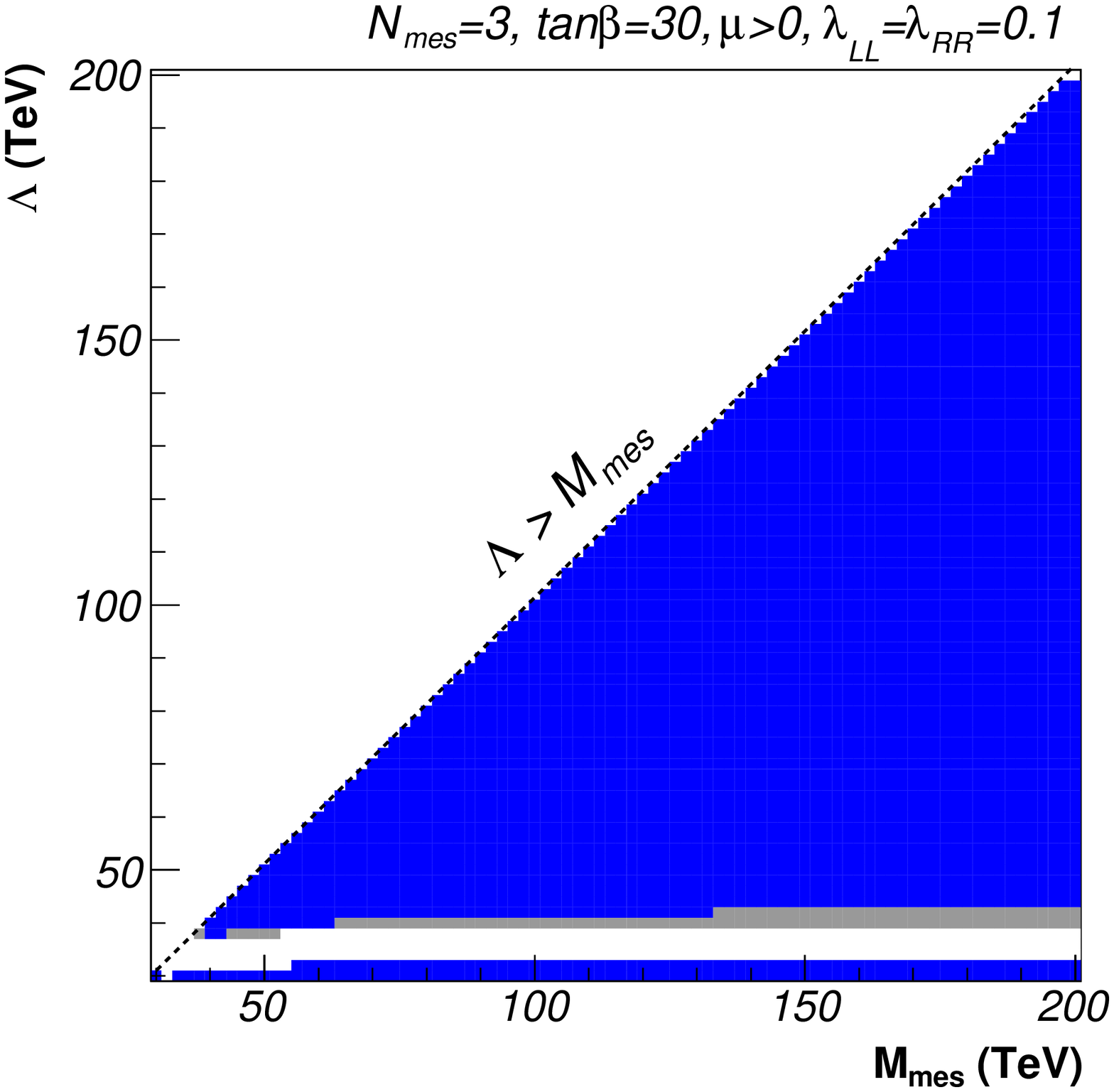}
	\includegraphics[scale=0.27]{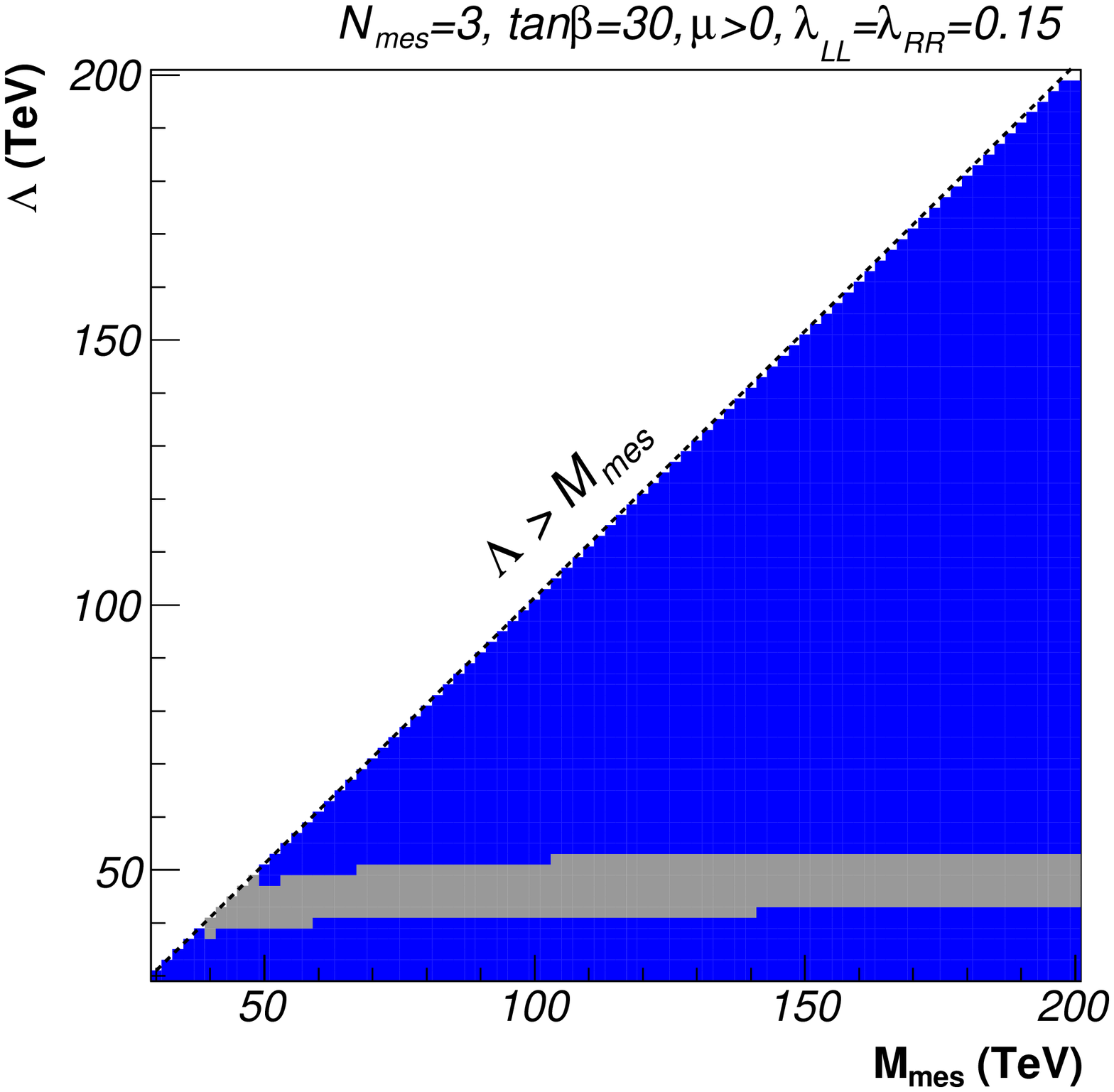}
	\includegraphics[scale=0.27]{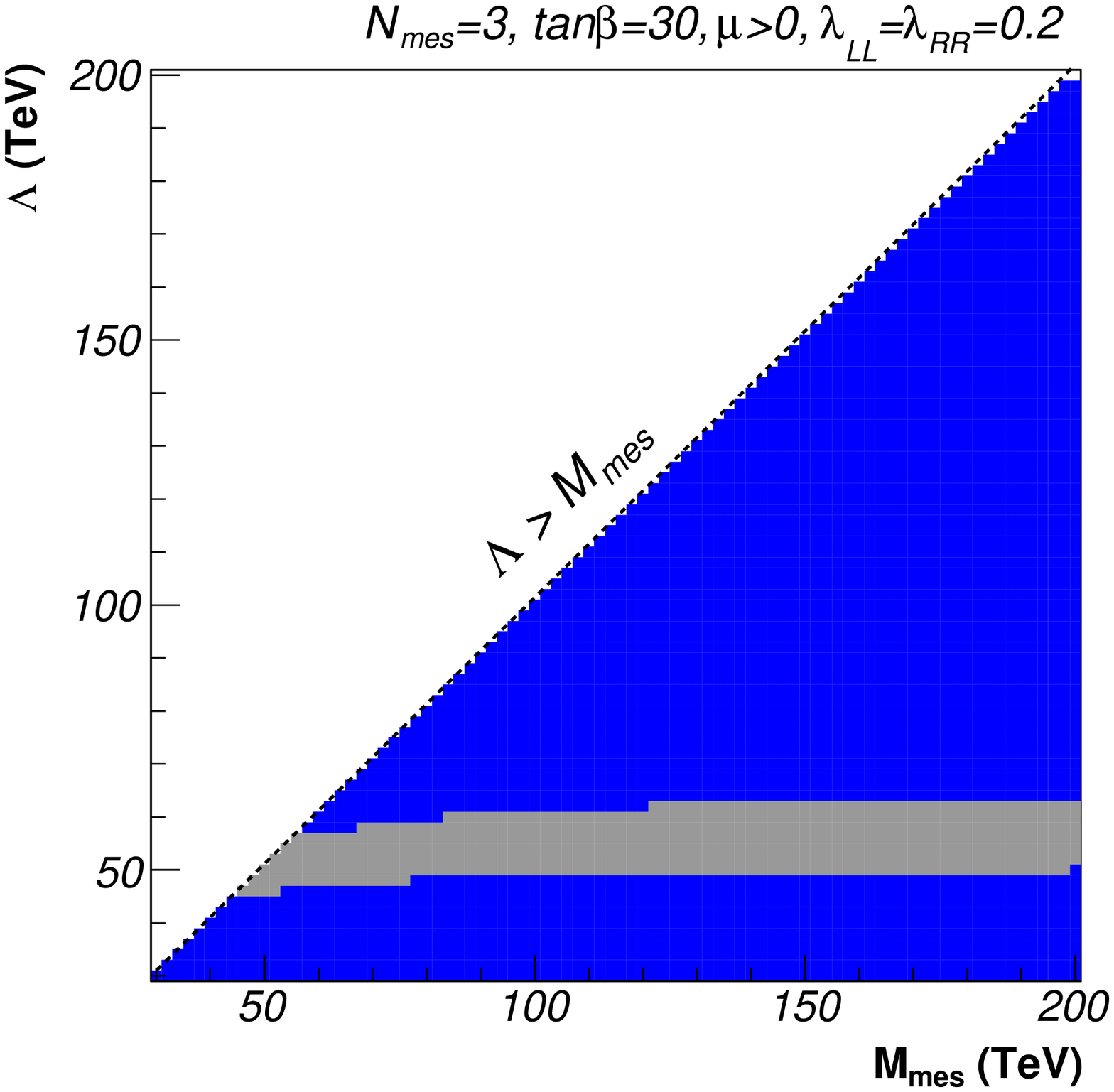}
\caption{Same as Fig.\ \ref{fig2} for $N_{\rm mes}=3$ and $\tan\beta=30$.}
\label{fig5}
\end{center}\end{figure}

\begin{figure}\begin{center}
	\includegraphics[scale=0.27]{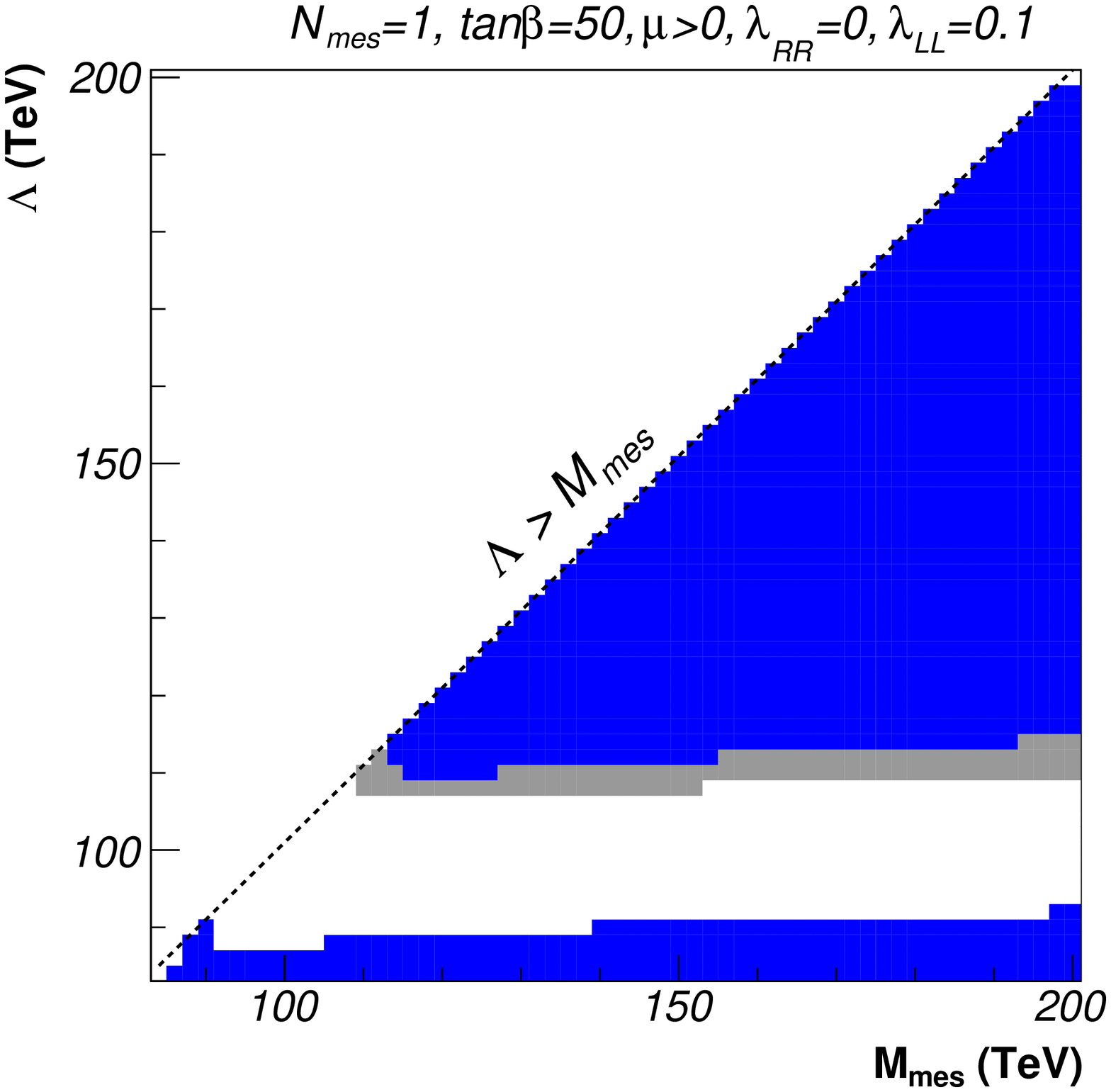}
	\includegraphics[scale=0.27]{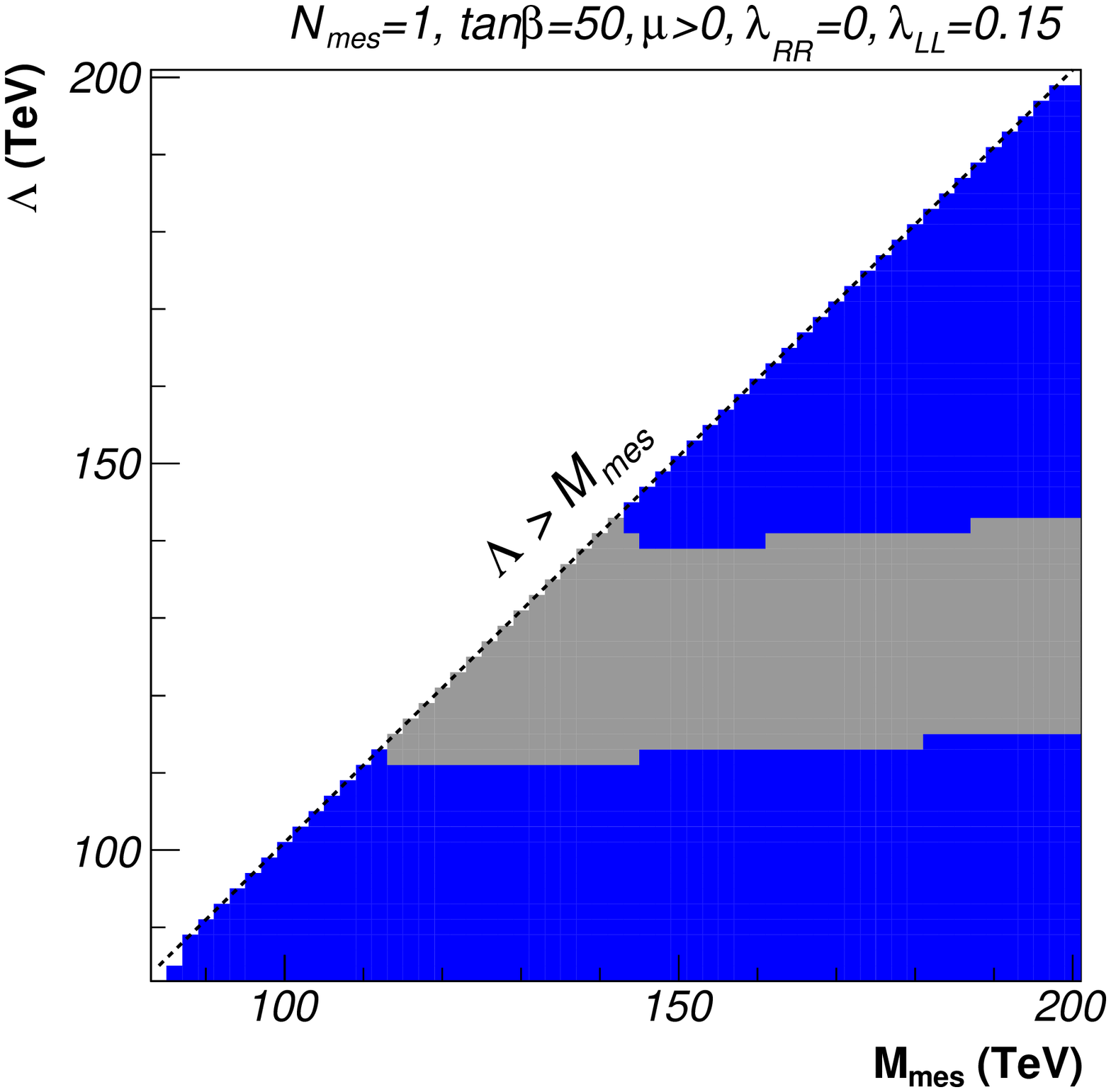}
	\includegraphics[scale=0.27]{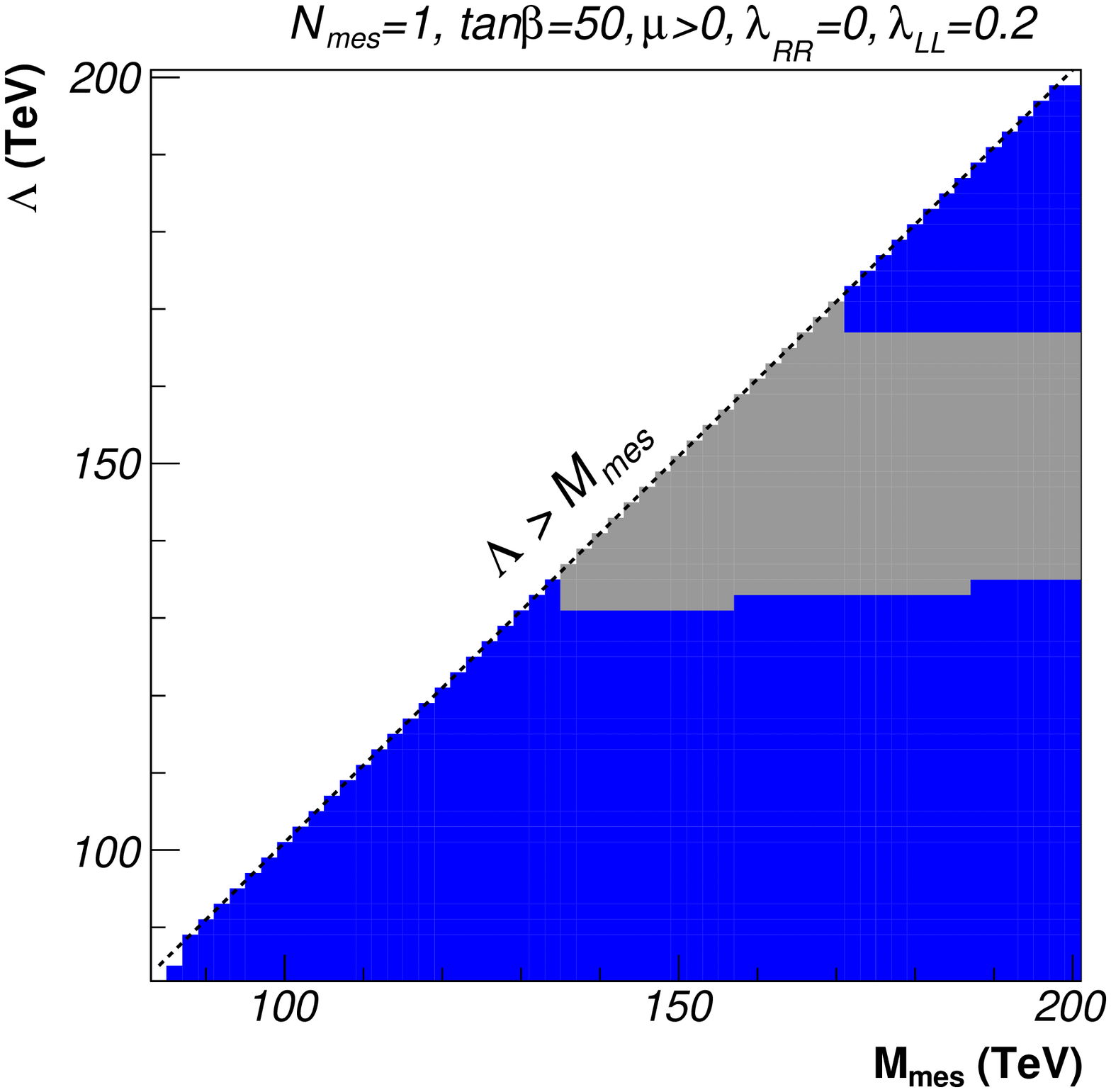}
	\includegraphics[scale=0.27]{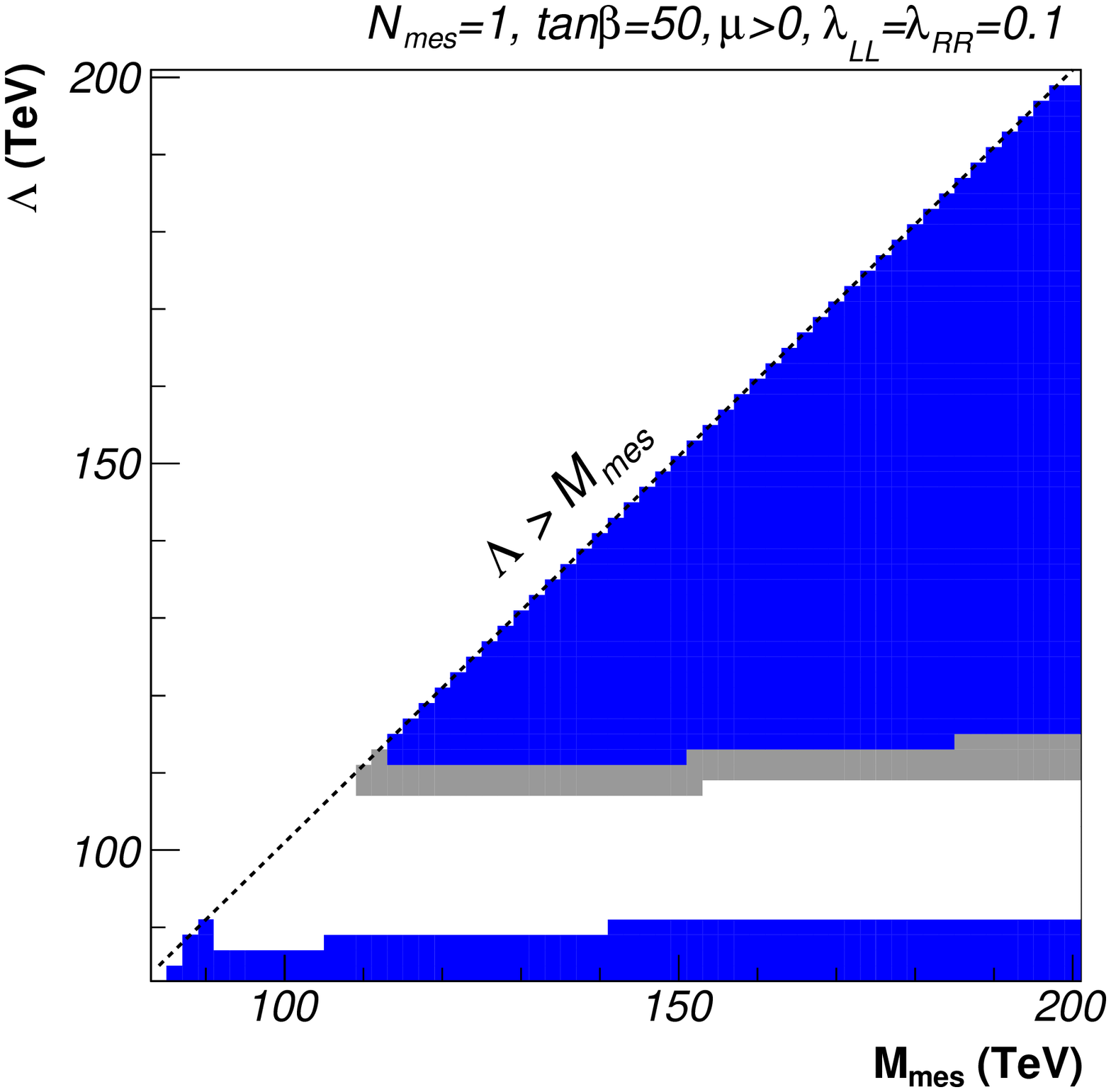}
	\includegraphics[scale=0.27]{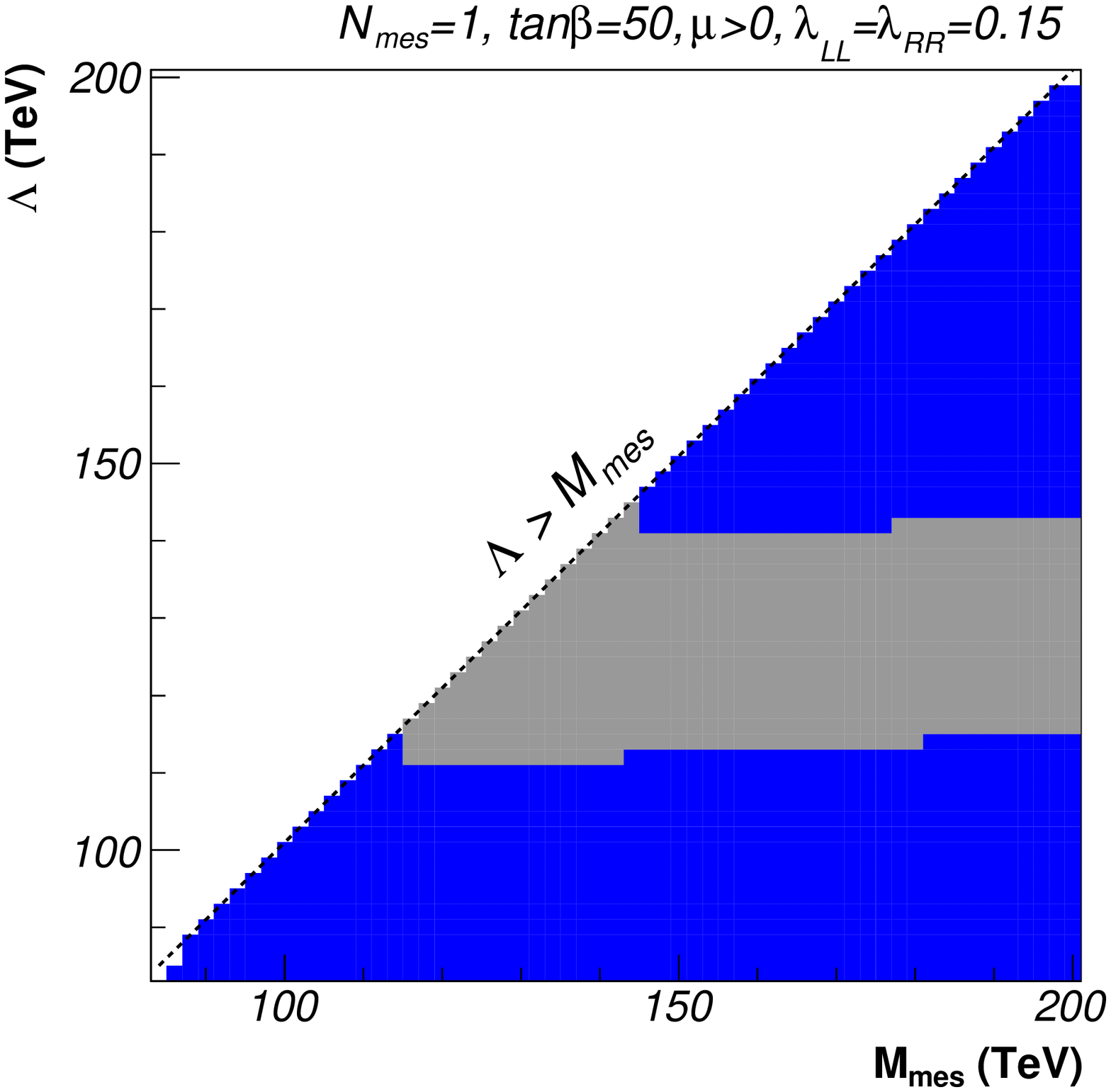}
	\includegraphics[scale=0.27]{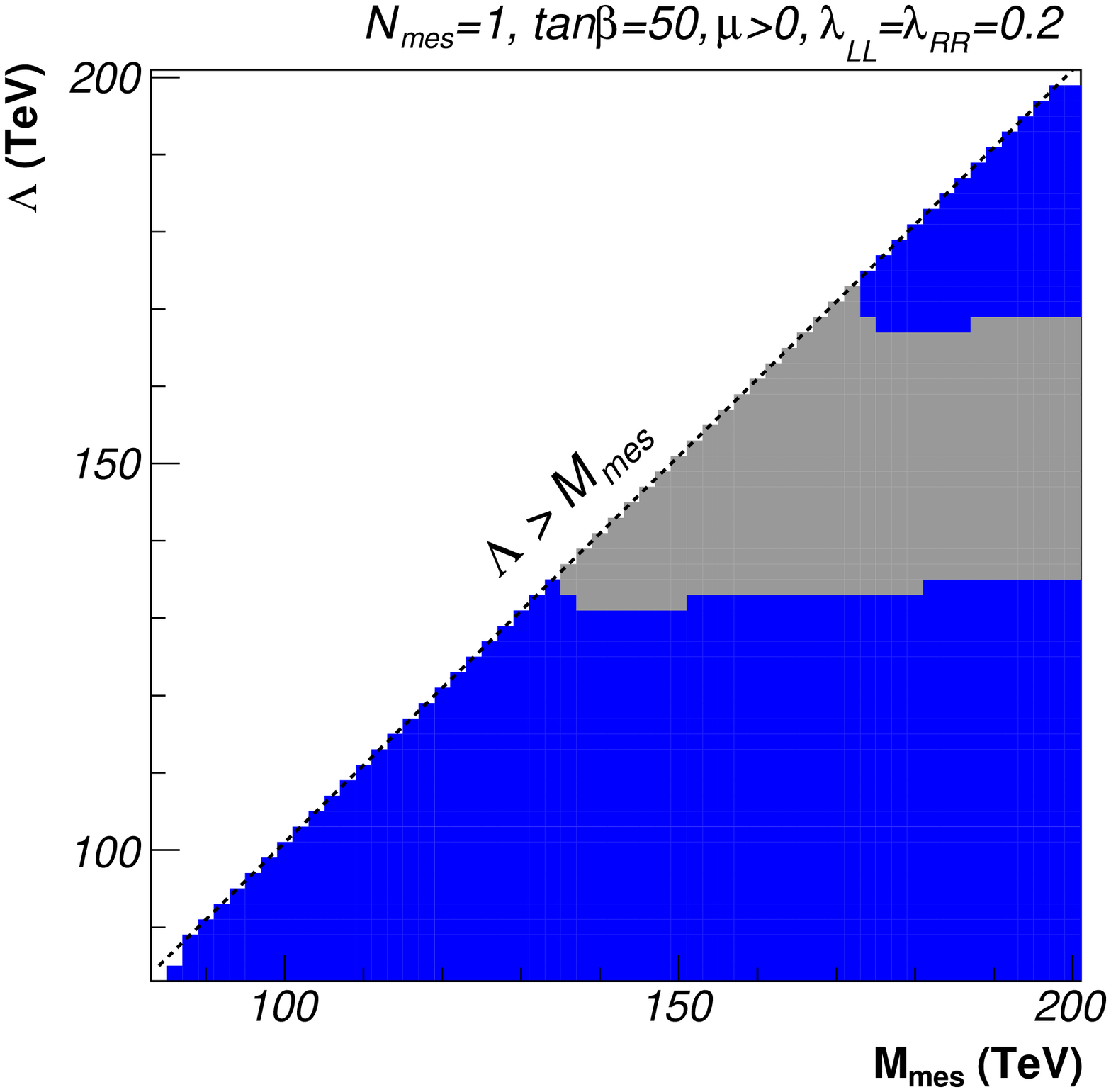}
\caption{Same as Fig.\ \ref{fig2} for $N_{\rm mes}=1$ and $\tan\beta=50$.}
\label{fig6}
\end{center}\end{figure}

\begin{figure}\begin{center}
	\includegraphics[scale=0.27]{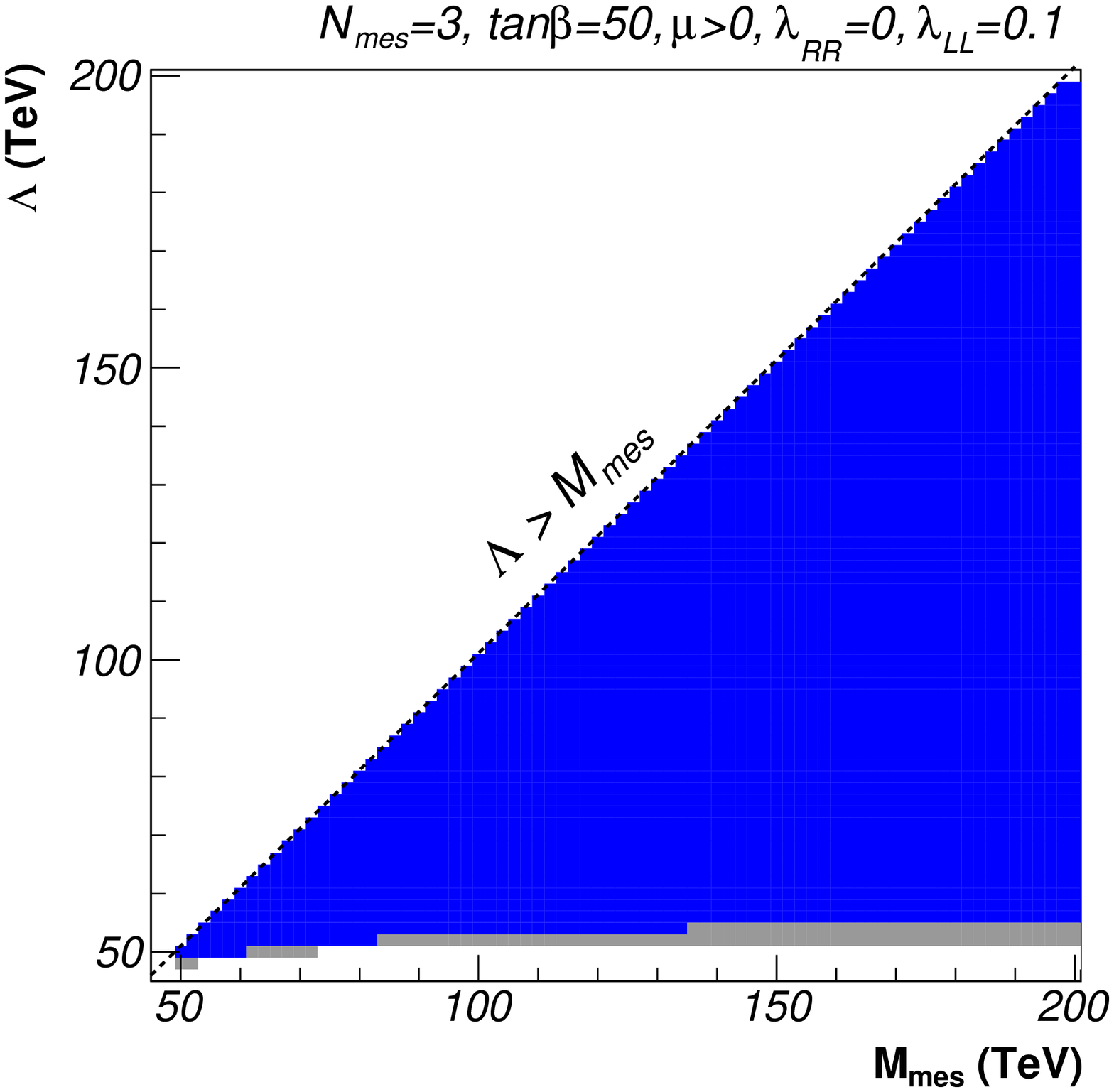}
	\includegraphics[scale=0.27]{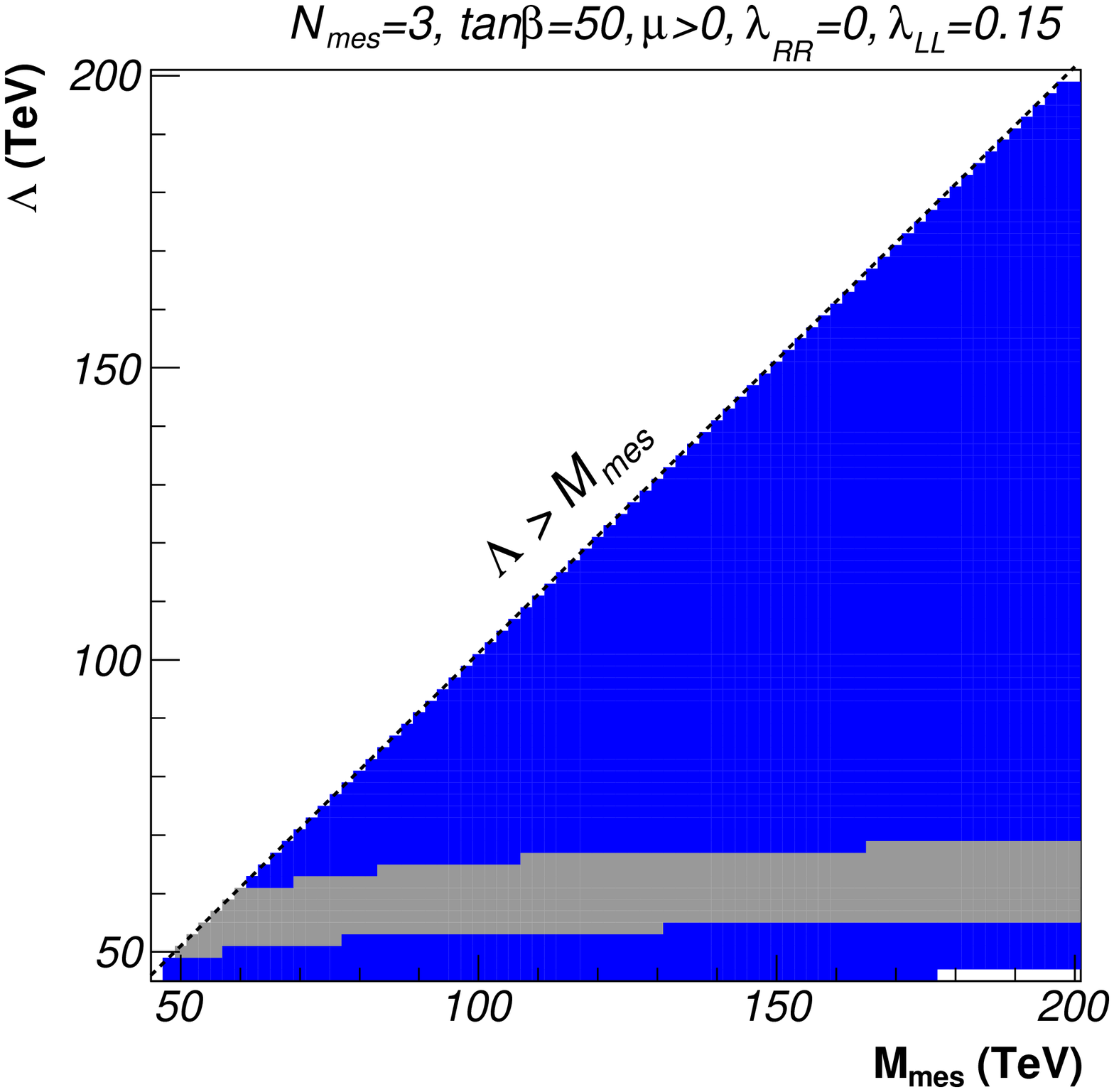}
	\includegraphics[scale=0.27]{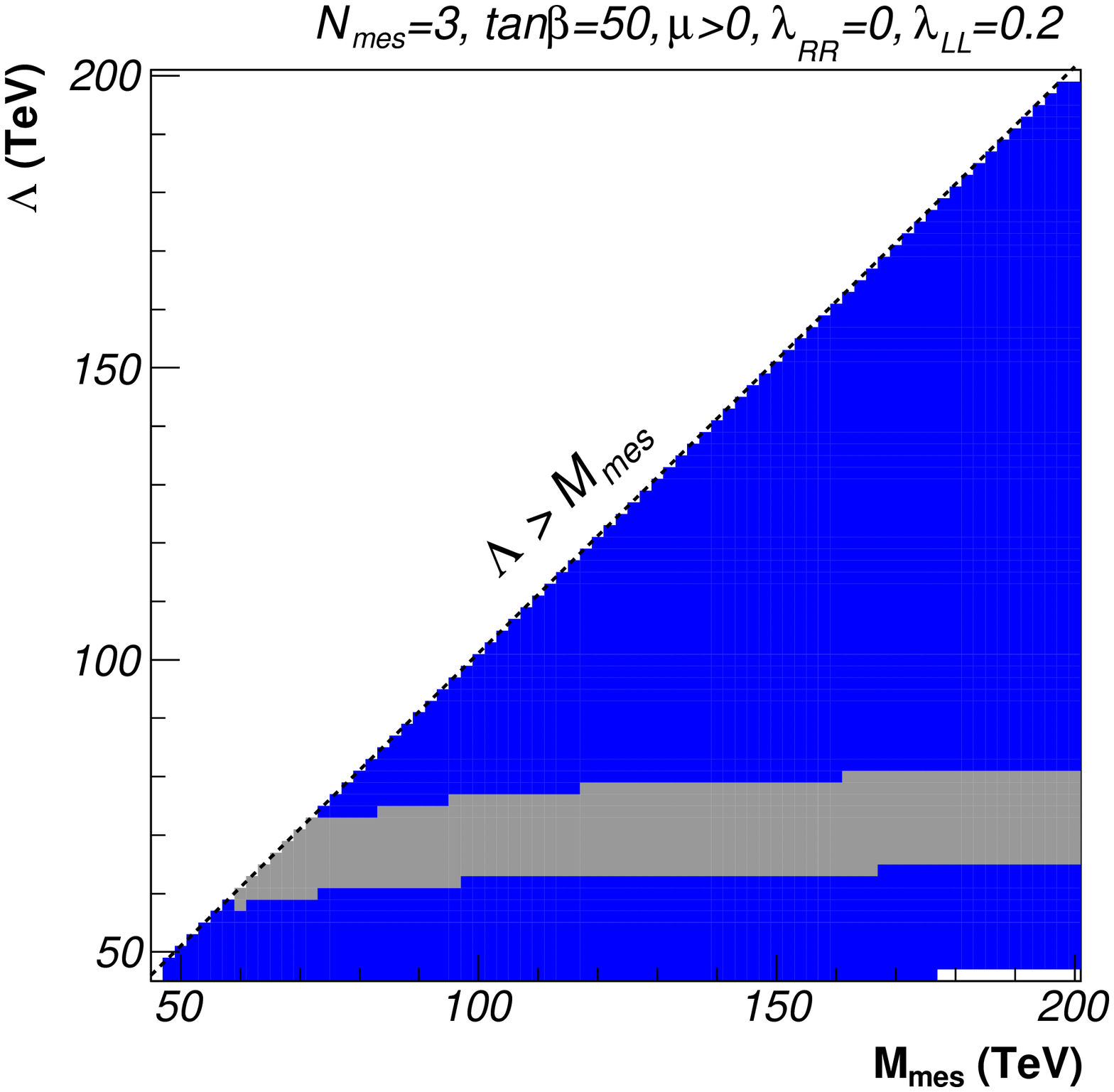}
	\includegraphics[scale=0.27]{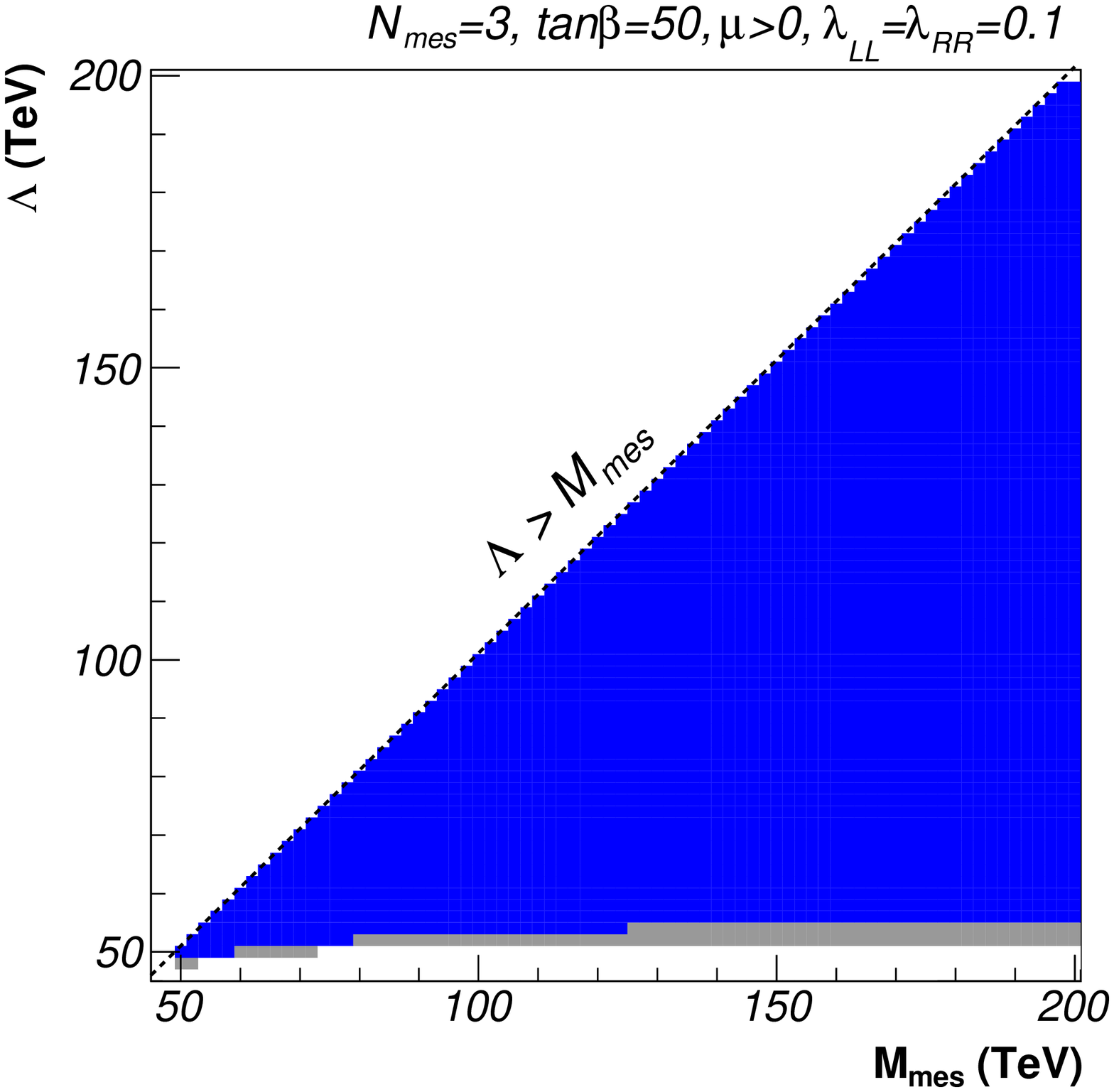}
	\includegraphics[scale=0.27]{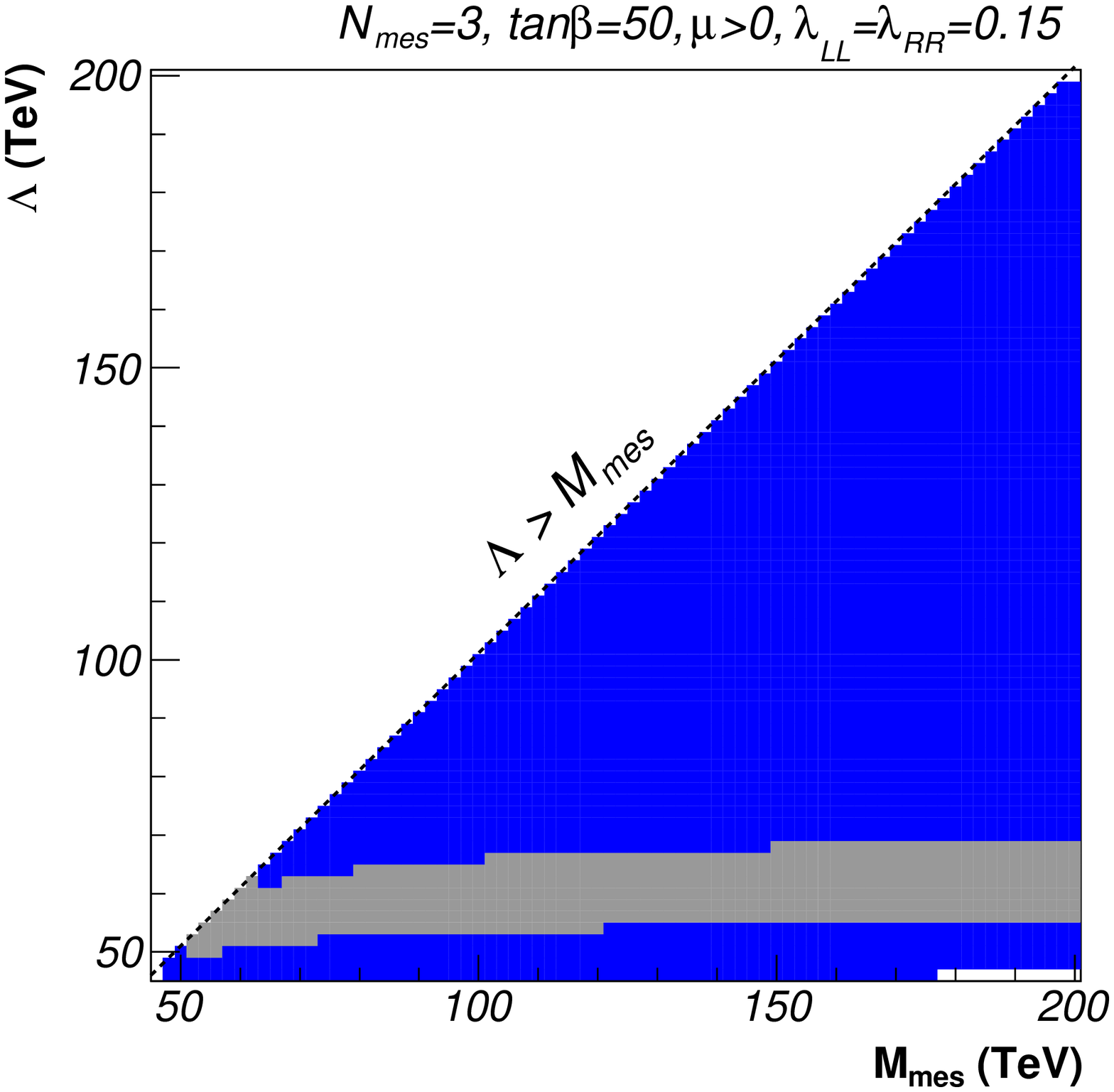}
	\includegraphics[scale=0.27]{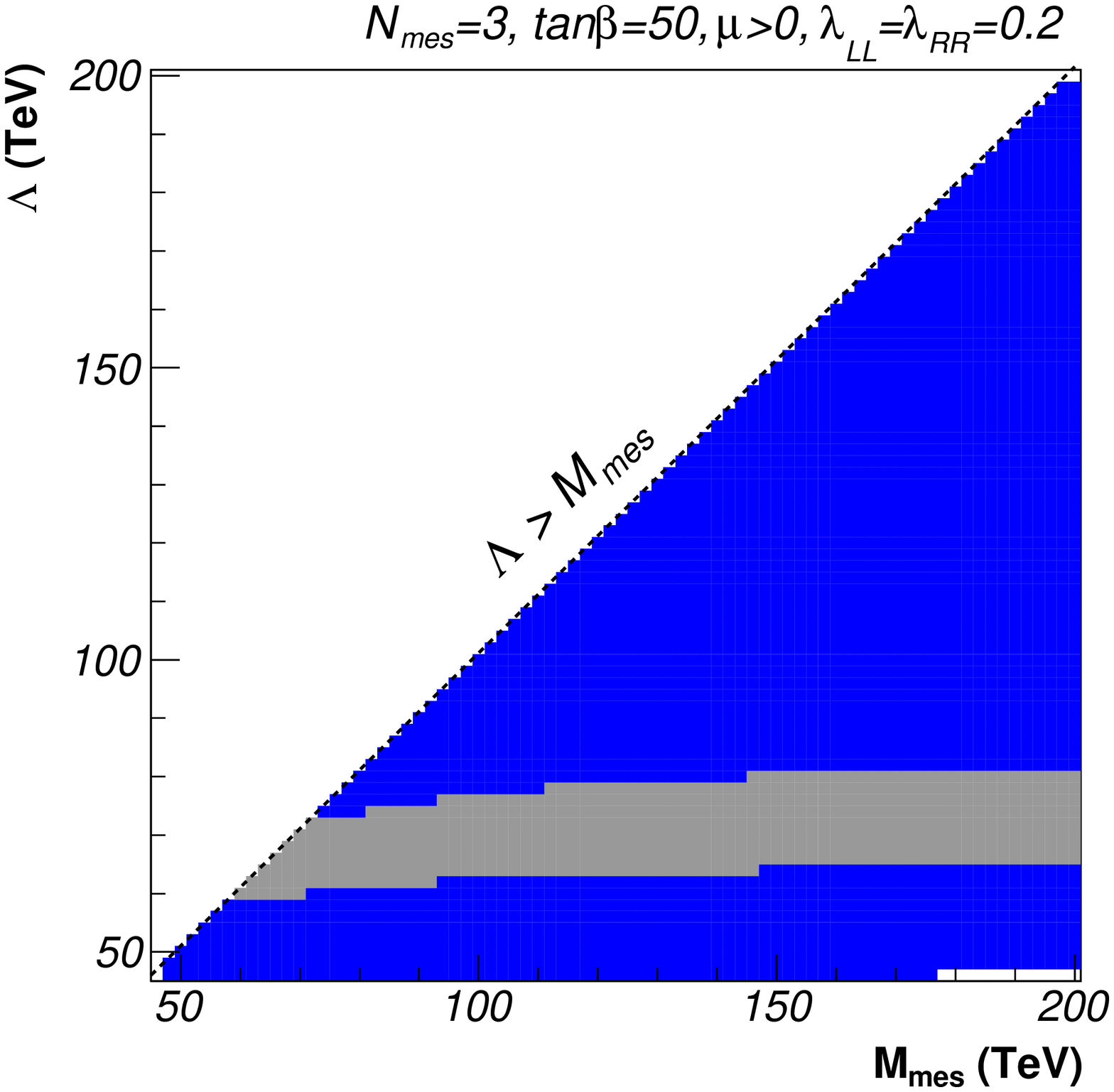}
\caption{Same as Fig.\ \ref{fig2} for $N_{\rm mes}=3$ and $\tan\beta=50$.}
\label{fig7}
\end{center}\end{figure}

In Figs.\ \ref{fig2} -- \ref{fig7} we show scans of the
$\Lambda$--$M_{\rm mes}$ plane for $\mu>0$ and the same values $N_{\rm mes} =
1,~3$ and $\tan\beta = 15,~30,~50$ as in Sec.\ \ref{sec2}. The region favoured by
the anomalous magnetic moment of the muon $a_{\mu}$ (light/grey) is quite
insensitive
to variations of the parameter $\lambda_{\rm LL}$, since dominant SUSY effects
come from induced quantum loops of gauginos and sleptons, while squarks
contribute only at the two-loop level, which reduces the dependence on squark
flavour violation. As expected, the $b\to s\gamma$ excluded region (dark/blue)
depends strongly on flavour mixing, while the constraint coming from
$\Delta\rho$ does not play a role for the moderate SUSY masses relevant to
our region of interest, so that the corresponding excluded regions are not
shown. Note that the difference between the two flavour violation scenarios
considered here is relatively small, as can be seen in Figs.\ \ref{fig2} --
\ref{fig7}. It becomes clear that, if we allow for
flavour mixing between the second and third generation squarks, windows in the
parameter space both favoured by $a_{\mu}$ and not excluded by the
stringent constraint from $b\to s\gamma$ make their appearance for small and
moderate SUSY masses.

Within these regions, we propose six benchmark scenarios permitting
non-minimal flavour violation and not yielding too high SUSY 
masses (``collider-friendly''), so that possible SUSY signals should be
observable at present and/or future hadron colliders. Our choices
are presented in Tab.\ \ref{tab1}, labeled starting at the point E due to our
four benchmark proposals for mSUGRA scenarios including NMFV
\cite{Bozzi:2007me}. Note that, in contrast to the mSUGRA case, these 
scenarios are not valid assuming cMFV ($\lambda_{\rm LL} = \lambda_{\rm RR} = 0$), so
that we indicate the allowed ranges for our flavour mixing parameter
$\lambda_{\rm LL}$. 

\begin{table}
\caption{GMSB benchmark points allowing for non-minimal flavour violation in the
left-left ($\lambda_{\rm RR}=0$) or both the left-left and
right-right ($\lambda_{\rm RR}=\lambda_{\rm LL}$) chiral squark sectors. We
also indicate the allowed range for the NMFV-parameter $\lambda_{\rm LL}$, the
nature of the next-to-lightest SUSY particle (NLSP), and the closest SPS
benchmark point (if relevant), which are valid for both flavour-violating
scenarios.}
\begin{center}
  \begin{tabular}{l||ccccc|c|c|c|}
     & $\Lambda$ [TeV] & $M_{\rm mes}$ [TeV] & $N_{\rm mes}$ & $\tan\beta$ &
     sgn($\mu$) & $\lambda_{\rm LL}$ & NLSP & SPS \\ 
  \hline \hline
  E & 65 & 90 & 1 & 15 & + & [0.14, 0.20] & $\tilde{\chi}_1^0$ & 8 \\
  F & 30 & 80 & 3 & 15 & + & [0.12, 0.18] & $\tilde{\tau}_1$ & 7 \\
  \hline
  G & 100 & 110 & 1 & 30 &+& [0.14, 0.20] & $\tilde{\tau}_1$ & -- \\
  H & 45 & 100 & 3 & 30 & +& [0.12, 0.18] & $\tilde{\tau}_1$ & -- \\
  \hline
  I & 130 & 140 & 1 & 50 &+& [0.14, 0.20] & $\tilde{\tau}_1$ & -- \\
  J & 60 & 100 & 3 & 50 & +& [0.14, 0.20] & $\tilde{\tau}_1$ & -- \\
  \hline
  \end{tabular}
\label{tab1}
\end{center}
\end{table}

Starting with $\tan\beta=15$ and $N_{\rm mes}=1$ (see Fig.\ \ref{fig2}), we
choose our benchmark point E in the region both favoured by the electroweak precision
constraints and corresponding to rather light SUSY particles. As for any GMSB
scenario, the gravitino is the lightest SUSY particle (LSP). The
next-to-lightest SUSY particle (NLSP) is the lightest neutralino with
$m_{\tilde{\chi}_1^0} = 95.4$ GeV, but the three lightest charged sleptons are
very close with similar masses around 100 GeV. The other sleptons,
sneutrinos, and gauginos have moderate masses of about 150 -- 300 GeV, while the
squarks and gluino are quite heavy with masses lying in the range of 700 -- 800
GeV. However, they are much lighter than those corresponding to the point SPS 8
with its larger values of $\Lambda$ and $M_{\rm mes}$, lie well above the
experimental limits obtained from direct searches assuming cMFV, and are
experimentally accessible at the LHC.

The point F (see Fig.\ \ref{fig3}) differs very little from the
point SPS 7, with the SUSY-breaking scale $\Lambda$ shifted from 40 to 30
TeV, so that it now lies in the preferred region with respect to the $b\to
s\gamma$ constraint. As for SPS 7, the three lightest sleptons have masses
around 100 GeV, the lightest being the stau with $m_{\tilde{\tau}_1} =
90.7$ GeV. The other sleptons, sneutrinos, and gauginos are a bit heavier
(120-200 GeV), and the squarks and gluino are rather heavy (600-700 GeV). 

The points G, H, I, and J (see Figs.\ \ref{fig4}-\ref{fig7}) all
have a stau NLSP with a mass between 99 and 160 GeV. The main difference in
their spectra is the number and nature of the particles that are closest in mass
to the NLSP. For the point G, these are two sleptons and the lightest neutralino,
whereas for the point H these are only the two sleptons. The points I and J do
not have any particles close to the NLSP in mass. For the four points G, H, I,
and J, the other sleptons and gauginos are rather light (200-600 GeV), while
the squarks and the gluino are very heavy (1-1.5 TeV).

\begin{figure}\begin{center}
	\includegraphics[scale=0.27]{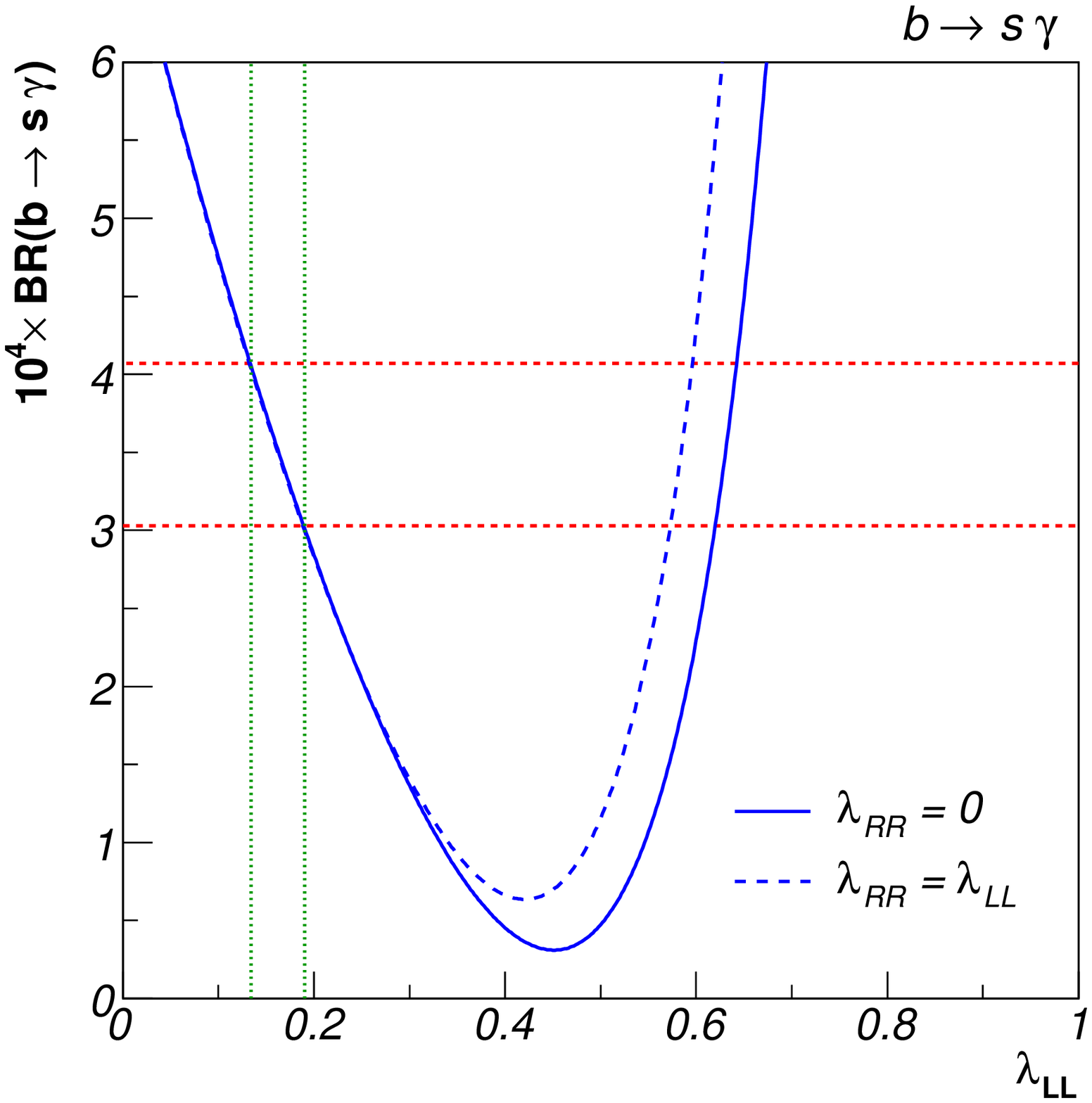}
	\includegraphics[scale=0.27]{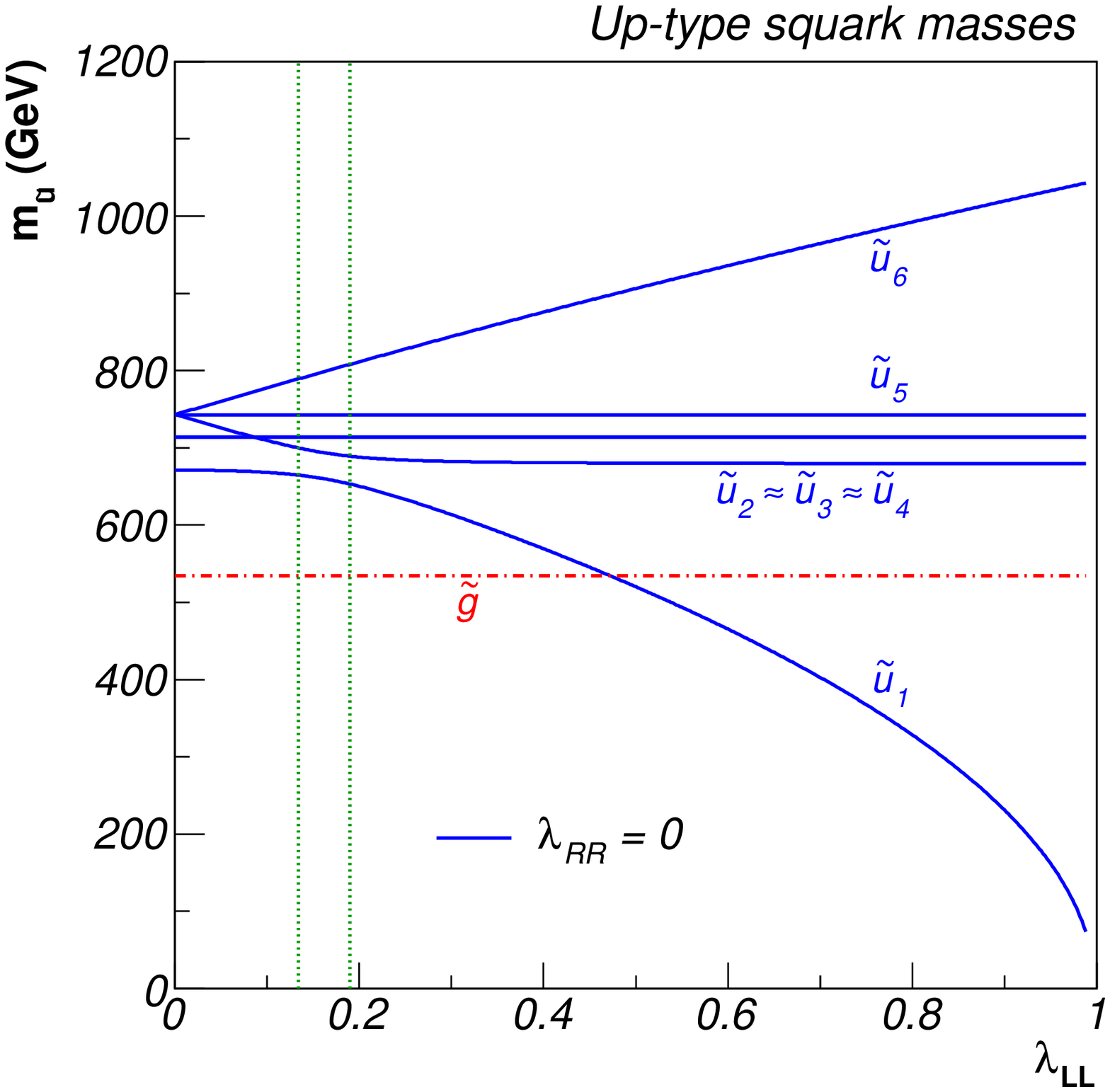}
	\includegraphics[scale=0.27]{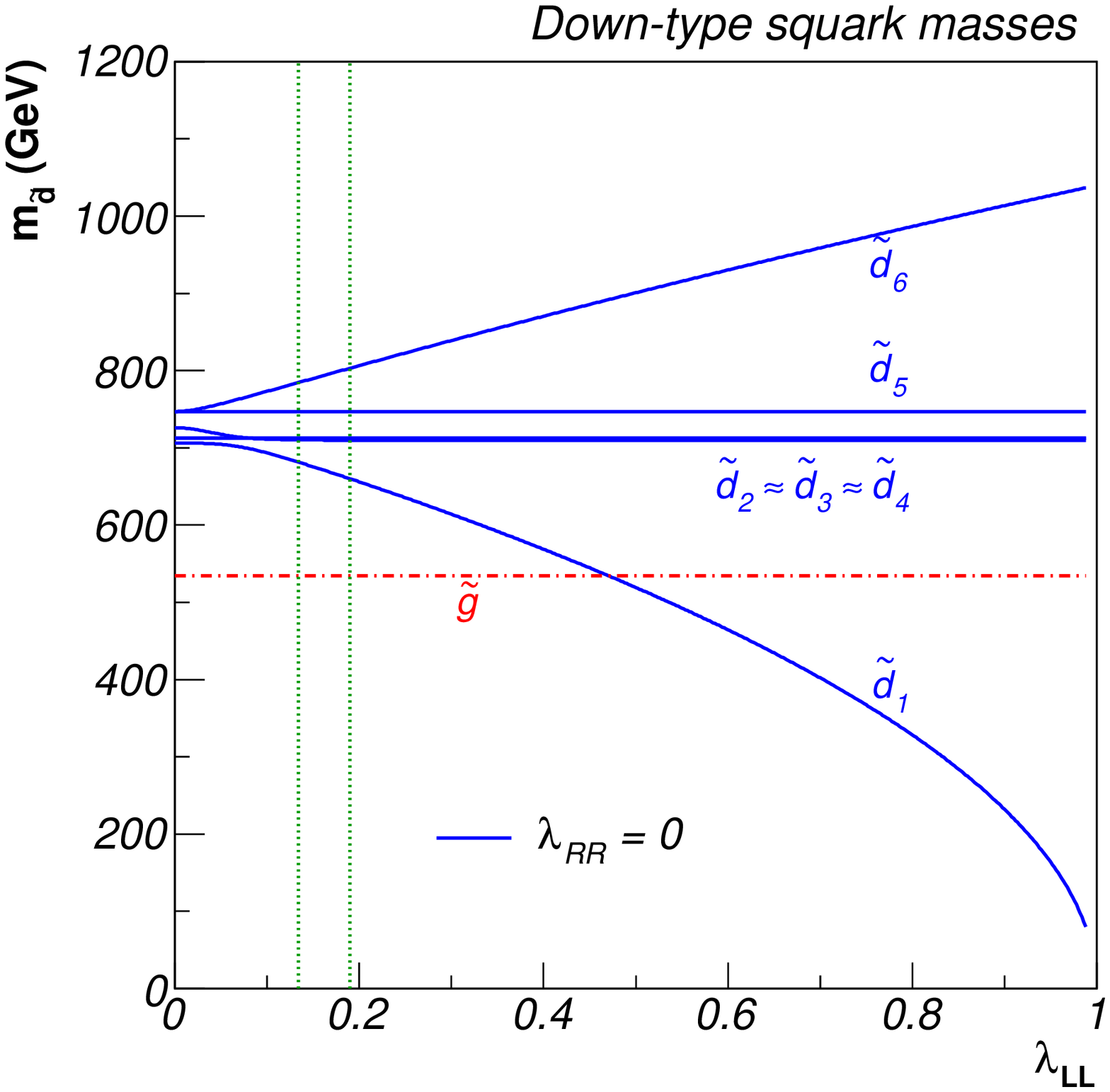}
	\includegraphics[scale=0.27]{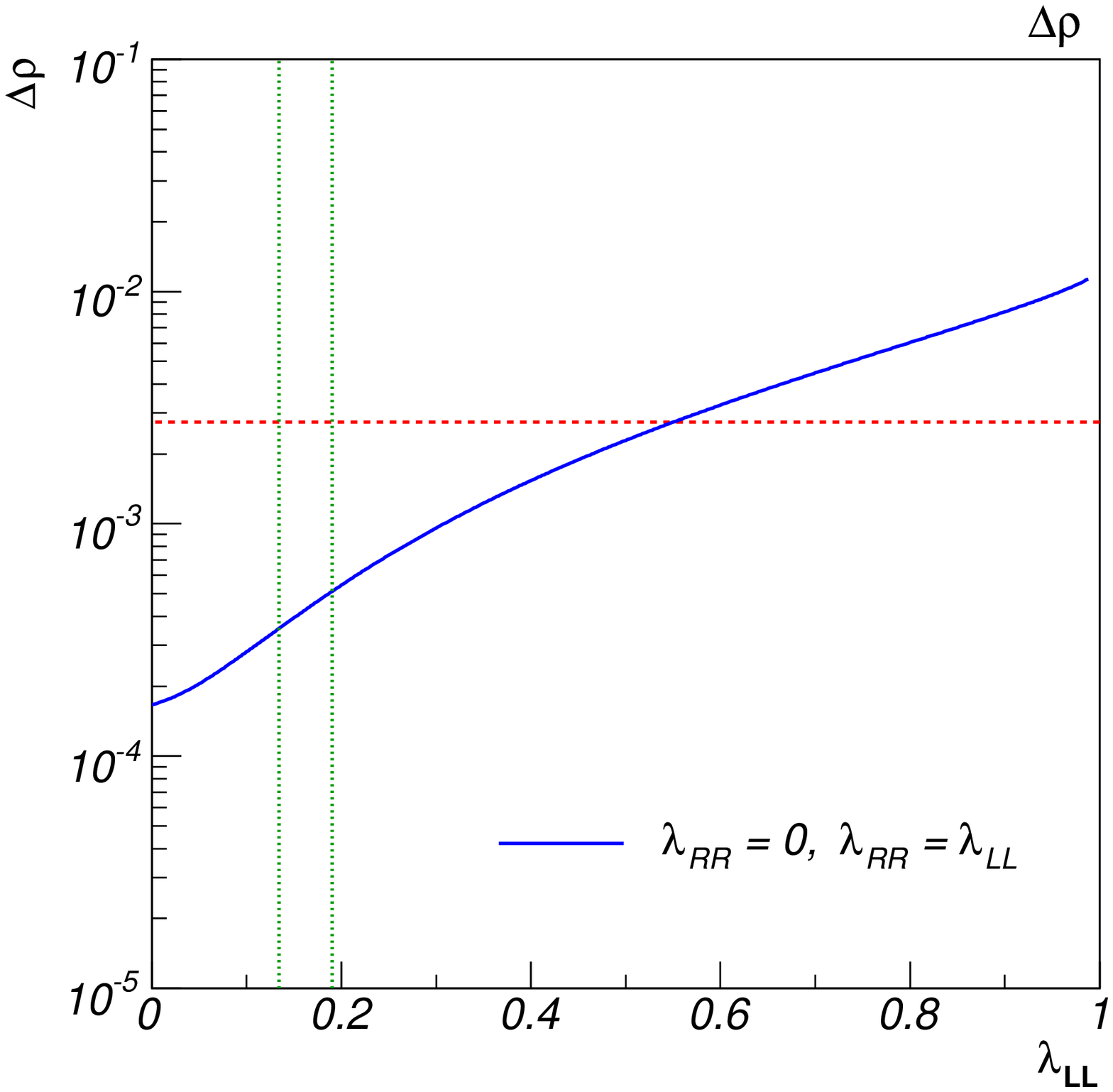}
	\includegraphics[scale=0.27]{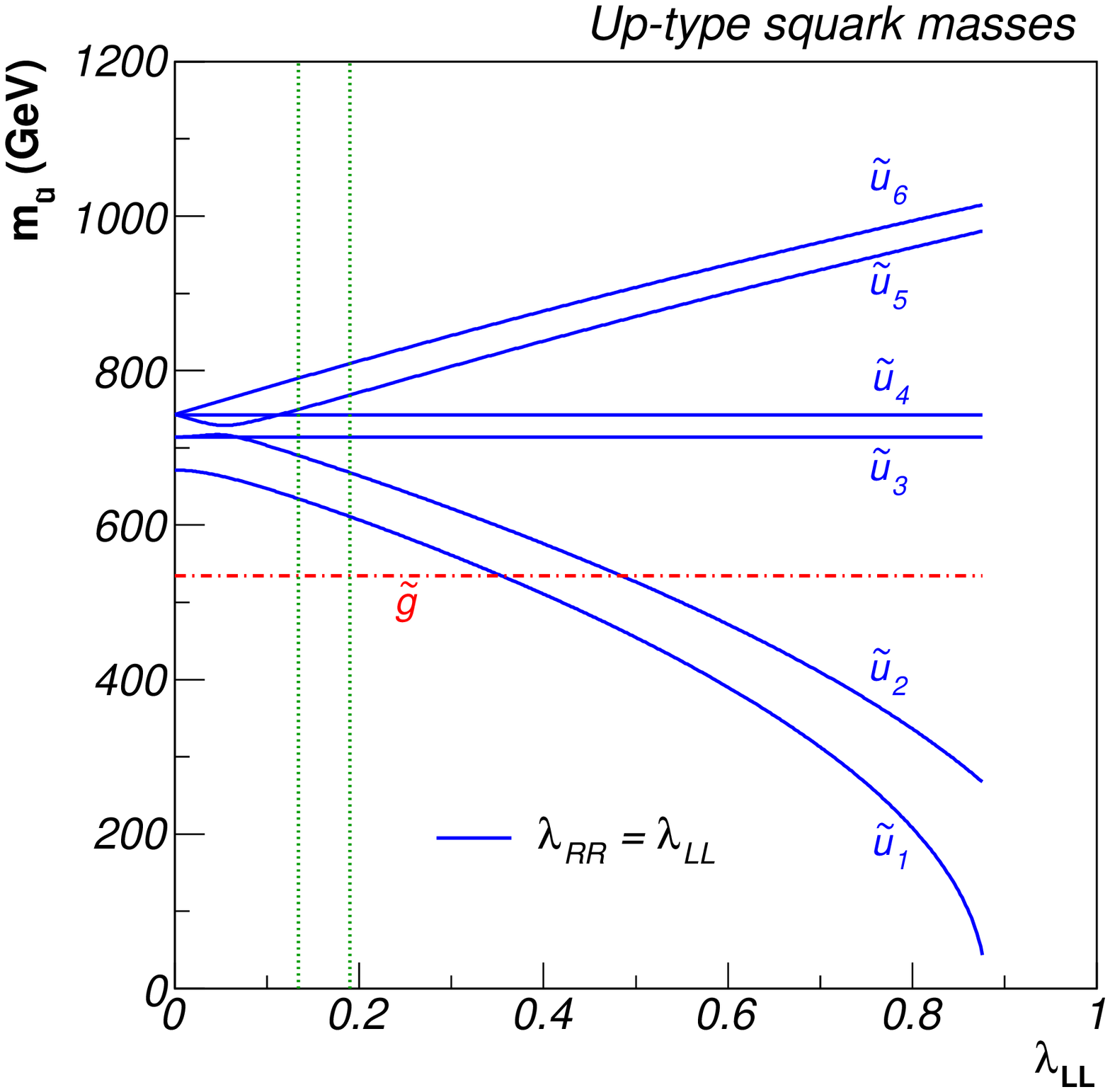}
	\includegraphics[scale=0.27]{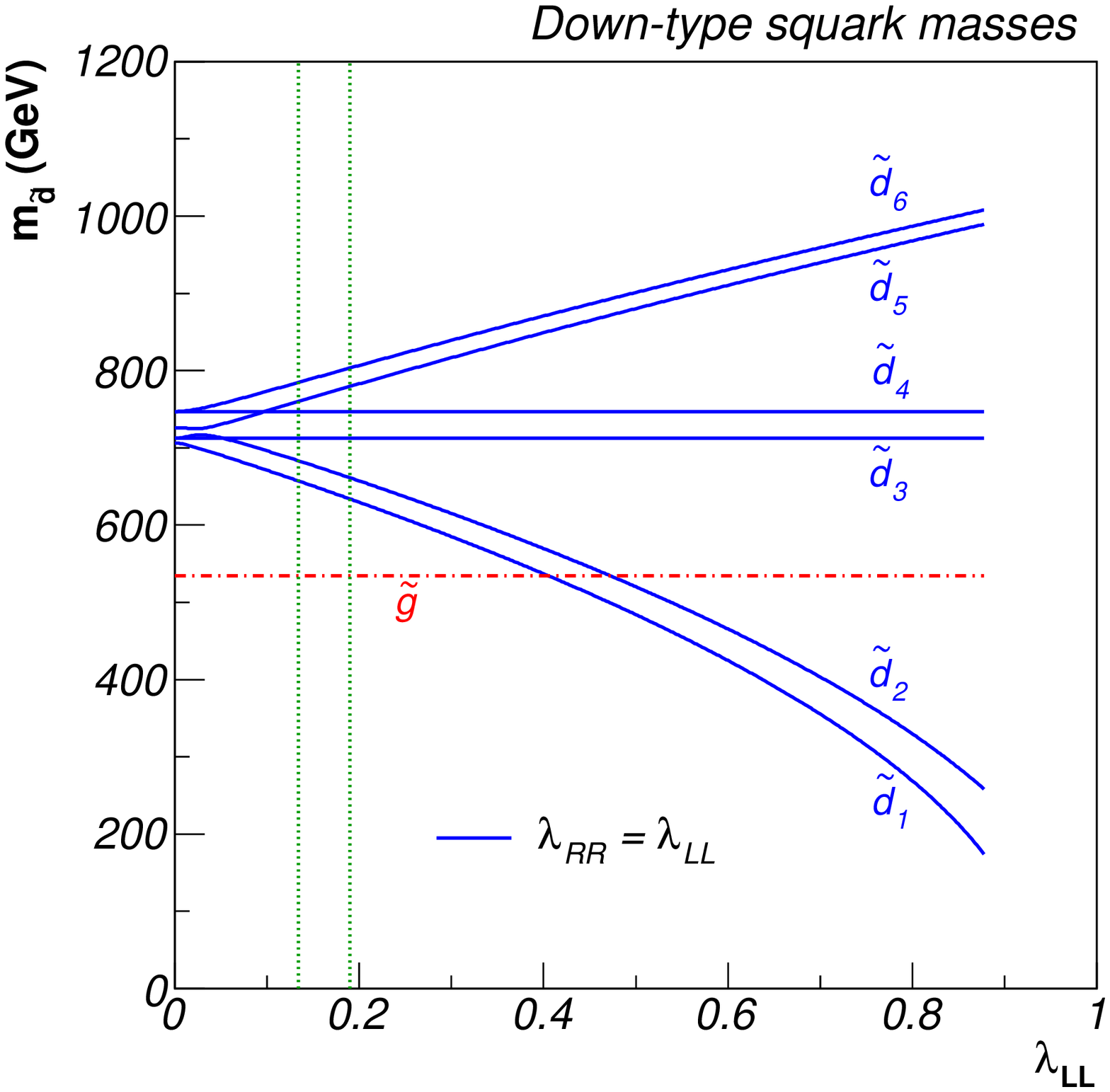}
\caption{Dependence of the precision variables ${\rm BR}(b\to s\gamma)$ and
$\Delta\rho$ as well as of the up- and down-type squark masses on the
NMFV-parameter $\lambda_{\rm LL}$ for flavour mixing in the left-left
($\lambda_{\rm RR}=0$) or both the left-left and right-right ($\lambda_{\rm
RR}=\lambda_{\rm LL}$) chiral squark sectors for our benchmark scenario E. The
experimentally allowed ranges within $2\sigma$ are indicated by horizontal
dashed lines. The vertical dotted lines indicate the allowed range for
$\lambda_{\rm LL}$ with respect to the most stringent constraint from $b\to
s\gamma$. For $\lambda_{\rm LL}=\lambda_{\rm RR}\geq0.9$ no physical solutions
are possible.}
\label{fig8}
\end{center}\end{figure}

\begin{figure}\begin{center}
	\includegraphics[scale=0.27]{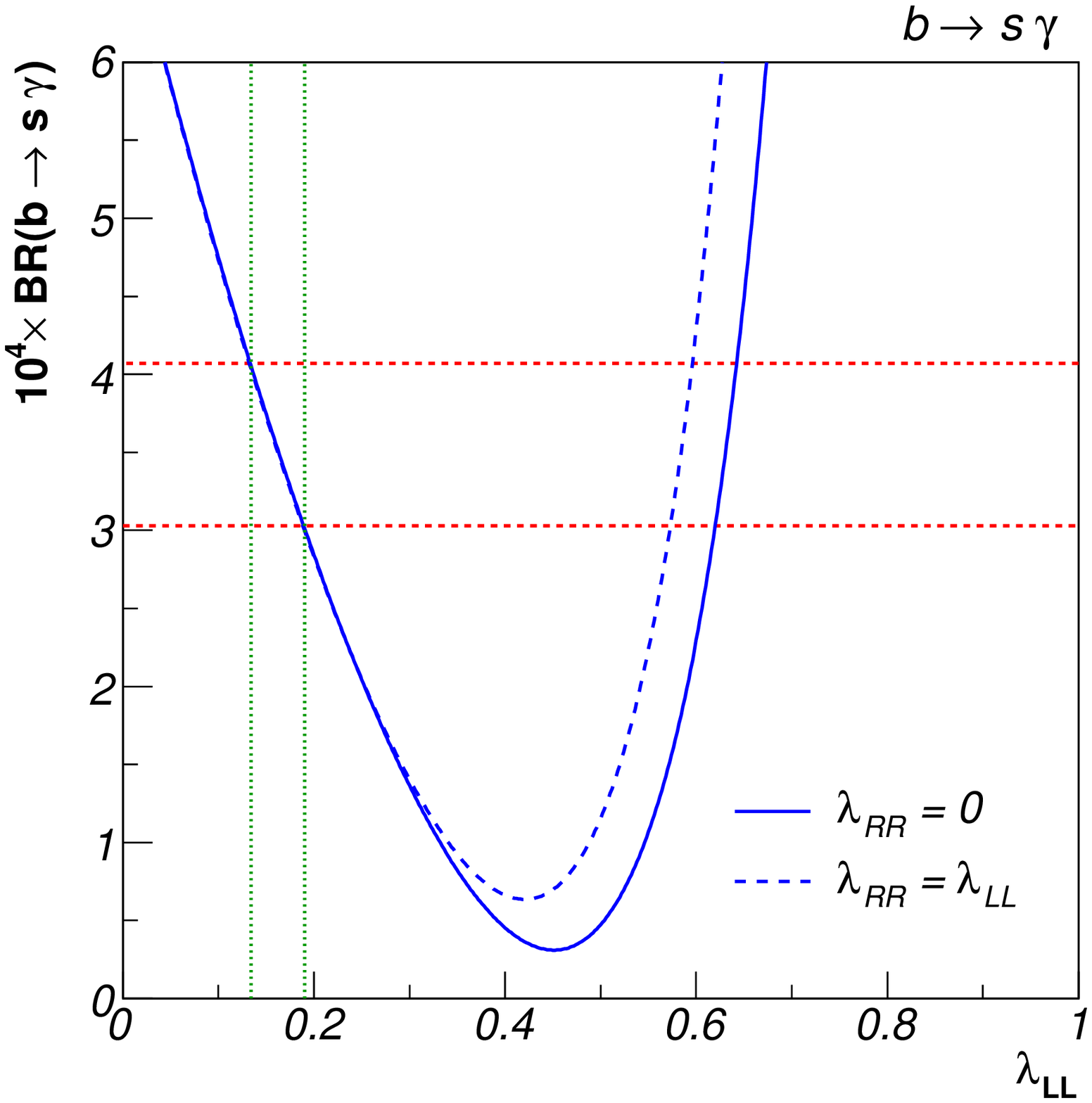}
	\includegraphics[scale=0.27]{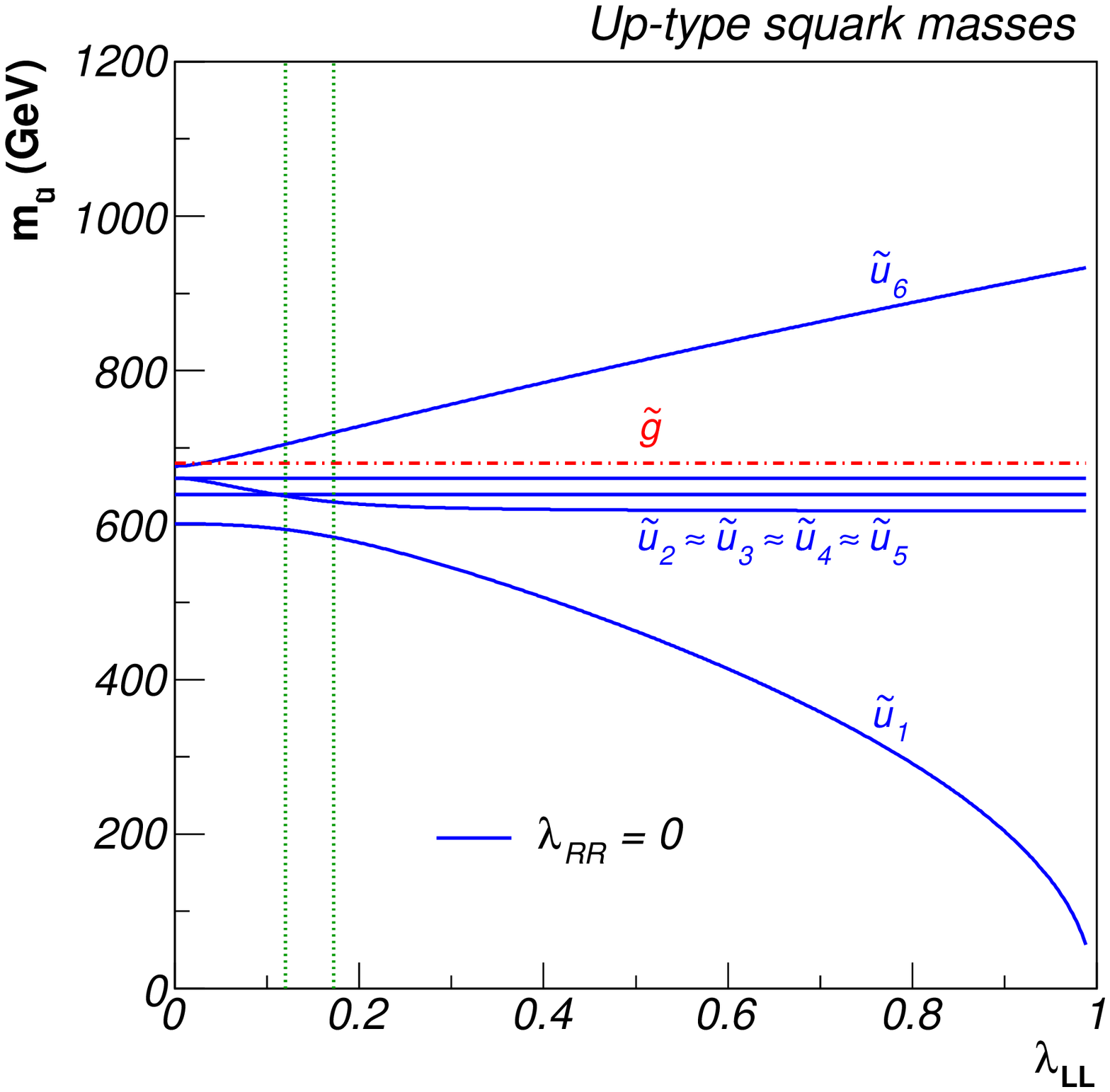}
	\includegraphics[scale=0.27]{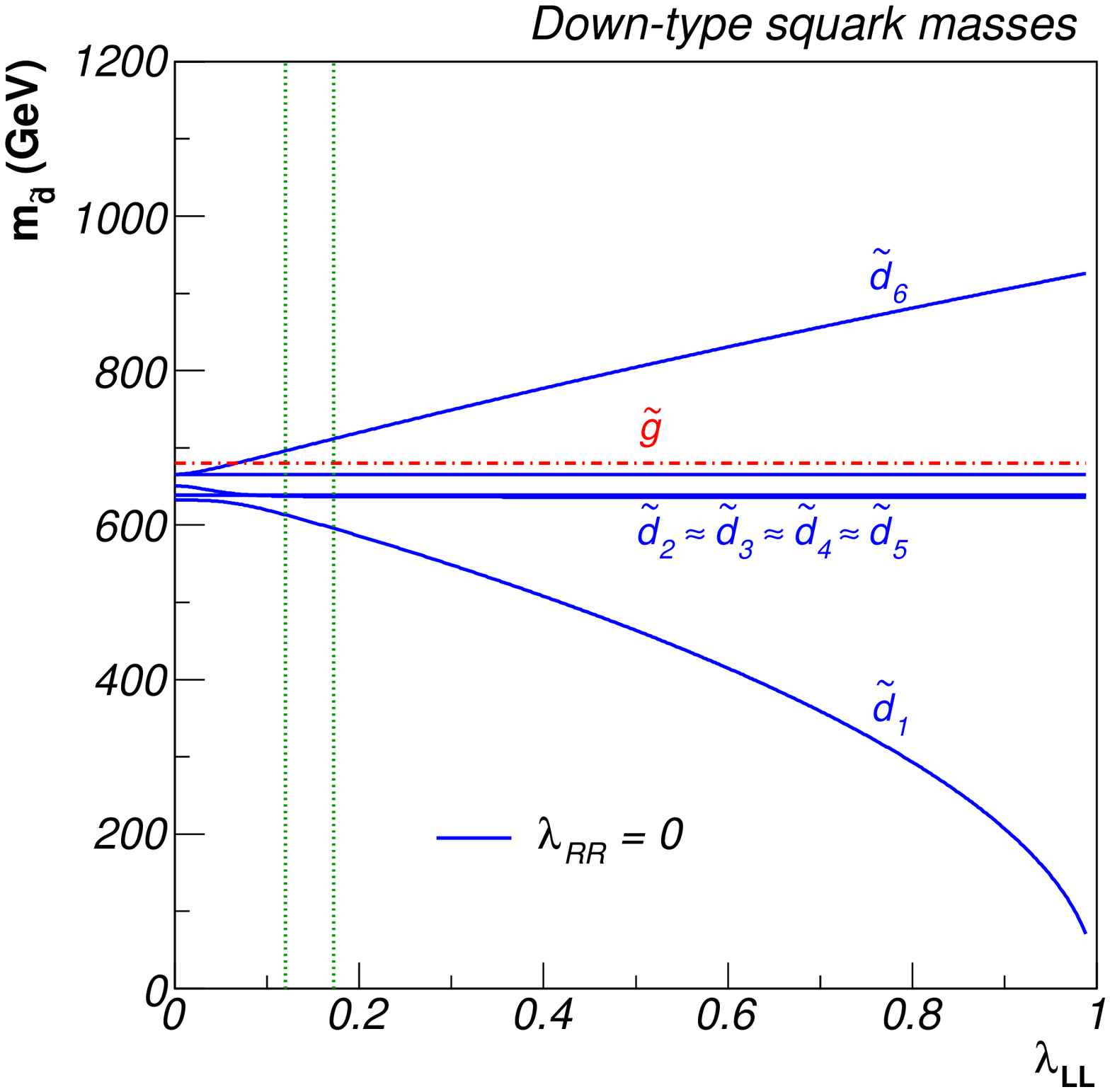}
	\includegraphics[scale=0.27]{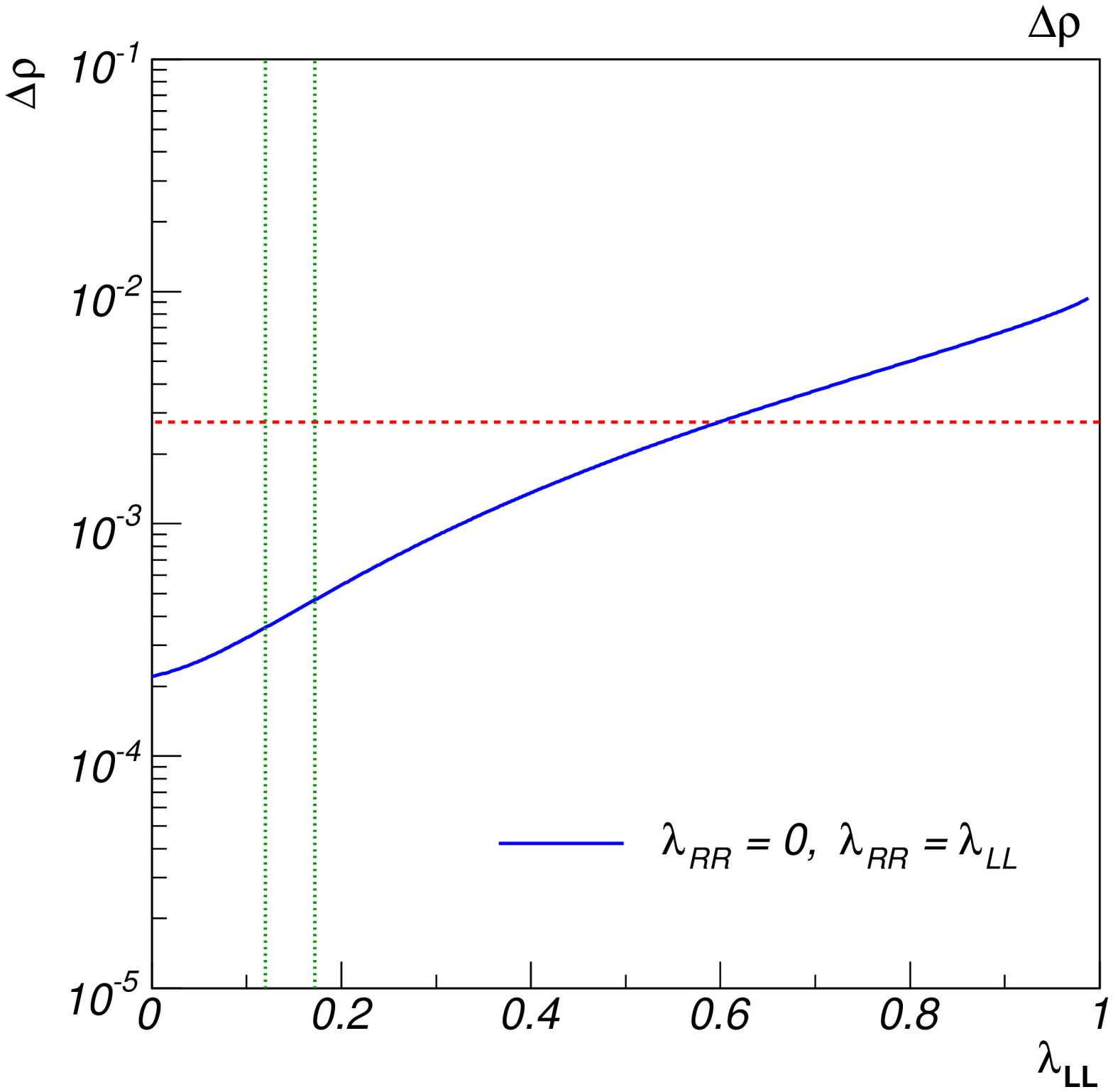}
	\includegraphics[scale=0.27]{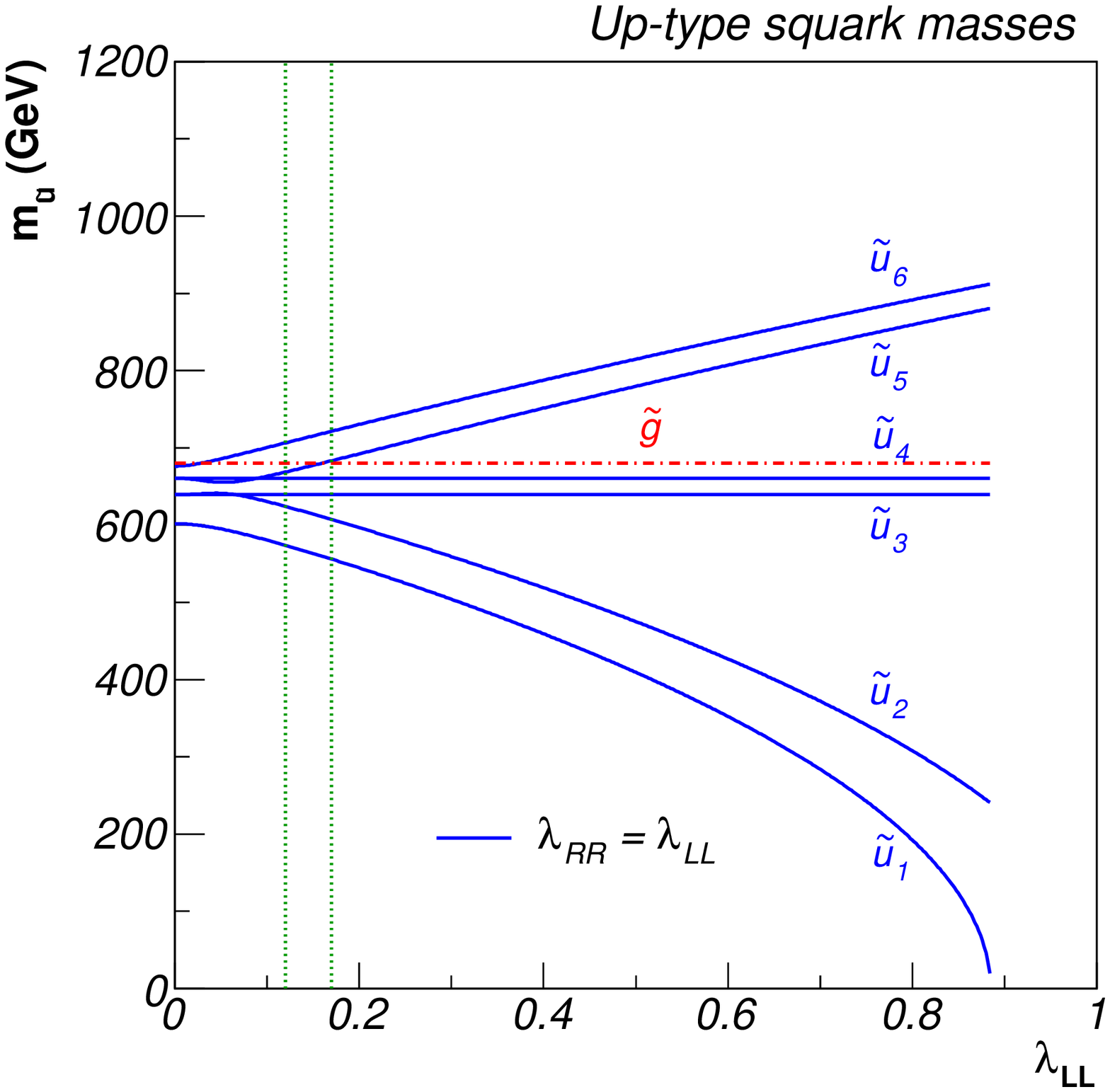}
	\includegraphics[scale=0.27]{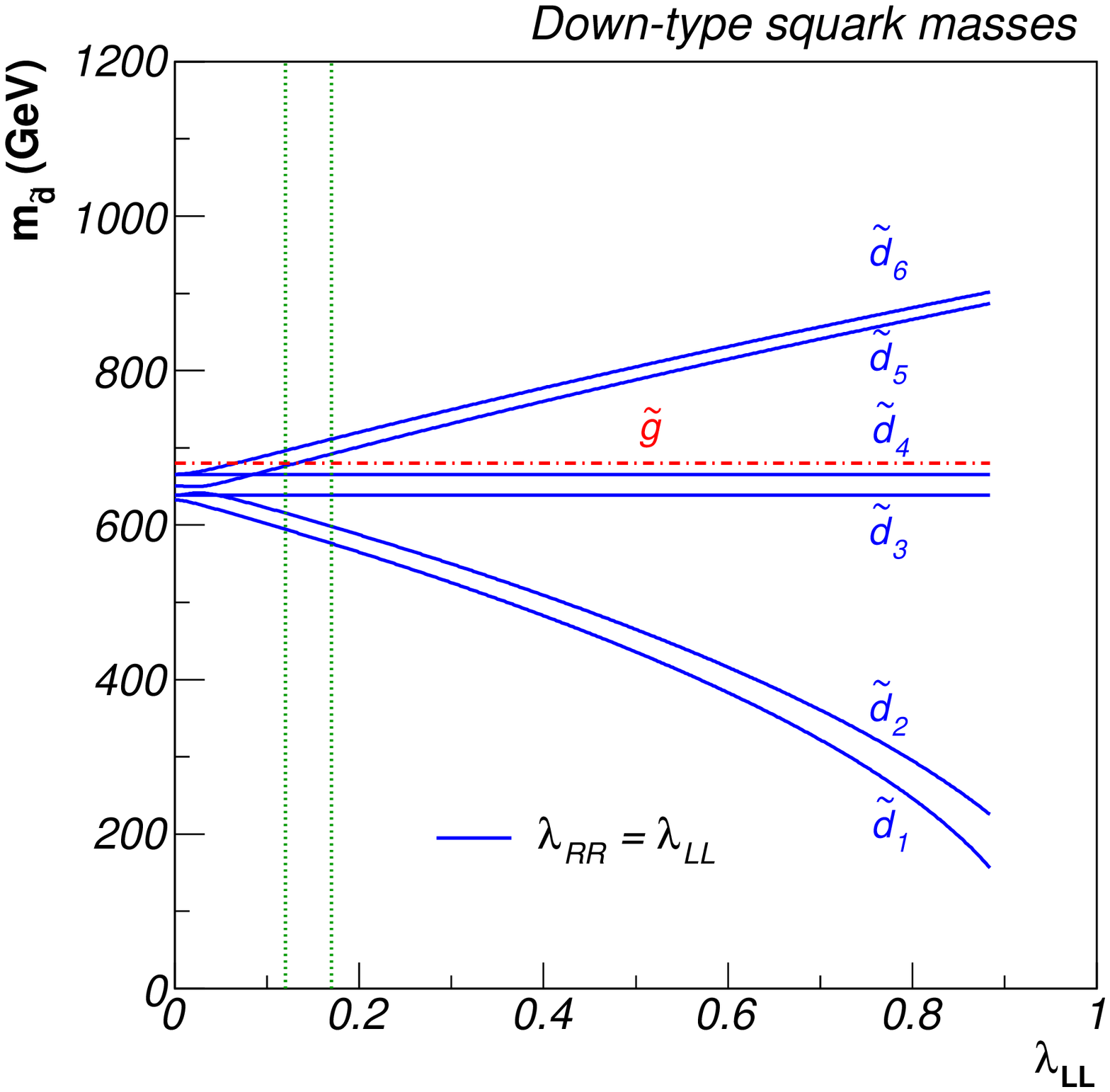}
\caption{Same as Fig.\ \ref{fig8} for our benchmark scenario F.}
\label{fig9}
\end{center}\end{figure}

\begin{figure}\begin{center}
	\includegraphics[scale=0.27]{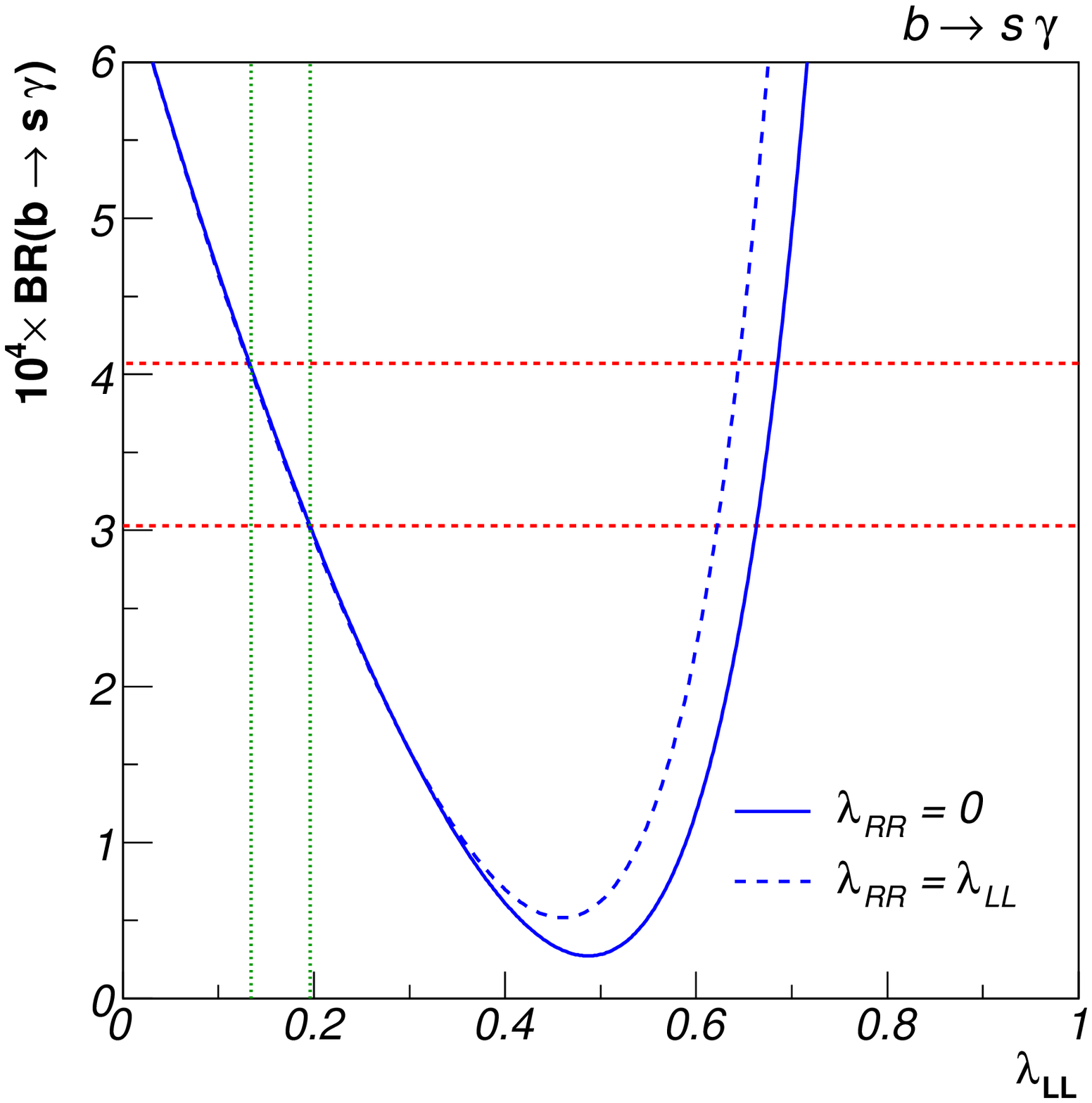}
	\includegraphics[scale=0.27]{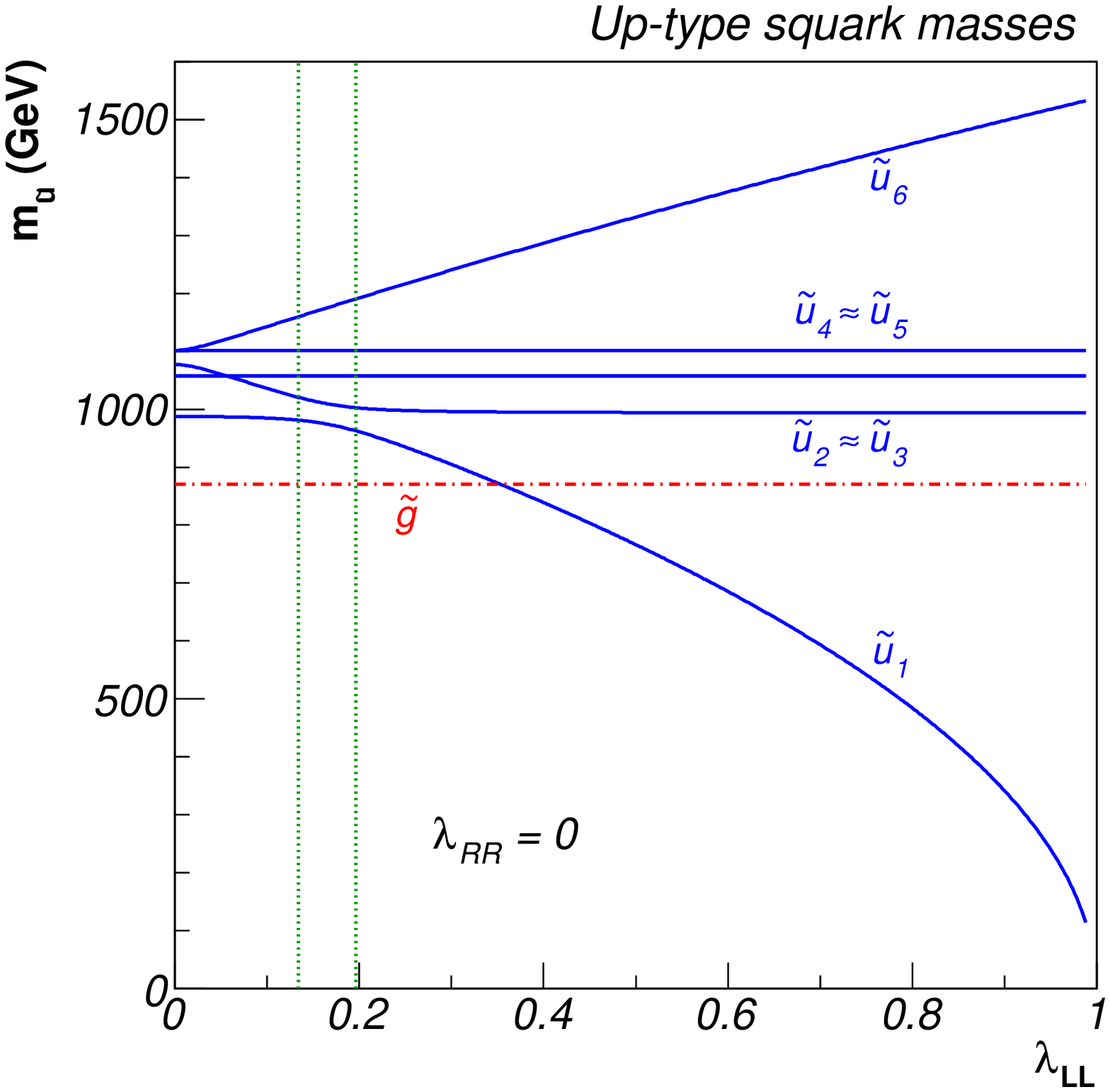}
	\includegraphics[scale=0.27]{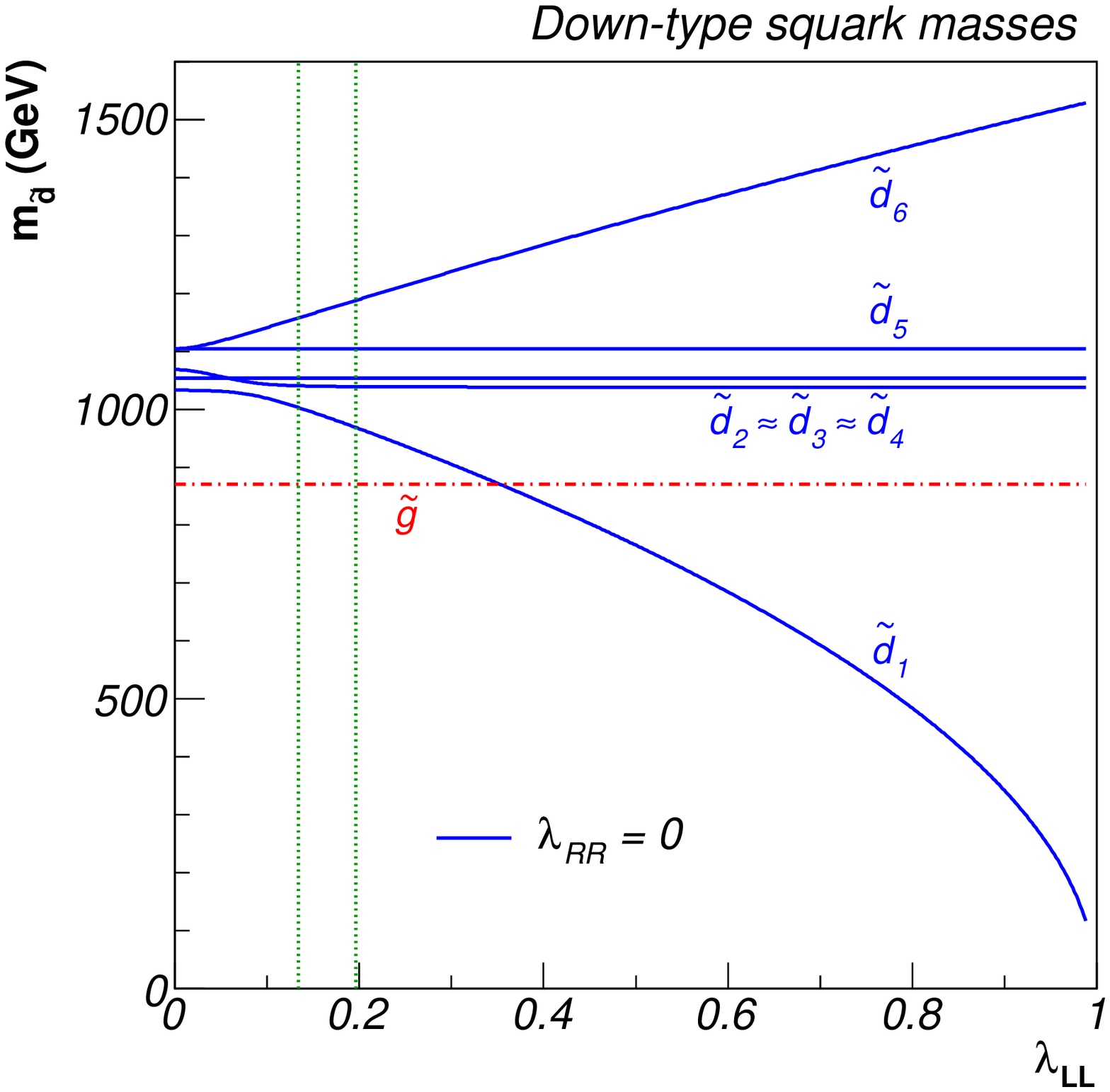}
	\includegraphics[scale=0.27]{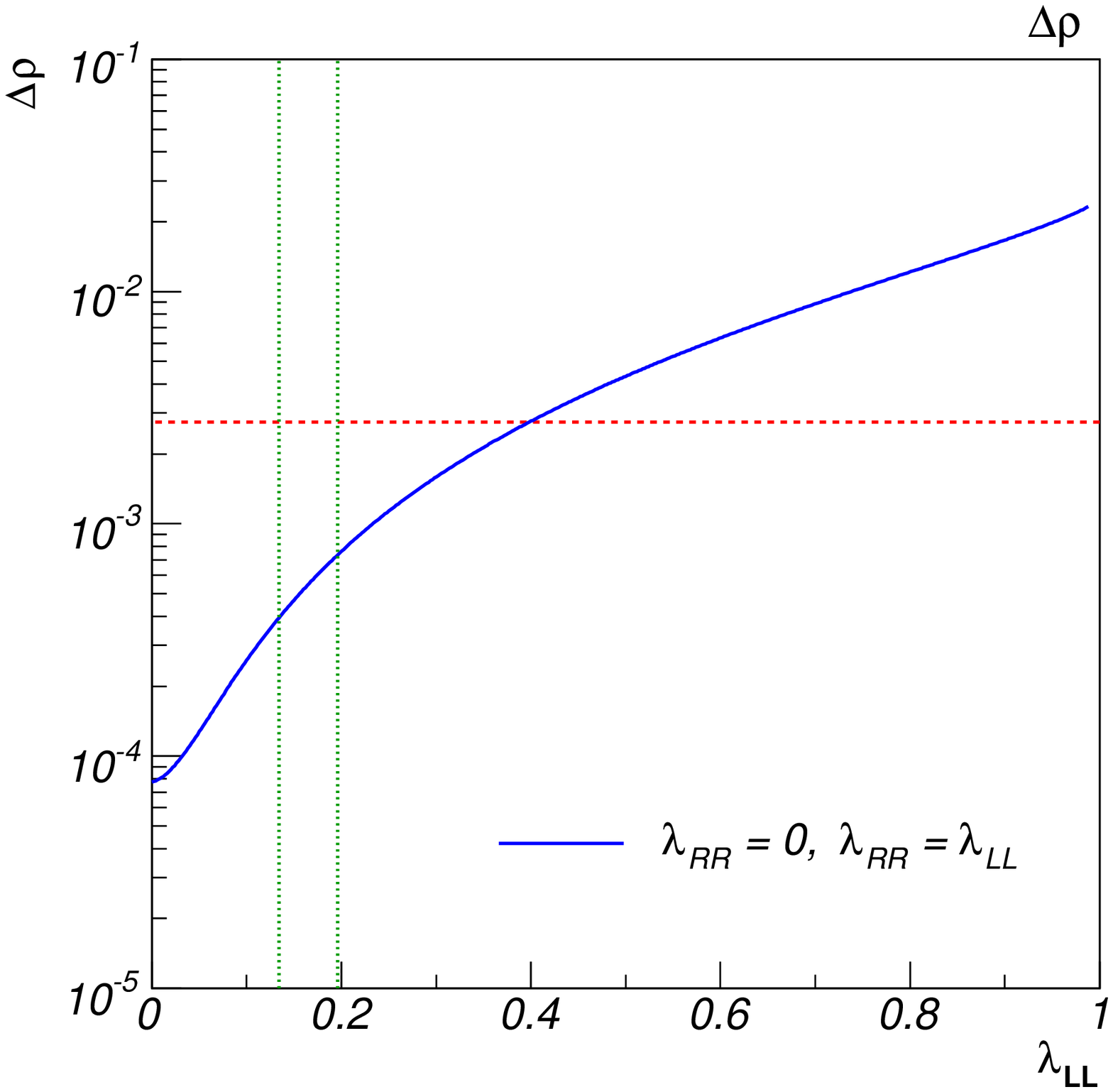}
	\includegraphics[scale=0.27]{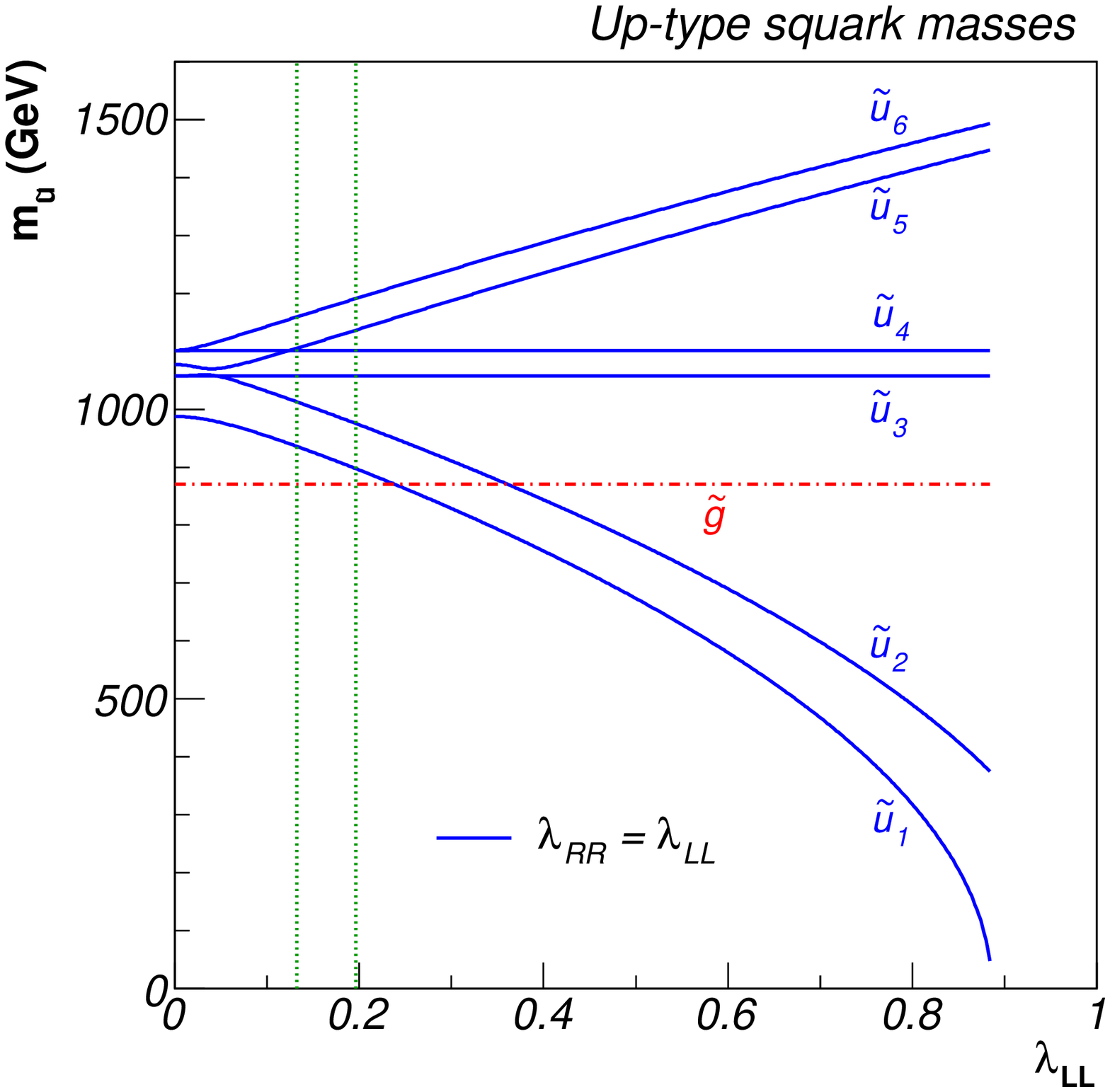}
	\includegraphics[scale=0.27]{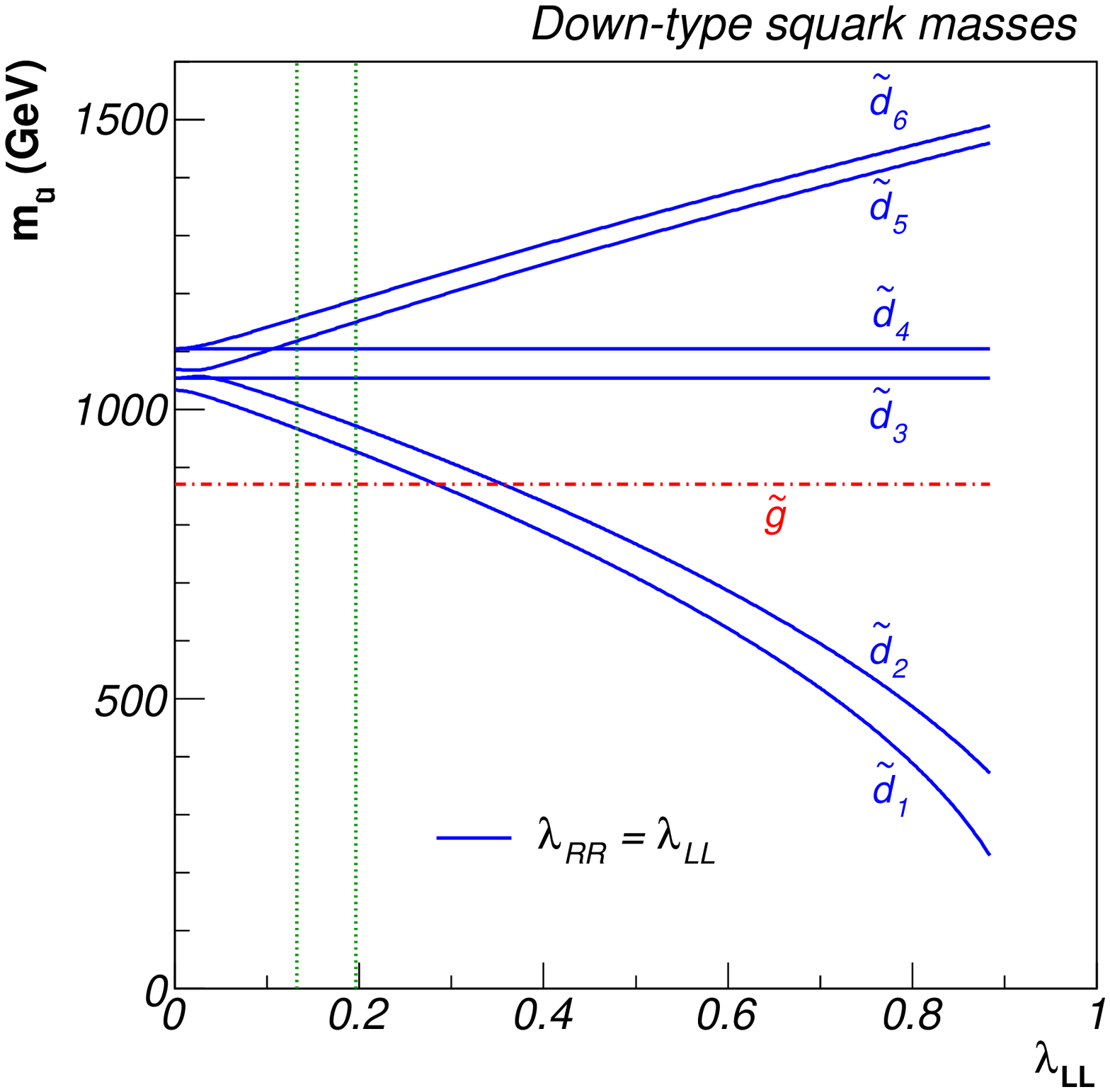}
\caption{Same as Fig.\ \ref{fig8} for our benchmark scenario G.}
\label{fig10}
\end{center}\end{figure}

\begin{figure}\begin{center}
	\includegraphics[scale=0.27]{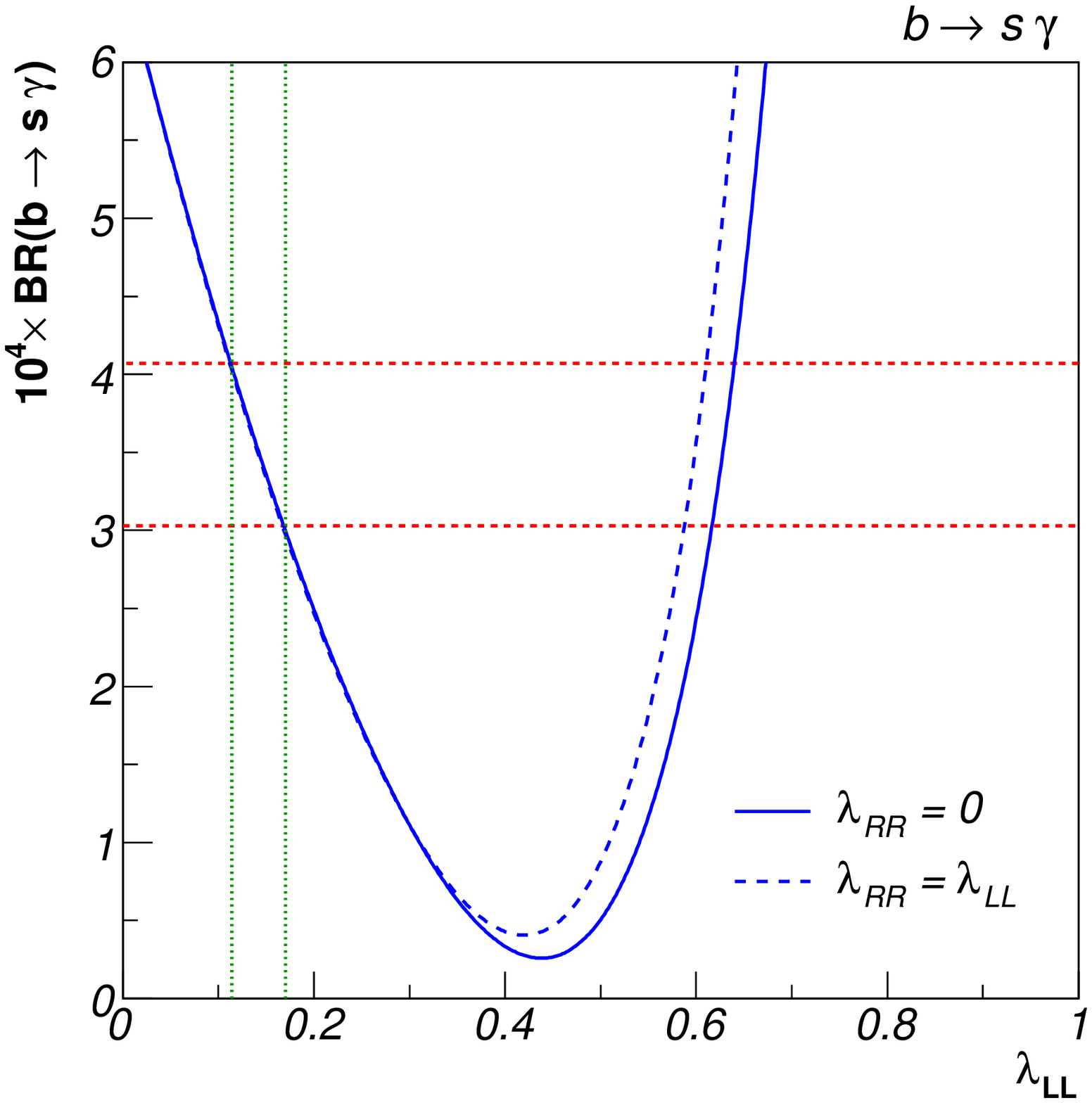}
	\includegraphics[scale=0.27]{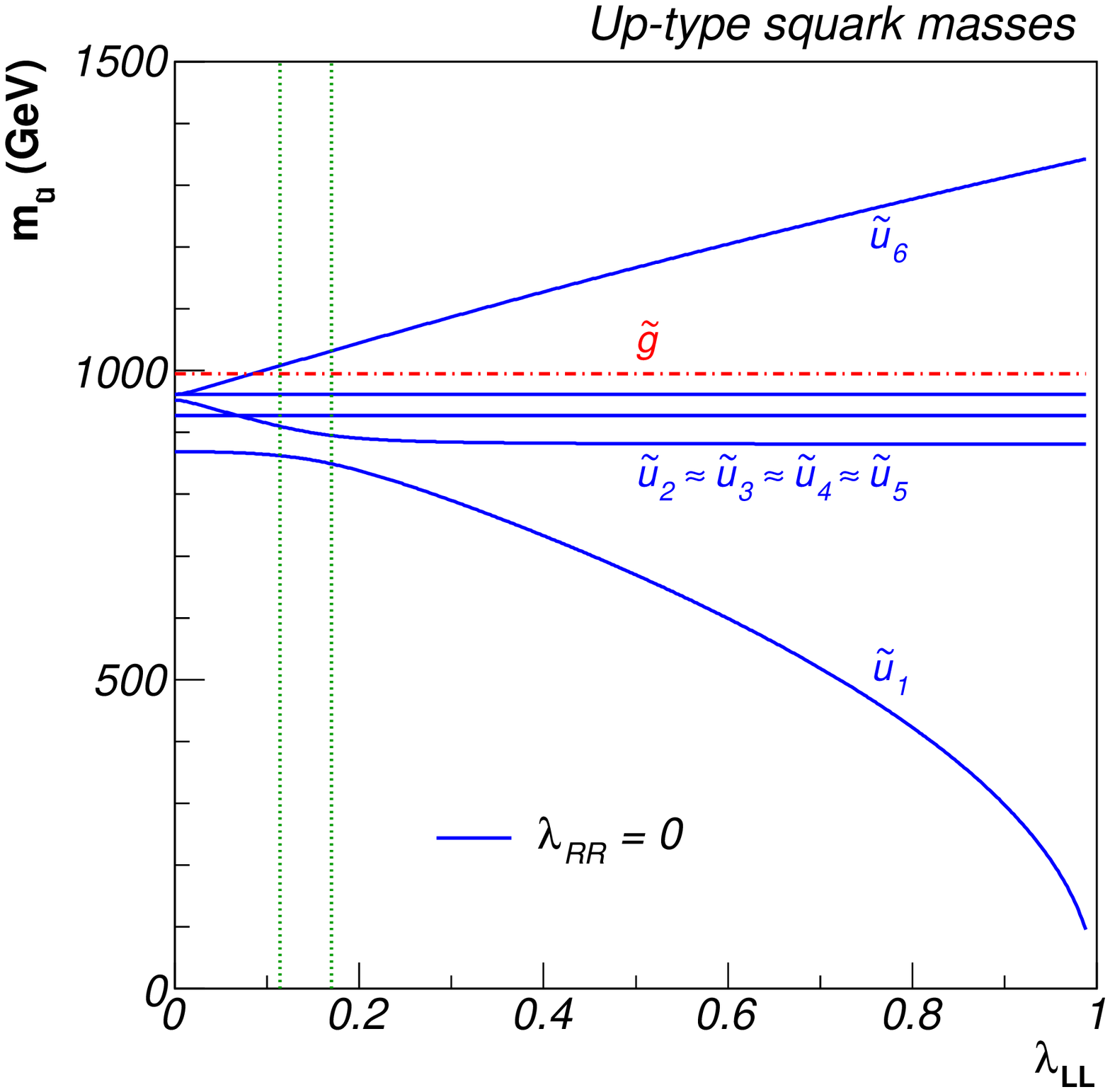}
	\includegraphics[scale=0.27]{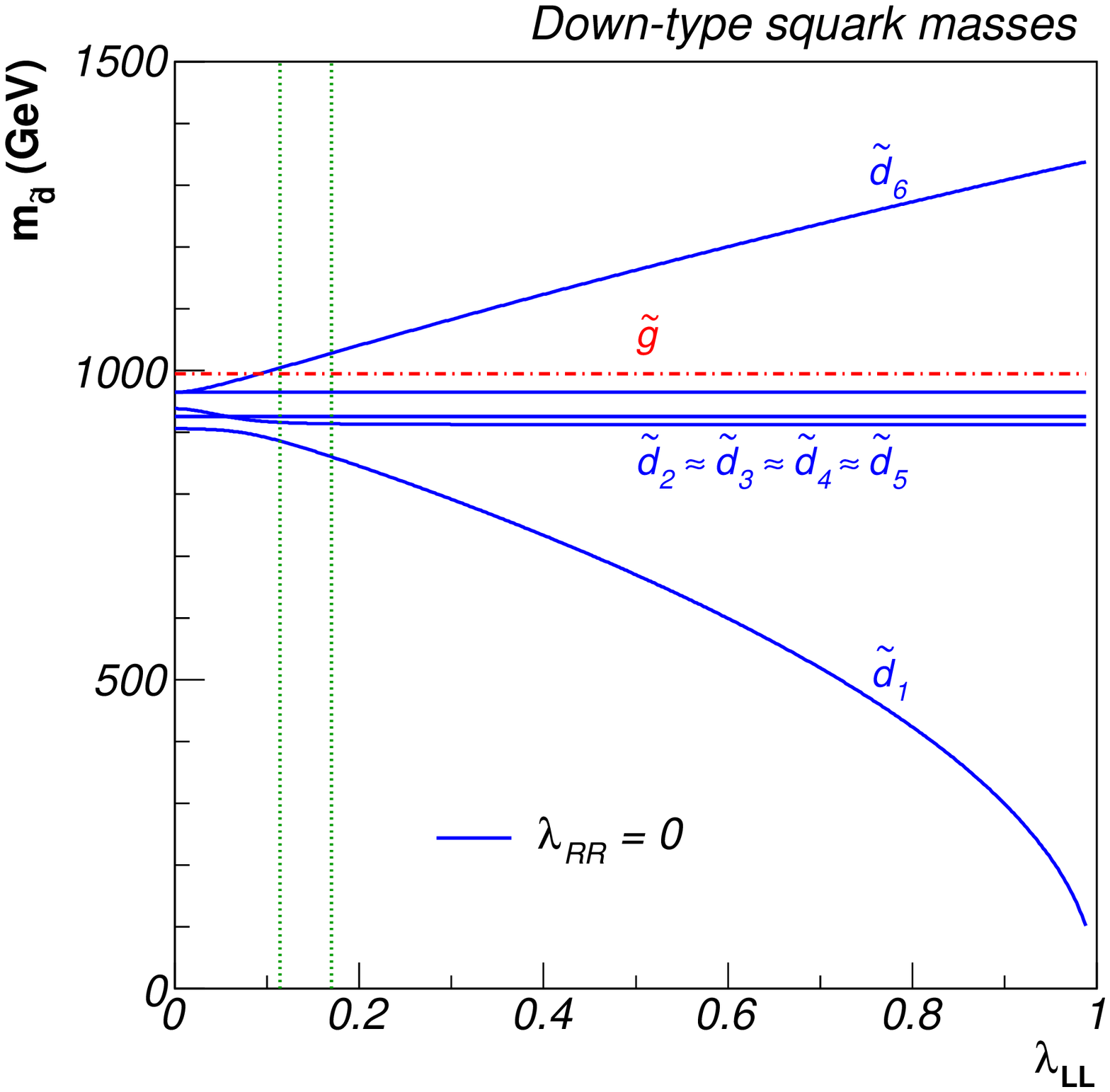}
	\includegraphics[scale=0.27]{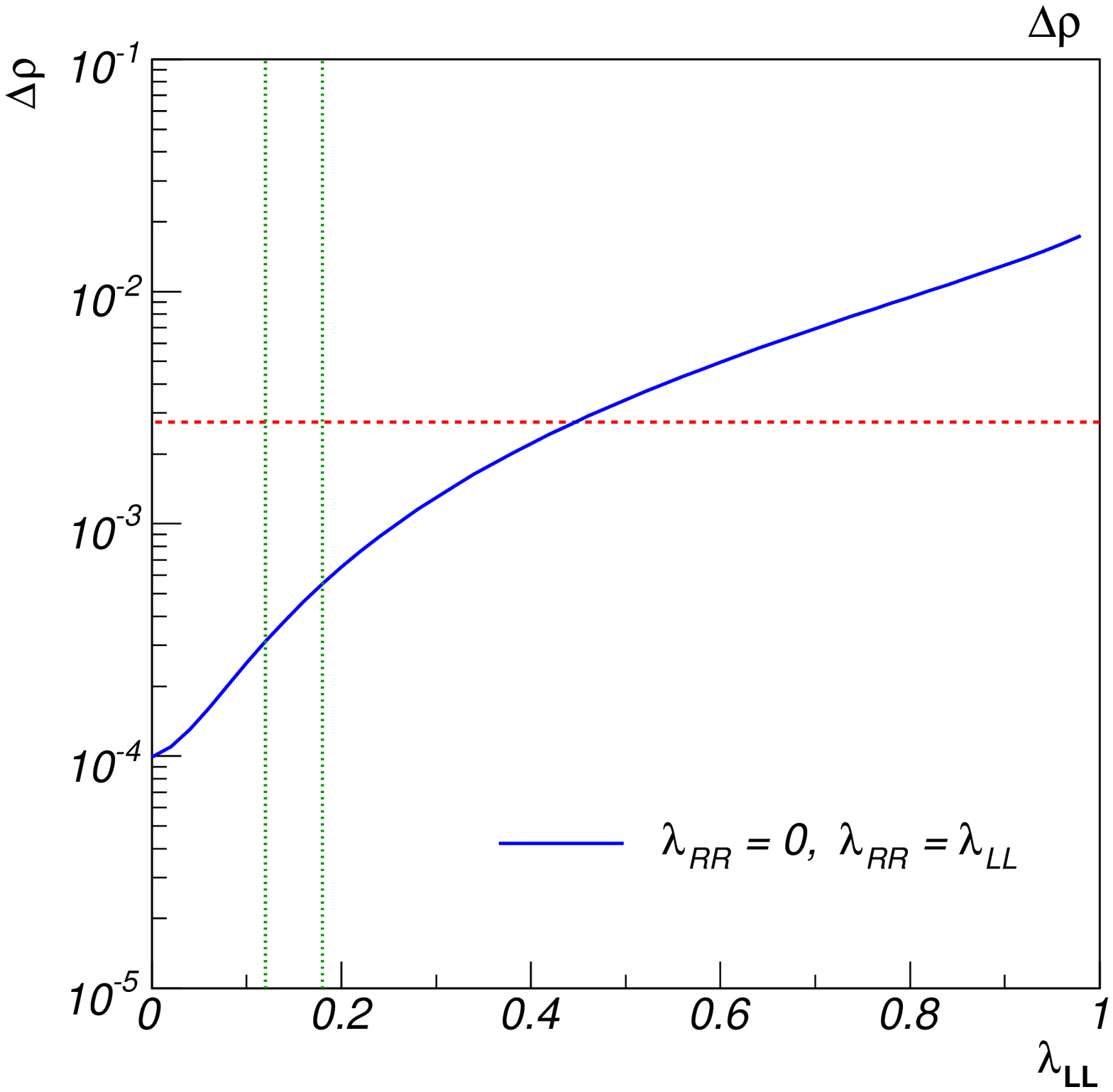}
	\includegraphics[scale=0.27]{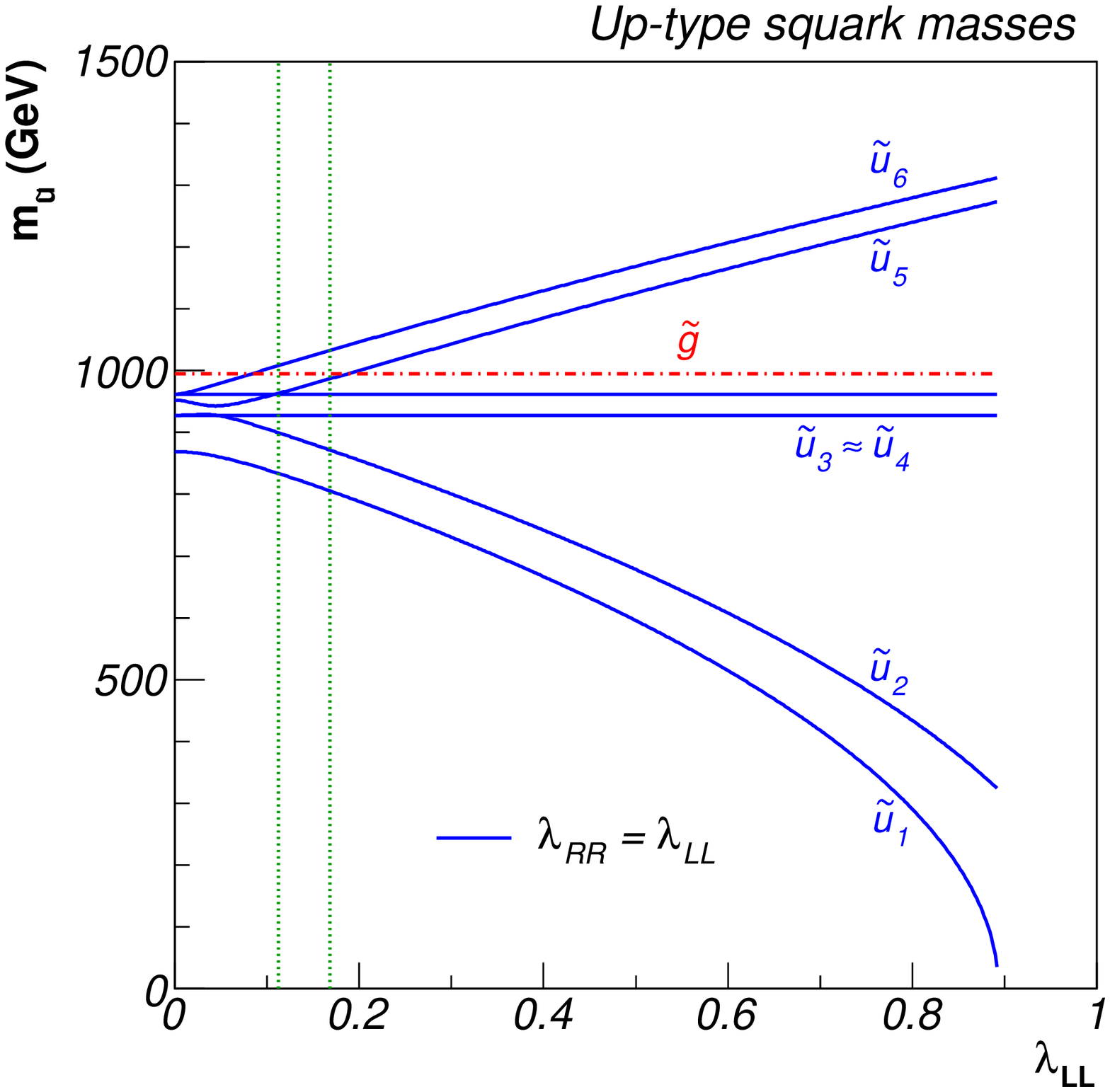}
	\includegraphics[scale=0.27]{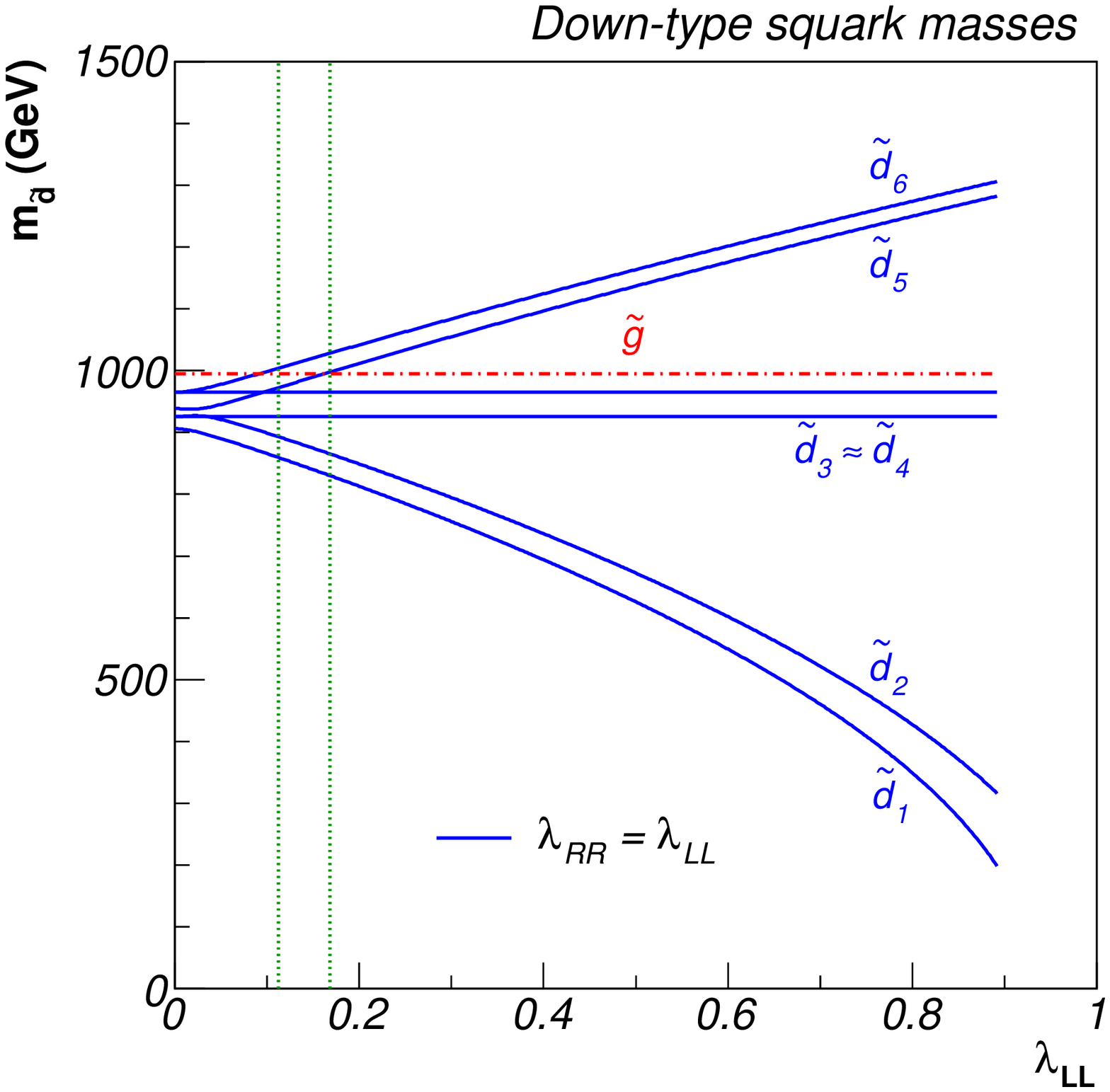}
\caption{Same as Fig.\ \ref{fig8} for our benchmark scenario H.}
\label{fig11}
\end{center}\end{figure}

\begin{figure}\begin{center}
	\includegraphics[scale=0.27]{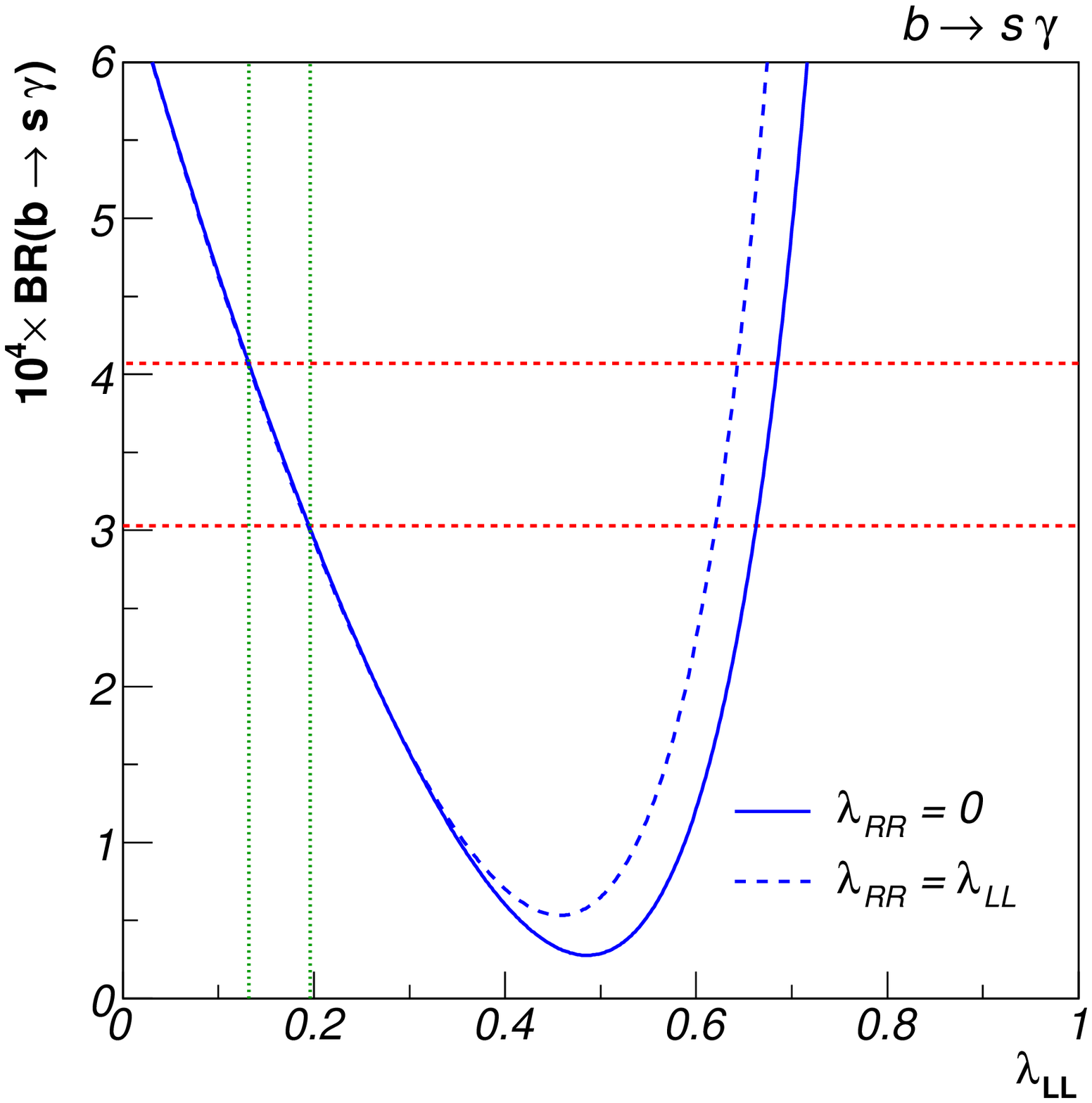}
	\includegraphics[scale=0.27]{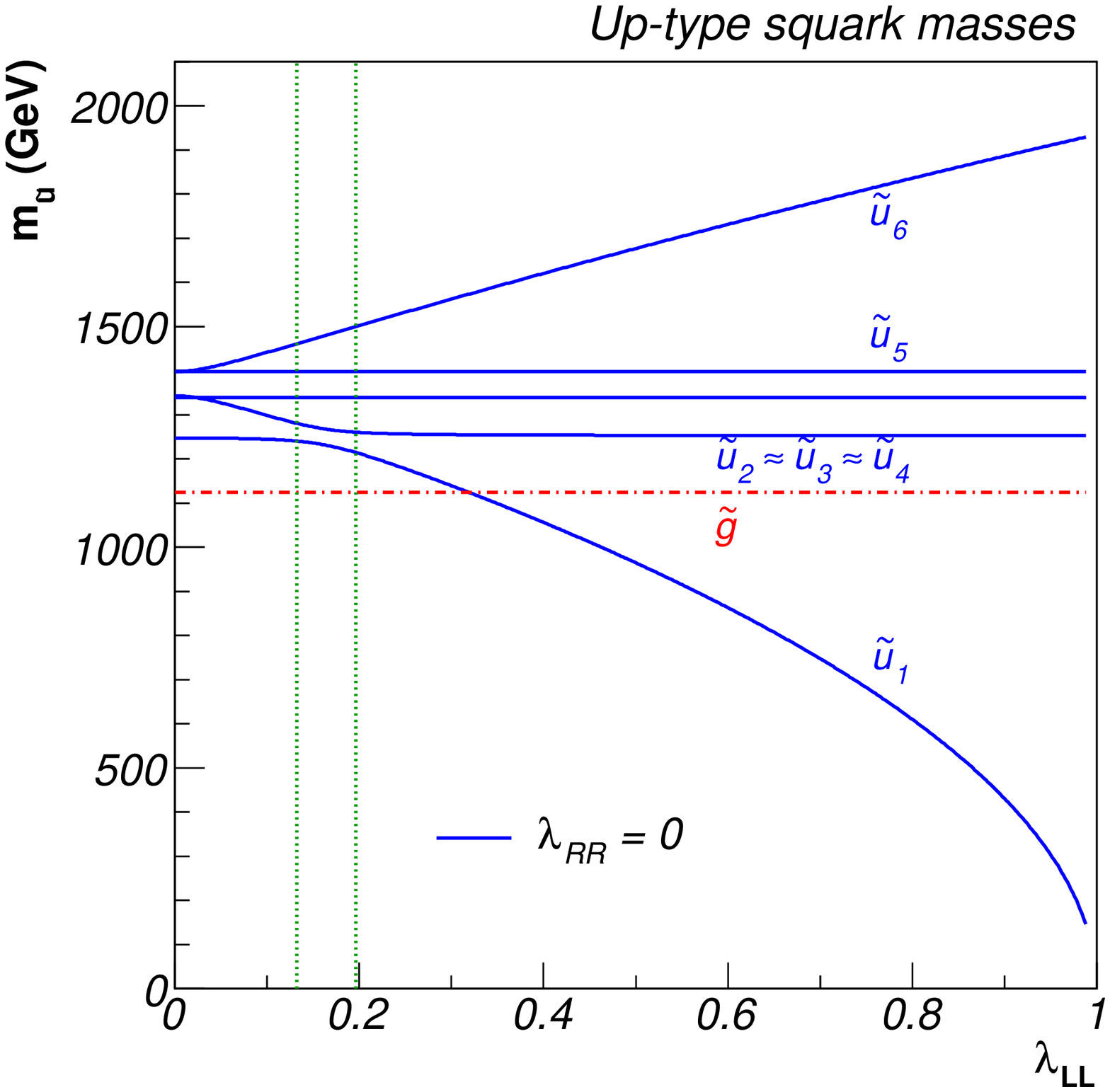}
	\includegraphics[scale=0.27]{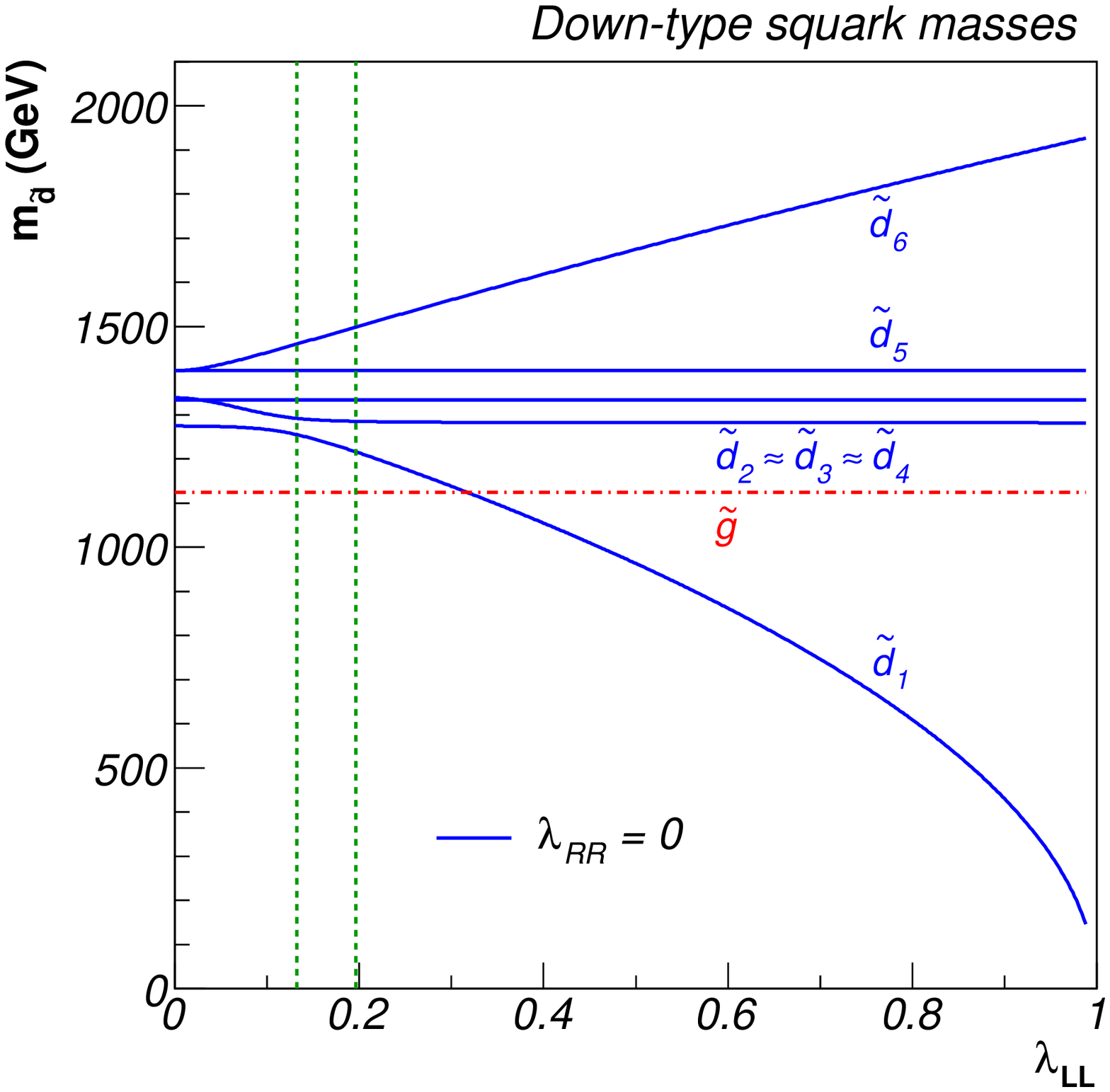}
	\includegraphics[scale=0.27]{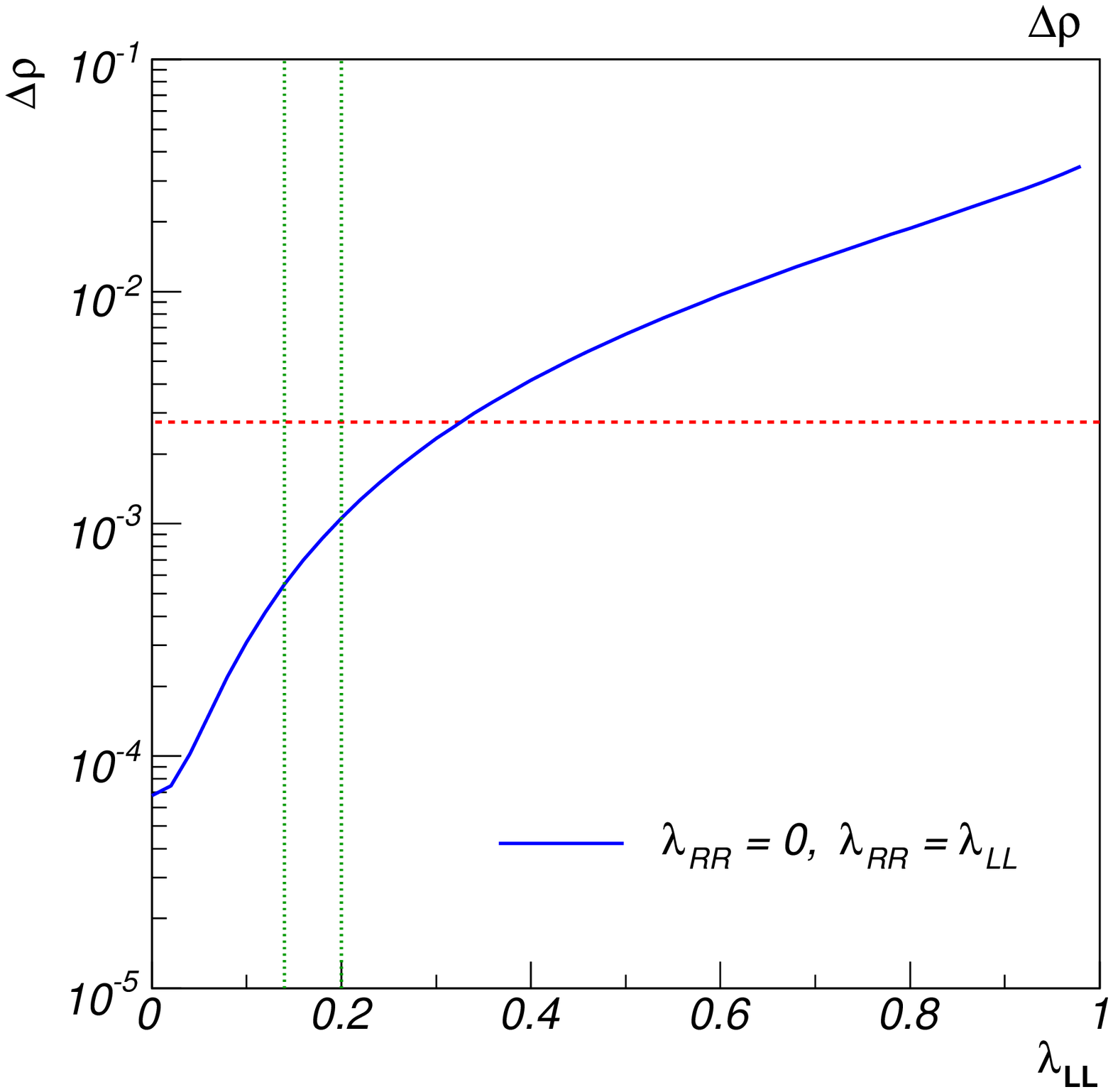}
	\includegraphics[scale=0.27]{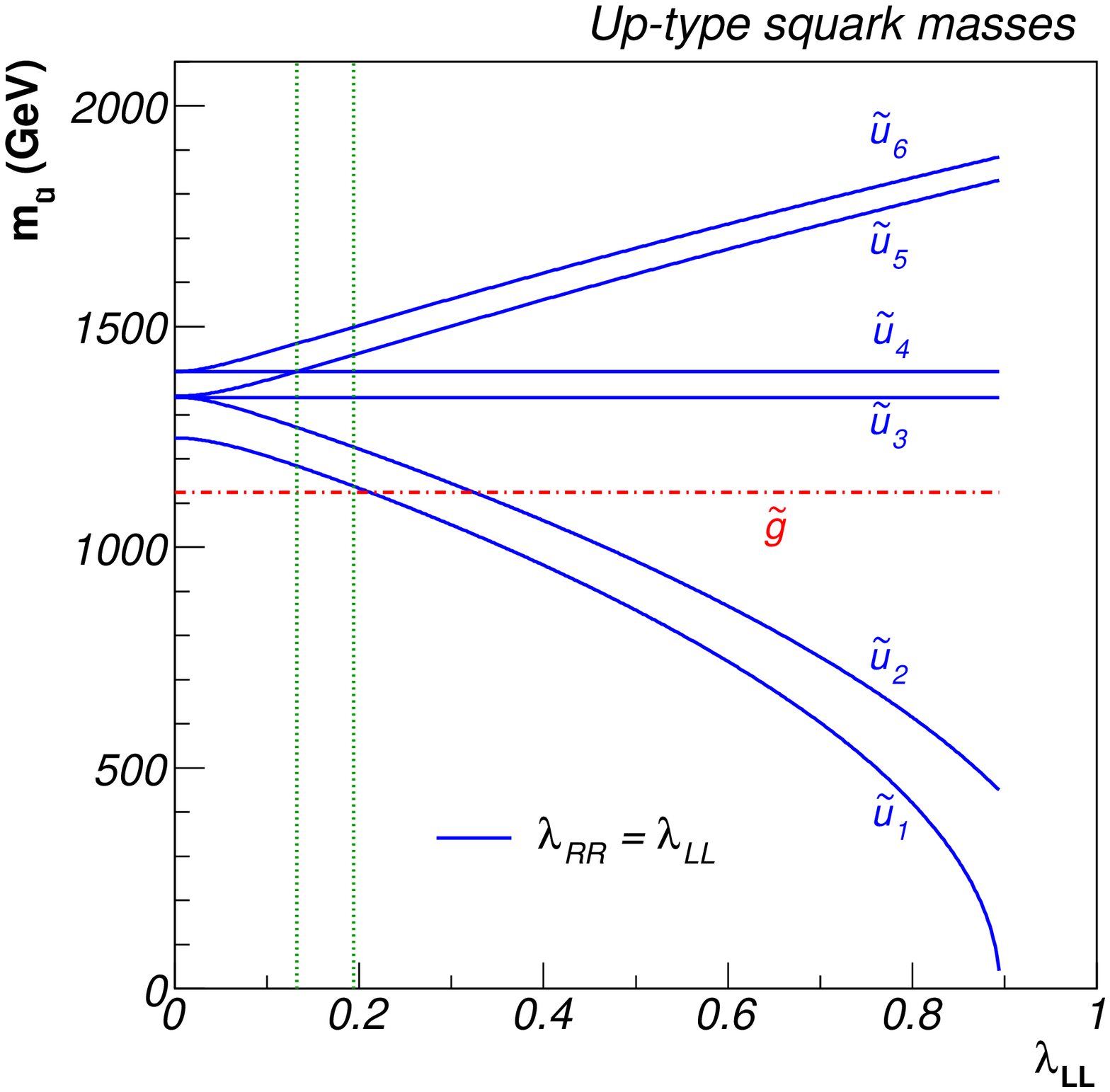}
	\includegraphics[scale=0.27]{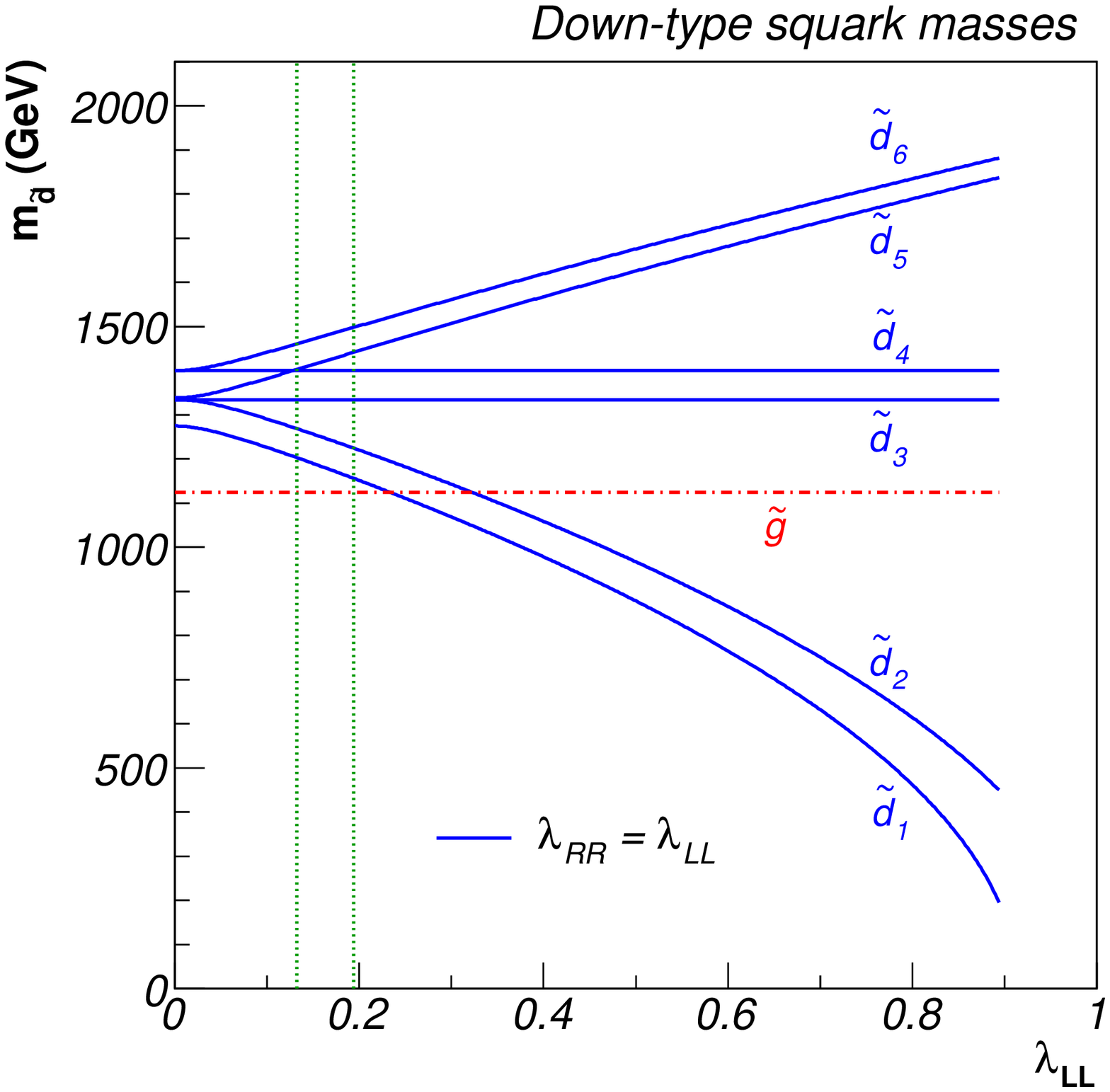}
\caption{Same as Fig.\ \ref{fig8} for our benchmark scenario I.}
\label{fig12}
\end{center}\end{figure}

\begin{figure}\begin{center}
	\includegraphics[scale=0.27]{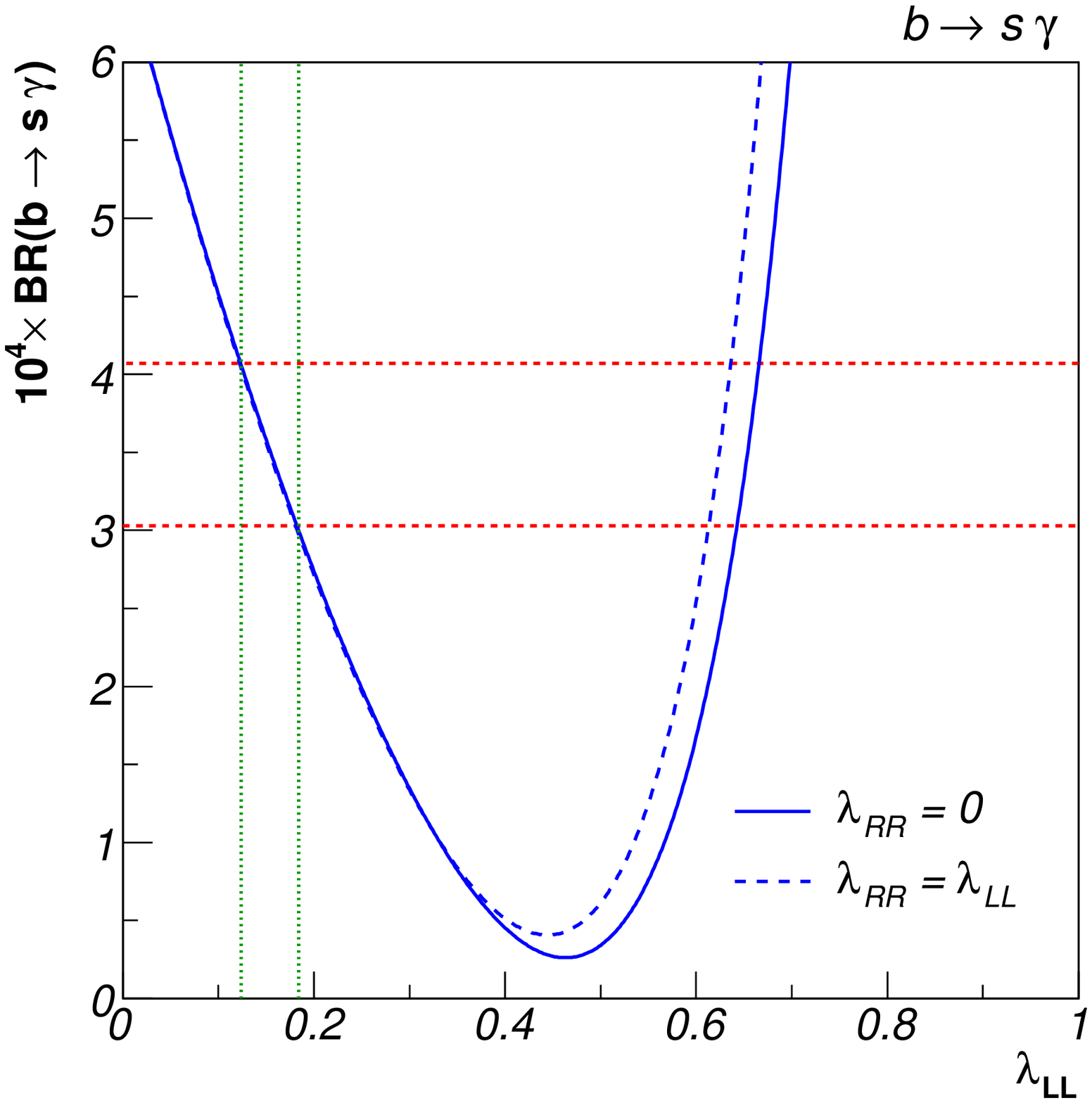}
	\includegraphics[scale=0.27]{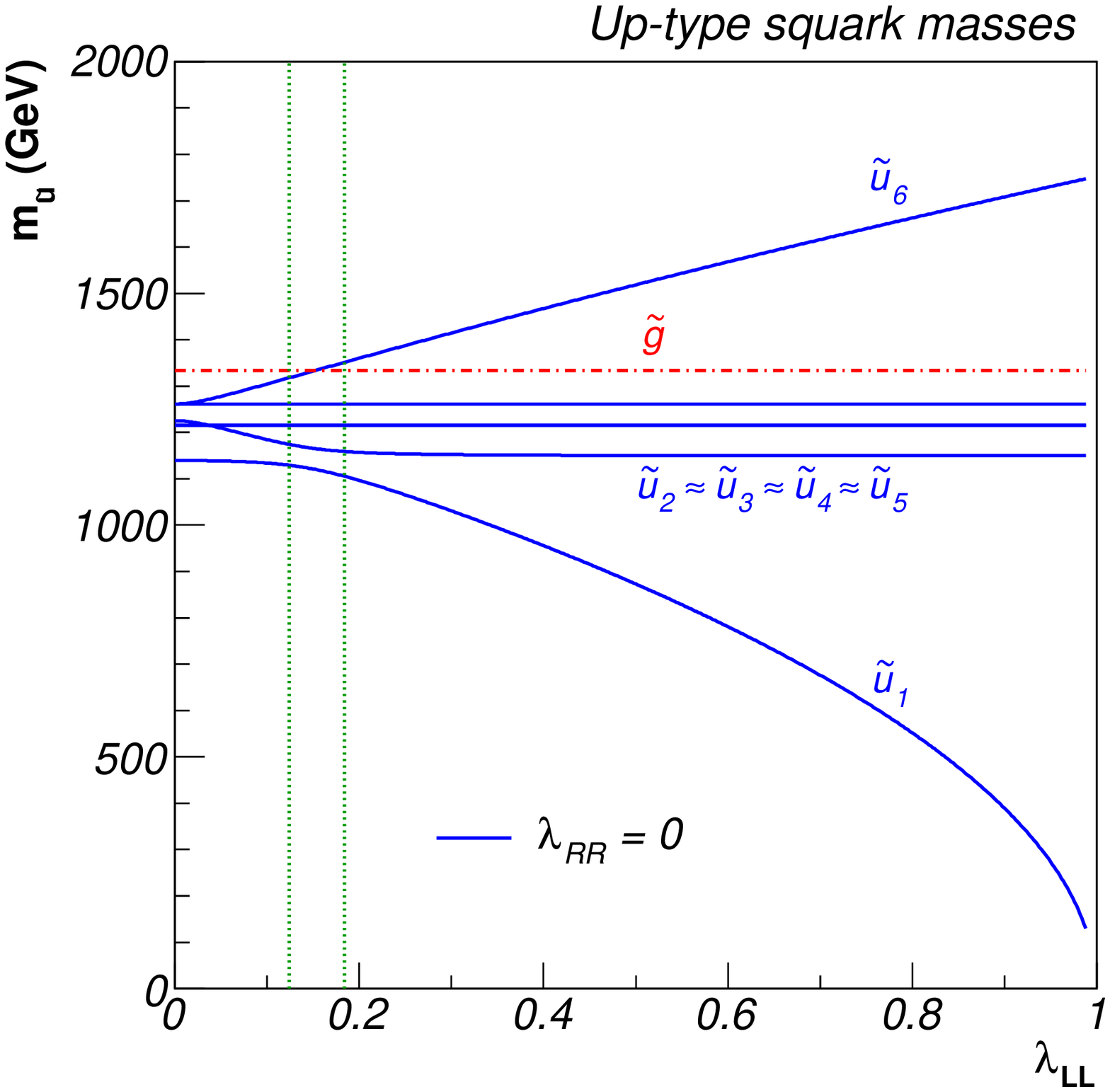}
	\includegraphics[scale=0.27]{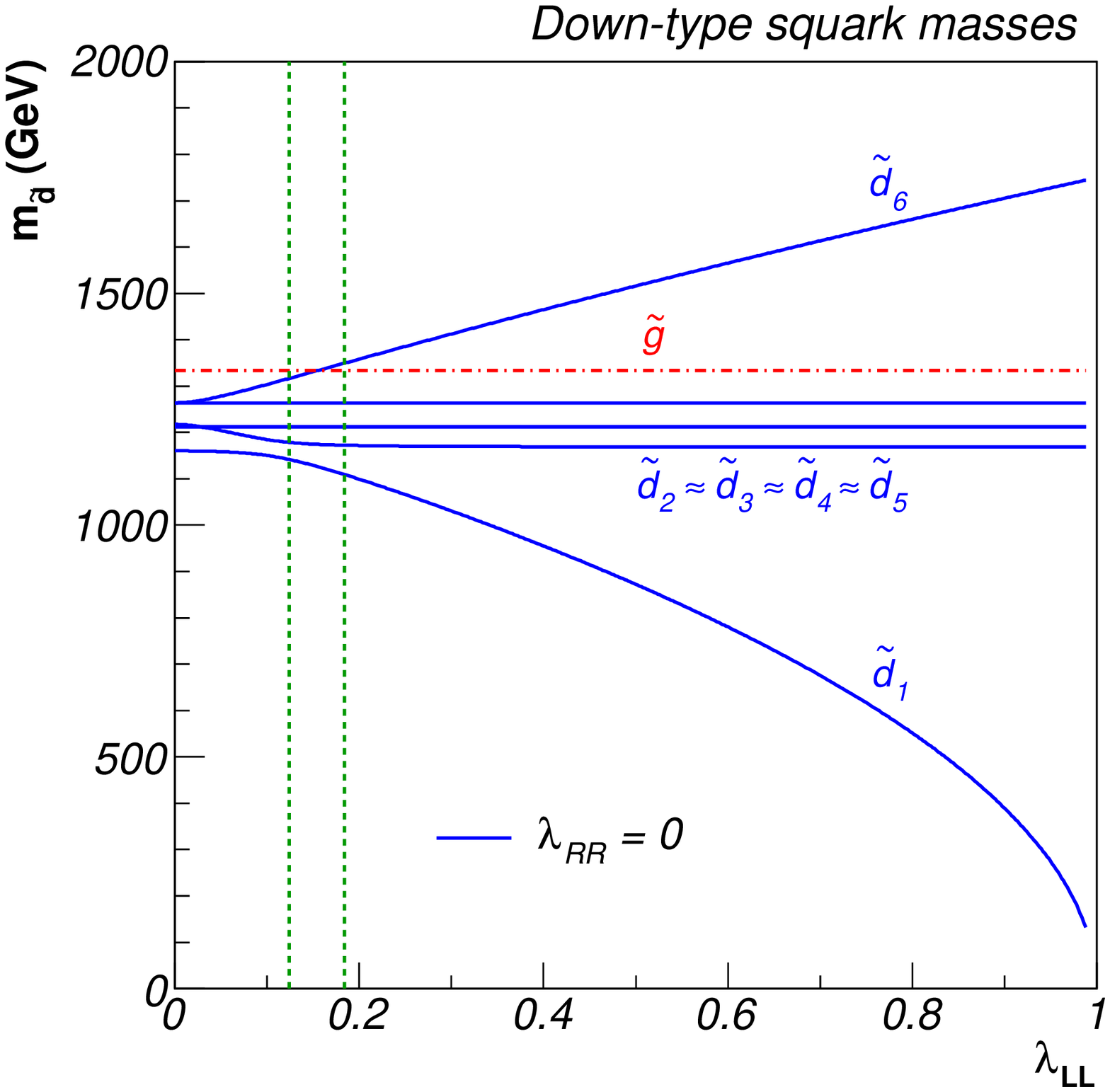}
	\includegraphics[scale=0.27]{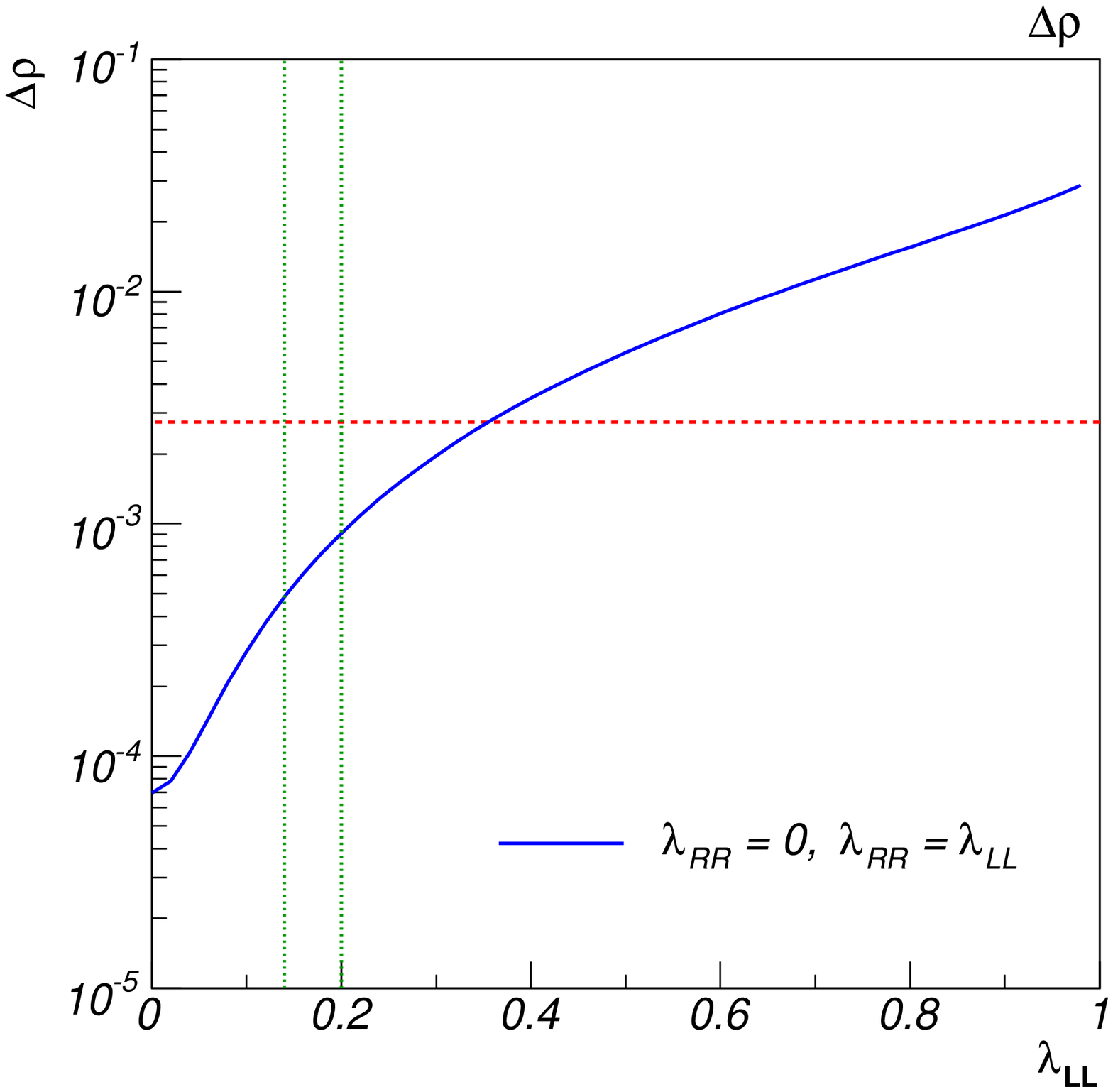}
	\includegraphics[scale=0.27]{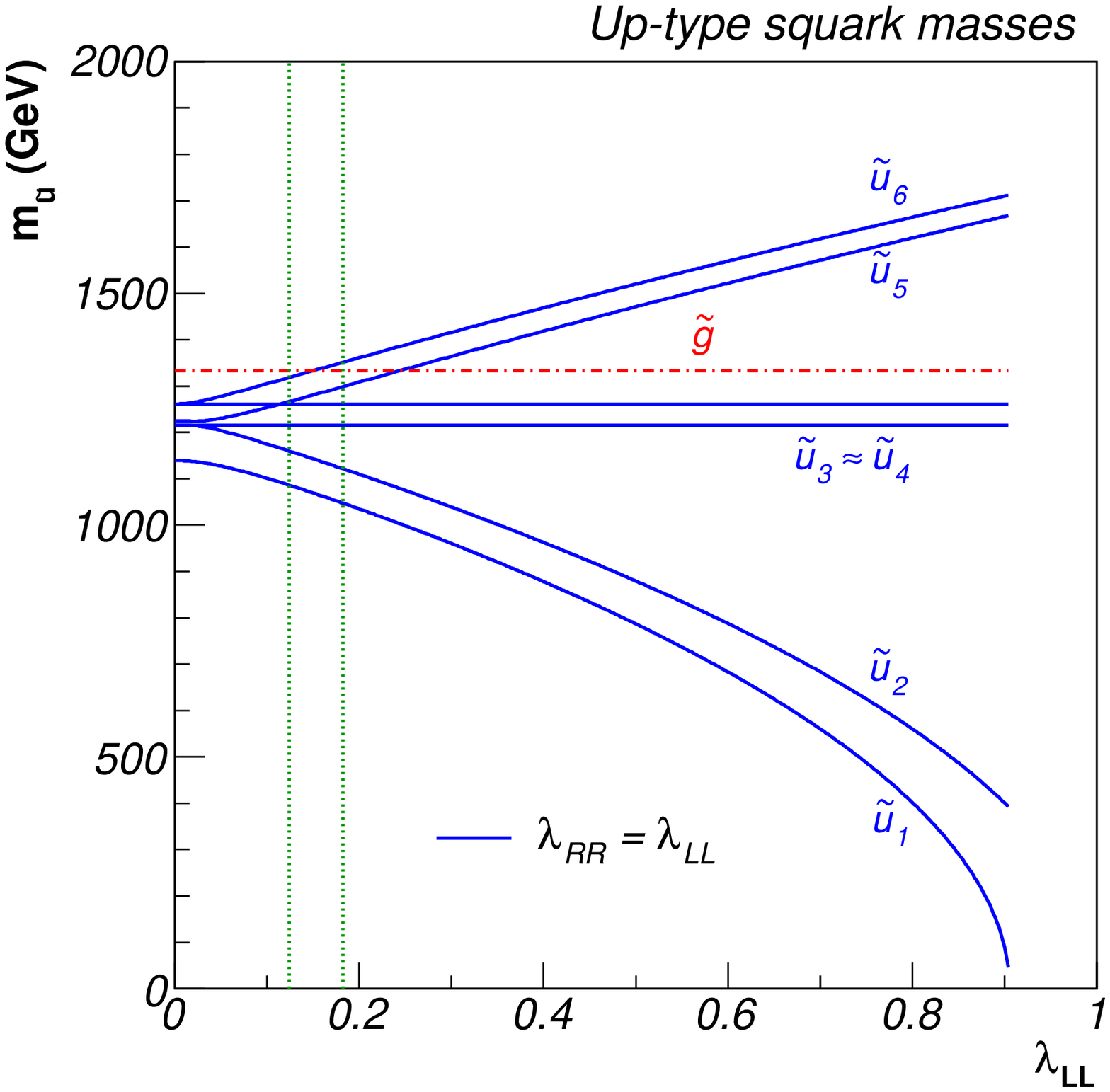}
	\includegraphics[scale=0.27]{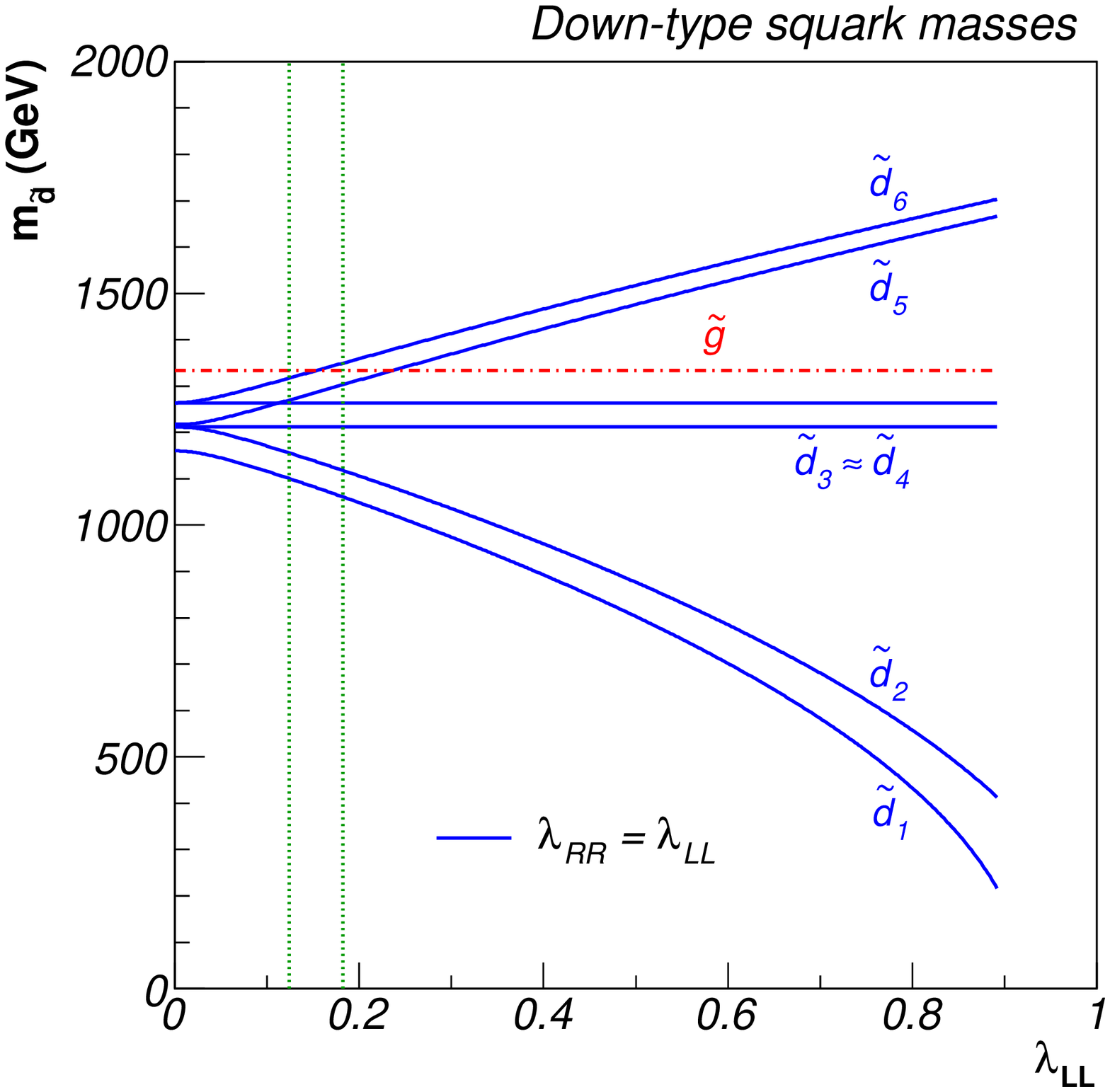}
\caption{Same as Fig.\ \ref{fig8} for our benchmark scenario J.}
\label{fig13}
\end{center}\end{figure}
We now study in detail the dependence of the electroweak precision and
low-energy observables as well as the mass spectra of the points E, F, G, H, I,
and J. In Figs.\ \ref{fig8} -- \ref{fig13}, we show the corresponding
branching ratio BR($b\to s\gamma$) (top left) and the observable $\Delta\rho$
(bottom left) as a function of the NMFV-parameter $\lambda_{\rm LL}$. We include
both flavour violation scenarios, the one for fundamental messengers with
$\lambda_{\rm RR}=0$ as well as the one with antisymmetric messengers, where
$\lambda_{\rm RR}=\lambda_{\rm LL}$. As already mentioned, the leptonic
observable $a_{\mu}$ depends only very weakly (at the two-loop level only) on
the squarks. As a consequence, we find values of $a_{\mu}$ independent of
$\lambda_{\rm LL}$ for our six benchmark scenarios, which are $a_{\mu}^{\rm
SUSY} = 37.7 \cdot 10^{-10}$, $41.3 \cdot 10^{-10}$, $31.4 \cdot 
10^{-10}$, $36.6 \cdot 10^{-10}$, $31.8 \cdot 10^{-10}$, and $34.2 \cdot
10^{-10}$ for the points E, F, G, H, I, and J, respectively. These values lie
well within 2$\sigma$ of the experimentally  favoured range of Eq.\
(\ref{eq.amu}), and even within 1$\sigma$ for the points E, G, H, I, and J.
For the inclusive branching ratio BR($b\to s\gamma$), the experimentally
allowed range within 2$\sigma$ is indicated by two horizontal dashed lines. The
good agreement between the measurements and the two-loop SM prediction in
combination with the strong dependence of the SUSY contribution on squark
flavour mixing only leave two allowed narrow intervals for our flavour violation
parameter, one being at relatively low values of $\lambda_{\rm LL} \sim 0.15$, the
second one at higher values of $\lambda_{\rm LL} \sim 0.5 - 0.7$. It is well
known that the latter is disfavoured by $b\to s\mu^+\mu^-$ data
\cite{Gambino:2004mv}. The remaining one is indicated by vertical dotted lines.
Note that the difference between the two scenarios is small for the relevant
values of $\lambda_{\rm LL} \lesssim 0.2$.
Concerning the observable $\Delta\rho$, the difference between the two
considered flavour mixing scenarios is not visible, so that only one curve
is shown. Again, the horizontal line indicates the favoured range within
2$\sigma$, where only the upper limit is visible on our logarithmic scale. In
contrast to BR($b\to s\gamma$), here the relatively large experimental errors
allow for values of $\lambda_{\rm LL} \lesssim 0.3 - 0.6$, depending on the
benchmark point. The vertical dashed lines indicate the allowed range for $\lambda_{\rm
LL}$ with respect to the more stringent constraint coming from $b\to s\gamma$.

The difference between the two flavour violation scenarios becomes more obvious
when we study the squark mass eigenvalues. The up- and down-type squark masses
are shown as a function of the flavour mixing parameter $\lambda_{\rm LL}$ in the
centre and right upper panels of Figs.\ \ref{fig8} -- \ref{fig13} for
mixing with fundamental messengers ($\lambda_{\rm RR}=0$) and in the centre and
right lower panels for antisymmetric messengers ($\lambda_{\rm RR} =
\lambda_{\rm LL}$). We observe here the same level-reordering phenomenon between
neighbouring states as already in the case of minimal supergravity, discussed
in Ref.\ \cite{Bozzi:2007me}. With increasing flavour violation the mass splitting
between the lightest and heaviest mass eigenstates becomes larger, while the
intermediate squark masses are practically unchanged. At the points, where two
levels should cross, we observe so-called ``avoided crossings'' of the mass
eigenvalues. This phenomenon of level-reordering is due to the fact that the
mass matrices depend on on single real parameter $\lambda_{\rm LL}$.
Unfortunately, many ``avoided crossings'' lie below, but some also within
the allowed ranges of the flavour-violating
parameter $\lambda_{\rm LL}$, indicated by green vertical lines for each of
the six benchmark scenarios. The level-reordering phenomenon is of similar
importance for up-type and down-type squarks. Concerning the difference between
our two implementations of flavour violation in the squark matrices, we observe
an important splitting for only the lightest and heaviest eigenstates in the
case of flavour mixing only in the left-left chiral sector. In contrast, for
flavour violation in both the left-left and right-right chiral squark sectors,
the two lightest and two heaviest mass eigenvalues give rise to an important
splitting, while only the remaining two masses are practically independent of
$\lambda_{\rm LL}$. This is a direct consequence of the fact that we have
introduced additional flavour mixing in two distinct sectors of the squark mass
matrices, and this will influence the squark and gaugino production cross sections
presented in Sec.\ \ref{sec5}. Note also that in the case of flavour mixing only
in the left-left chiral sector, ``avoided'' level crossings occur among
the $\tilde{q}_{1,2}$, $\tilde{q}_{3,4}$, and $\tilde{q}_{5,6}$ mass eigenstates,
whereas in the case of flavour mixing in both the left-left and right-right chiral
squark sectors, we rather observe the mass flips among the $\tilde{q}_{2,3}$ and
$\tilde{q}_{4,5}$ mass eigenstates, respectively.

\section{Cosmological Constraints \label{sec4}}

To be cosmologically viable, a supersymmetric model should include a convincing
candidate for the cold dark matter (CDM) in our Universe. This particle has to
be stable, electrically neutral, and a colour singlet \cite{Goldberg:1983nd,
Ellis:1983ew}. Furthermore its relic density has to lie within the range
\beq
    0.094 \le \Omega_{\rm CDM}h^2 \le 0.136 
\label{eq.wmap}
\eeq
at 95\% (2$\sigma$) confidence level. Here, $h$ denotes the present Hubble
expansion rate in units of 100 km\! s$^{-1}$\! Mpc$^{-1}$. This limit has
been obtained from the three-year data of the WMAP satellite, combined
with recent SDSS and SNLS survey and Baryon Acoustic Oscillation data, and
interpreted within an eleven-parameter inflationary model \cite{Hamann:2006pf},
which is more general than the usual six-parameter ``vanilla'' concordance model
of cosmology. Note that this range is well compatible with the older,
independently obtained range of $0.094 \le \Omega_{\rm CDM}h^2 \le 0.129$ of
Ref.\ \cite{Ellis:2003cw}.

A natural candidate in SUSY models is the lightest supersymmetric particle
(LSP), which is the gravitino in GMSB models. Depending on its mass, the
gravitino can account either for cold ($\mG \gtrsim 100$ keV), for warm ($1\
{\rm keV} \lesssim \mG \lesssim 100$ keV), or for hot ($\mG \lesssim 1$ keV)
dark matter. 
Today's gravitino abundance in the Universe has two contributions. First,
gravitinos are produced by thermal scattering in the very early Universe. The
corresponding energy density \cite{Bolz:2000fu,Pradler:2006qh,Rychkov:2007uq} 
\beq
  \Omega_{\tilde{G}}^{\rm th}h^2 ~\simeq~ 0.27 \left( \frac{T_{\rm R}}{10^{10}{\
  \text{GeV}}} \right) \left( \frac{100{\ \text{GeV}}}{\mG} \right) \biggr(
  \frac{m_{\tilde{g}}}{1\ {\text{TeV}}} \biggr)^2
\label{eq.thgrav}
\eeq
involves the gluino mass $m_{\tilde{g}}$ at low energy and the reheating
temperature $T_{\rm R}$. The latter is the temperature of the Universe after
inflation, for which at present no stringent constraints exist. Values of
$T_{\rm R} \gtrsim 10^9$ GeV are preferred in scenarios that feature leptogenesis
in order to explain the cosmic baryon asymmetry \cite{Buchmuller:2004nz}. As the
resummation method leading to Eq.\ (15) may become unreliable below $T_{\rm R}
\simeq10^7$ GeV, we use the more accurate result given in \cite{Pradler:2006hh}
and do not make practical use of the low-temperature region (see below). Note that
the thermal gravitino relic density may also be affected by late-time entropy
production coming, e.g., from the decay of the messengers.

\begin{figure}\begin{center}
	\includegraphics[scale=0.29]{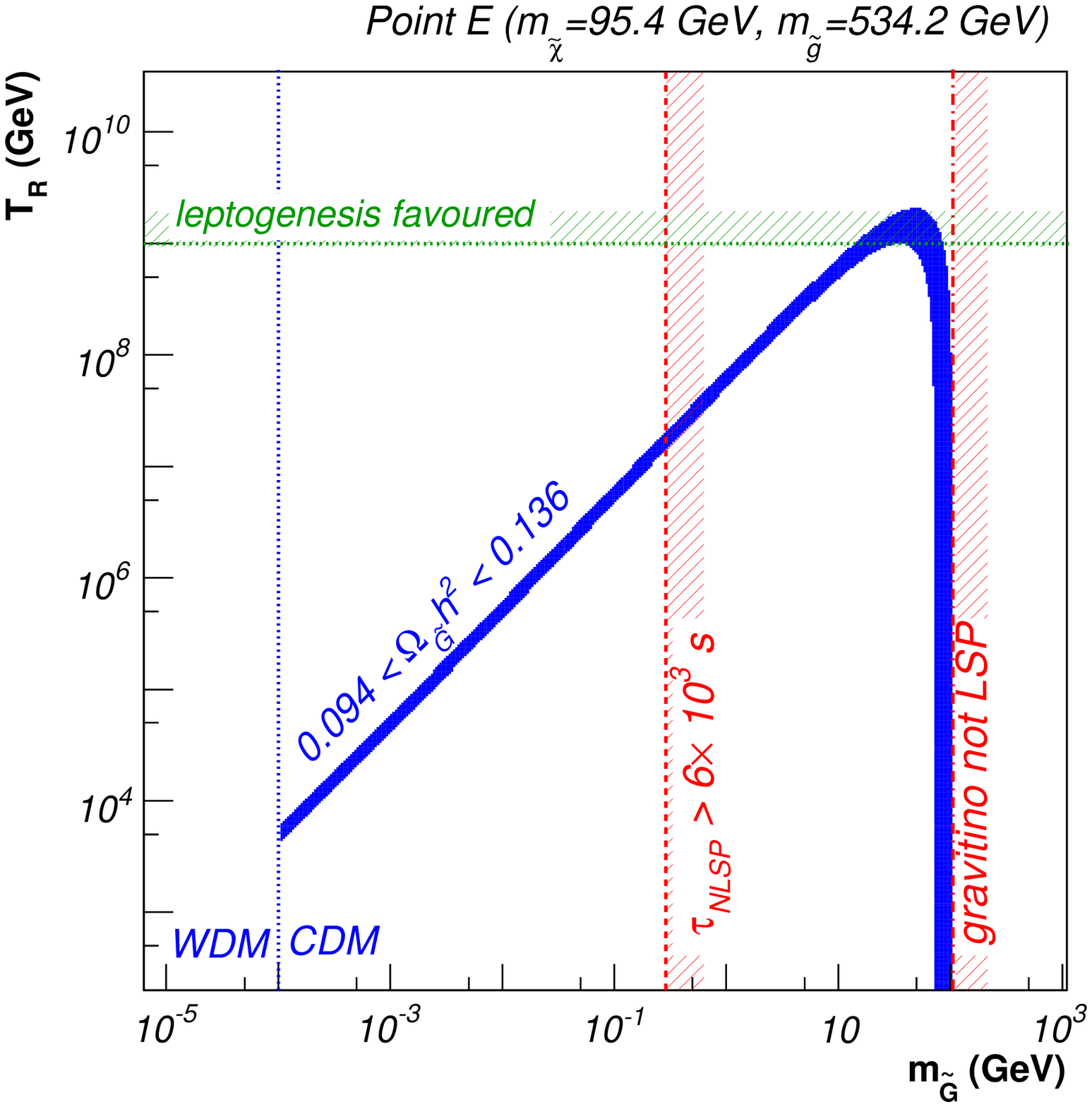}
	\includegraphics[scale=0.29]{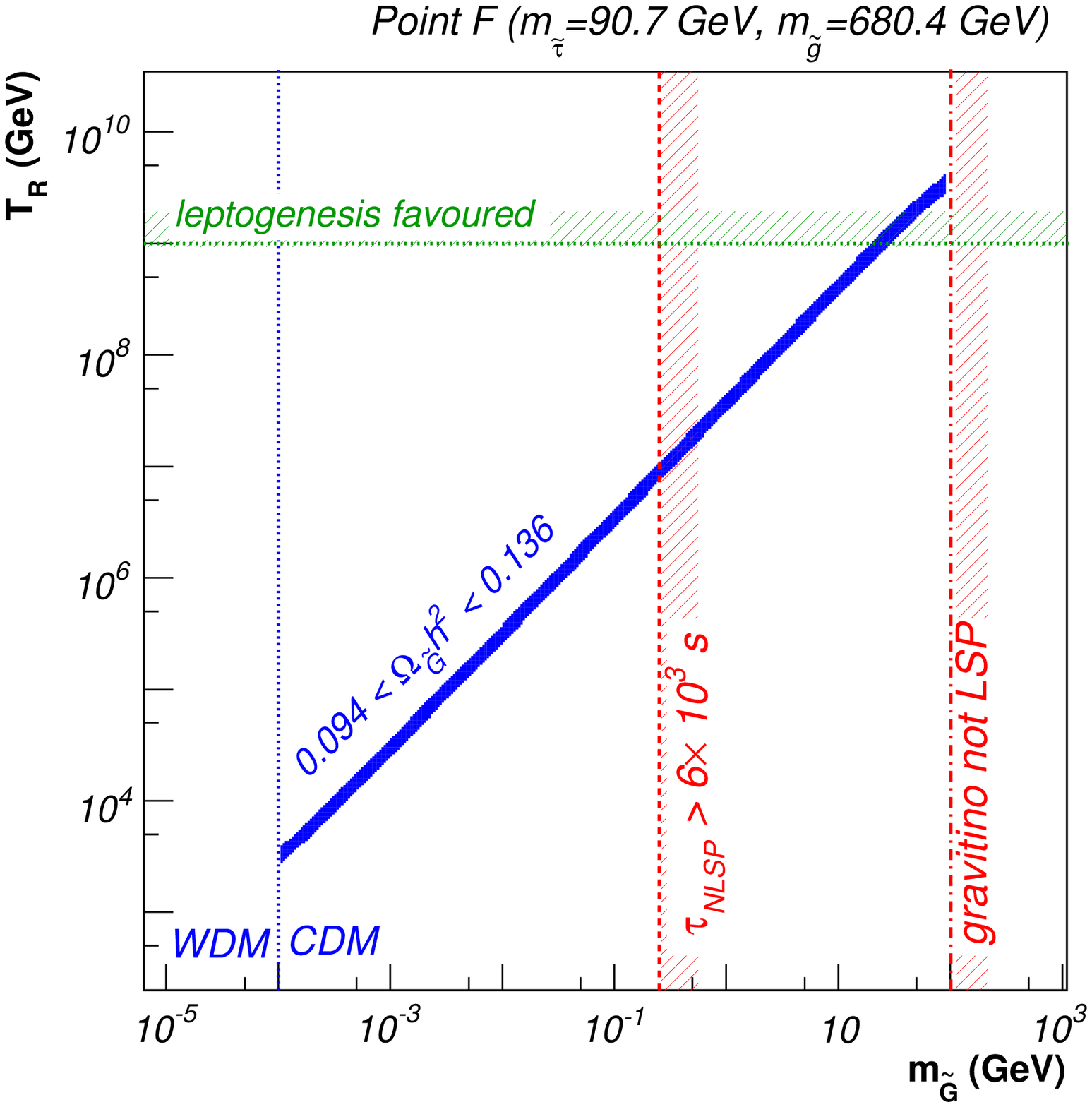}
	\includegraphics[scale=0.29]{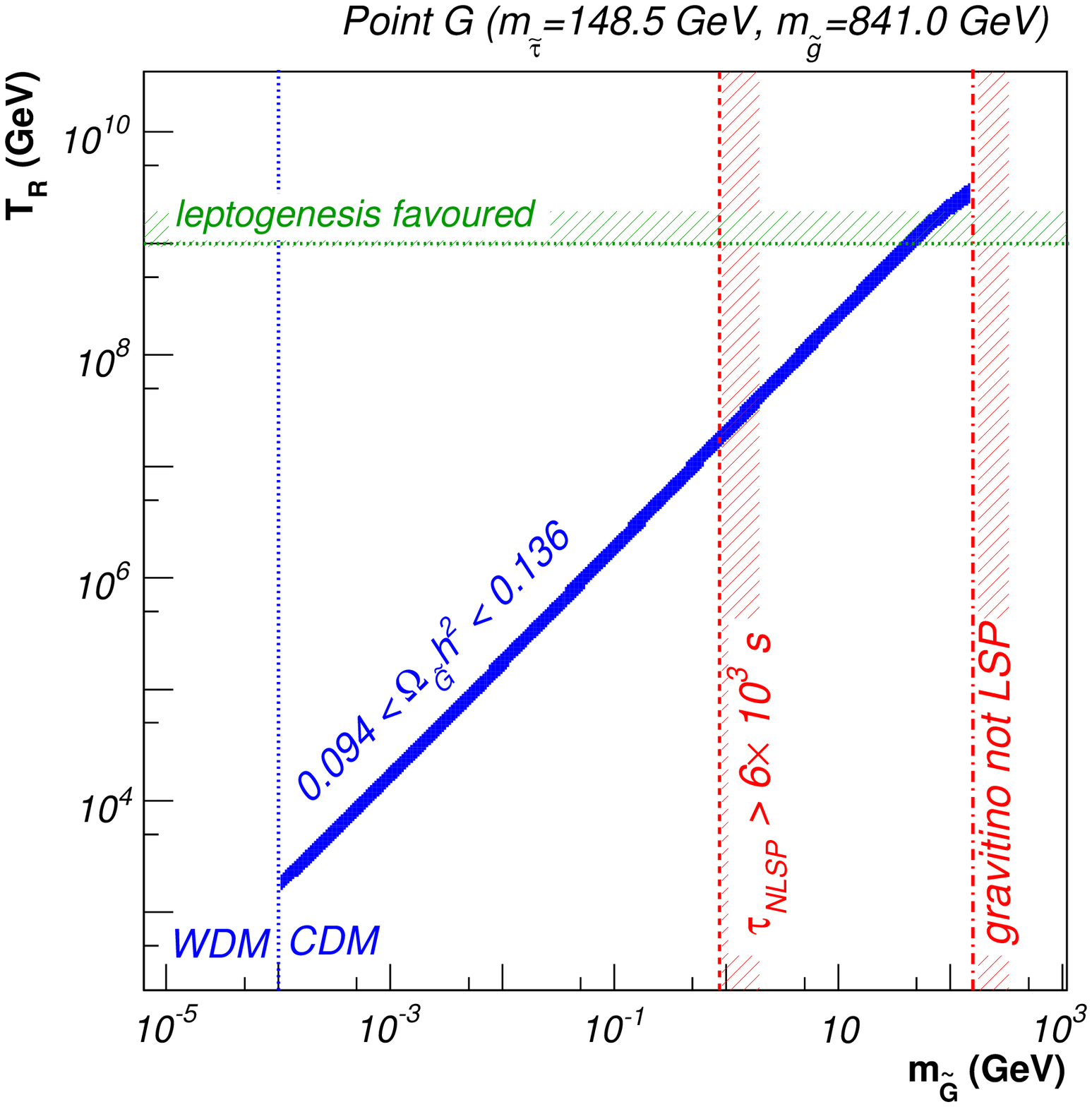}
	\includegraphics[scale=0.29]{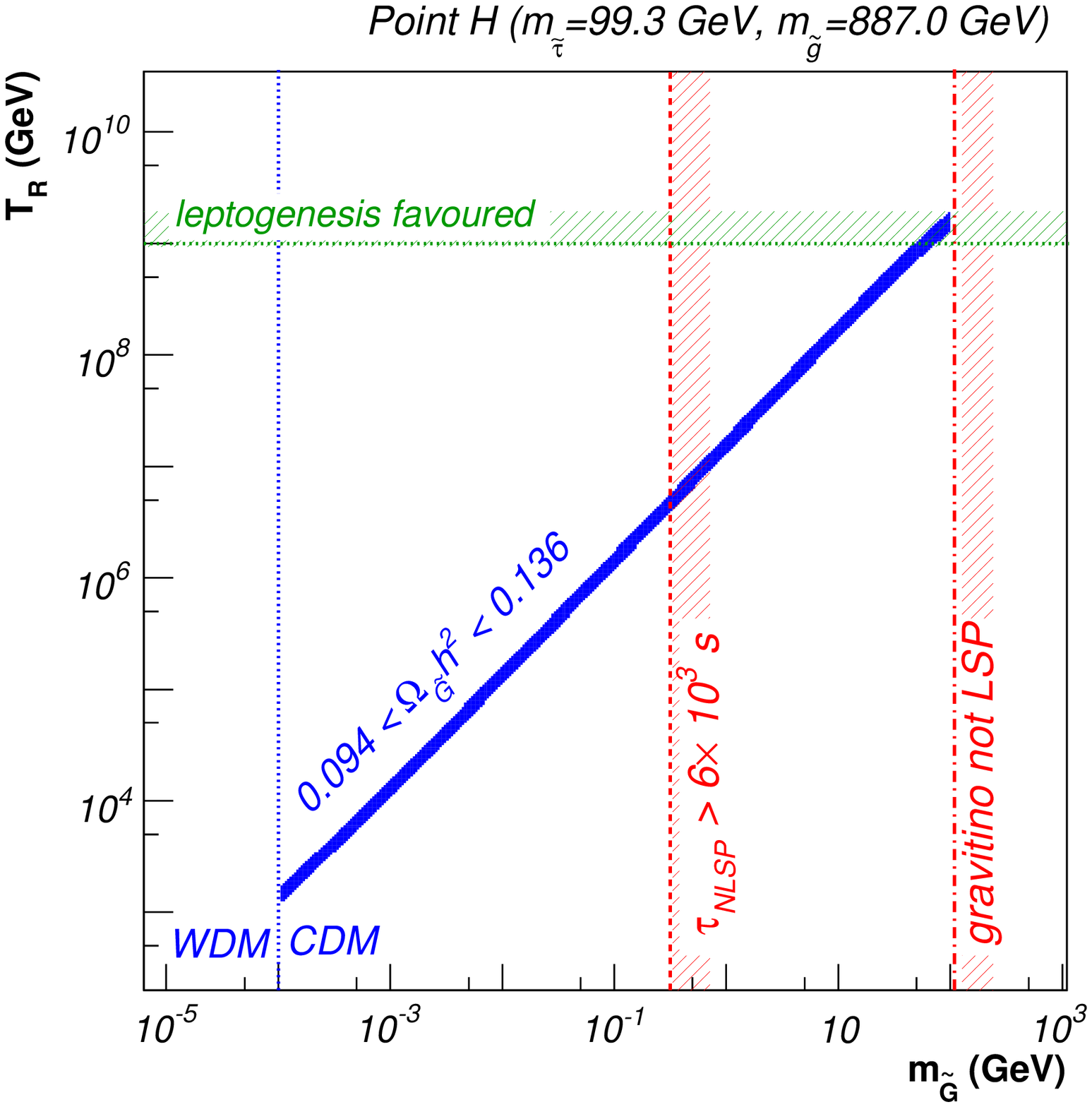}
	\includegraphics[scale=0.29]{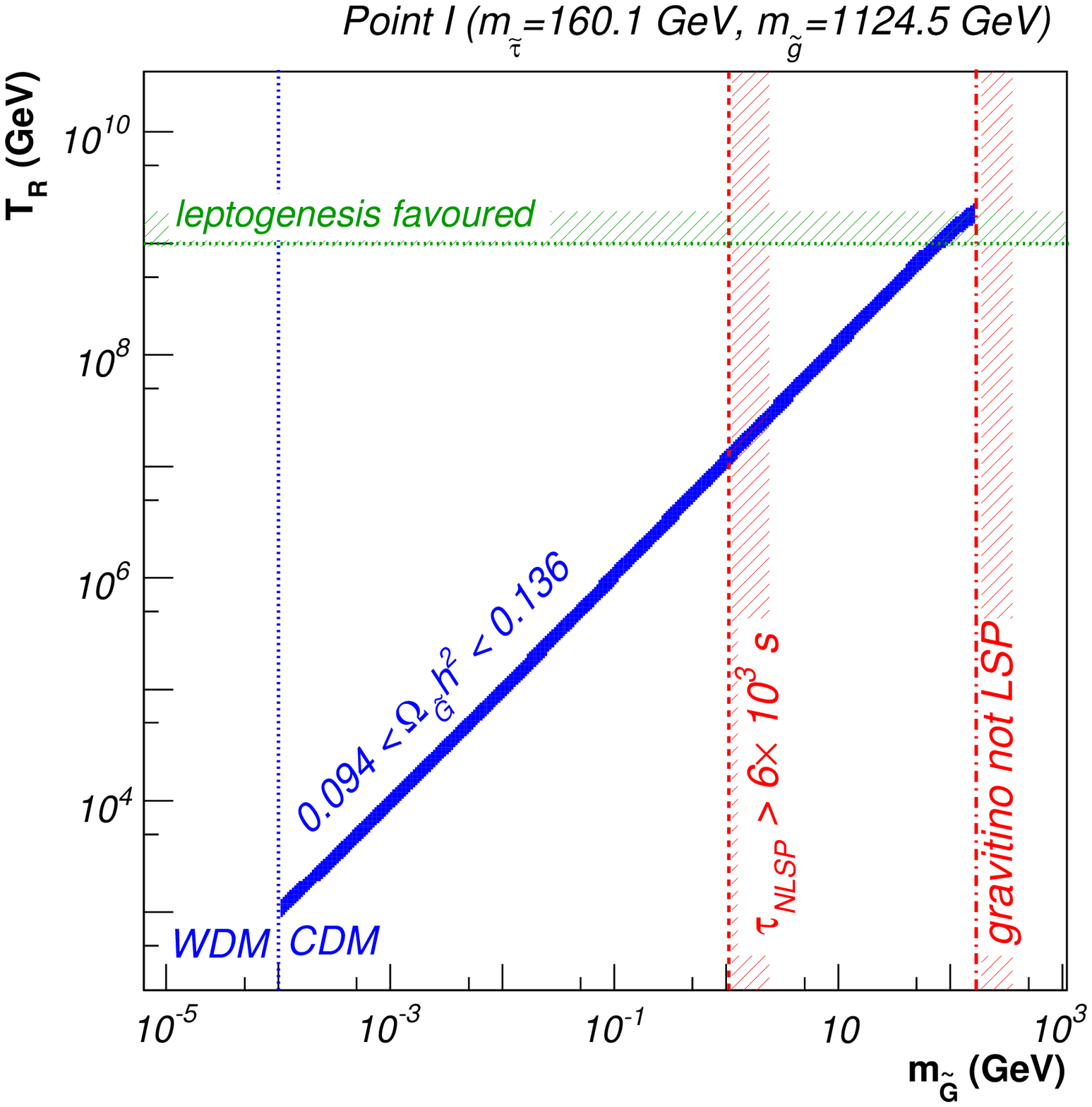}
	\includegraphics[scale=0.29]{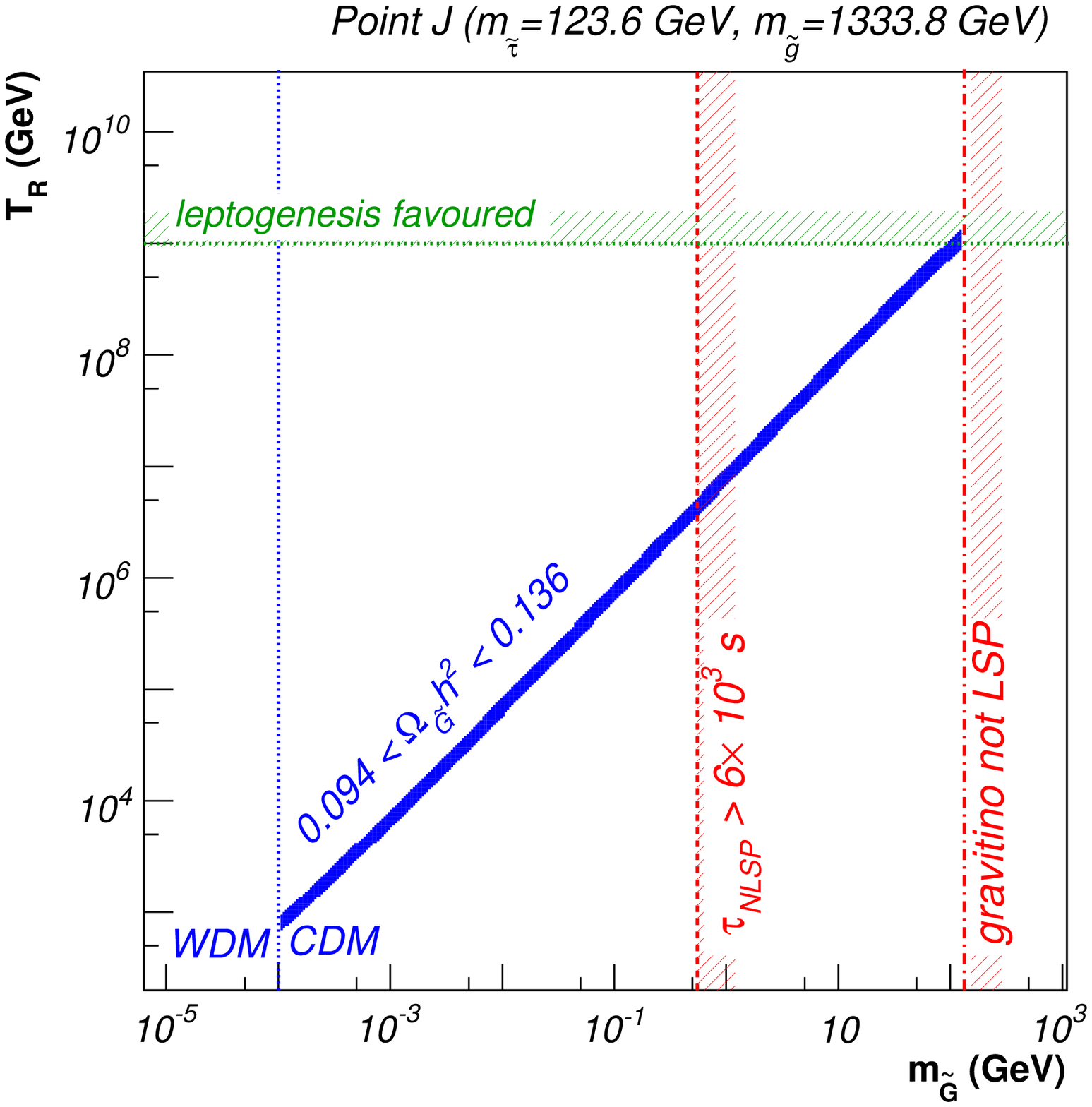}
\caption{The cosmologically favoured regions in the $\mG$--$T_{\rm R}$ plane for
our benchmark scenarios E, F, G, H, I, and J with respect to WMAP data
(dark/blue),
the NLSP lifetime (vertical red line), and leptogenesis (horizontal green line).
Also indicated are the limits between gravitino warm (WDM) or cold (CDM) dark
matter and the regions where the gravitino would not be the LSP.}
\label{fig14}
\end{center}\end{figure}

Second, there is non-thermal production through decay of the next-to-lightest
supersymmetric particle (NLSP) into the gravitino. As each NLSP will decay into
its Standard Model partner and one gravitino, the resulting gravitino energy
density can be obtained through
\beq
  \Omega_{\tilde{G}}^{\rm non-th}h^2 ~=~ 
  \frac{\mG}{m_{\rm NLSP}}~ \Omega^{\rm th}_{\rm NLSP}h^2 ,  
\eeq
where $\Omega^{\rm th}_{\rm NLSP}h^2$ is the thermal freeze-out relic density
the NLSP would have if it did not decay. Note that for low values of $\mG$
and/or high reheating temperatures $T_{\rm R}$ thermal production dominates,
whereas for high values of $\mG$ it is negligible with respect to the
contribution from NLSP decay. The NLSP would-be relic density $\Omega_{\rm
NLSP}^{\rm th}h^2$ is calculated by solving the Boltzmann equation and can be
evaluated numerically for any type of NLSP using the public code {\tt
micrOMEGAs} \cite{Belanger:2001fz}, whereas the programme {\tt DarkSUSY}
\cite{Gondolo:2004sc} is only adapted to the neutralino case. We therefore
always use {\tt micrOMEGAs} in this work.
 
Another constraint for scenarios with gravitino dark matter arises from the fact
that the NLSP spoils the abundances of light elements in our Universe, if it
does not decay rapidly enough \cite{Pradler:2006qh}. The lifetime of a
supersymmetric particle decaying into its Standard Model partner and a gravitino
is given by the inverse of the corresponding decay rate. Neglecting here flavour
violation and any SUSY particle mixing, which have only little impact
\cite{Bozzi:2007me}, we obtain for the lifetime of the NLSP
\beq
  \tau_{\rm NLSP} ~\simeq~ \big( 6.1 \cdot 10^3 {\text s} \big) 
                    \biggr( \frac{1\ {\text{TeV}}}{m_{\rm NLSP}} \biggr)^5
                    \biggr( \frac{\mG}{100\ {\text{GeV}}} \biggr)^2 ,
\eeq
where we have inserted the value of the reduced Planck mass $M_{\rm P} = (8\pi
G_{\rm N})^{-1/2}$ and $G_{\rm N} = 6.7097\cdot 10^{-39}$ GeV$^{-2}$
\cite{Yao:2006px}.
In order to preserve the abundances of the light elements,
that are well explained by
primordial nucleosynthesis, the lifetime of the NLSP should be shorter than
$\tau_{\rm NLSP} \lesssim 6 \cdot 10^3$ seconds \cite{Pospelov:2008ta}. 
As a consequence, the latter constraint favours scenarios having a light
gravitino, which might enter in conflict with the thermal production favouring a
rather high reheating temperature and therefore a rather high gravitino mass, as
can be seen from Eq.\ (\ref{eq.thgrav}).

In the case of gravitino cold dark matter, we compute the gravitino energy
density $\Omega_{\tilde{G}}h^2$ in our Universe as described above, taking into
account the contributions from thermal production in 
the early Universe and from NLSP decay. In Fig.\ \ref{fig14}, we compare the
obtained gravitino relic density to the 2$\sigma$ range of the cold dark matter
relic density of Eq.\ (\ref{eq.wmap}) as a function of the gravitino mass $\mG$
and the reheating temperature $T_{\rm R}$ for our benchmark scenarios E to J.
For each of the six scenarios, we also indicate the upper limit on the gravitino
mass coming from the constraint on the NLSP lifetime, the limit between warm
(WDM) and cold dark matter (CDM), as well as the region where the gravitino
would become heavier than the NLSP. Concerning the reheating temperature, we
indicate the favoured region with respect to leptogenesis above $T_{\rm R} \sim
10^9$ GeV.

Note that the contribution to $\Omega_{\tilde{G}}h^2$ from NLSP decay is only
relevant for our point E with its neutralino NLSP and a rather important
neutralino energy density $\Omega_{\rm NLSP}^{\rm th}h^2 = 0.1275$. The fact
that this value lies already within the interval favoured by WMAP opens an
allowed band around $\mG \approx m_{\tilde{\chi}_1^0} = 95.4$ GeV, as can be
seen in the first panel of Fig.\ \ref{fig14}. For the other points, the
annihilation cross section of the charged stau NLSP is more important, so that
the resulting relic NLSP density is quite low ($\Omega_{\rm NLSP}^{\rm th}h^2
\sim 0.003 - 0.012$) and its values lie below the lower limit 0.094 of the WMAP
2$\sigma$ range.

From the graphs in Fig.\ \ref{fig14} it becomes clear that for the chosen
``collider-friendly'' benchmark points, we cannot fulfill all three 
cosmological constraints at the same time. For instance, if we want a scenario
featuring leptogenesis, i.e.\ having $T_{\rm R} > 10^9$ GeV, the lifetime of the
next-to-lightest SUSY particle (NLSP) would be too long for not spoiling the
light element abundances. We therefore relax the less stringent constraint,
which is the one coming from leptogenesis.

If we then impose the constraint due to the lifetime of the NLSP, our six
benchmark scenarios all lead to an upper limit on the gravitino mass of the
order of $\mG \lesssim 10^{-1} - 1$ GeV. 
For simplicity, we propose the same value $\mG = 10^{-1}$~GeV for all points. 
This respects the limit due to the NLSP lifetime, allows for gravitino cold dark
matter with relic gravitino density that agrees with current WMAP data, and this
in combination with relatively high values of $T_{\rm R} \sim 10^7$ GeV for the
reheating temperature.

\section{Supersymmetric Particle Production at the LHC \label{sec5}}

\begin{figure}
  \begin{center}
     \includegraphics[scale=1]{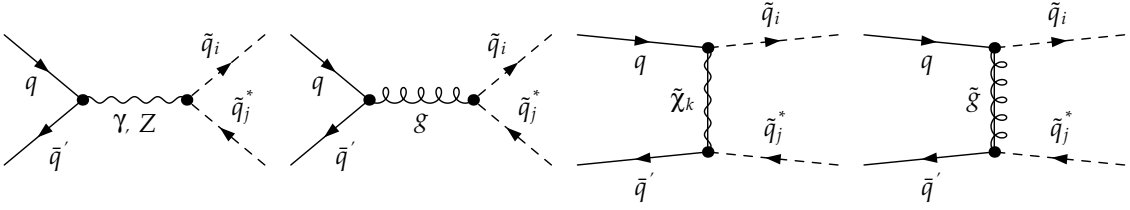}
  \end{center}
  \vspace*{-5mm}
  \caption{Tree-level Feynman diagrams for the production of neutral
           squark-antisquark pairs in quark-antiquark collisions.}
  \label{fig27}
\end{figure}

In this Section, we present numerical predictions for the production cross
sections of squark-antisquark pairs, squark pairs, the associated production of
squarks and gauginos, and gaugino pairs in NMFV SUSY at the LHC, i.e.\ for
$pp$-collisions at $\sqrt{S}=14$ TeV centre-of-momentum energy. Total
unpolarized hadronic cross sections
\beq
	\sigma ~=~ \int_{4m^2/S}^1 {\rm d}\tau \int_{-1/2 \ln\tau}^{1/2 \ln\tau}
	{\rm d}y \int_{t_{\rm min}}^{t_{\rm max}}{\rm d}t~f_{a/A}(x_a,M^2_a)~
	f_{b/B}(x_b,M^2_b) \frac{{\rm d}\hat{\sigma}}{{\rm d}t}
\eeq
are obtained through convolving the relevant partonic cross sections ${\rm
d}\hat{\sigma}/{\rm d}t$ with universal parton densities $f_{a/A}$ and $f_{b/B}$
of partons $a$, $b$ in the hadrons $A$, $B$, which depend on the longitudinal
momentum fractions of the two partons $x_{a,b} = \sqrt{\tau} e^{\pm y}$ and on
the unphysical factorization scales $M_{a,b}$. For consistency with our leading
order (LO) QCD calculation in the collinear approximation, where all quark
masses but the top mass are neglected with respect to the centre-of-momentum
energy $\sqrt{S}$, we employ the LO set of the latest CTEQ6 global parton
density fit \cite{Pumplin:2002vw}, which includes $n_f=5$ ``light'' quark
flavours and the gluon, but no top-quark density. Whenever it occurs, i.e.\ for
gluon initial states and gluon or gluino exchanges, the strong coupling constant
$\alpha_s(\mu_R)$ is calculated with the corresponding LO value of $\Lambda_{\rm
LO}^{n_f=5} = 165$ MeV. We identify the renormalization scale $\mu_R$ with the
factorization scales $M_a = M_b$ and set the scales to the average mass of the
produced SUSY particles $m$. 

\begin{figure}
  \begin{center}
    \includegraphics[scale=0.28]{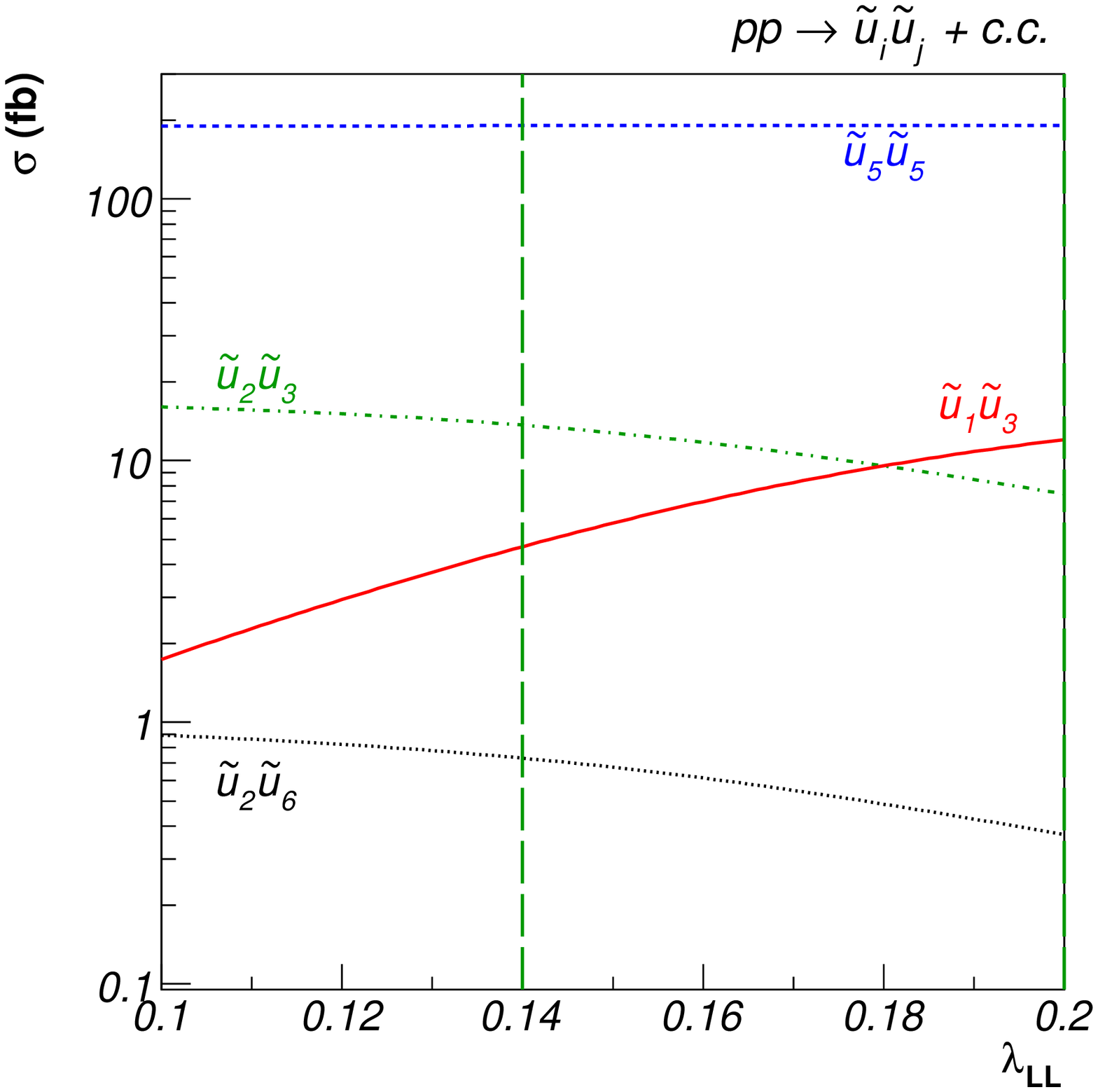} 
    \includegraphics[scale=0.28]{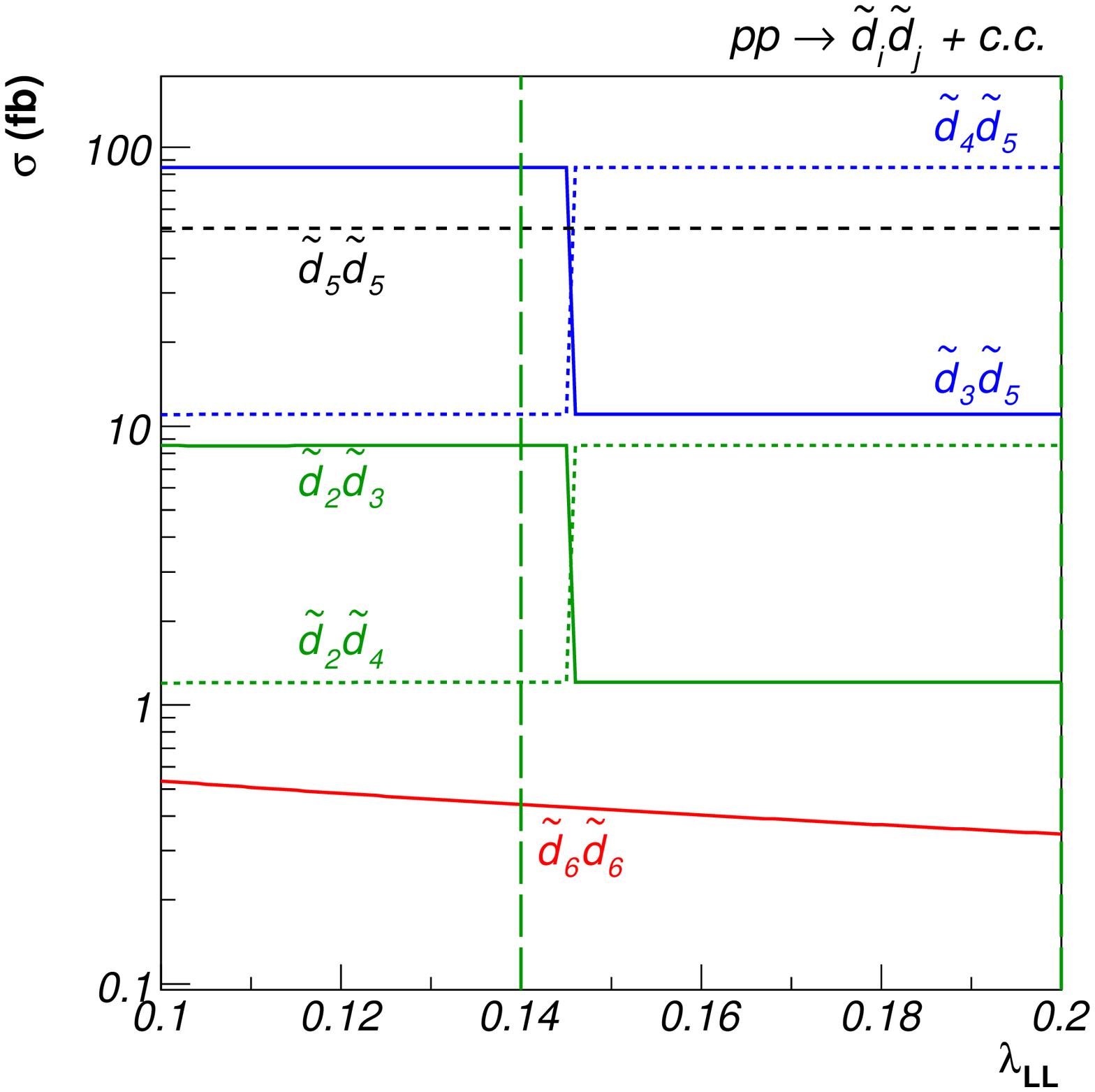} 
    \includegraphics[scale=0.28]{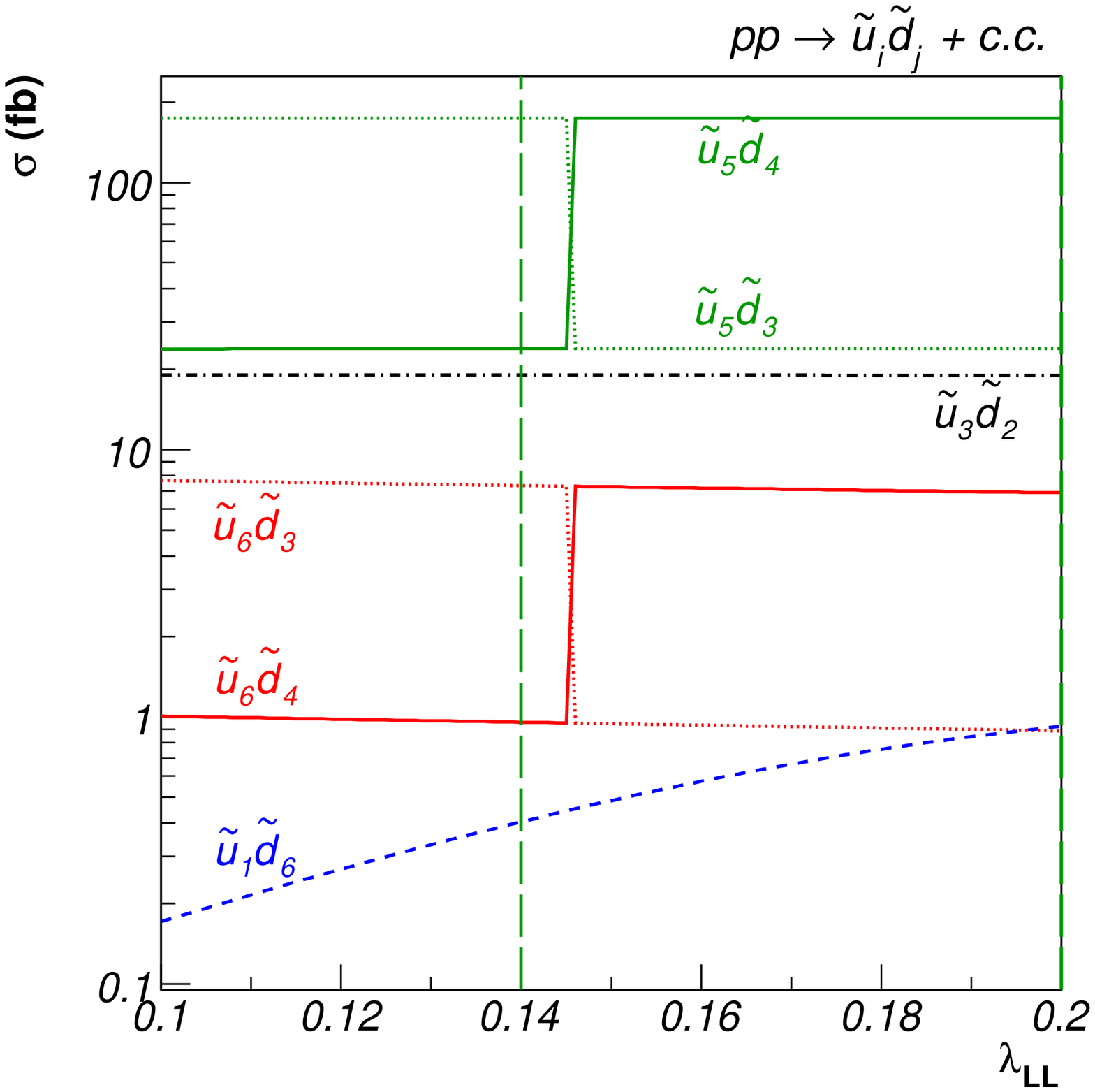} 
    \includegraphics[scale=0.28]{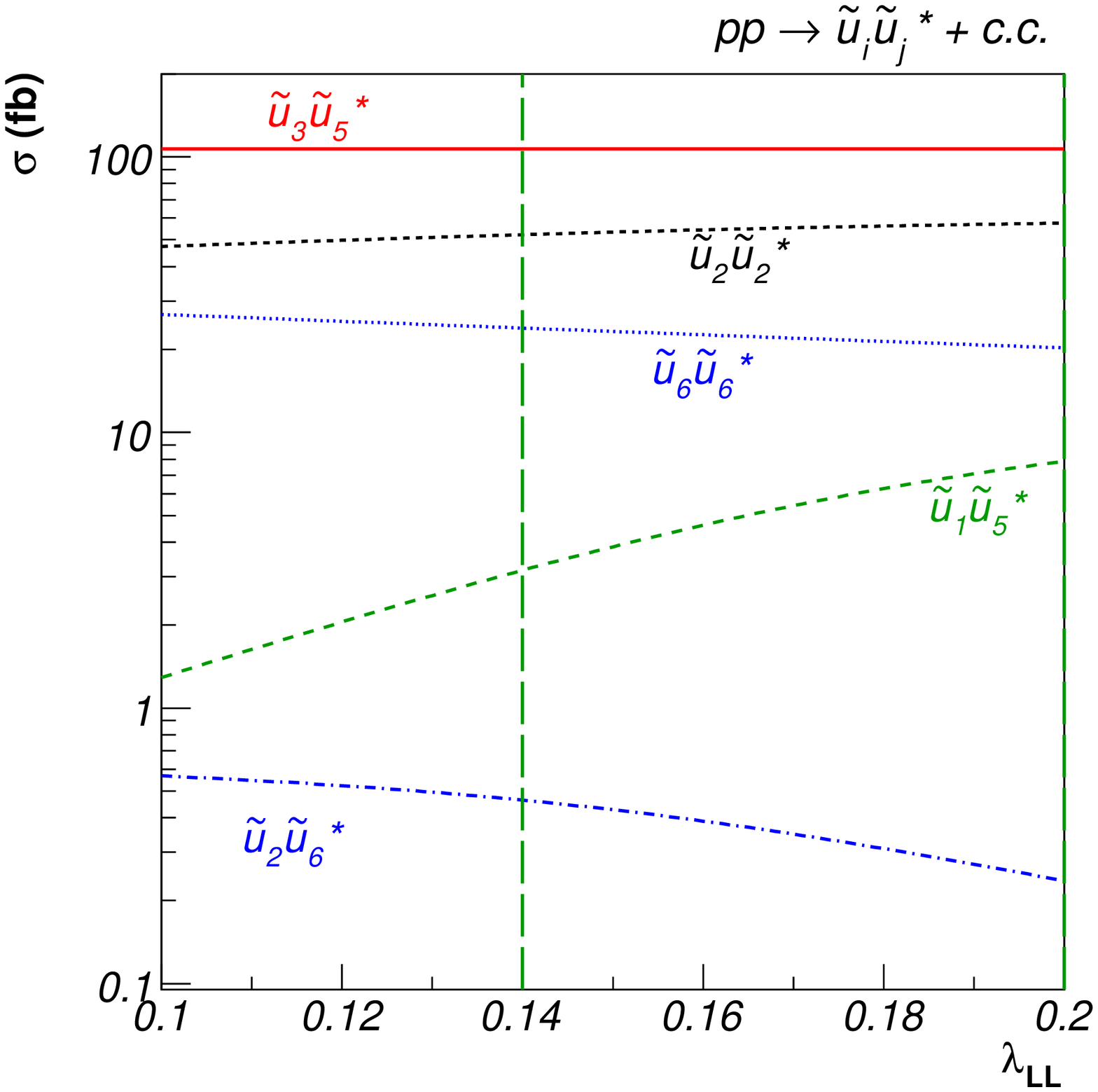}
    \includegraphics[scale=0.28]{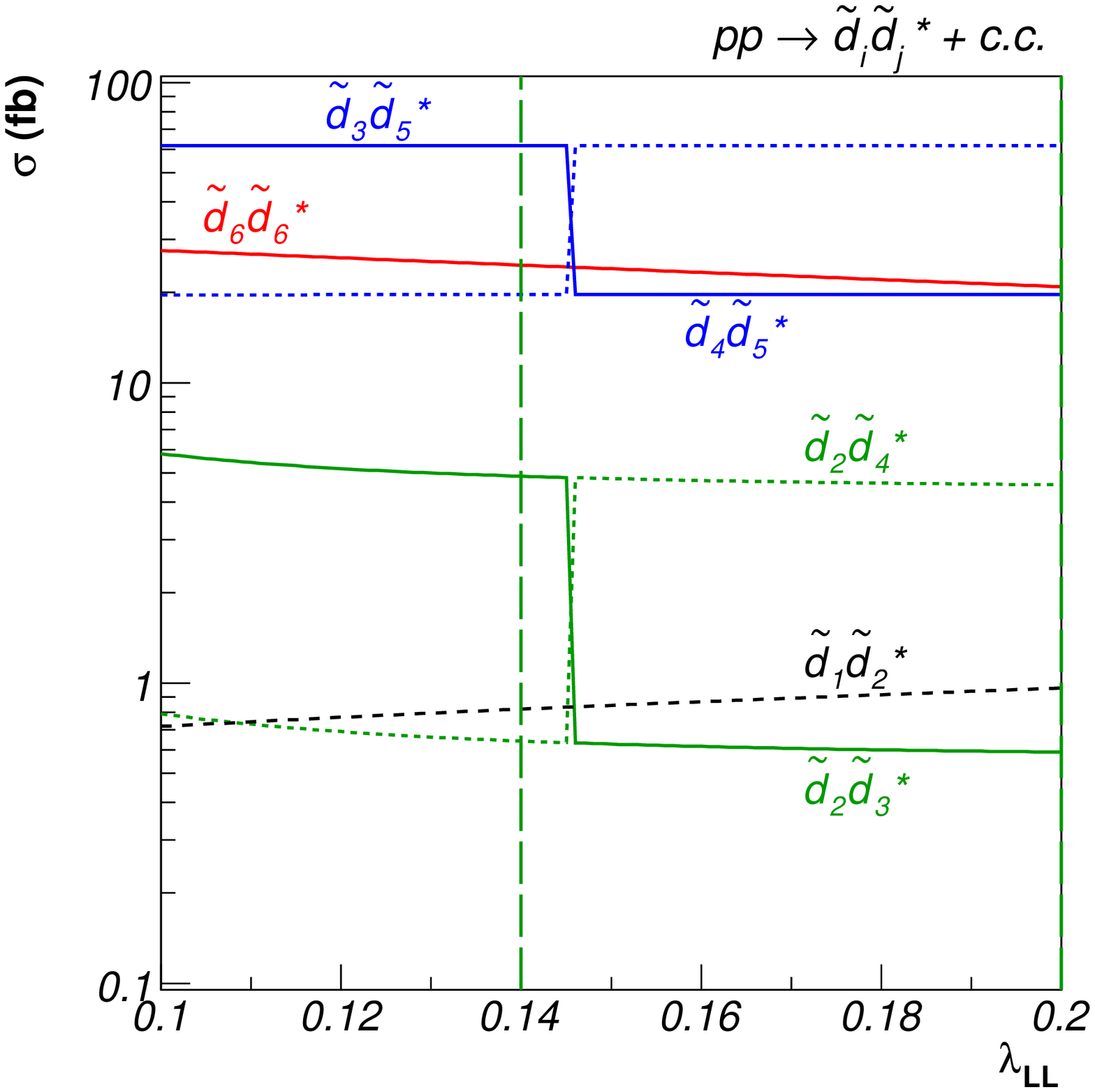}
    \includegraphics[scale=0.28]{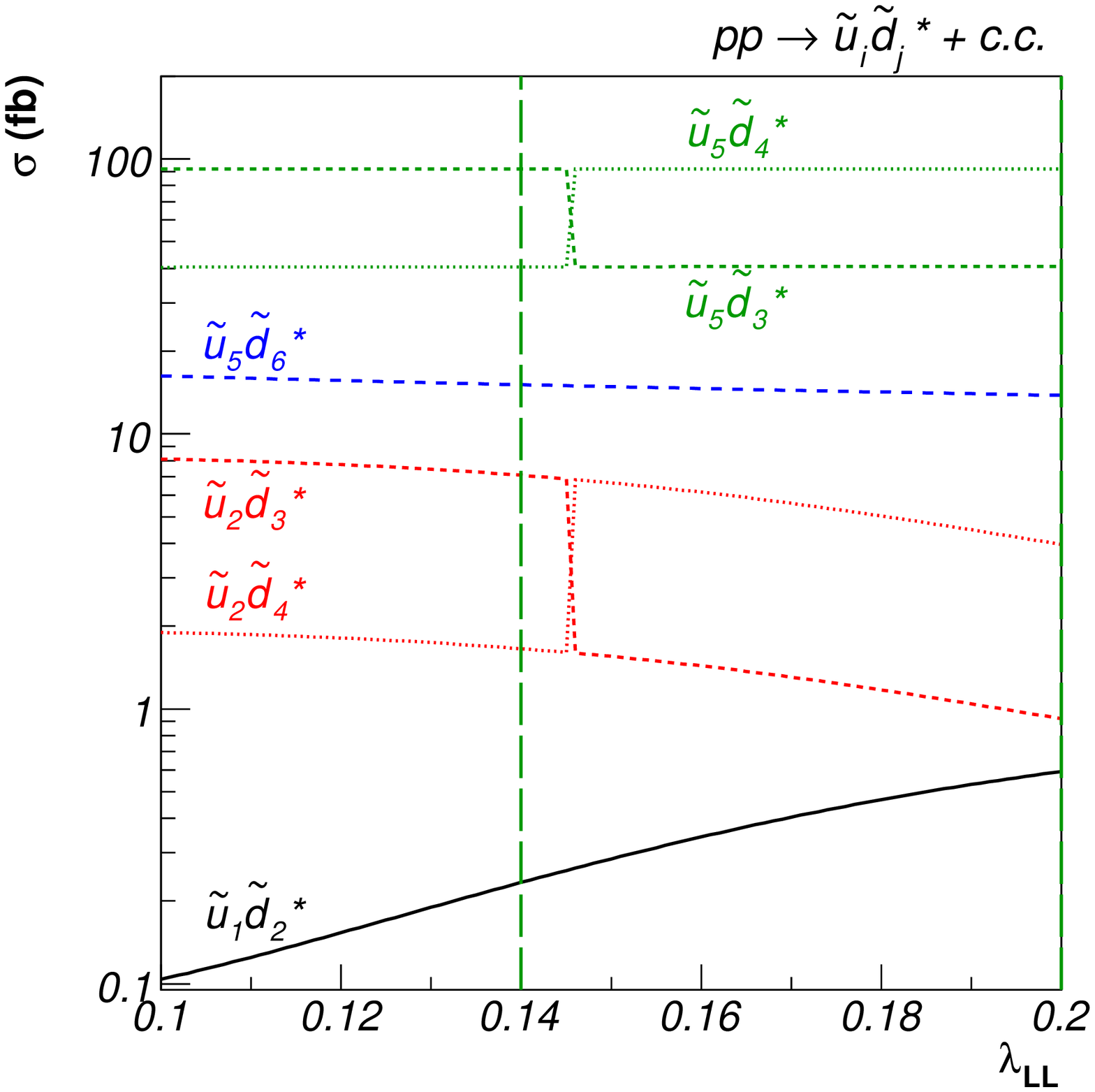} 
    \includegraphics[scale=0.28]{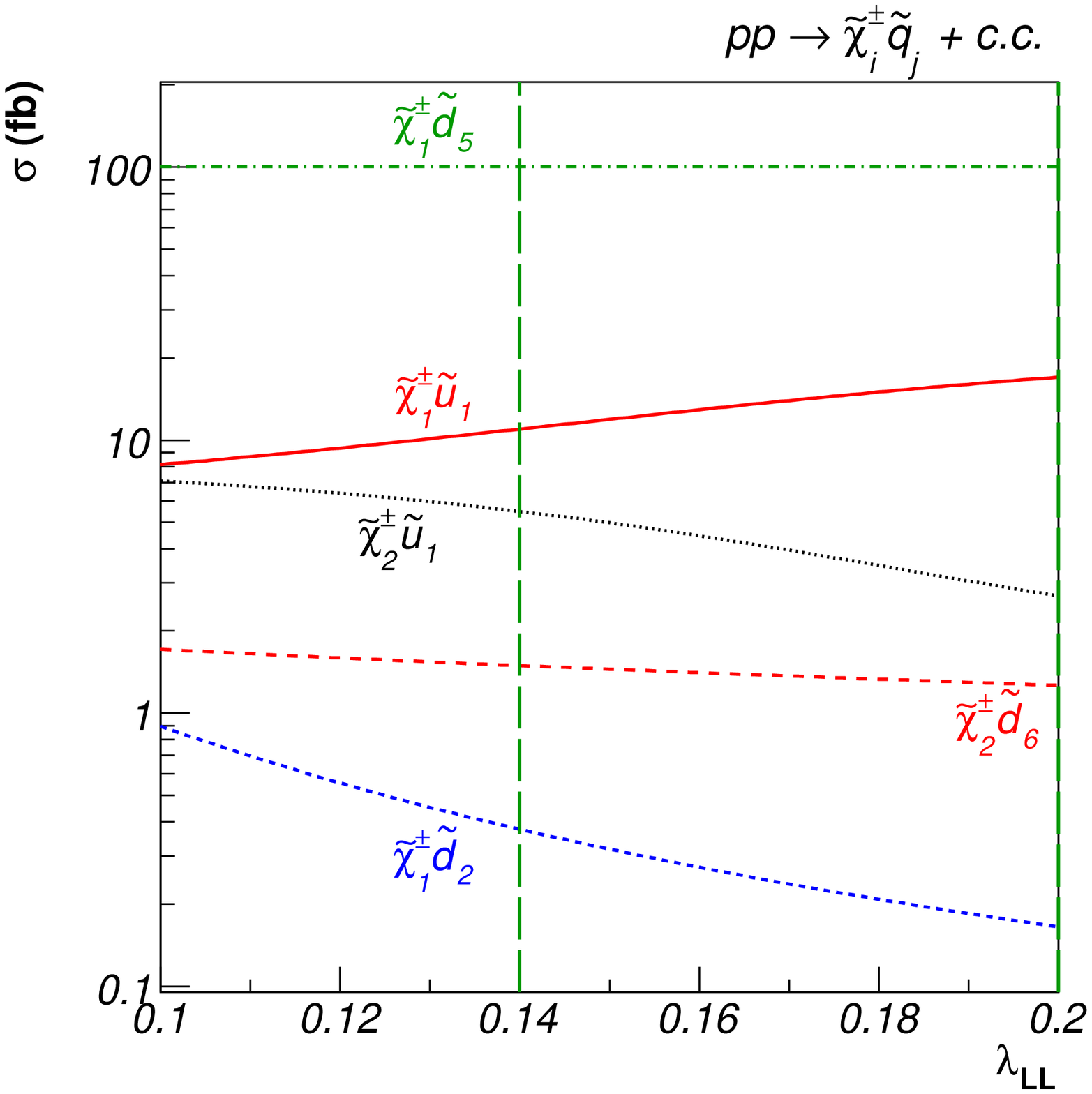} 
    \includegraphics[scale=0.28]{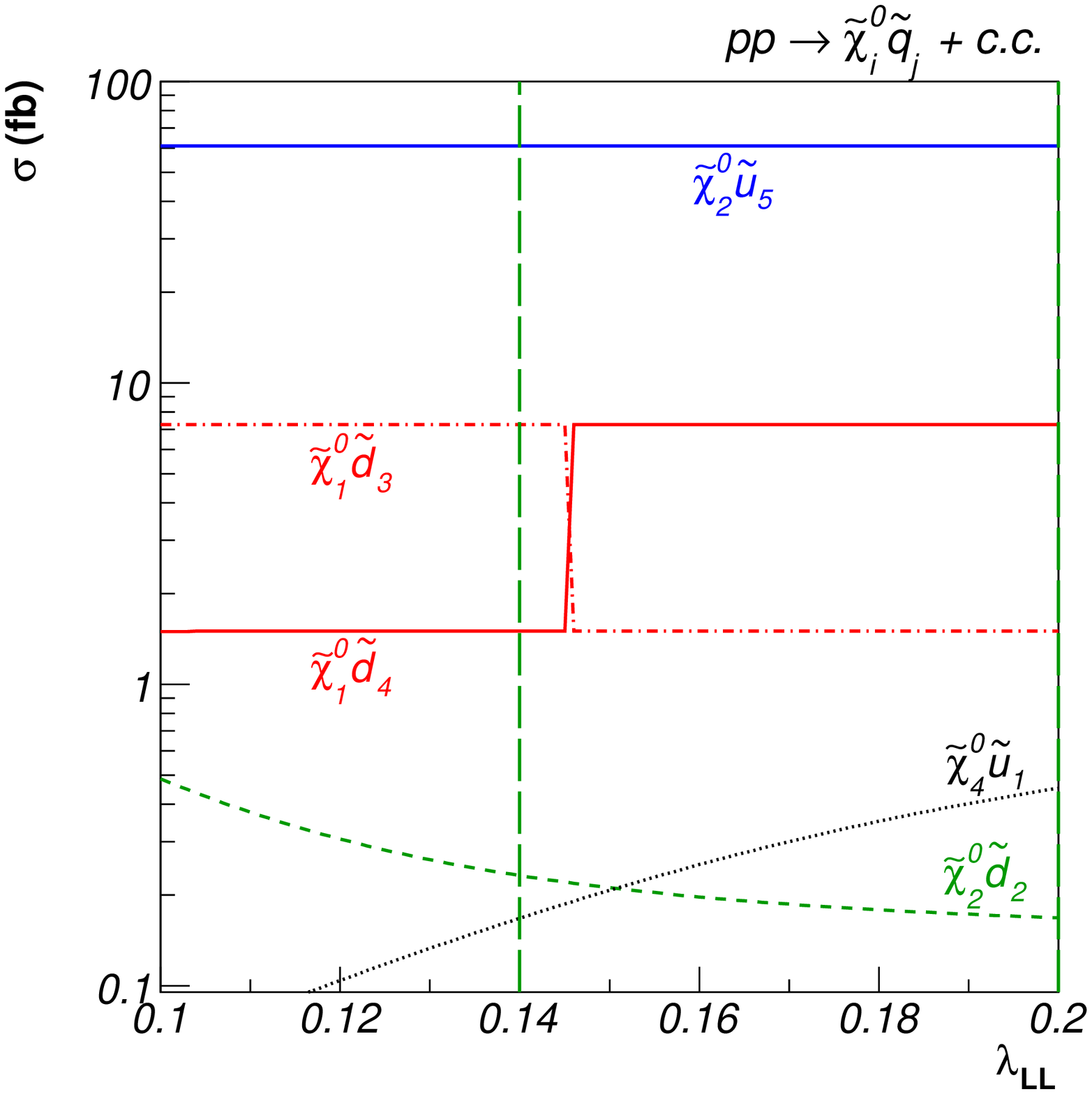} 
    \includegraphics[scale=0.28]{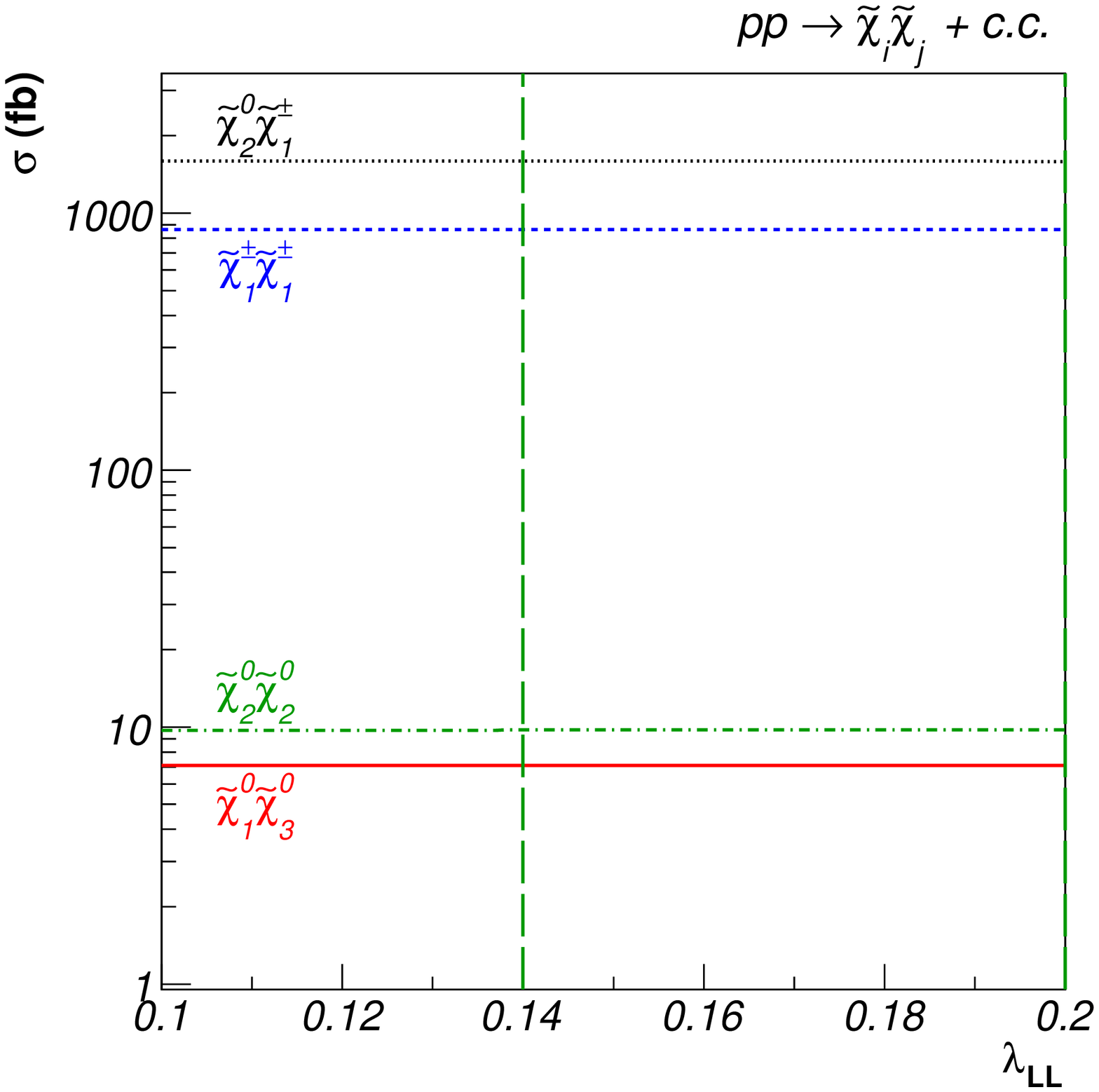} 
  \end{center}
  \vspace*{-5mm}
\caption{Examples of cross sections for charged squark-squark pair production,
 neutral and charged squark-antisquark pair production, associated production of
 squarks with charginos and neutralinos, and gaugino-pair production at the
 LHC in our benchmark scenario E with flavour violation in the left-left
 chiral sector ($\lambda_{\rm RR}=0$).}
\label{fig15}
\end{figure}

\begin{figure}
  \begin{center}
    \includegraphics[scale=0.28]{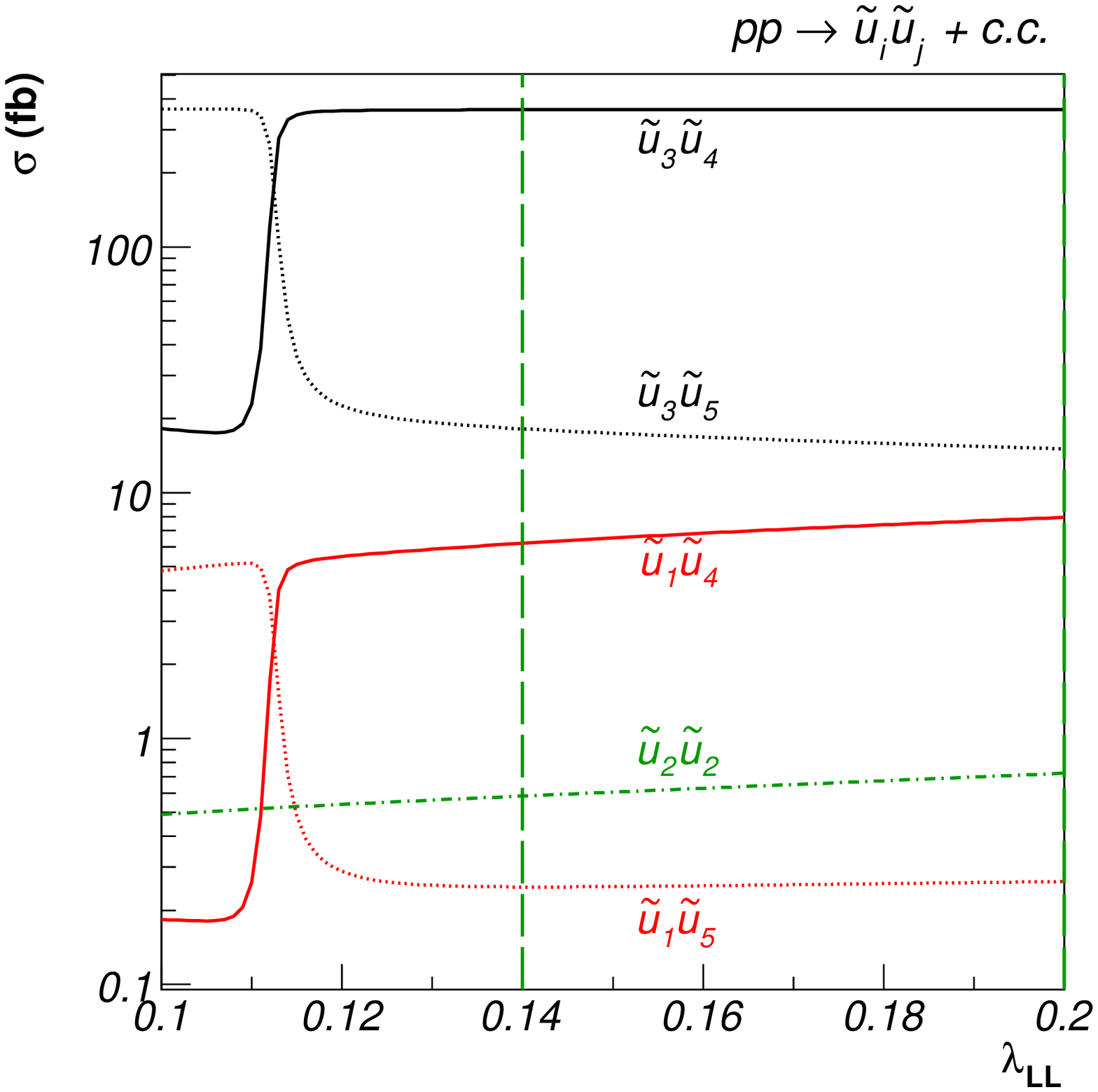} 
    \includegraphics[scale=0.28]{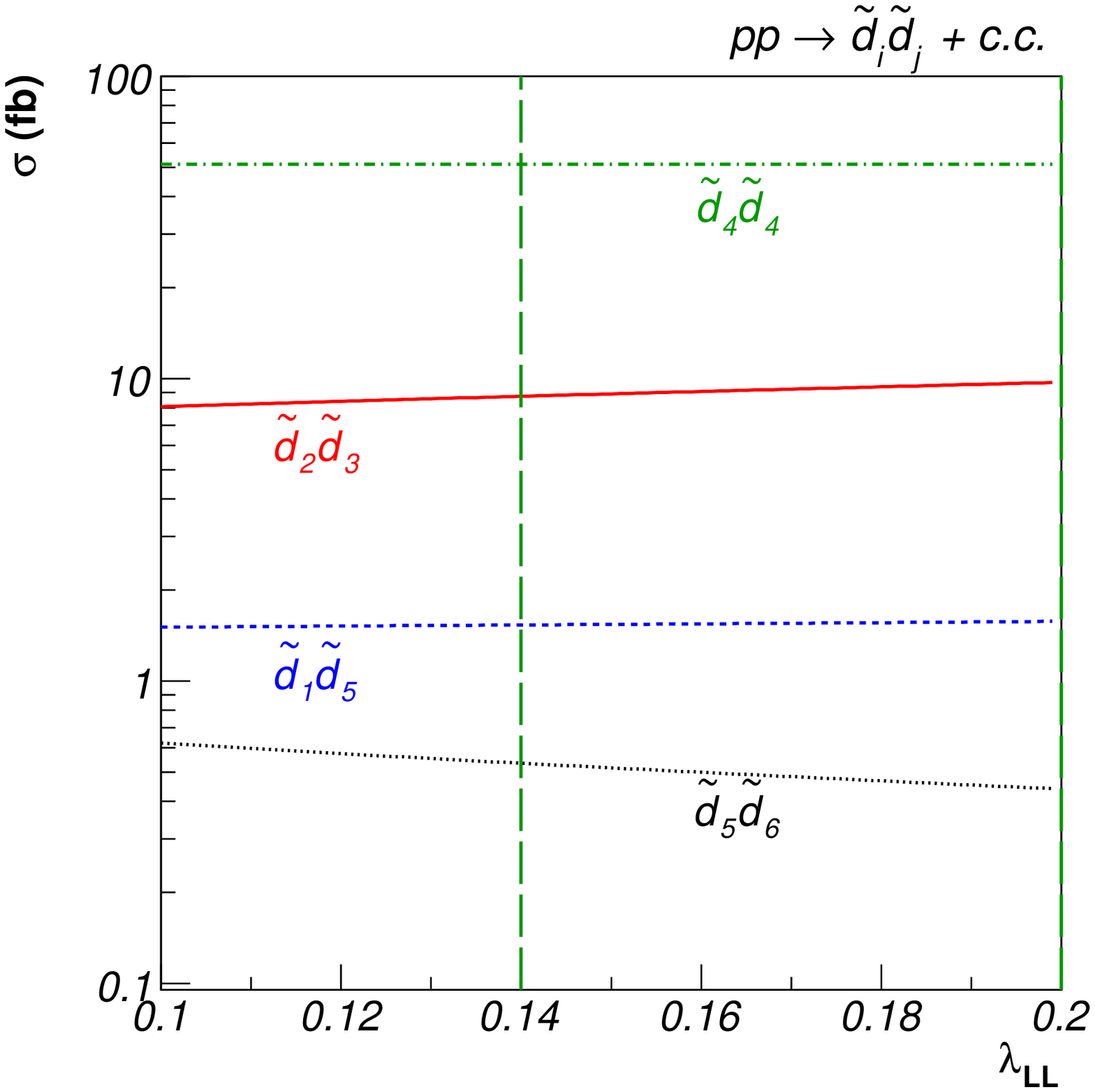} 
    \includegraphics[scale=0.28]{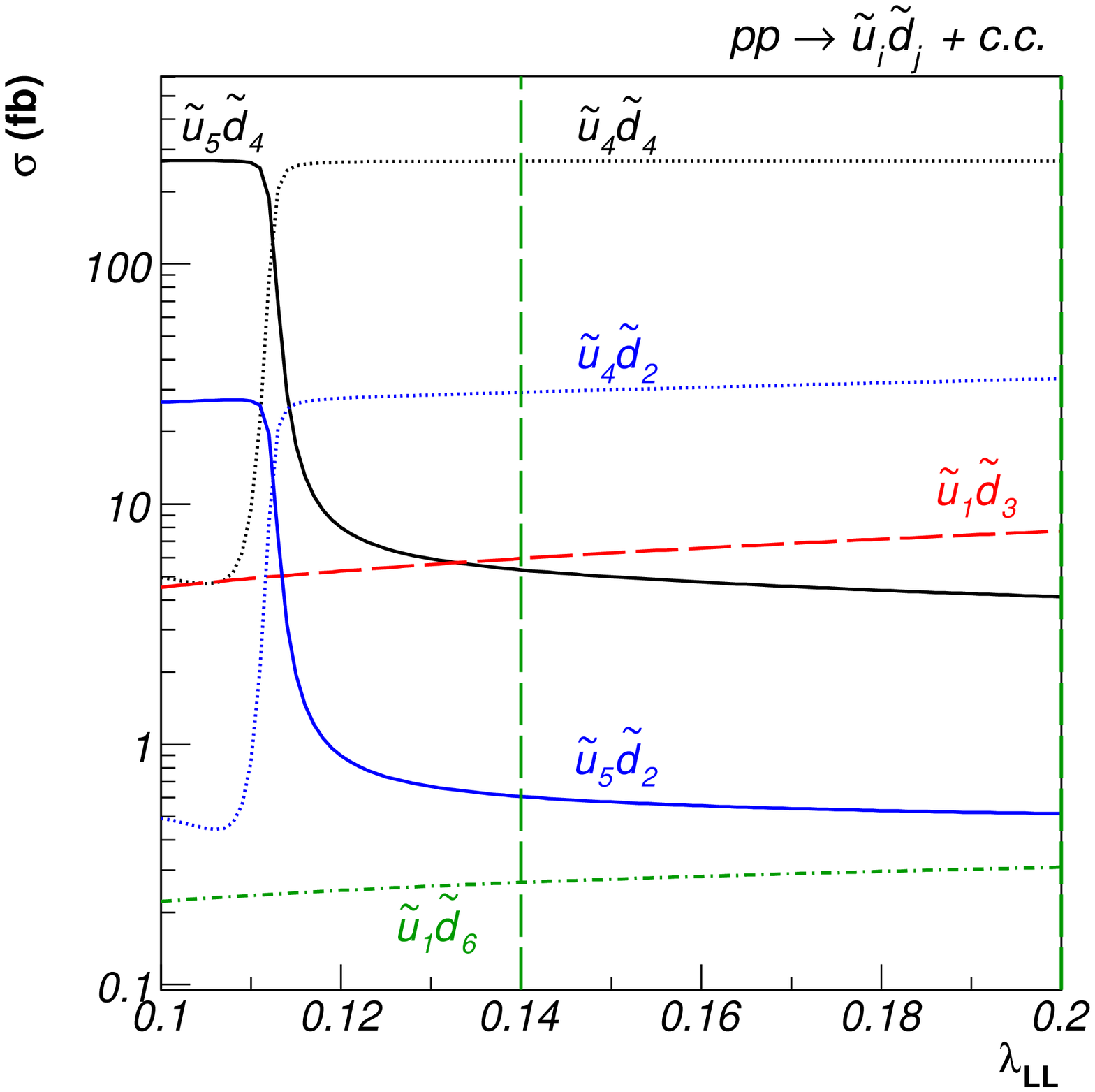} 
    \includegraphics[scale=0.28]{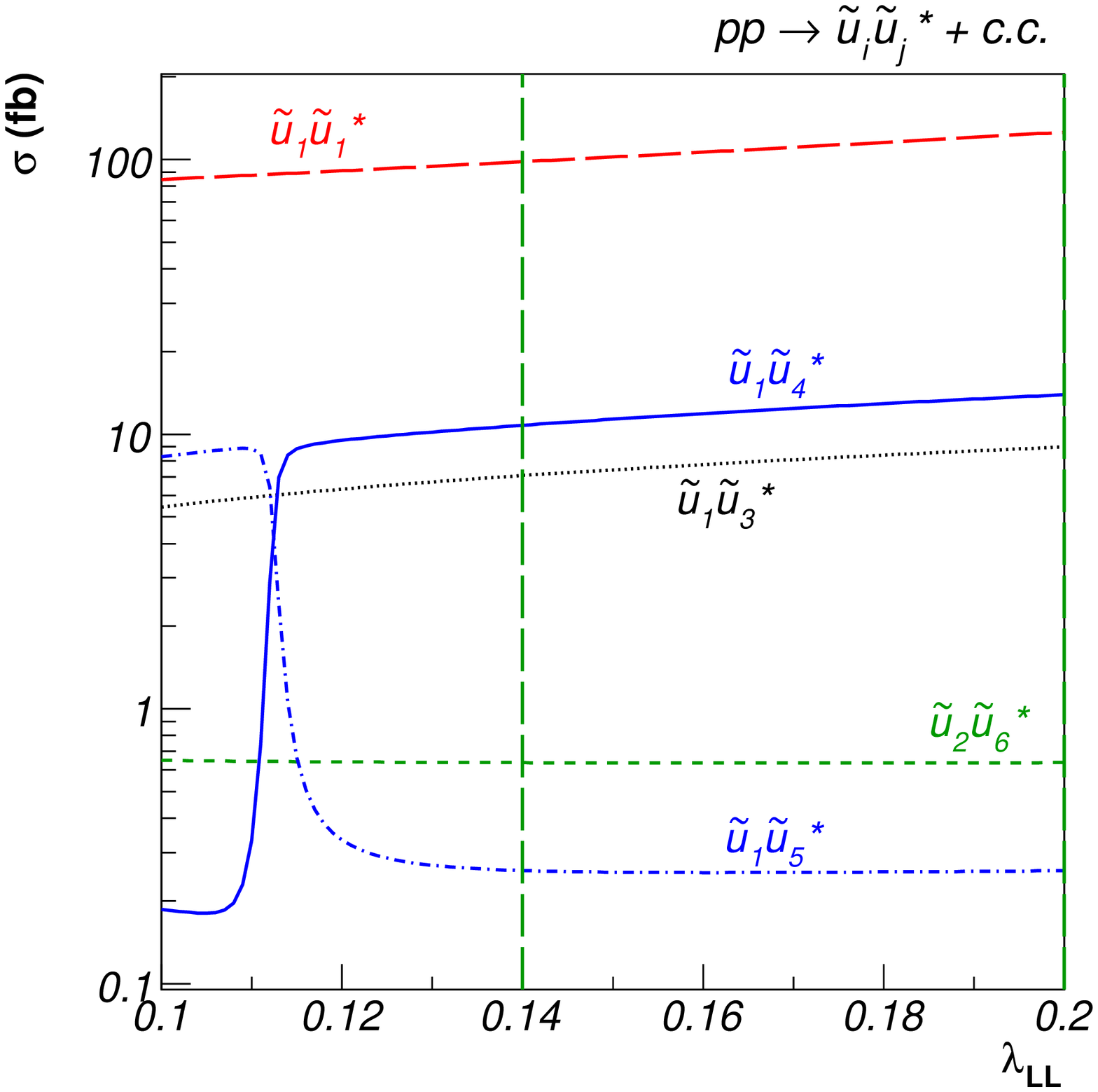}
    \includegraphics[scale=0.28]{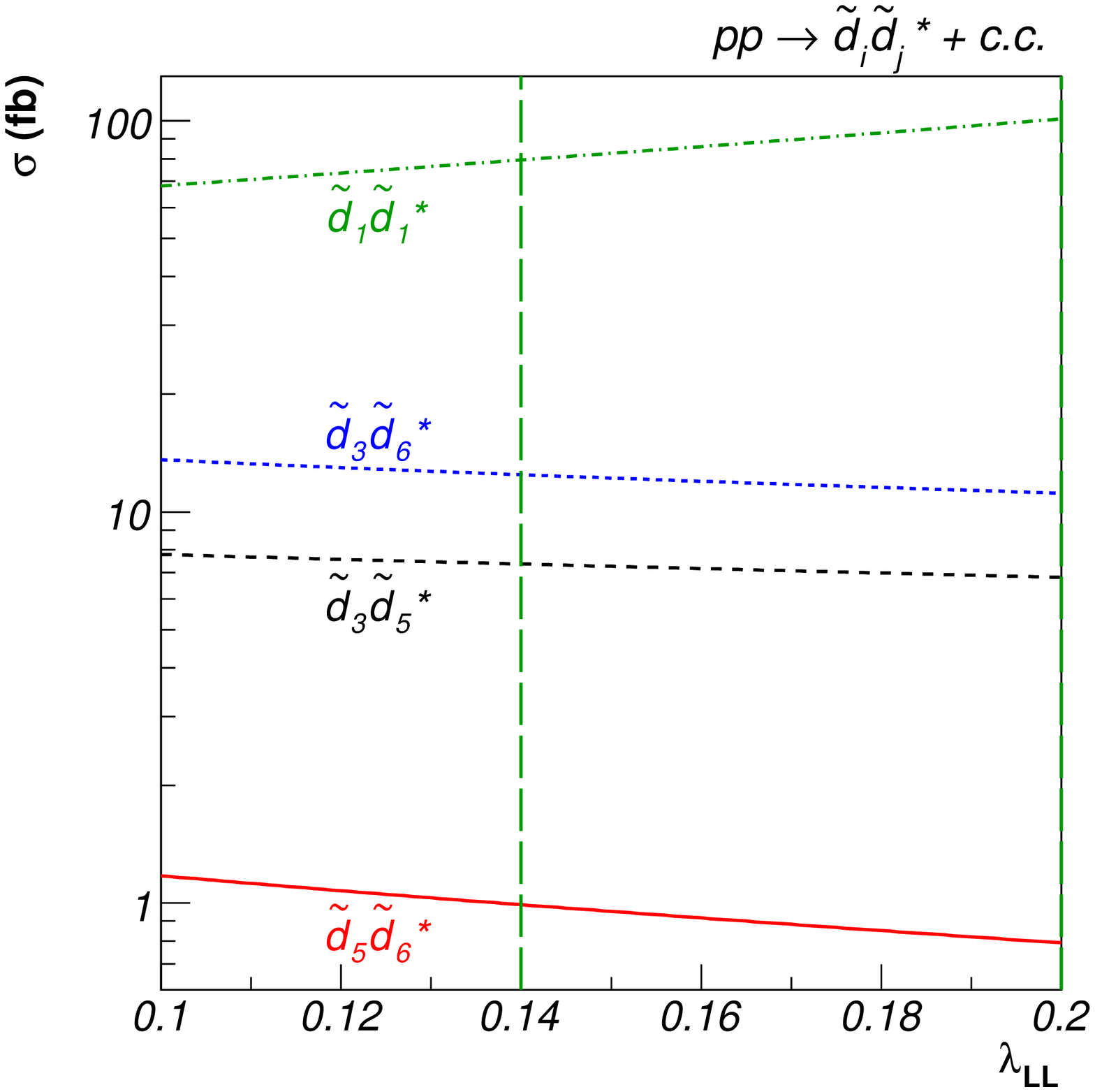}
    \includegraphics[scale=0.28]{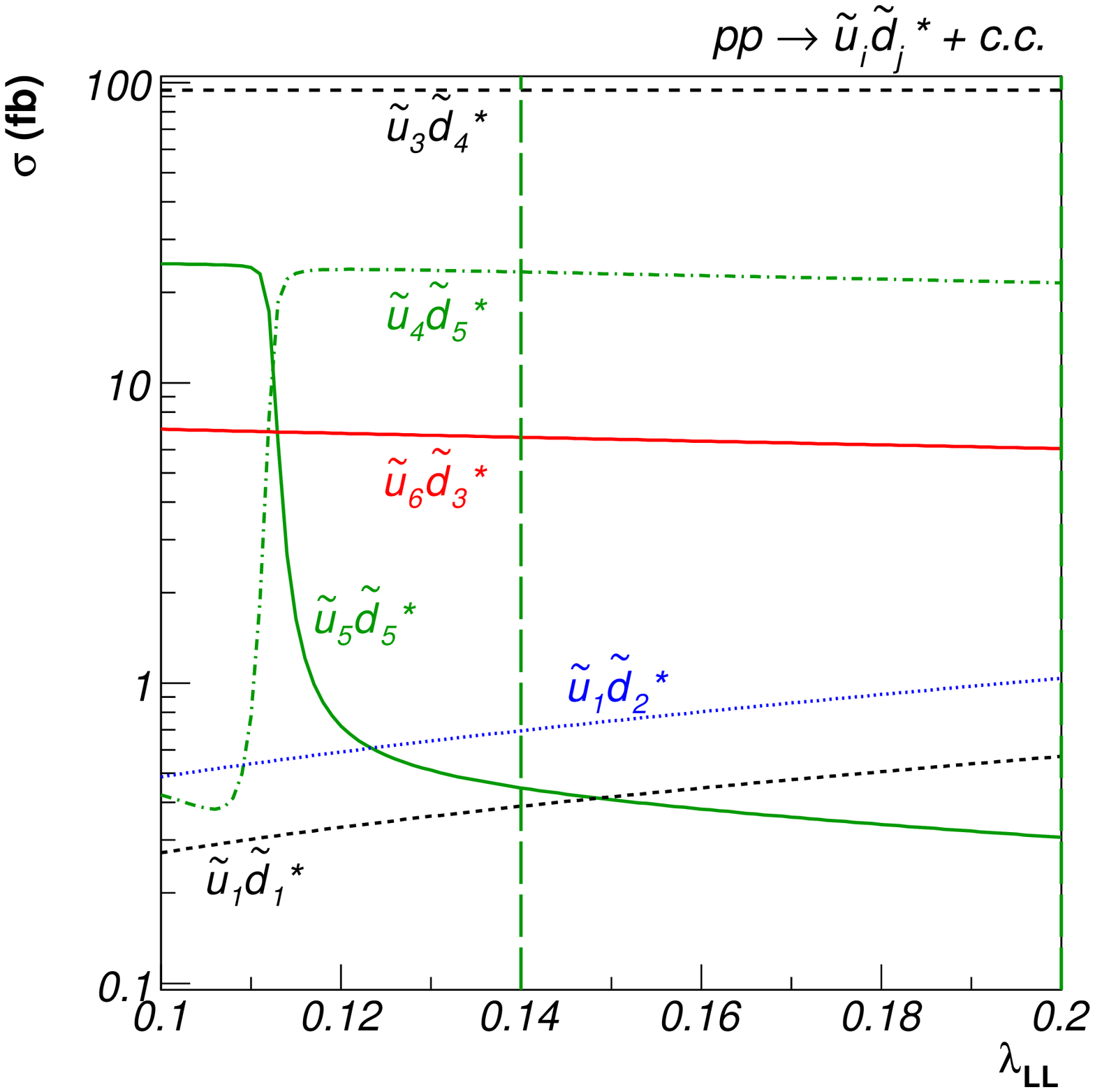}
    \includegraphics[scale=0.28]{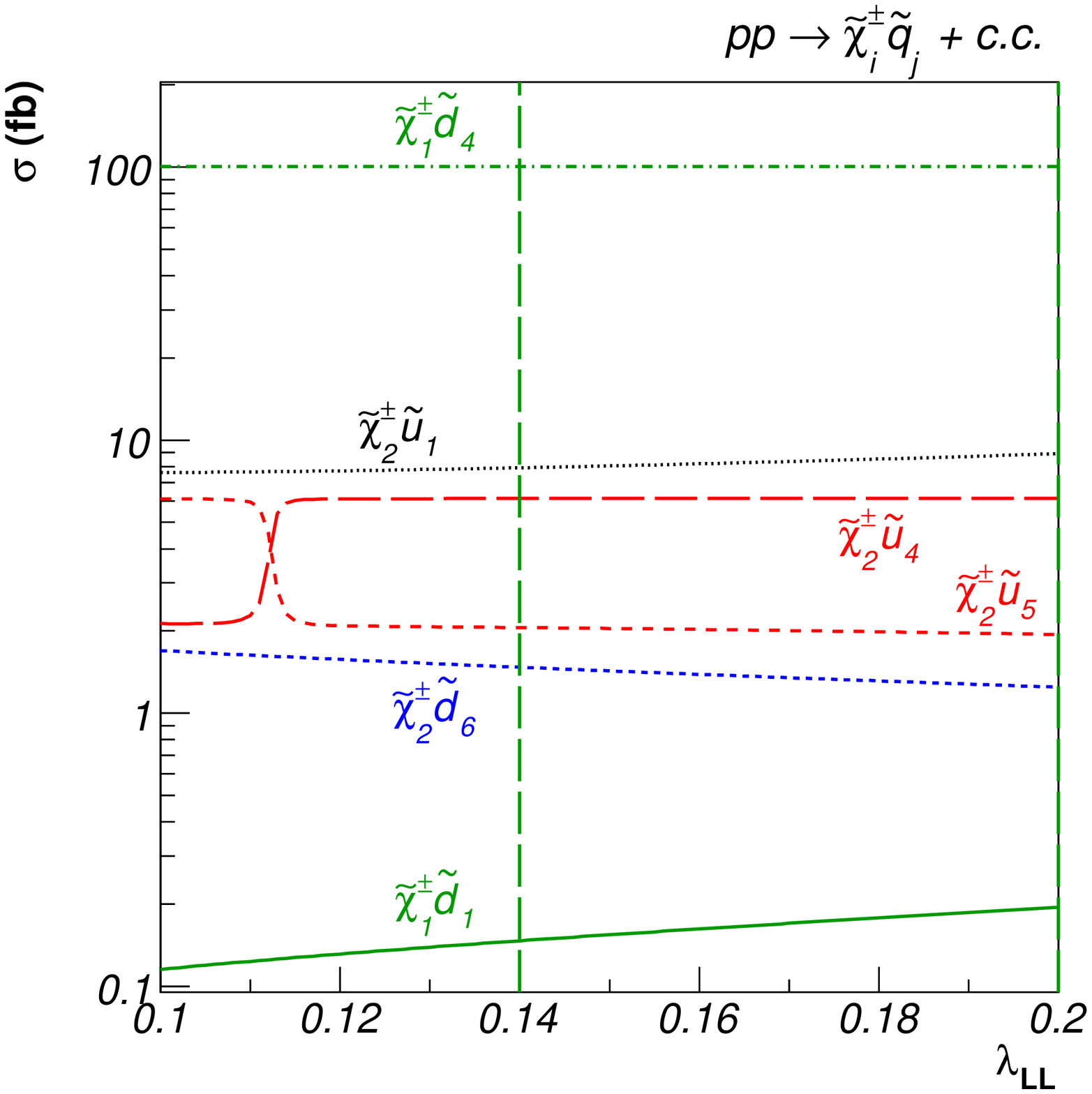} 
    \includegraphics[scale=0.28]{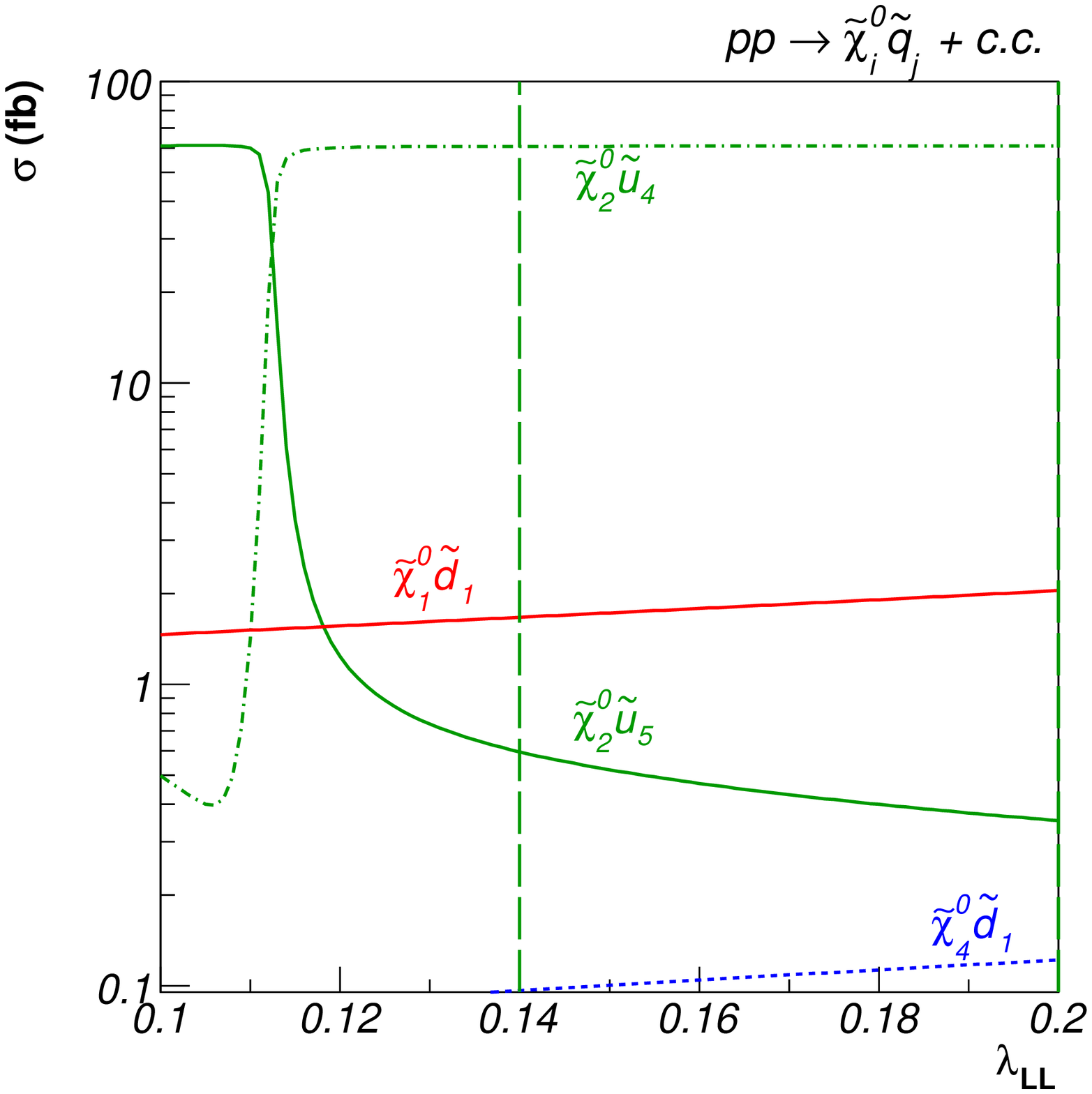} 
    \includegraphics[scale=0.28]{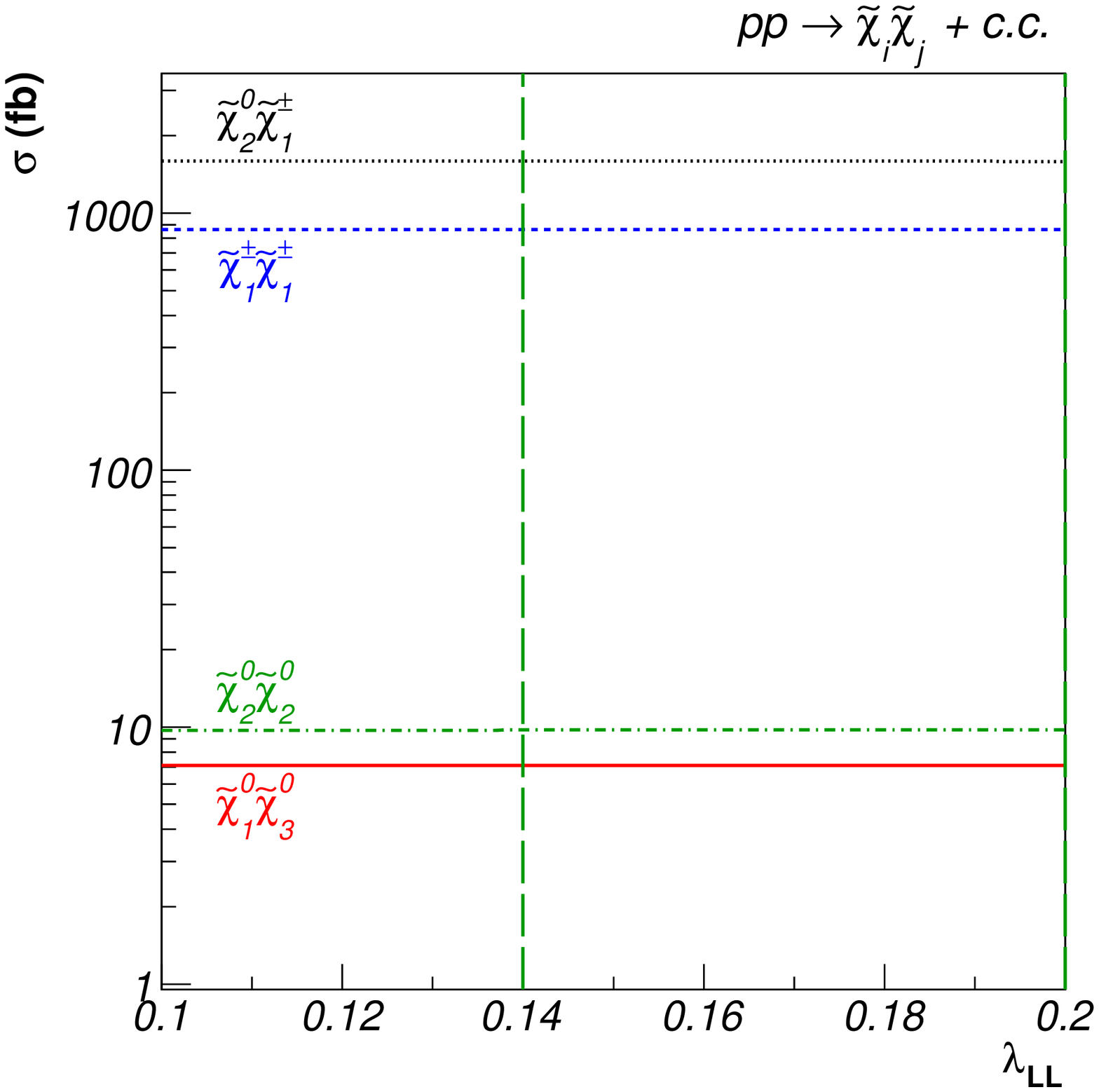} 
  \end{center}
  \vspace*{-5mm}
\caption{Same as Fig.\ \ref{fig15} for our benchmark scenario E with flavour
violation in the left-left and right-right chiral sectors ($\lambda_{\rm
RR}=\lambda_{\rm LL}$).} 
\label{fig16}
\end{figure}

\begin{figure}
  \begin{center}
    \includegraphics[scale=0.28]{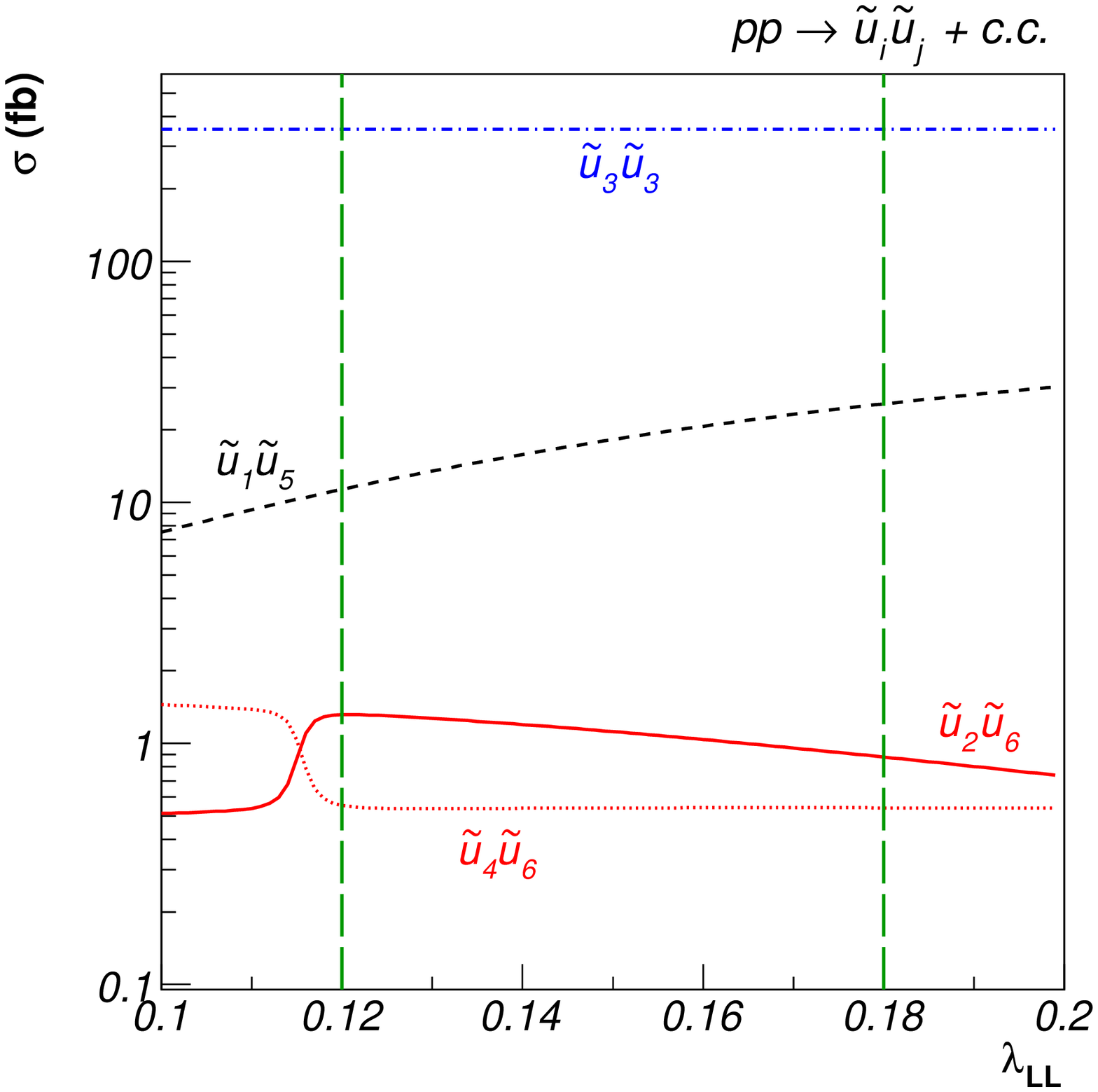} 
    \includegraphics[scale=0.28]{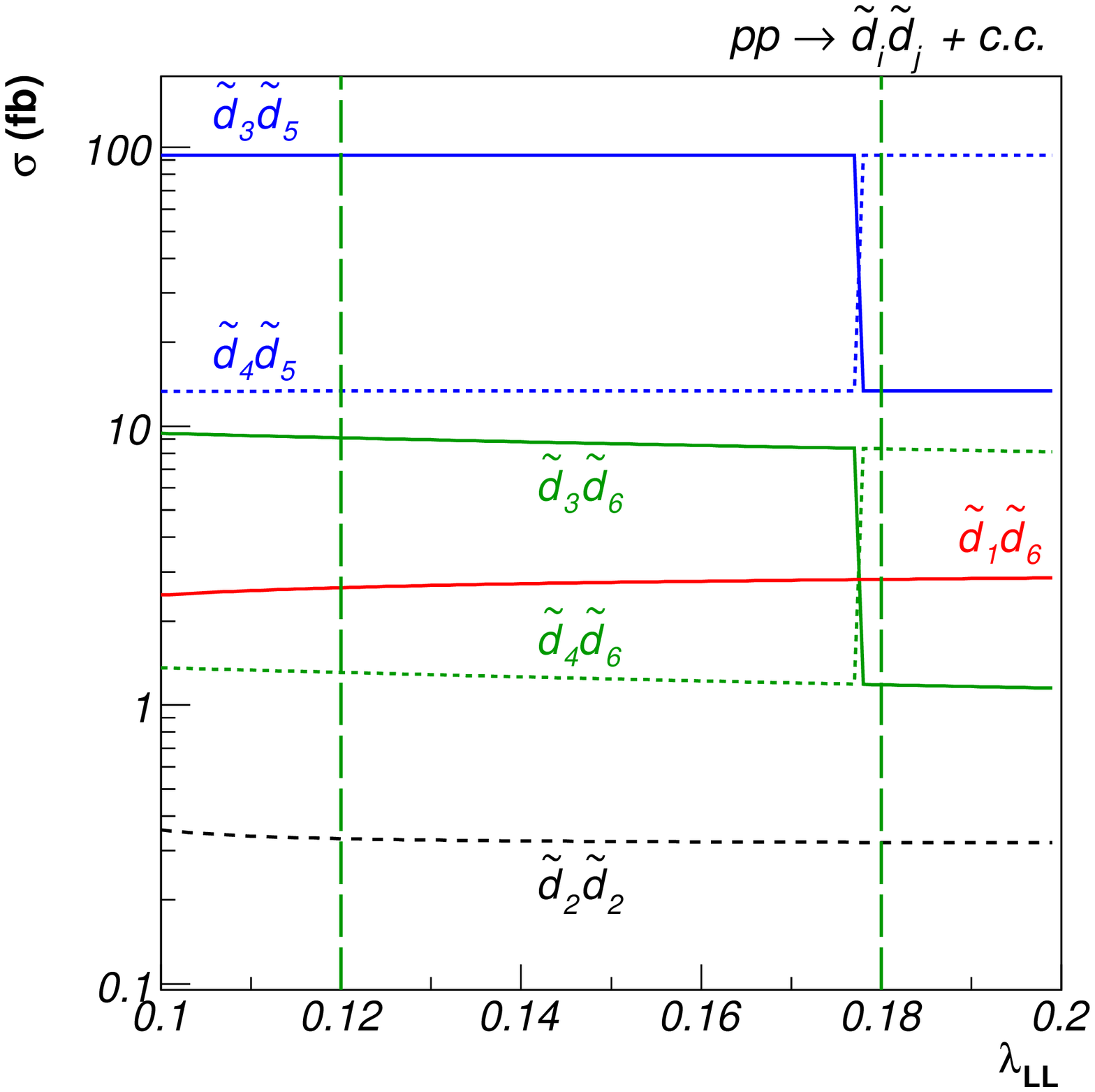} 
    \includegraphics[scale=0.28]{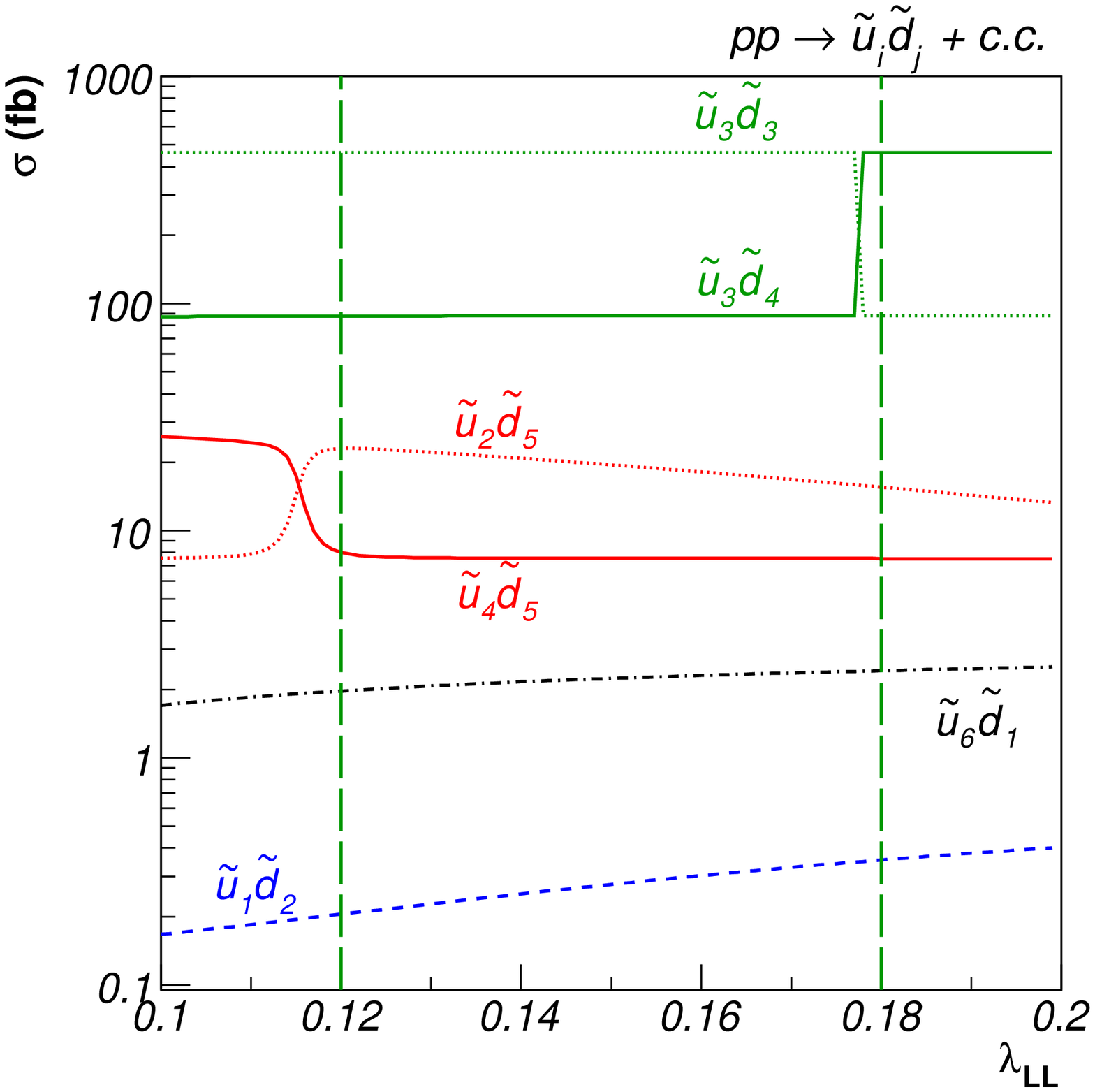} 
    \includegraphics[scale=0.28]{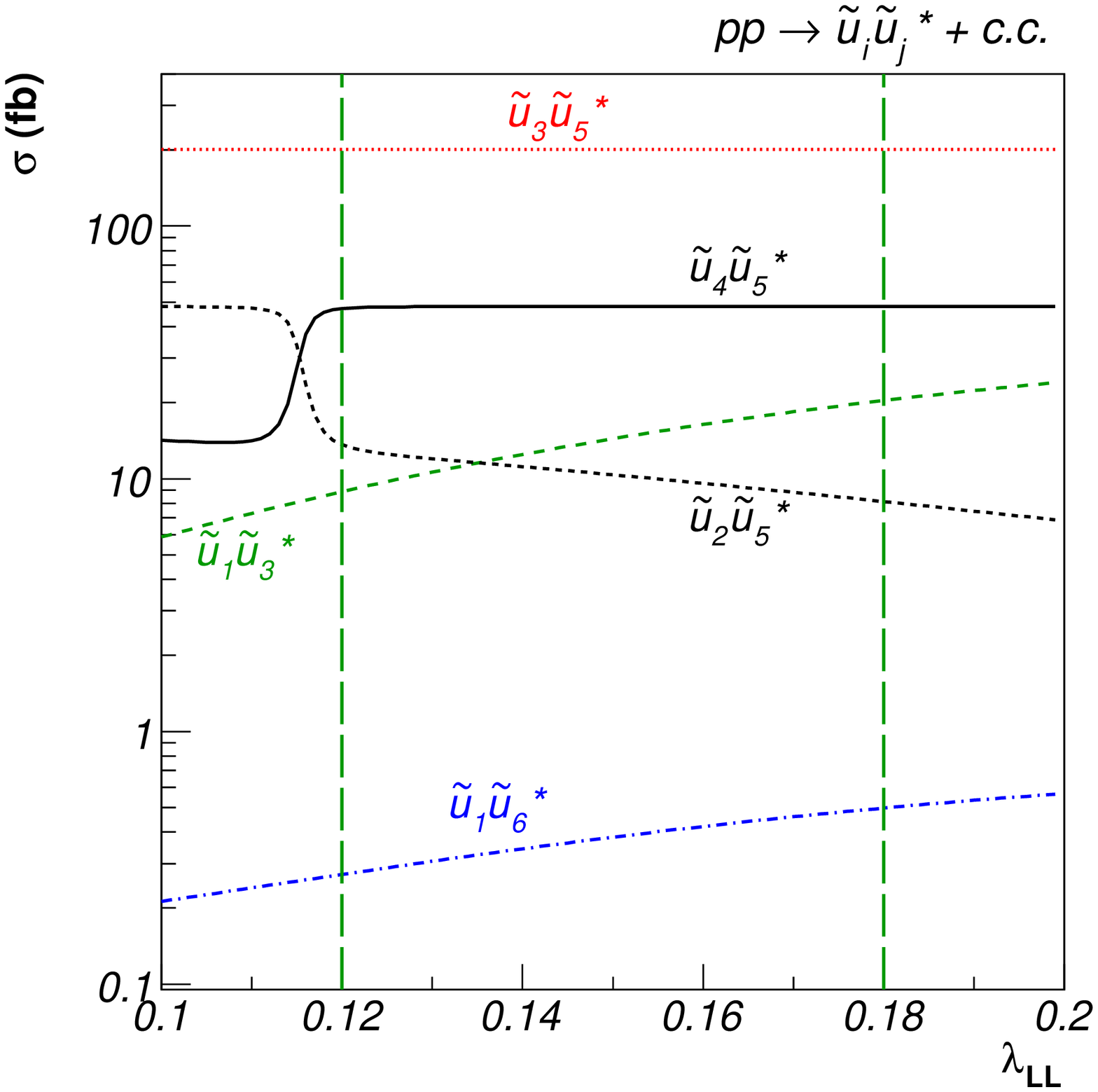} 
    \includegraphics[scale=0.28]{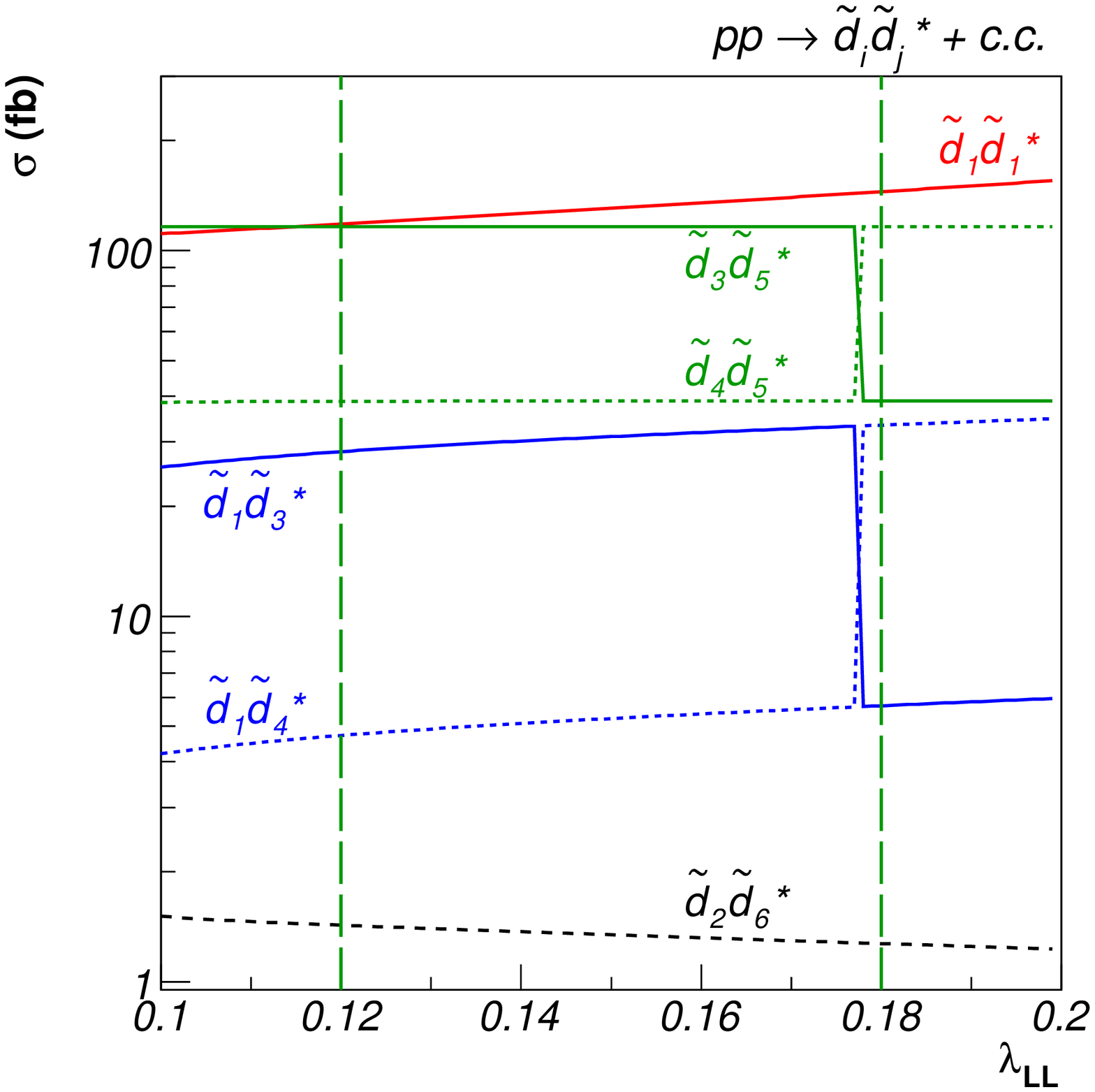} 
    \includegraphics[scale=0.28]{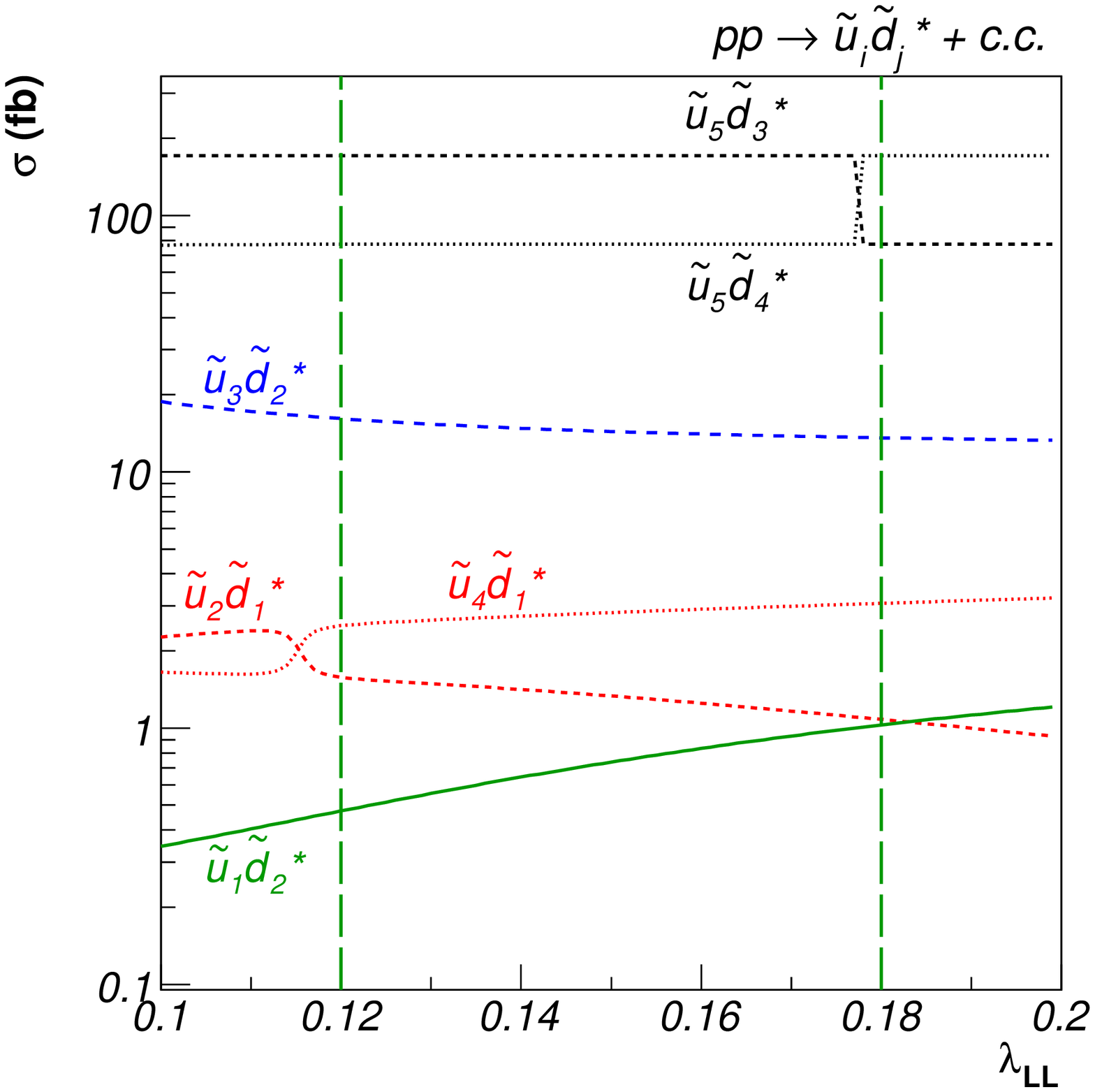} 
    \includegraphics[scale=0.28]{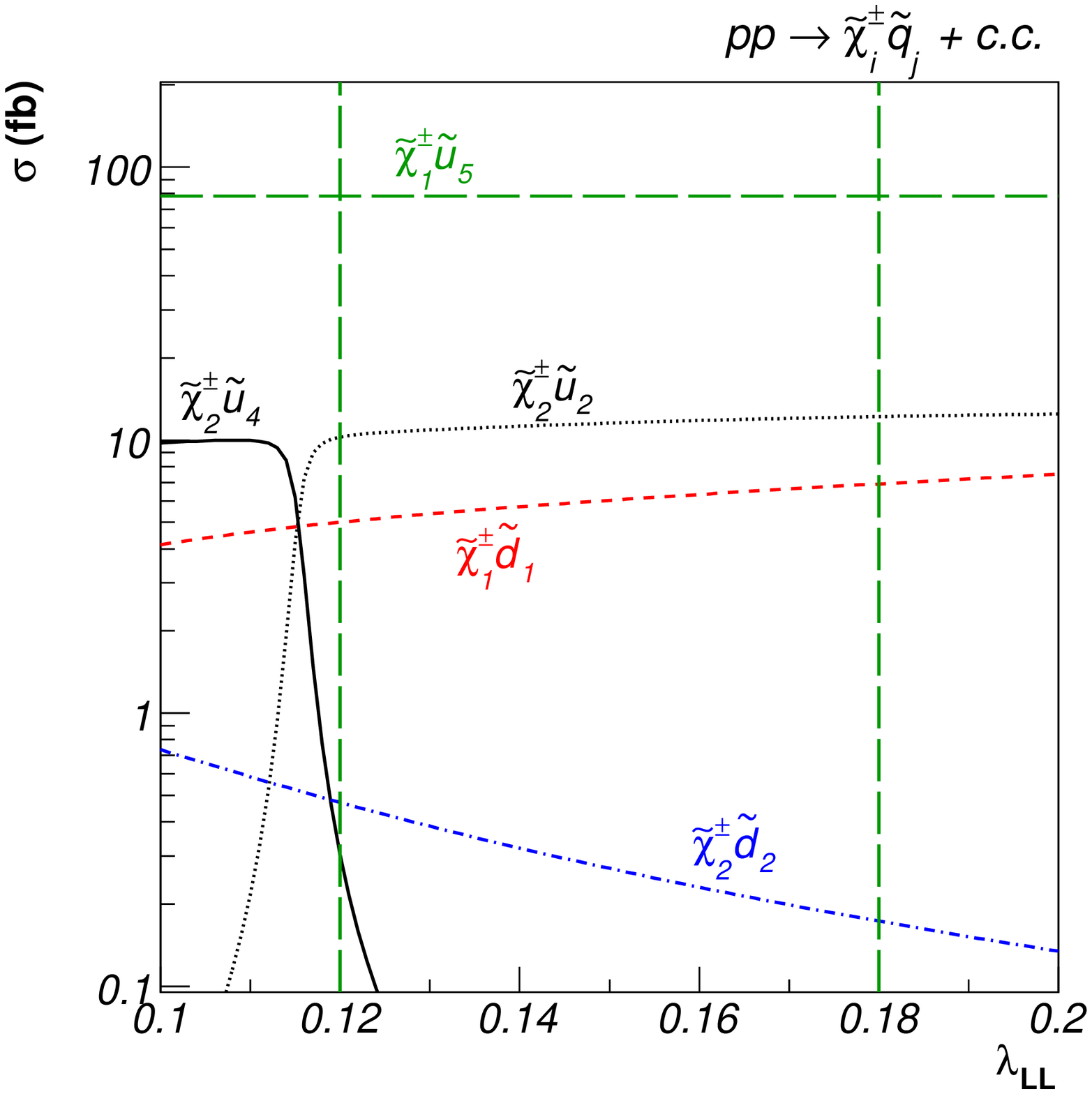} 
    \includegraphics[scale=0.28]{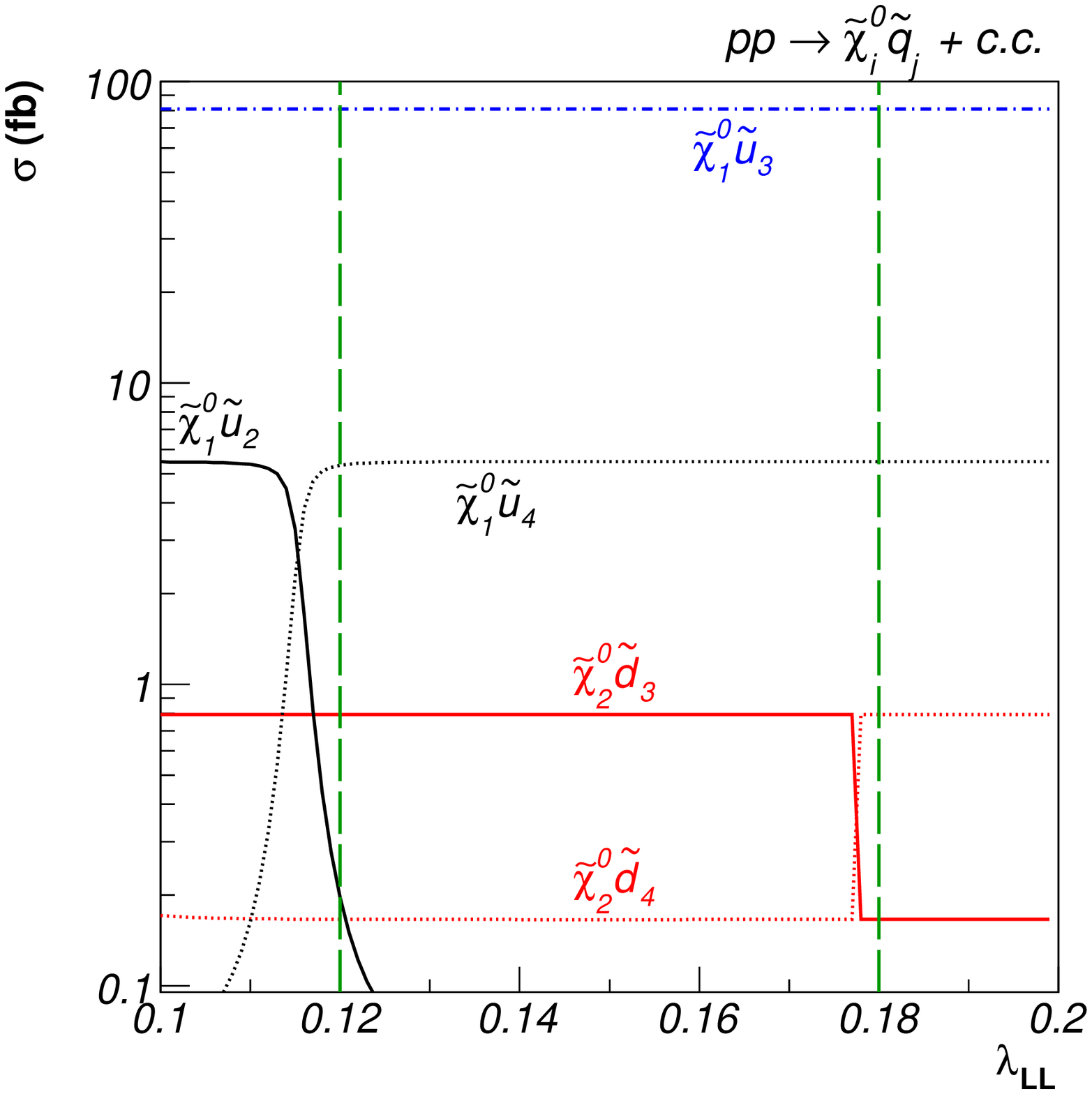} 
    \includegraphics[scale=0.28]{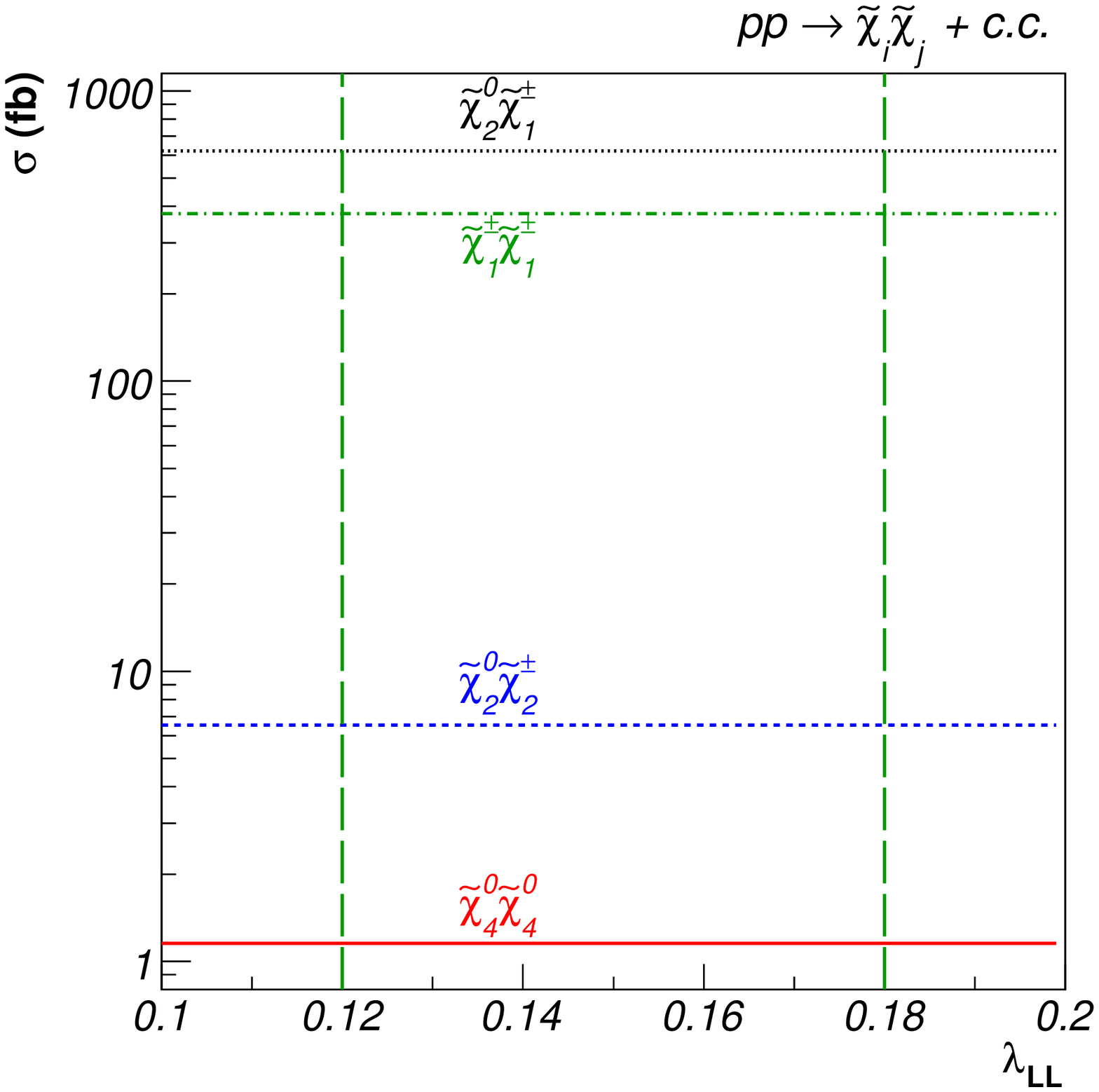} 
  \end{center}
  \vspace*{-5mm}
\caption{Same as Fig.\ \ref{fig15} for our benchmark scenario F.}
\label{fig17}
\end{figure}

\begin{figure}
  \begin{center}
    \includegraphics[scale=0.28]{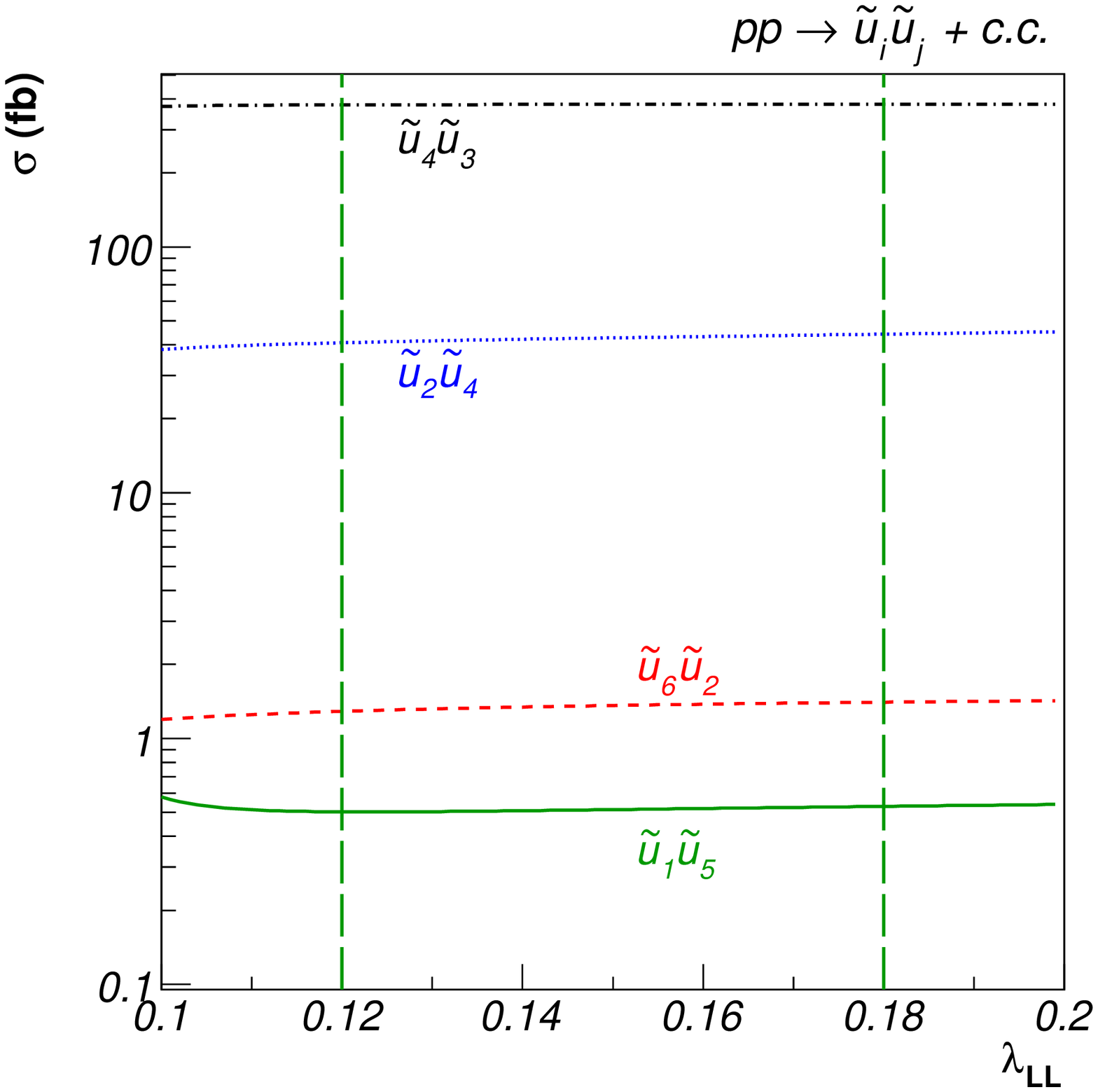} 
    \includegraphics[scale=0.28]{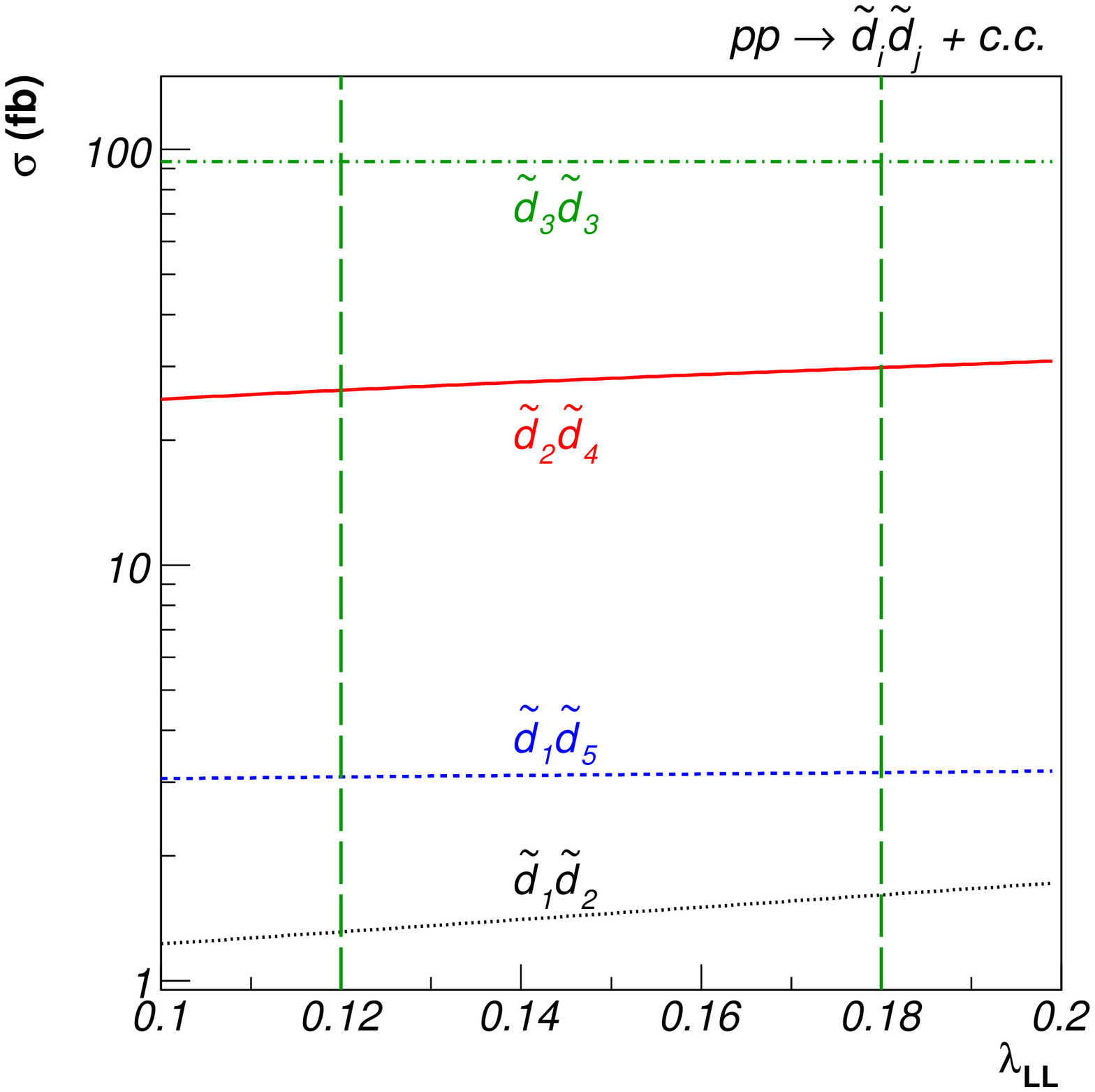} 
    \includegraphics[scale=0.28]{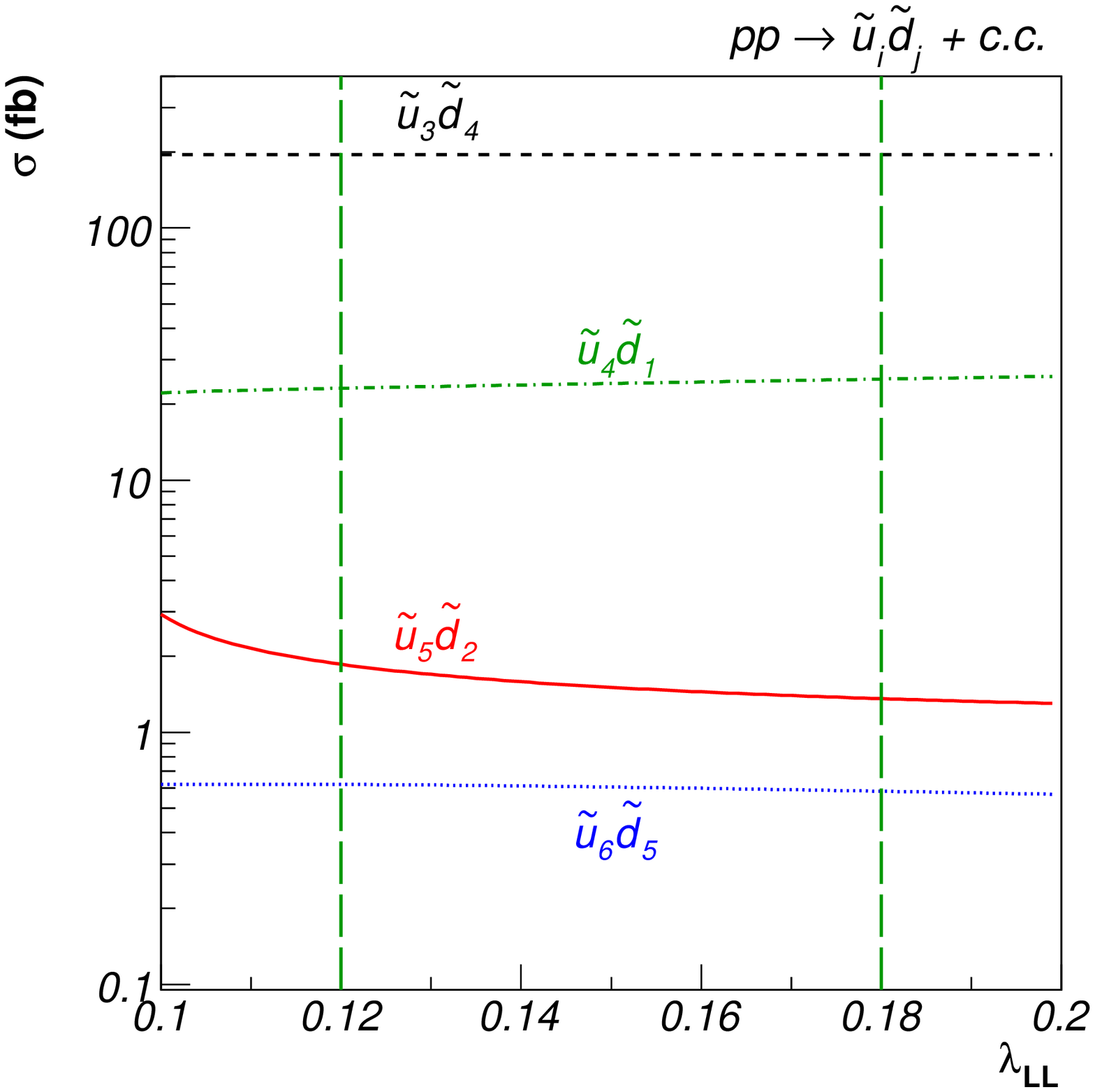} 
    \includegraphics[scale=0.28]{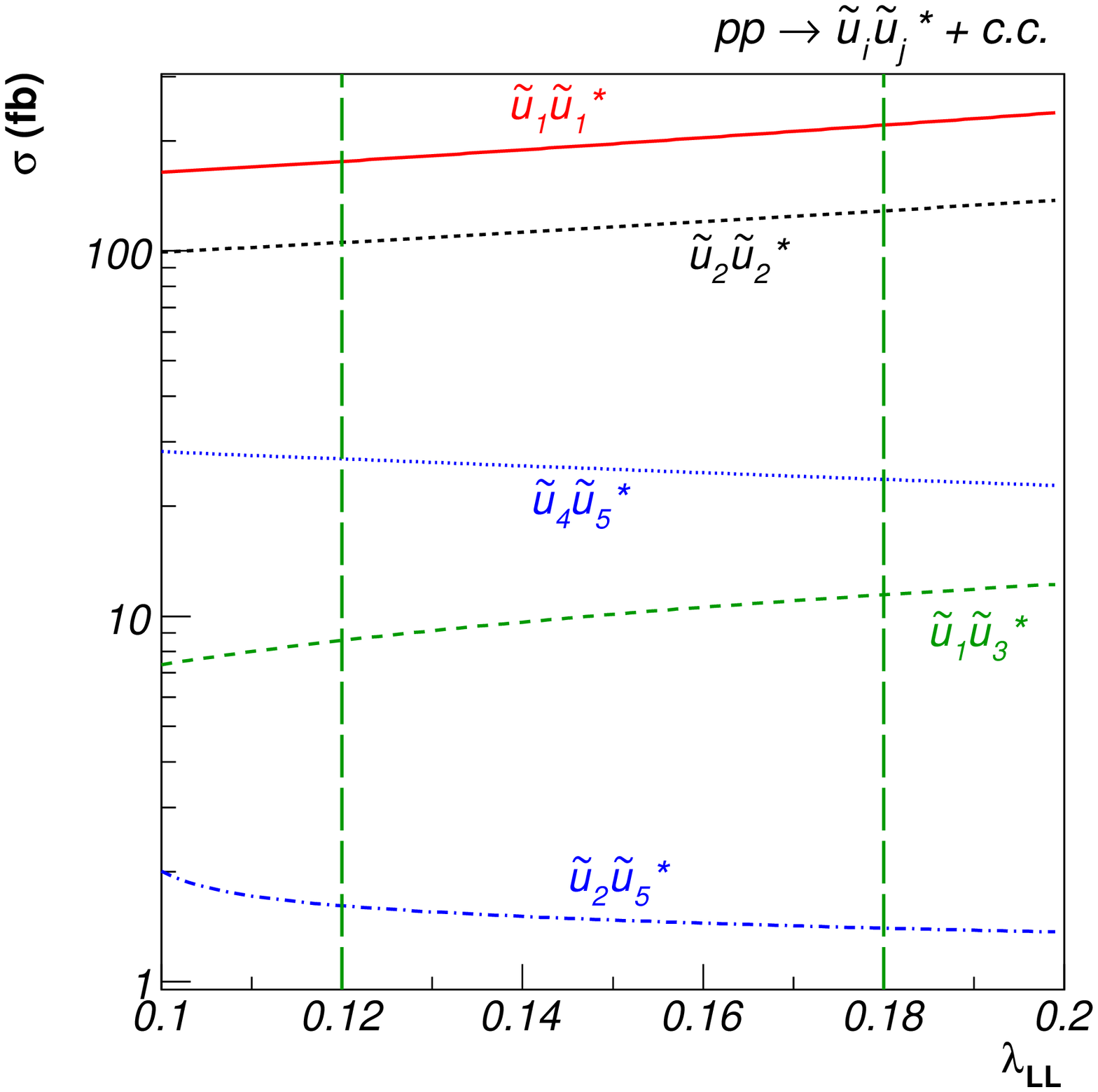} 
    \includegraphics[scale=0.28]{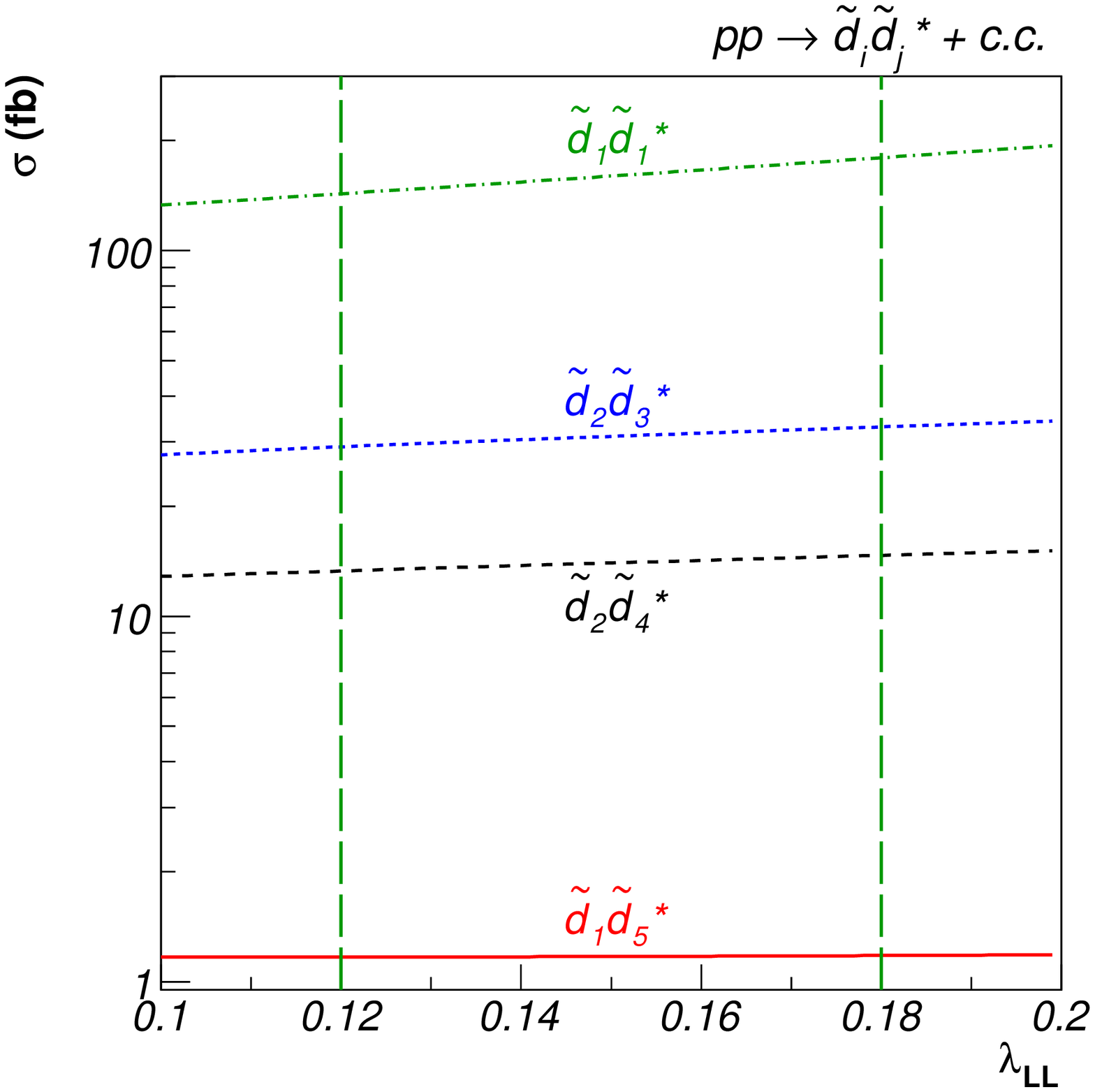} 
    \includegraphics[scale=0.28]{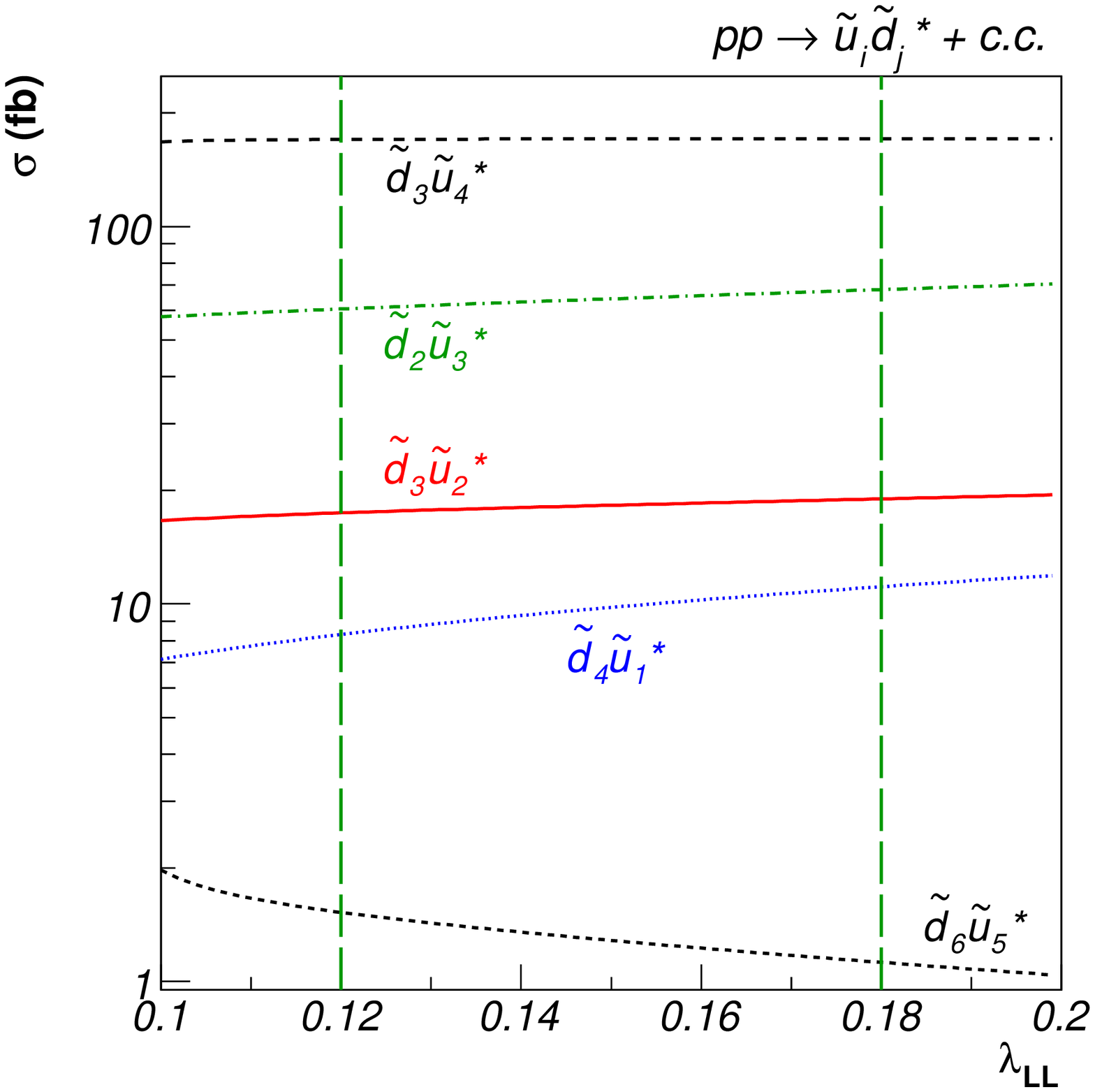} 
    \includegraphics[scale=0.28]{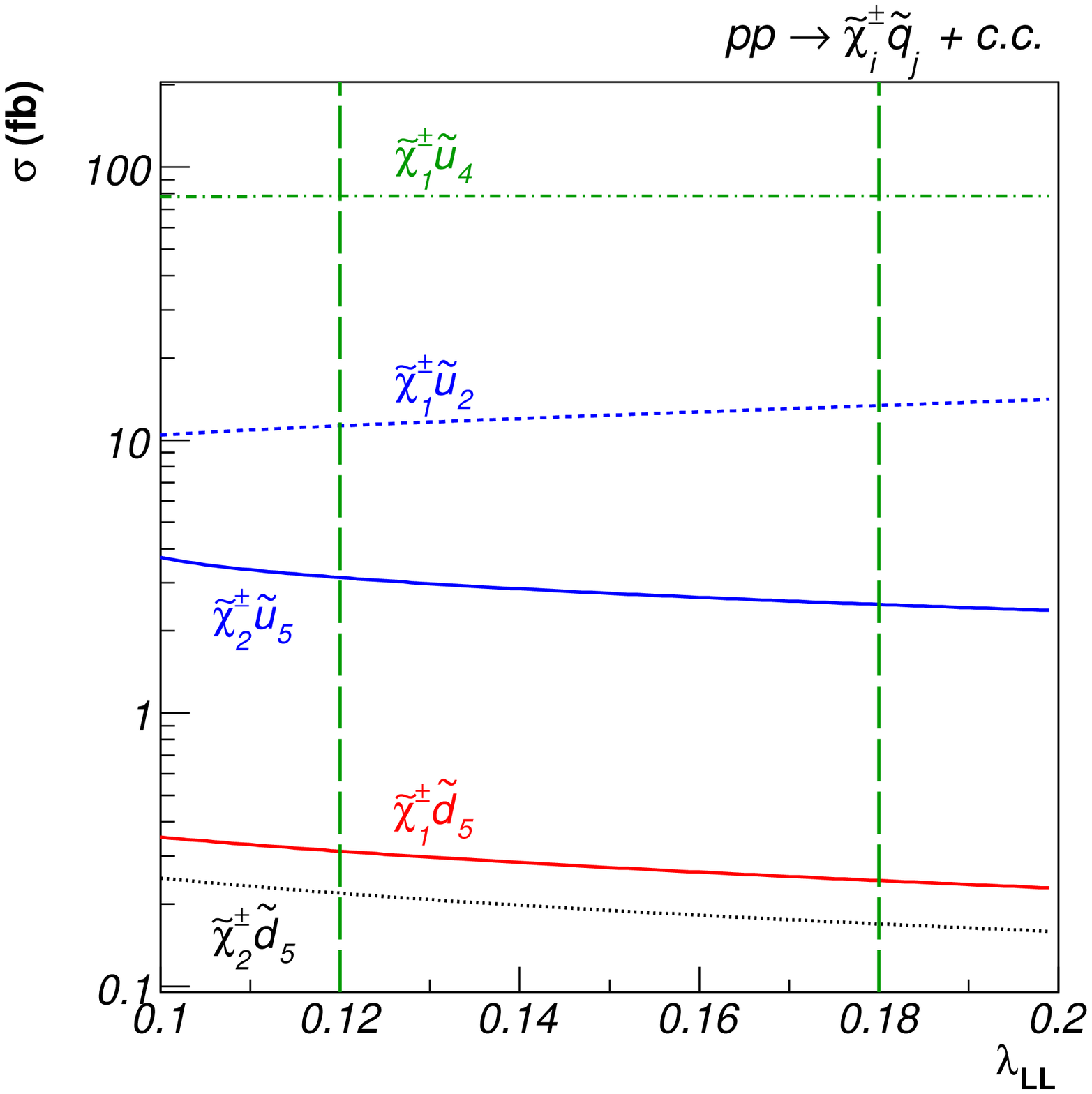} 
    \includegraphics[scale=0.28]{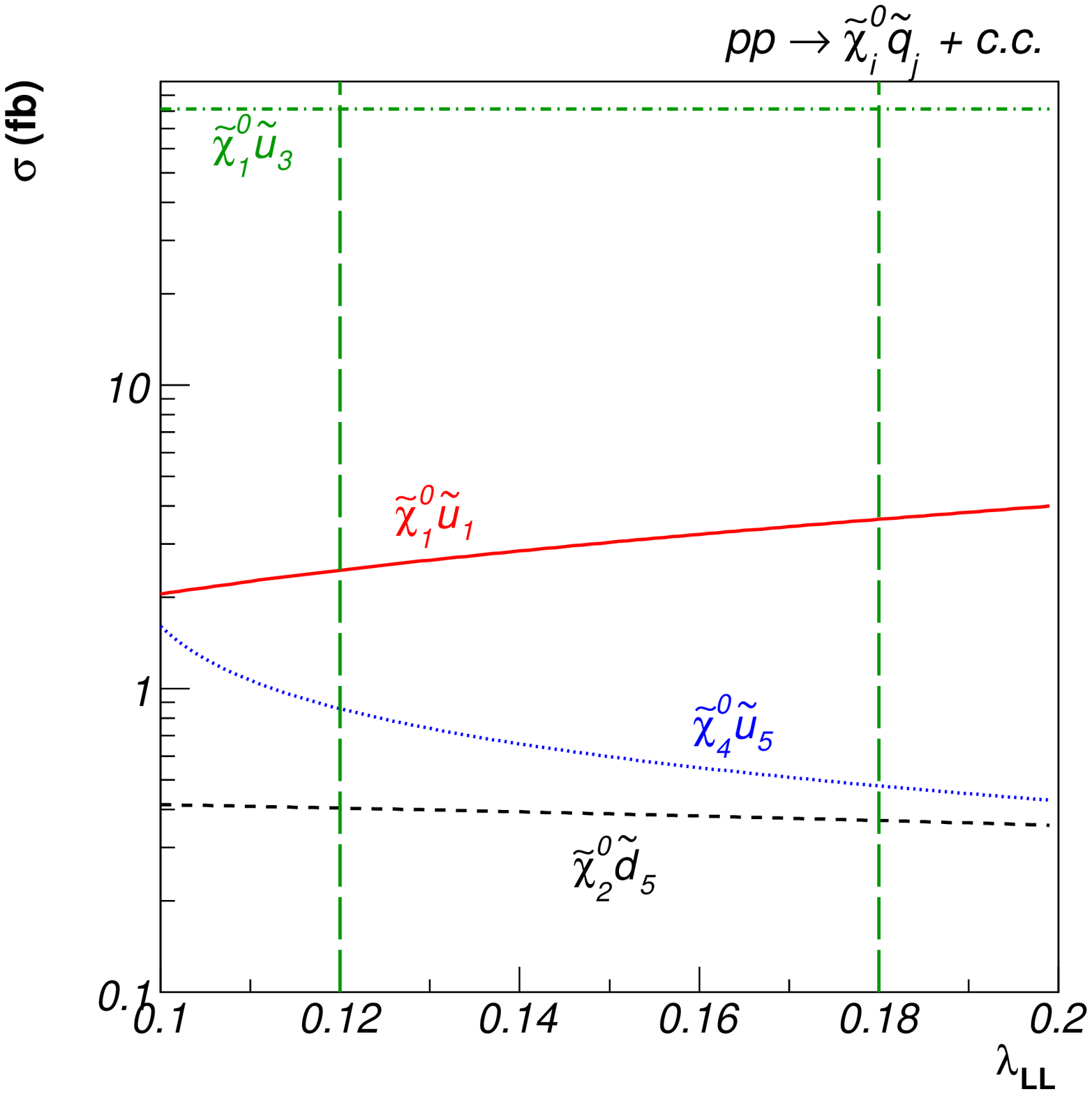} 
    \includegraphics[scale=0.28]{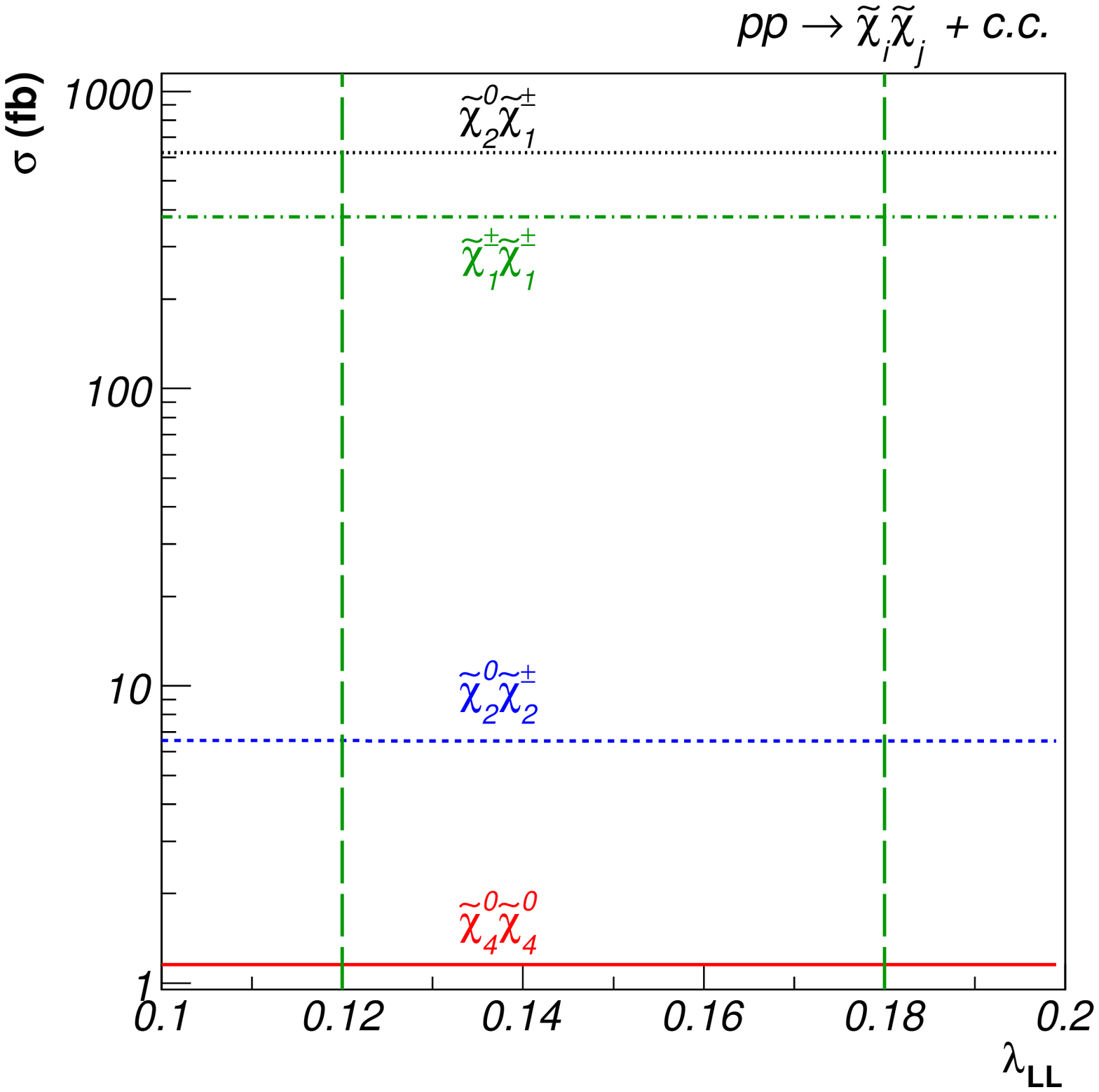} 
  \end{center}
  \vspace*{-5mm}
\caption{Same as Fig.\ \ref{fig16} for our benchmark scenario F.}
\label{fig18}
\end{figure}

\begin{figure}
  \begin{center}
    \includegraphics[scale=0.28]{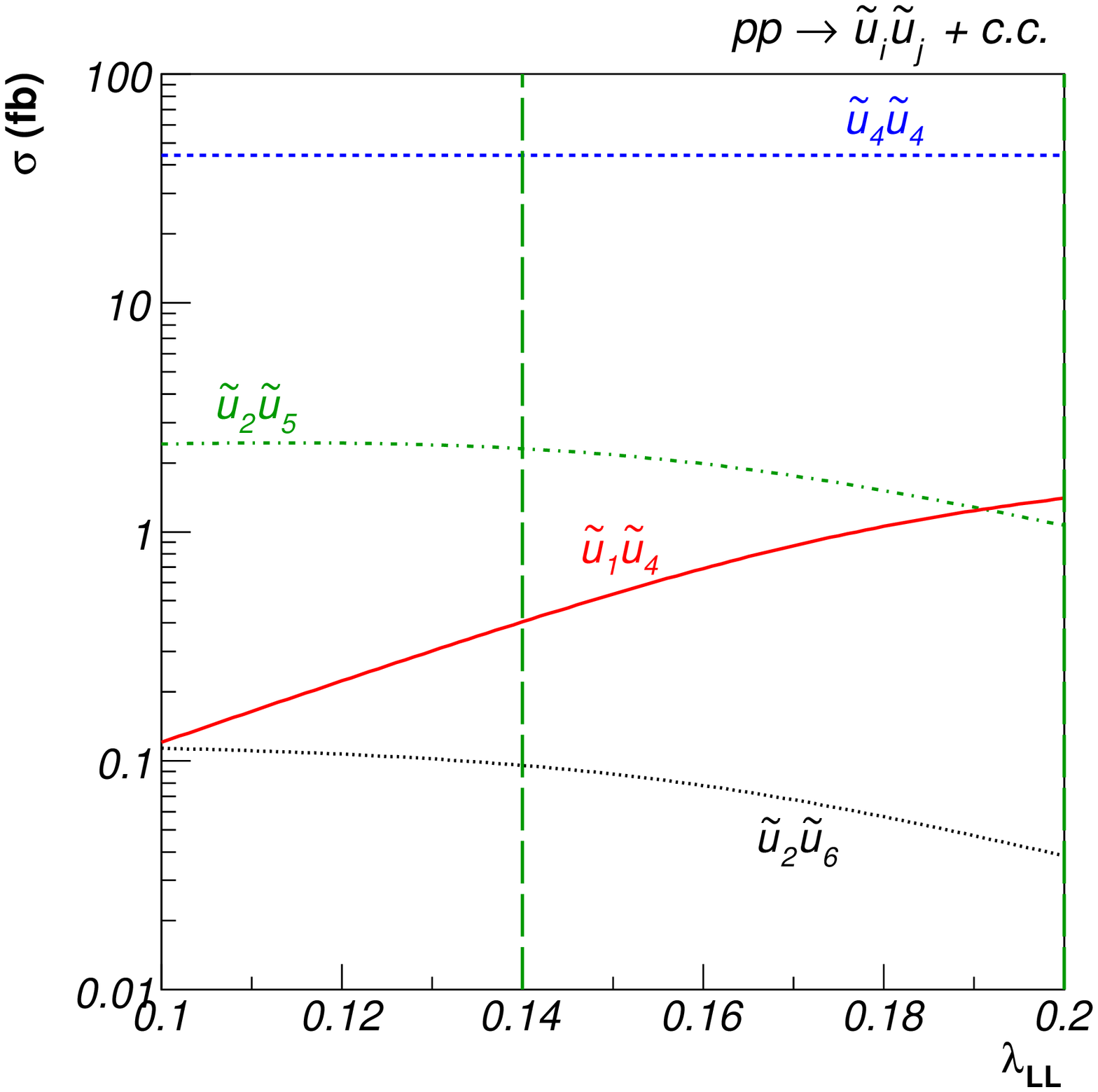} 
    \includegraphics[scale=0.28]{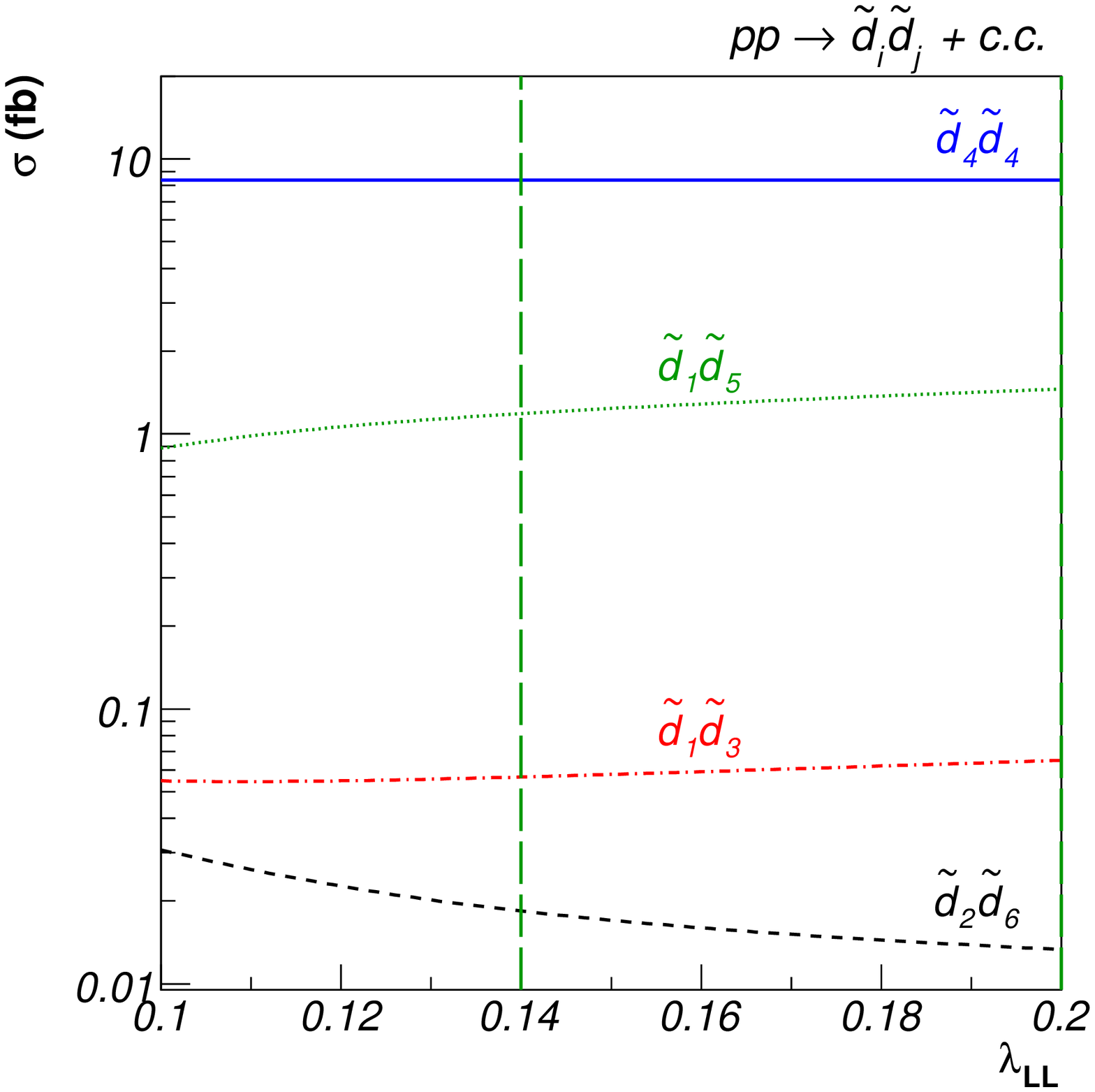} 
    \includegraphics[scale=0.28]{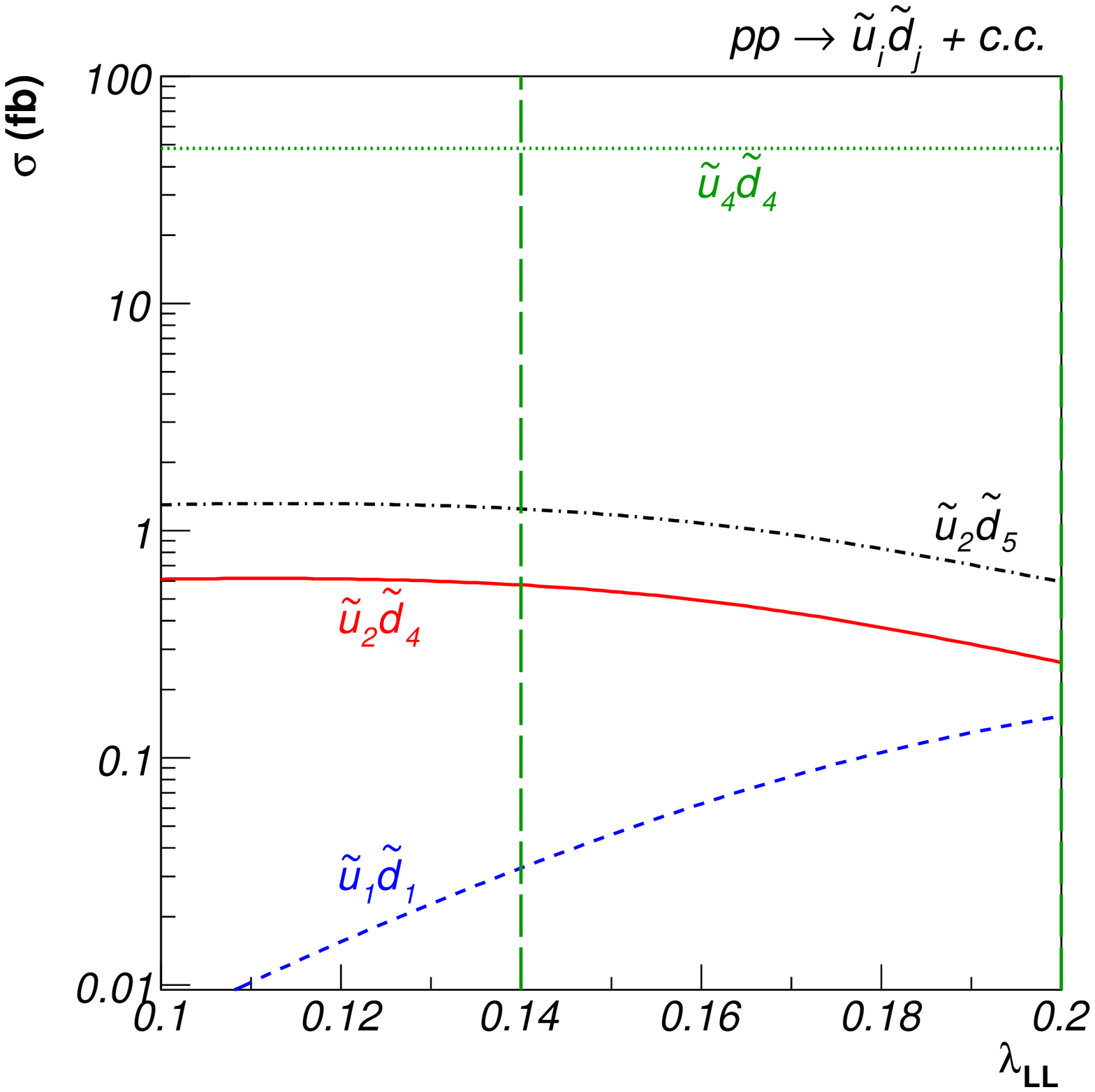} 
    \includegraphics[scale=0.28]{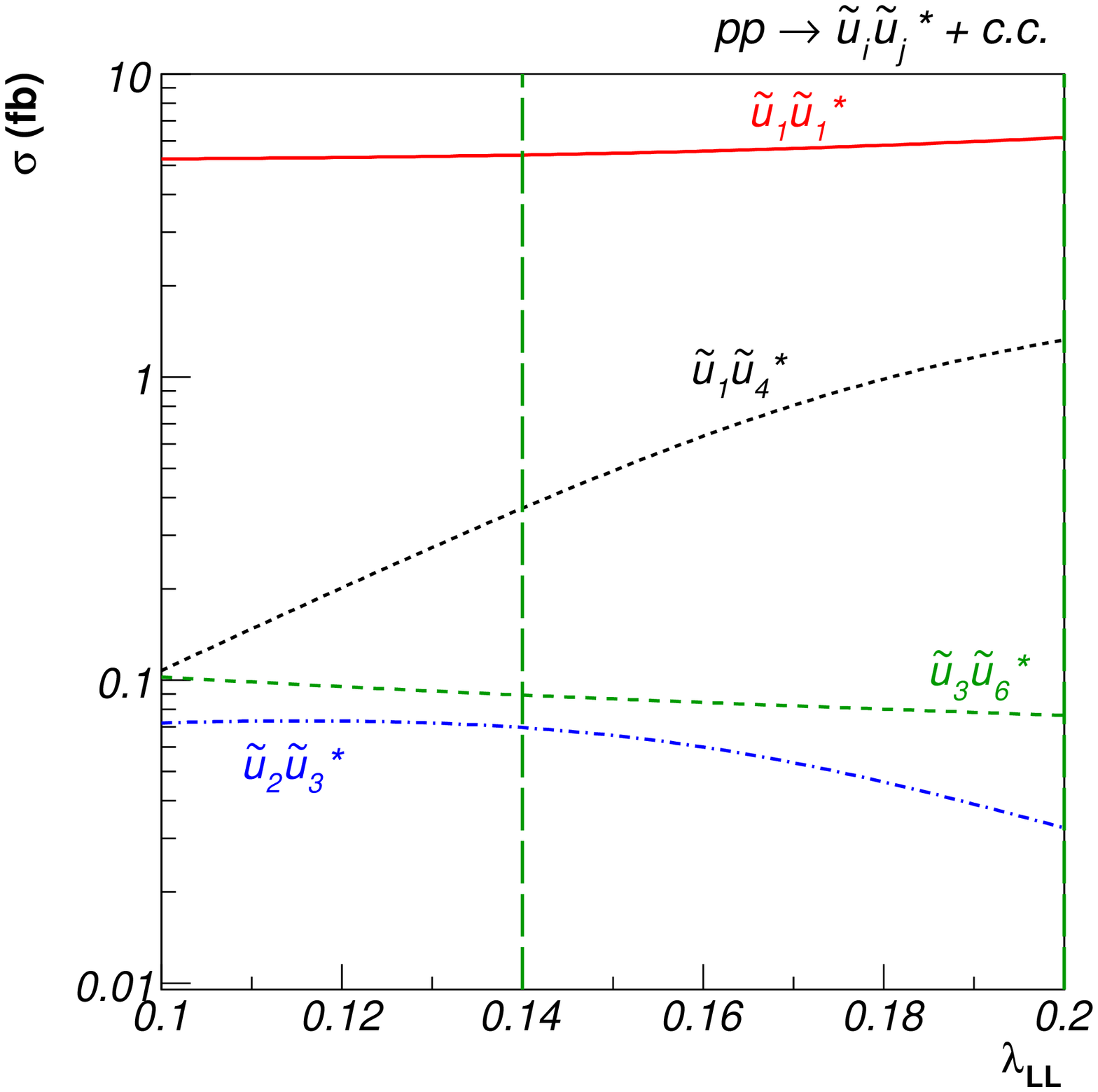} 
    \includegraphics[scale=0.28]{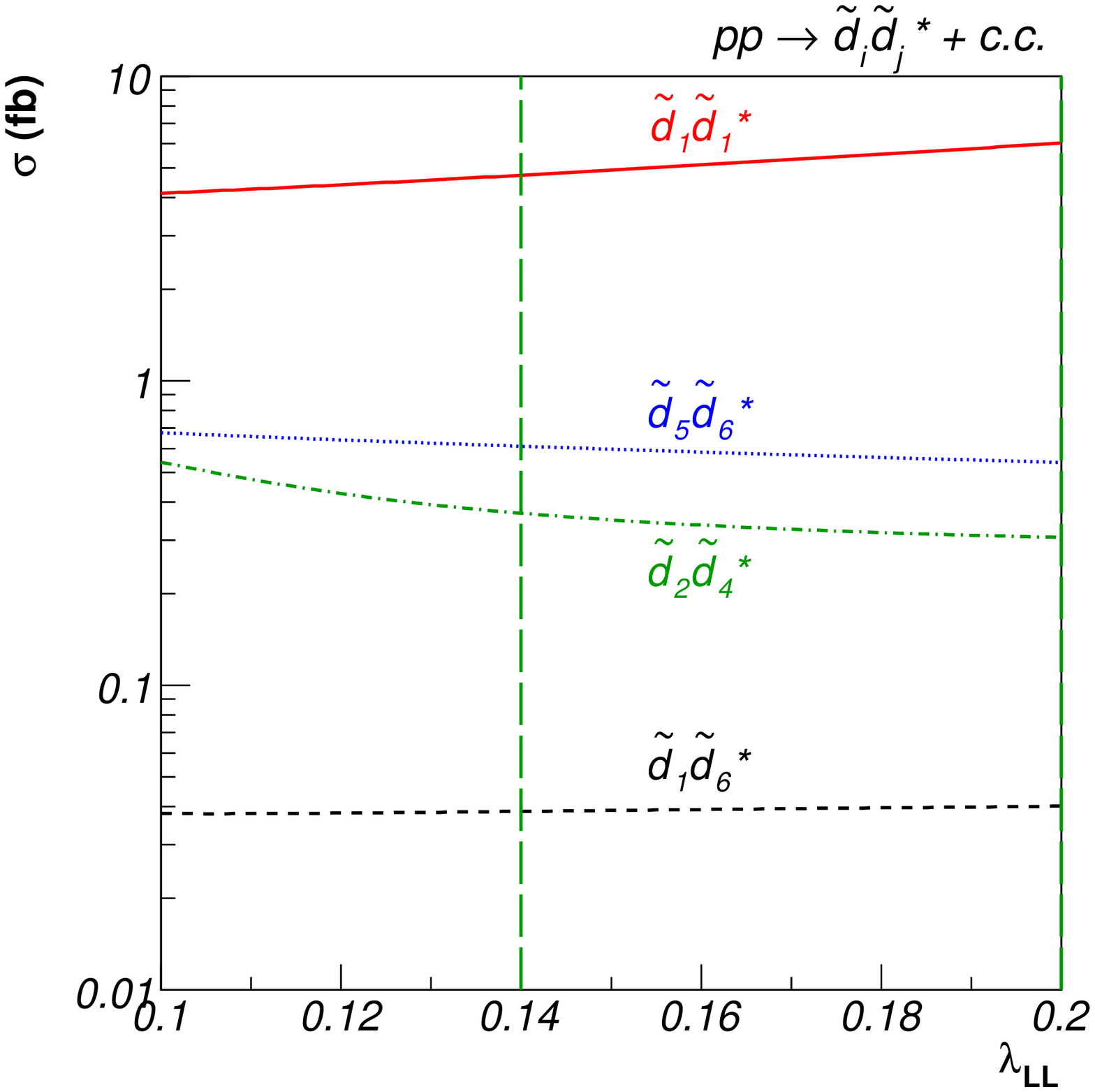} 
    \includegraphics[scale=0.28]{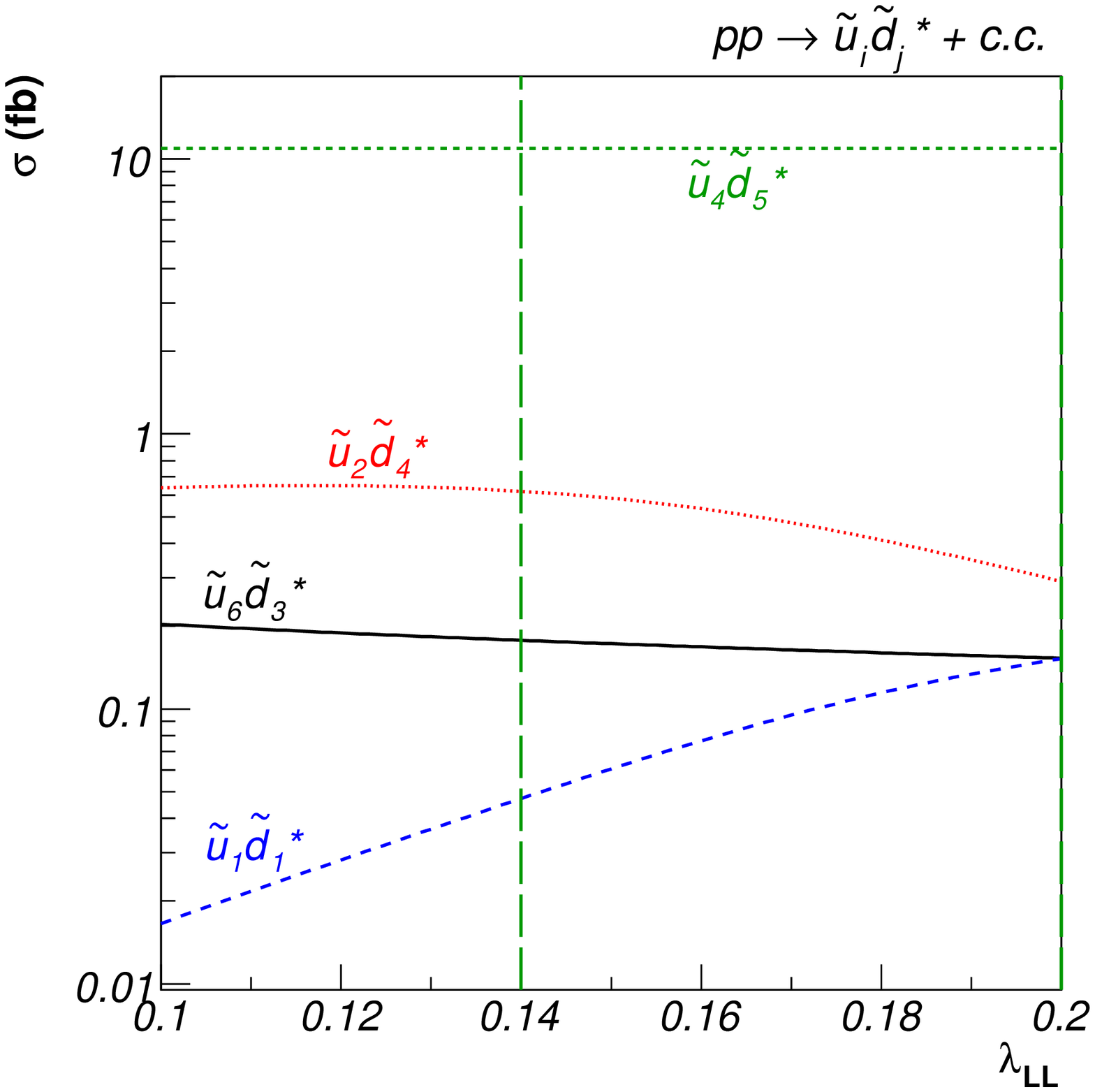} 
    \includegraphics[scale=0.28]{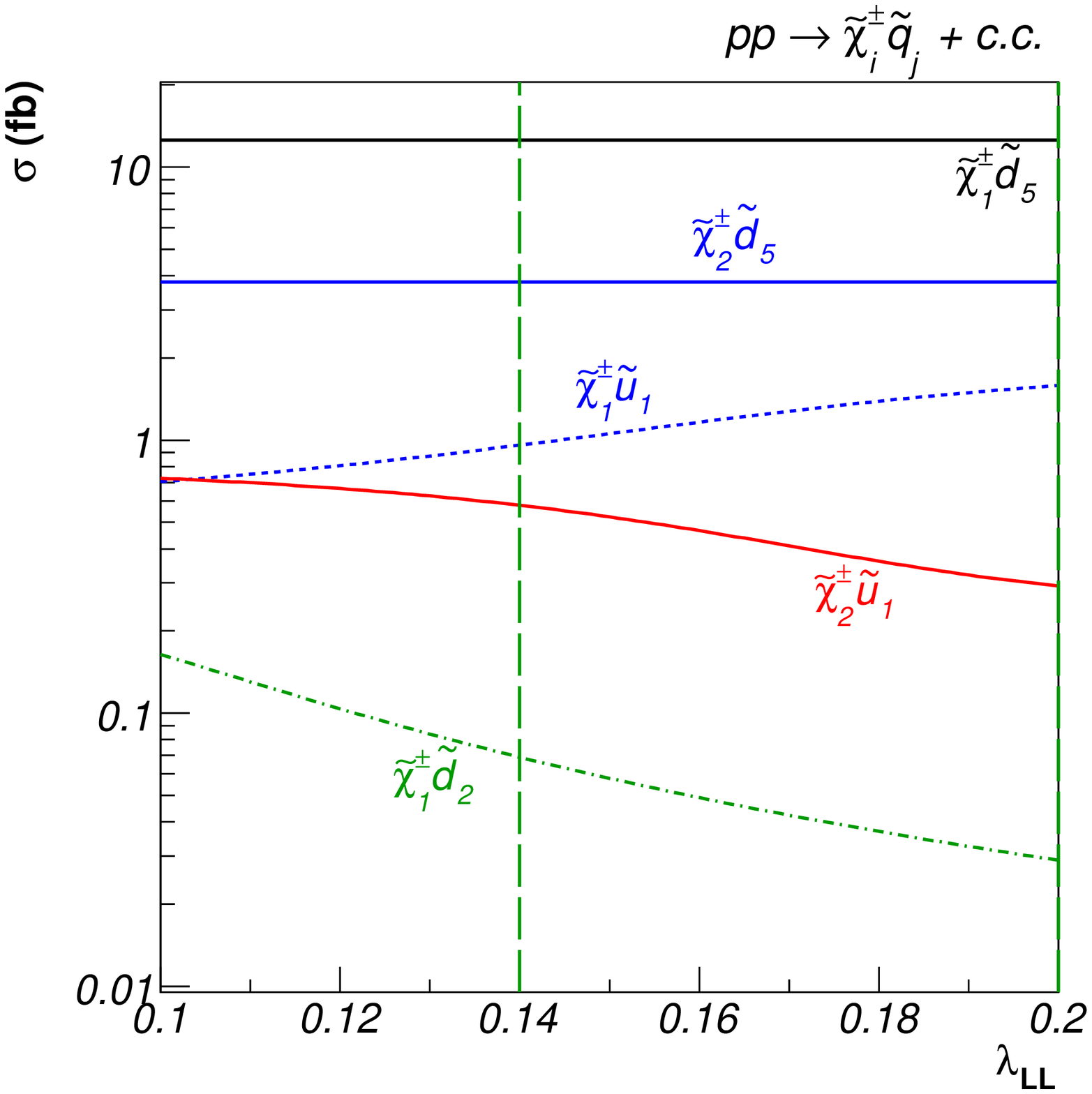} 
    \includegraphics[scale=0.28]{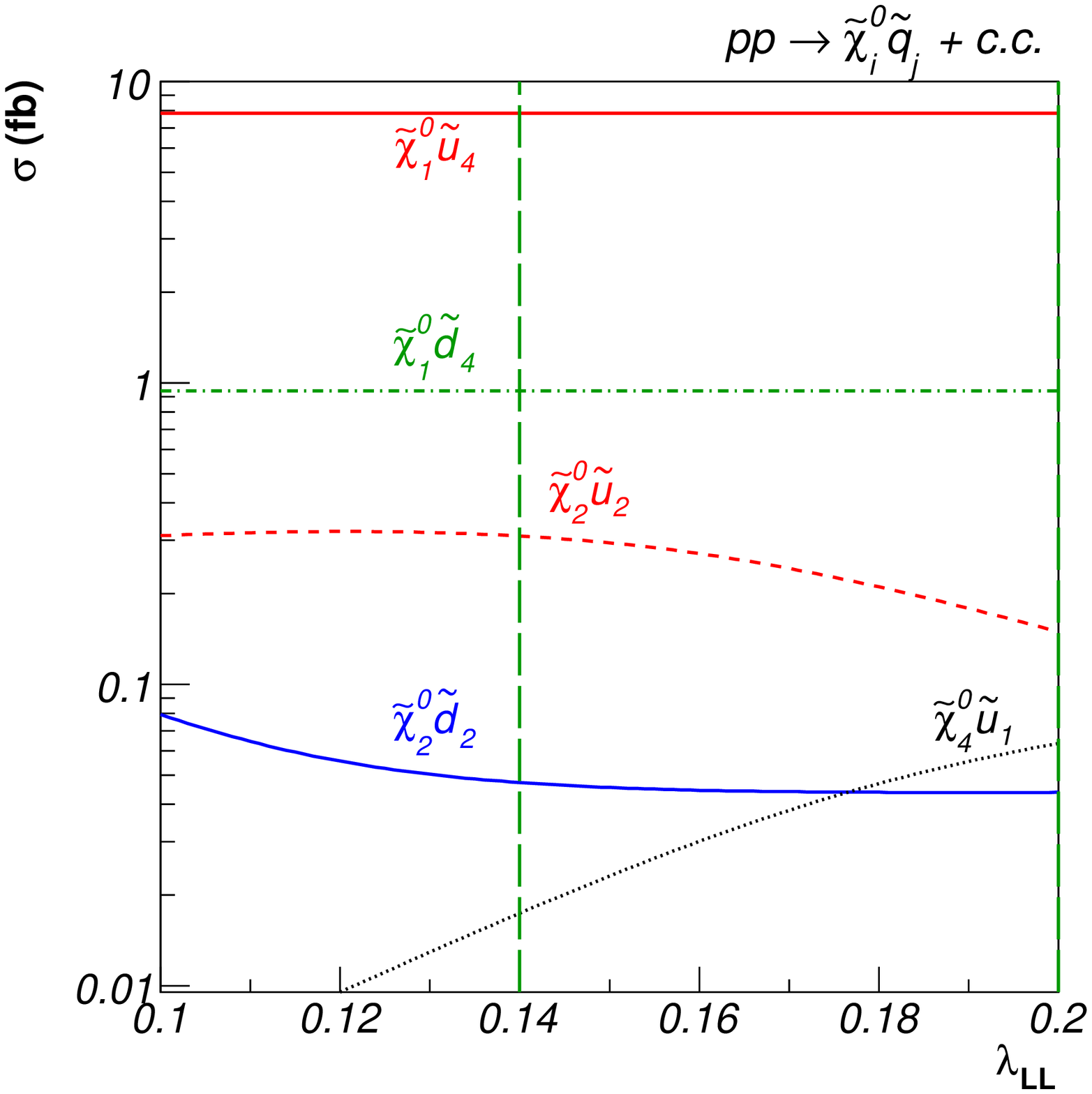} 
    \includegraphics[scale=0.28]{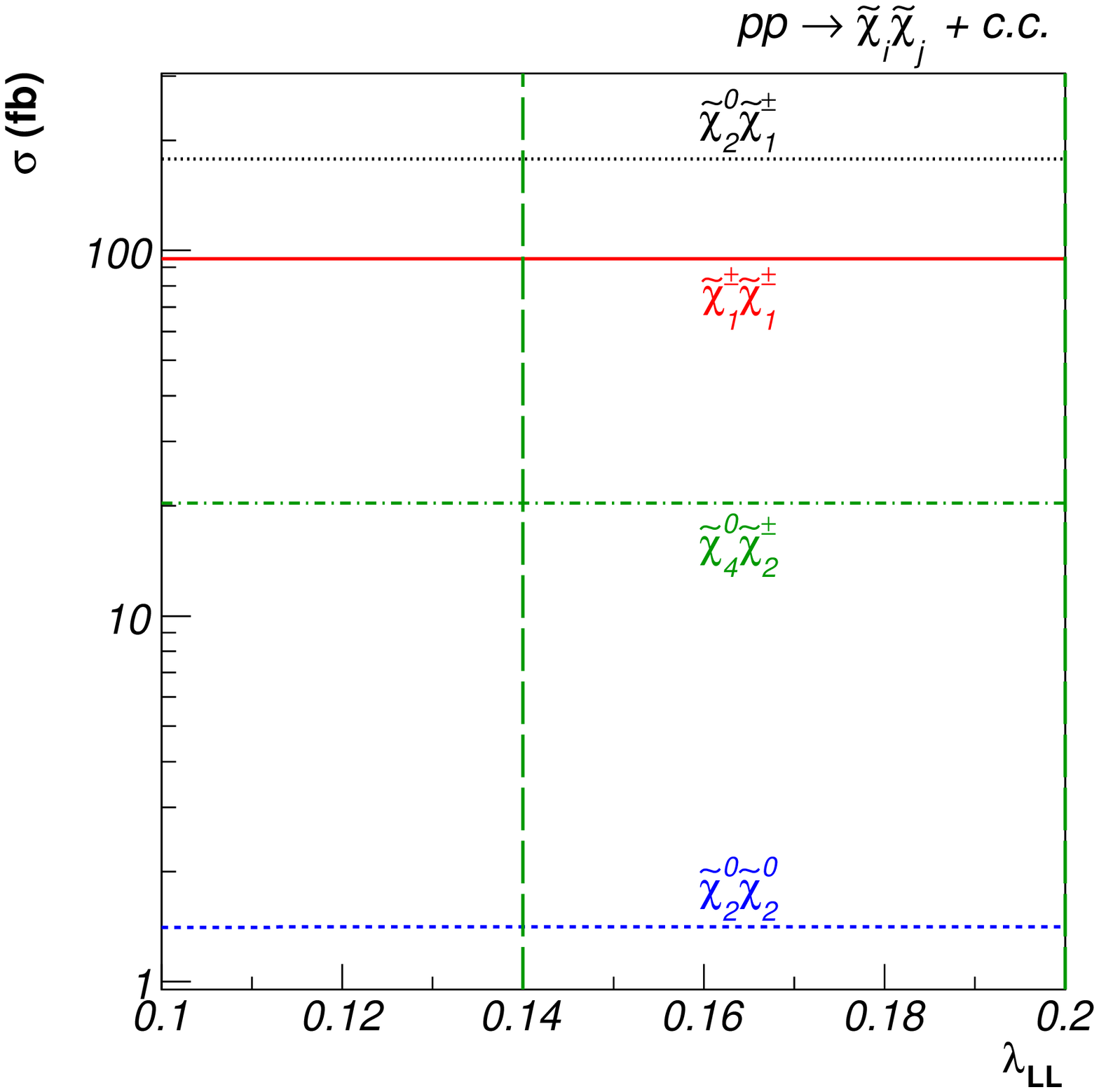} 
  \end{center}
  \vspace*{-5mm}
\caption{Same as Fig.\ \ref{fig15} for our benchmark scenario G.}
\label{fig19}
\end{figure}

\begin{figure}
  \begin{center}
    \includegraphics[scale=0.28]{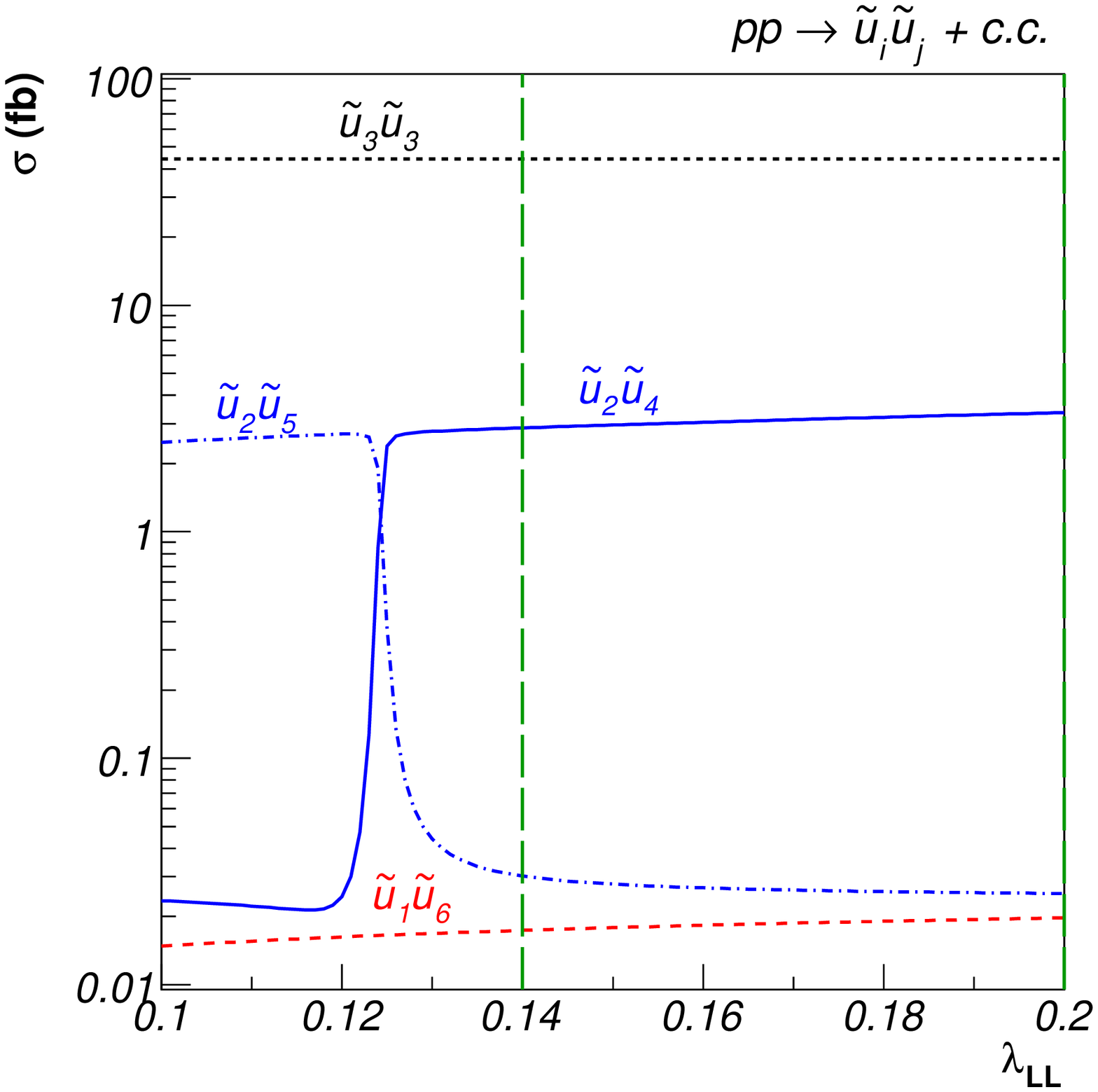} 
    \includegraphics[scale=0.28]{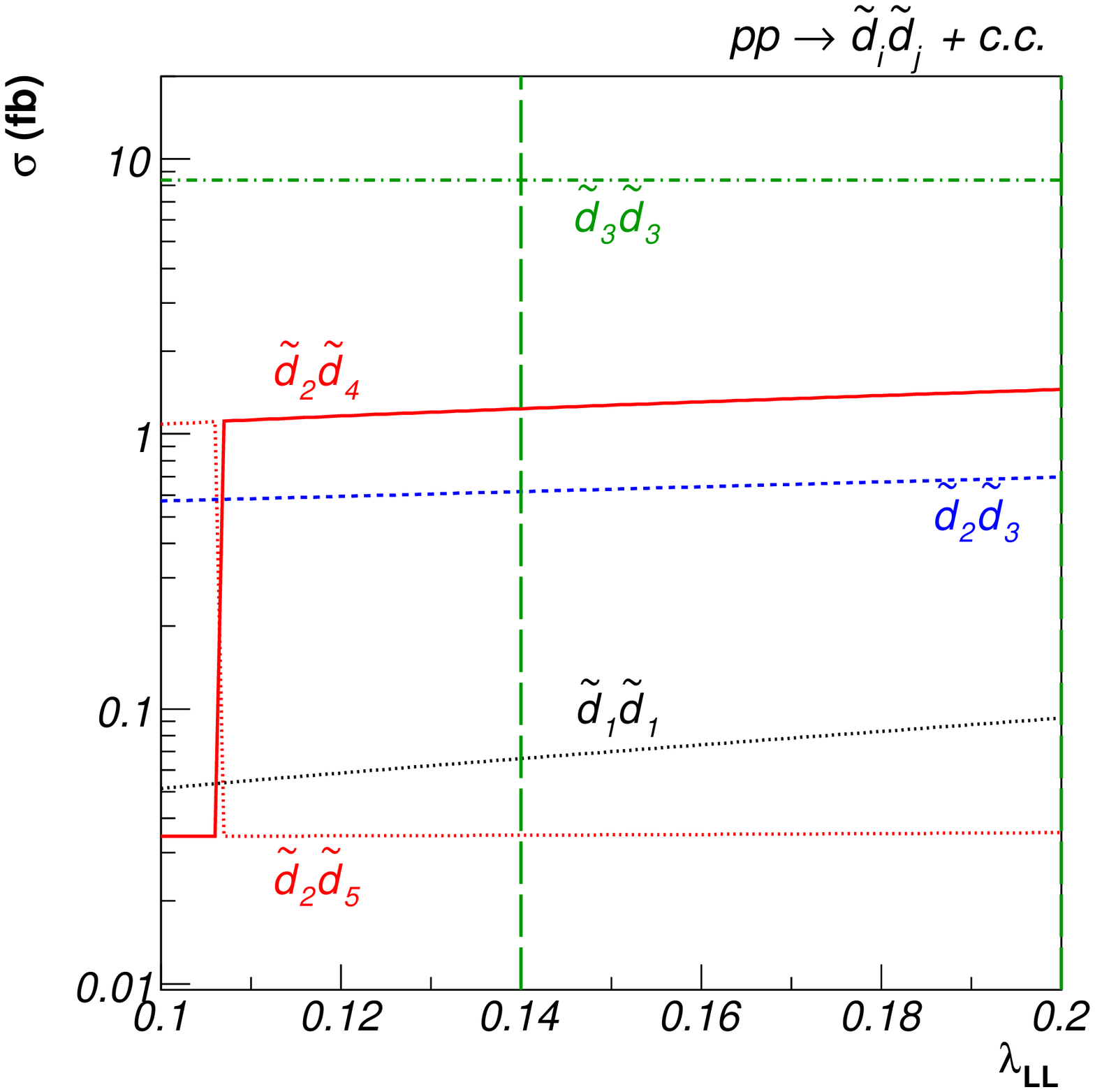} 
    \includegraphics[scale=0.28]{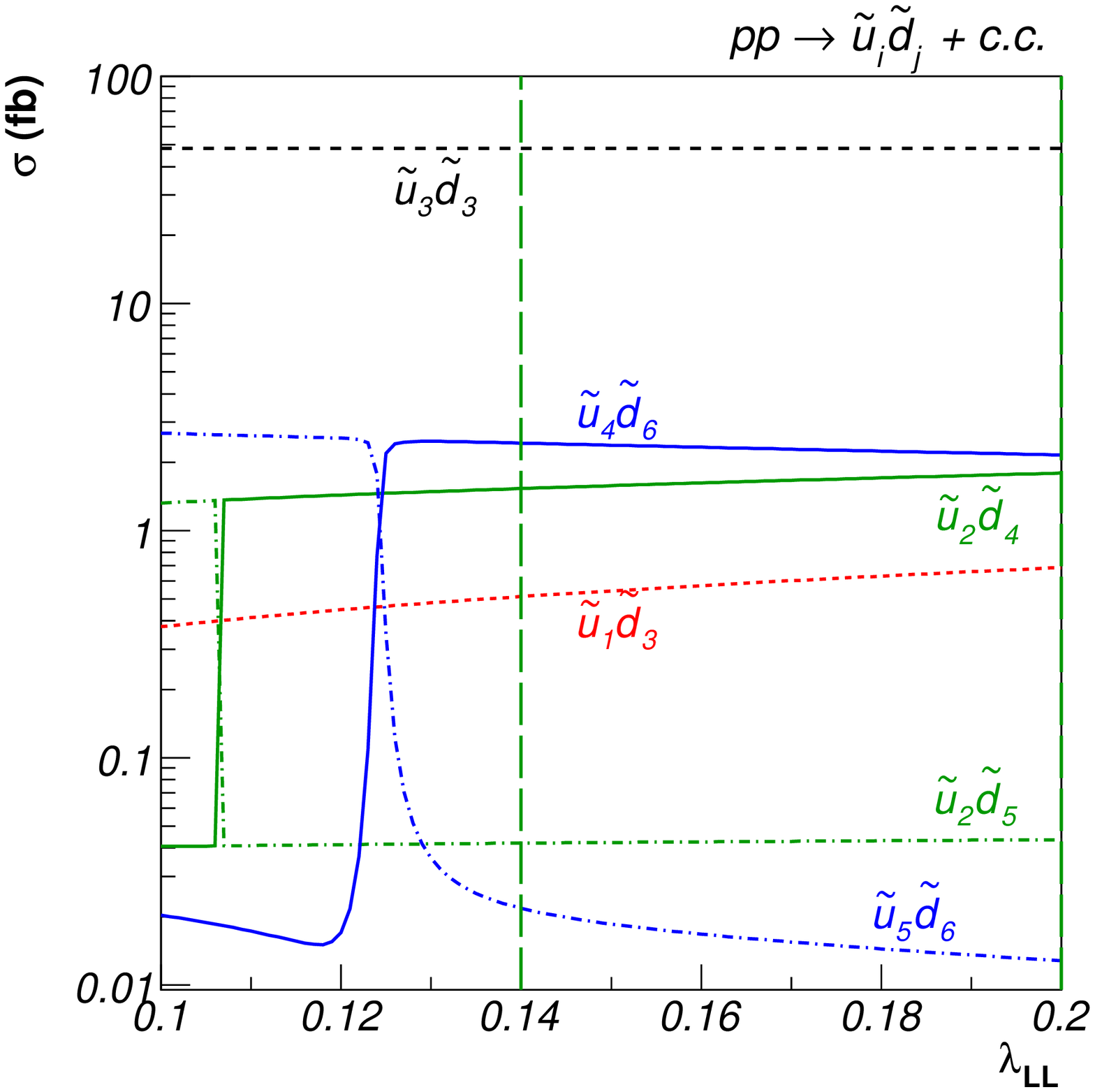} 
    \includegraphics[scale=0.28]{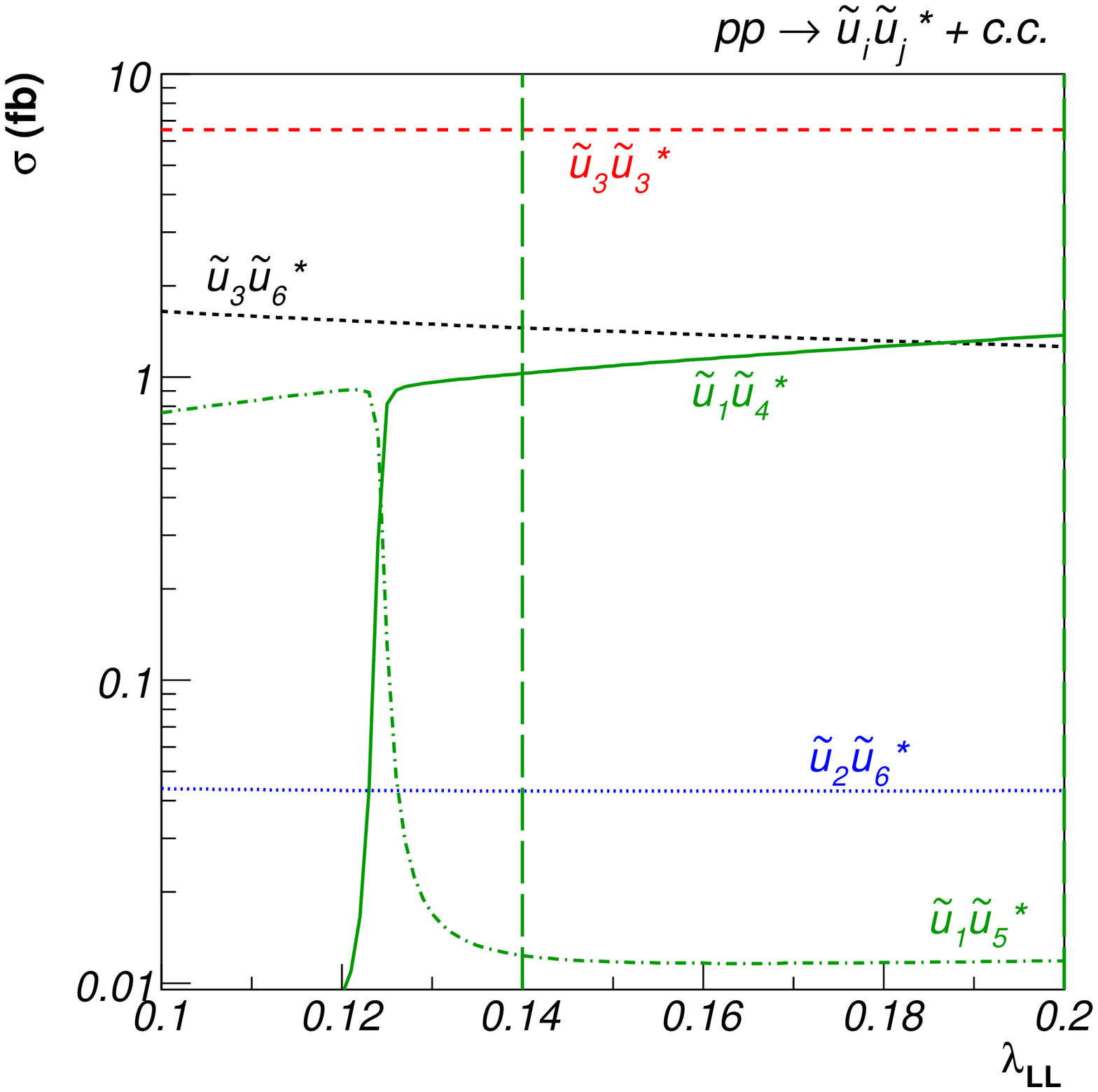} 
    \includegraphics[scale=0.28]{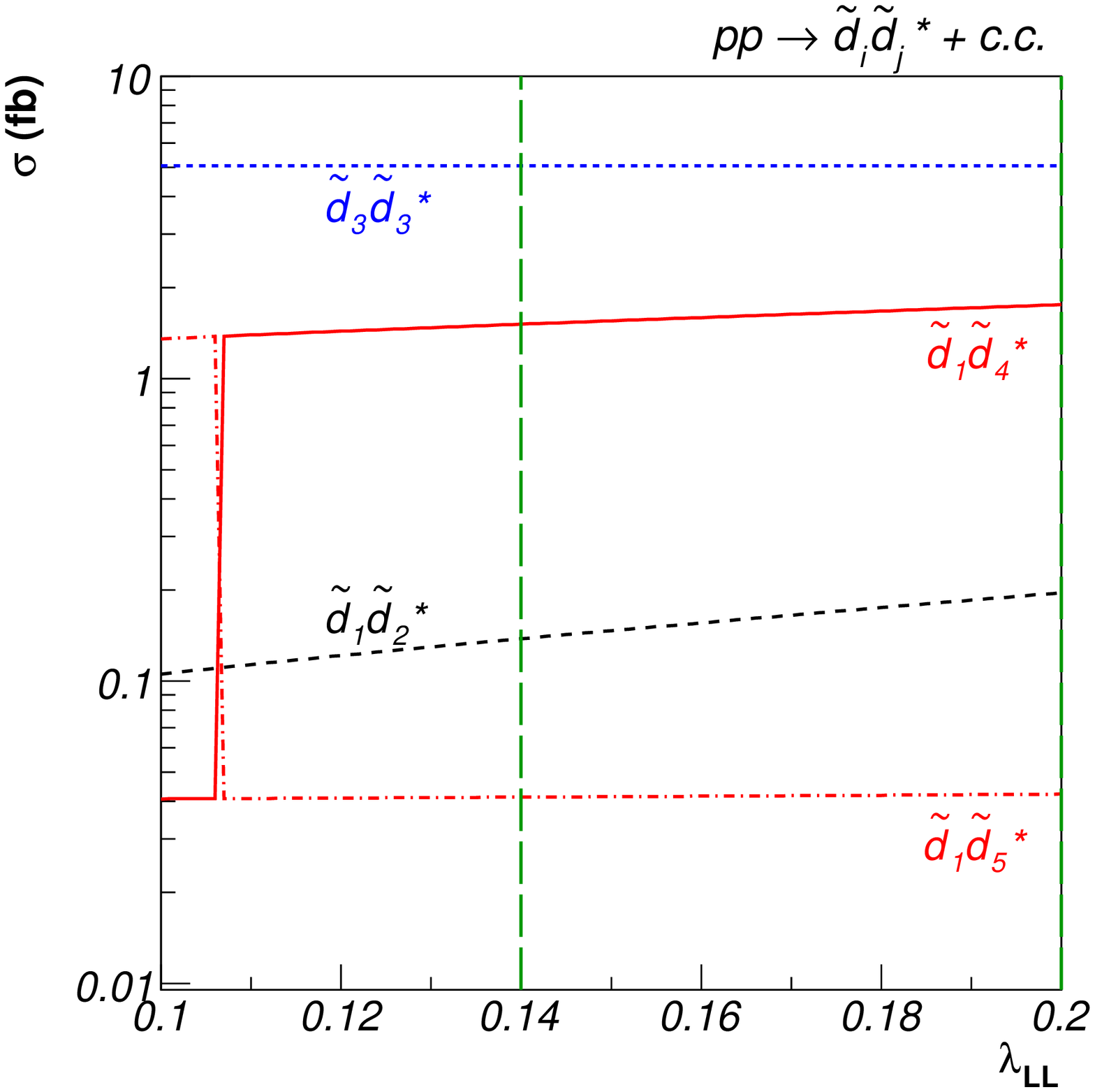} 
    \includegraphics[scale=0.28]{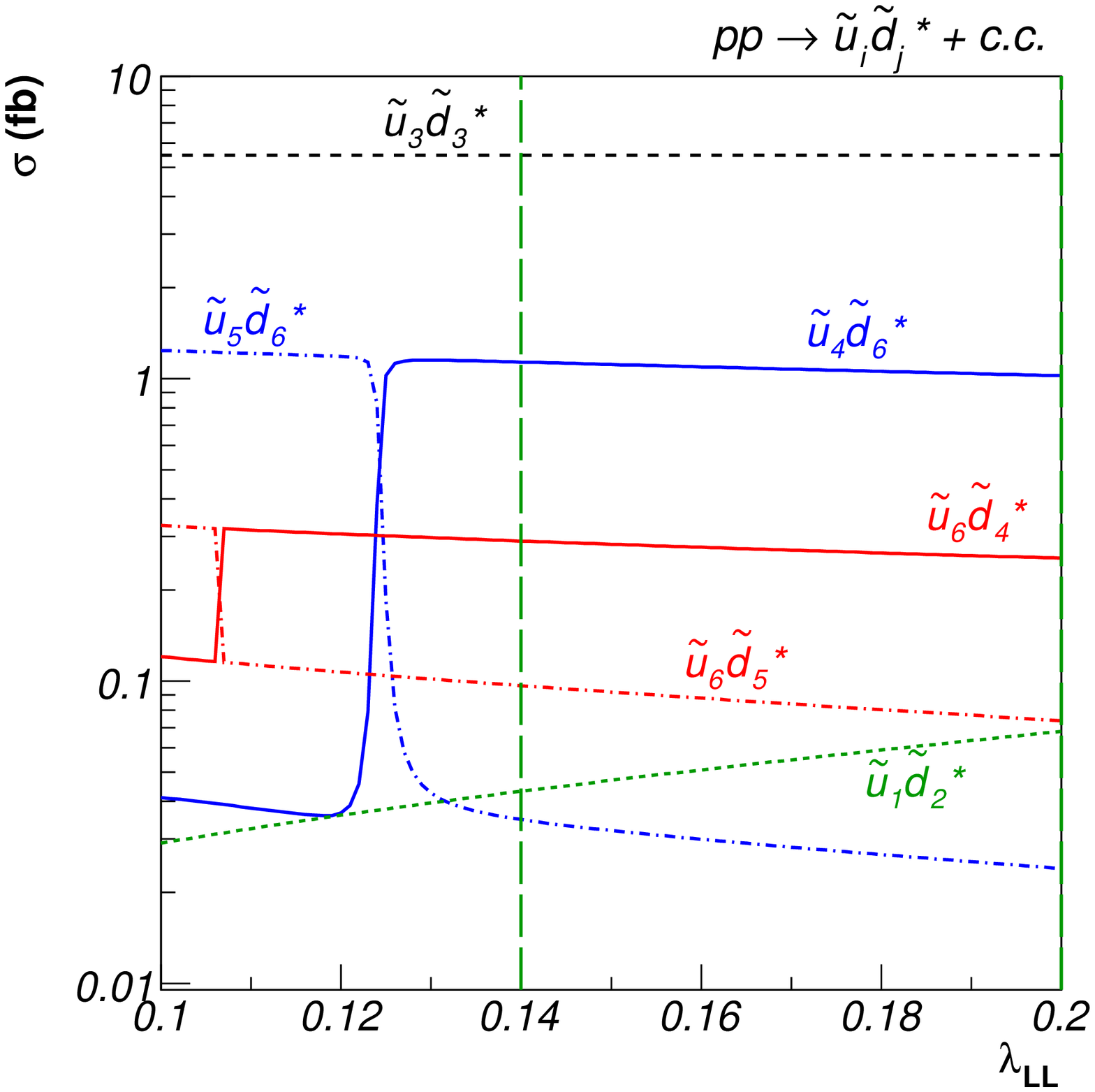} 
    \includegraphics[scale=0.28]{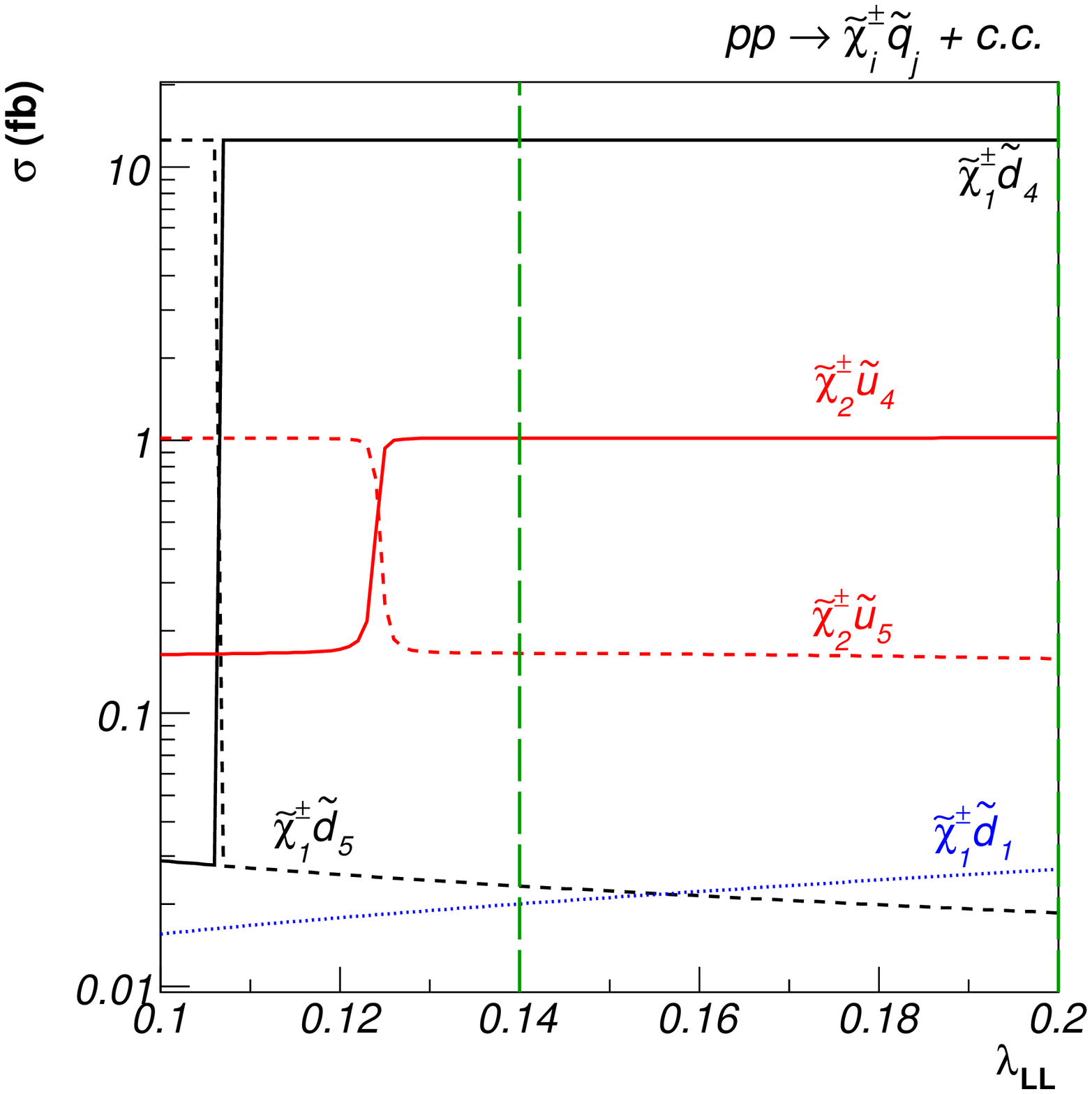} 
    \includegraphics[scale=0.28]{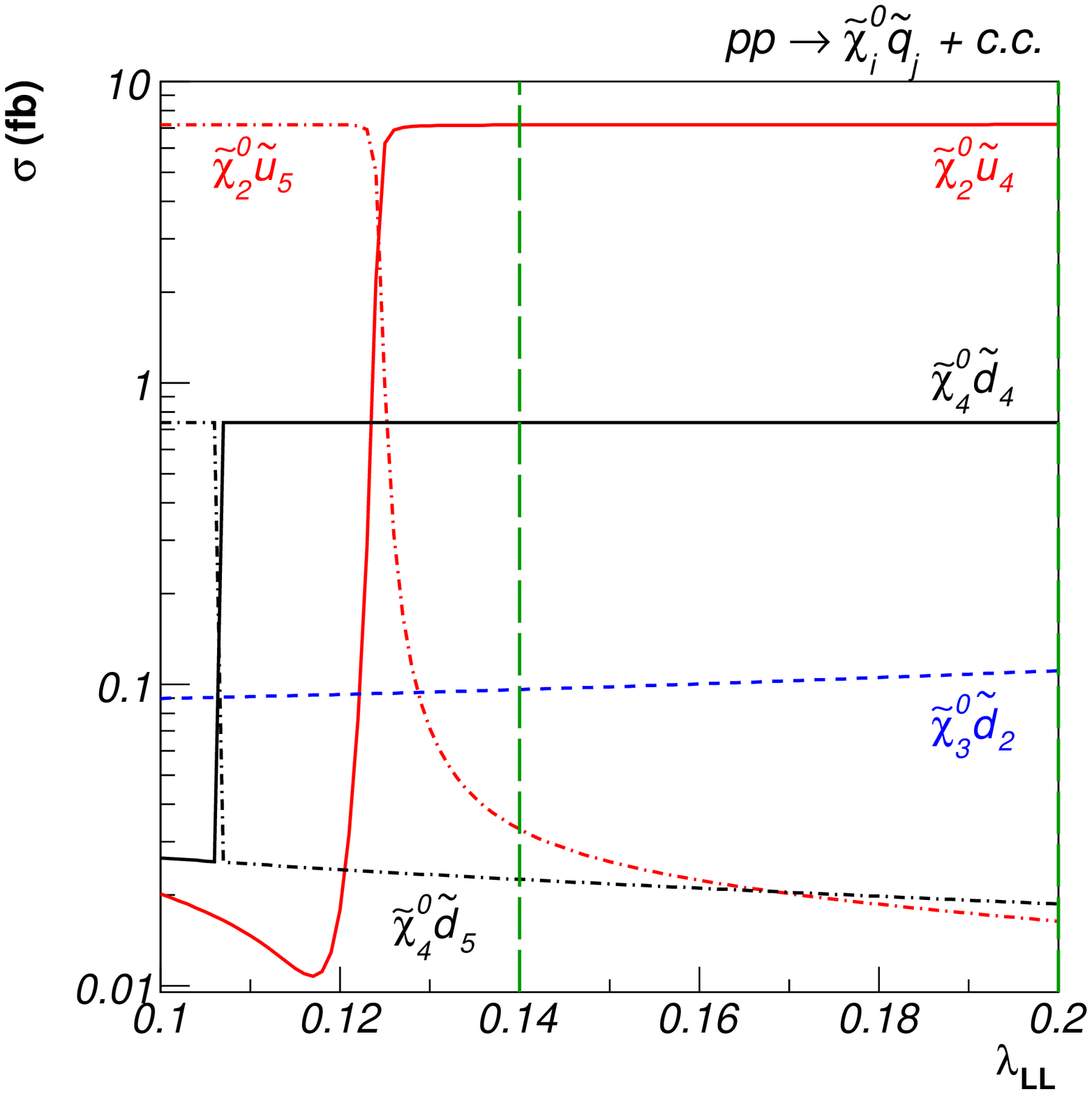} 
    \includegraphics[scale=0.28]{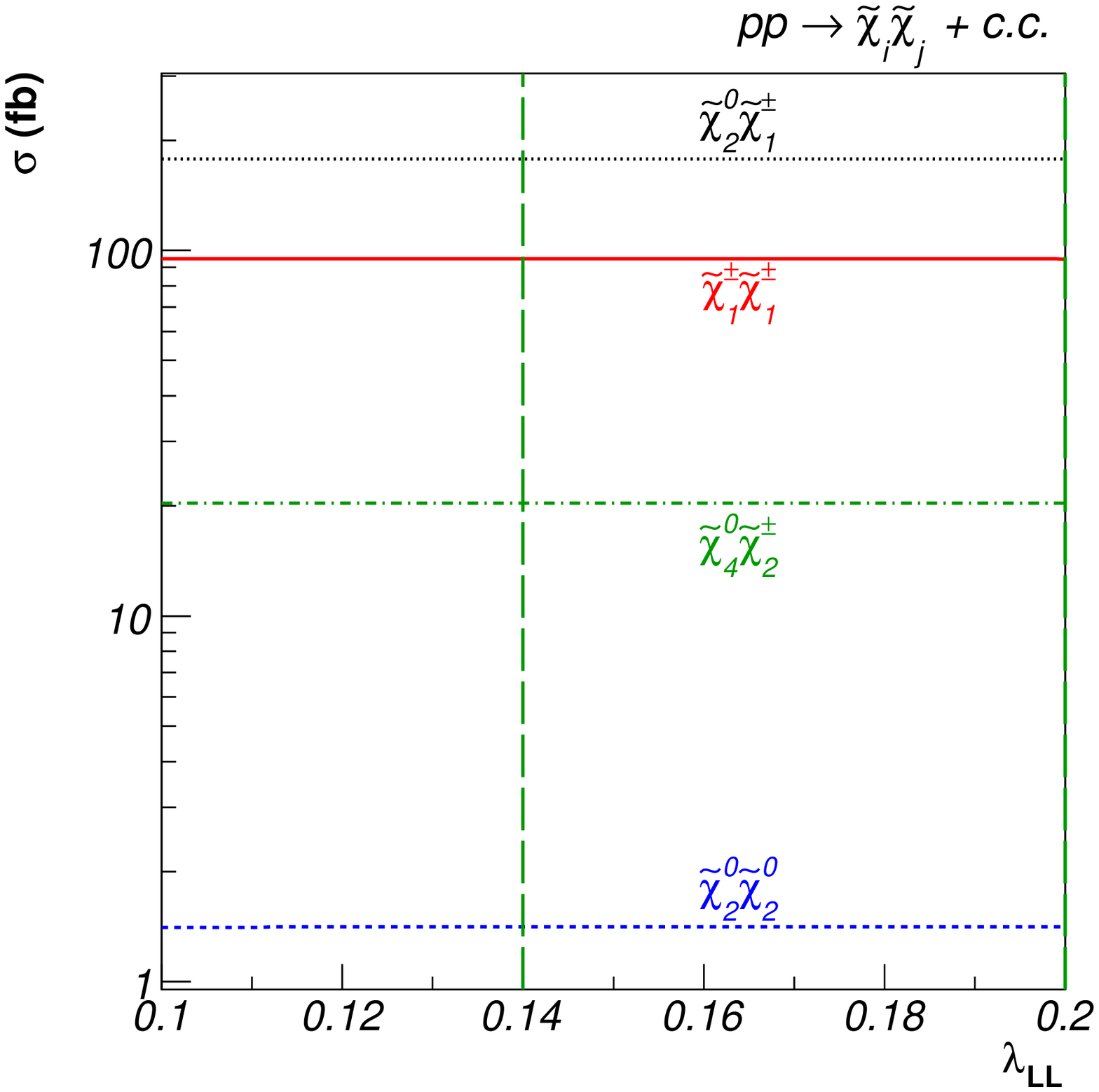} 
  \end{center}
  \vspace*{-5mm}
\caption{Same as Fig.\ \ref{fig16} for our benchmark scenario G.}
\label{fig20}
\end{figure}

\begin{figure}
  \begin{center}
    \includegraphics[scale=0.28]{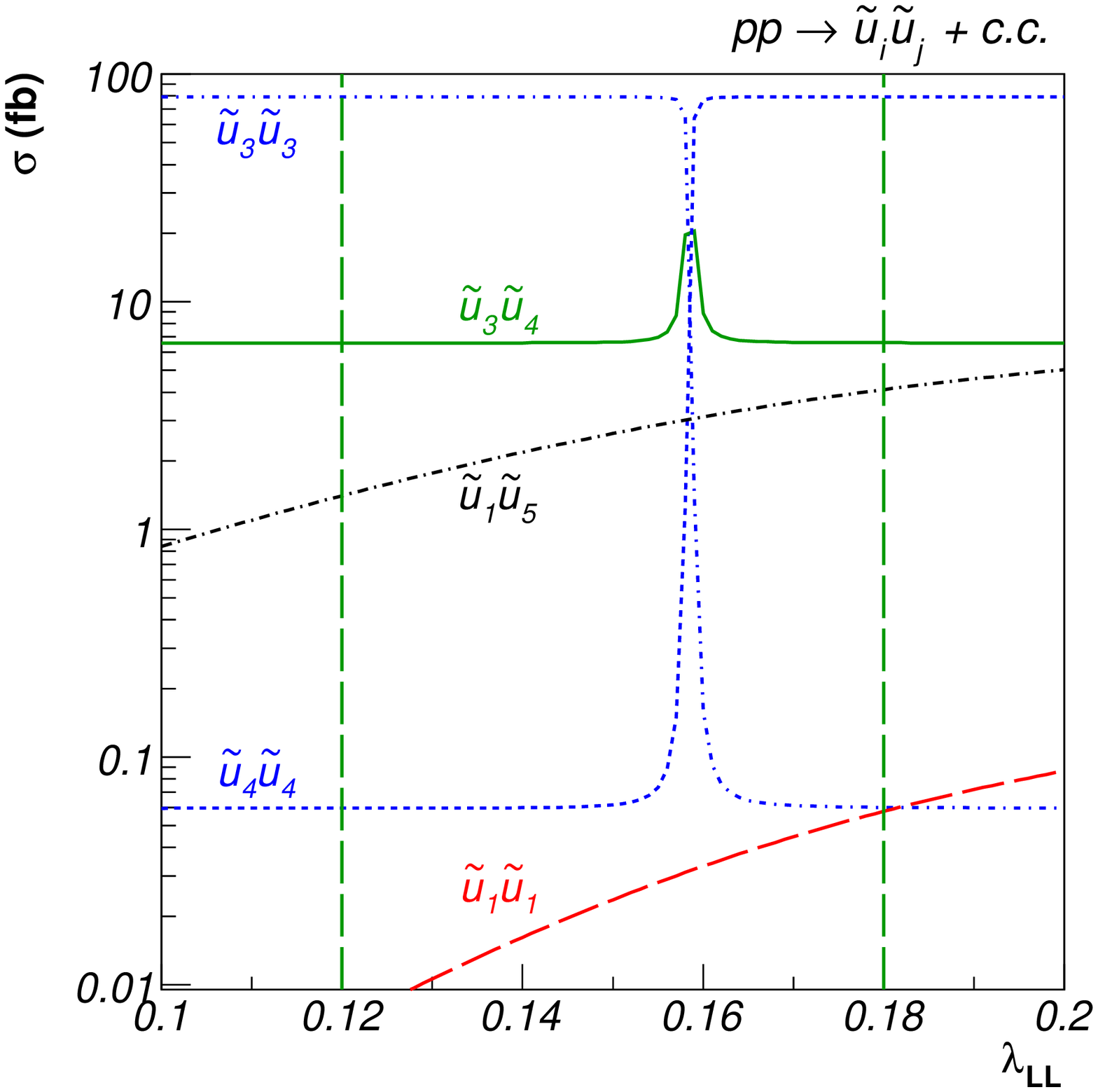} 
    \includegraphics[scale=0.28]{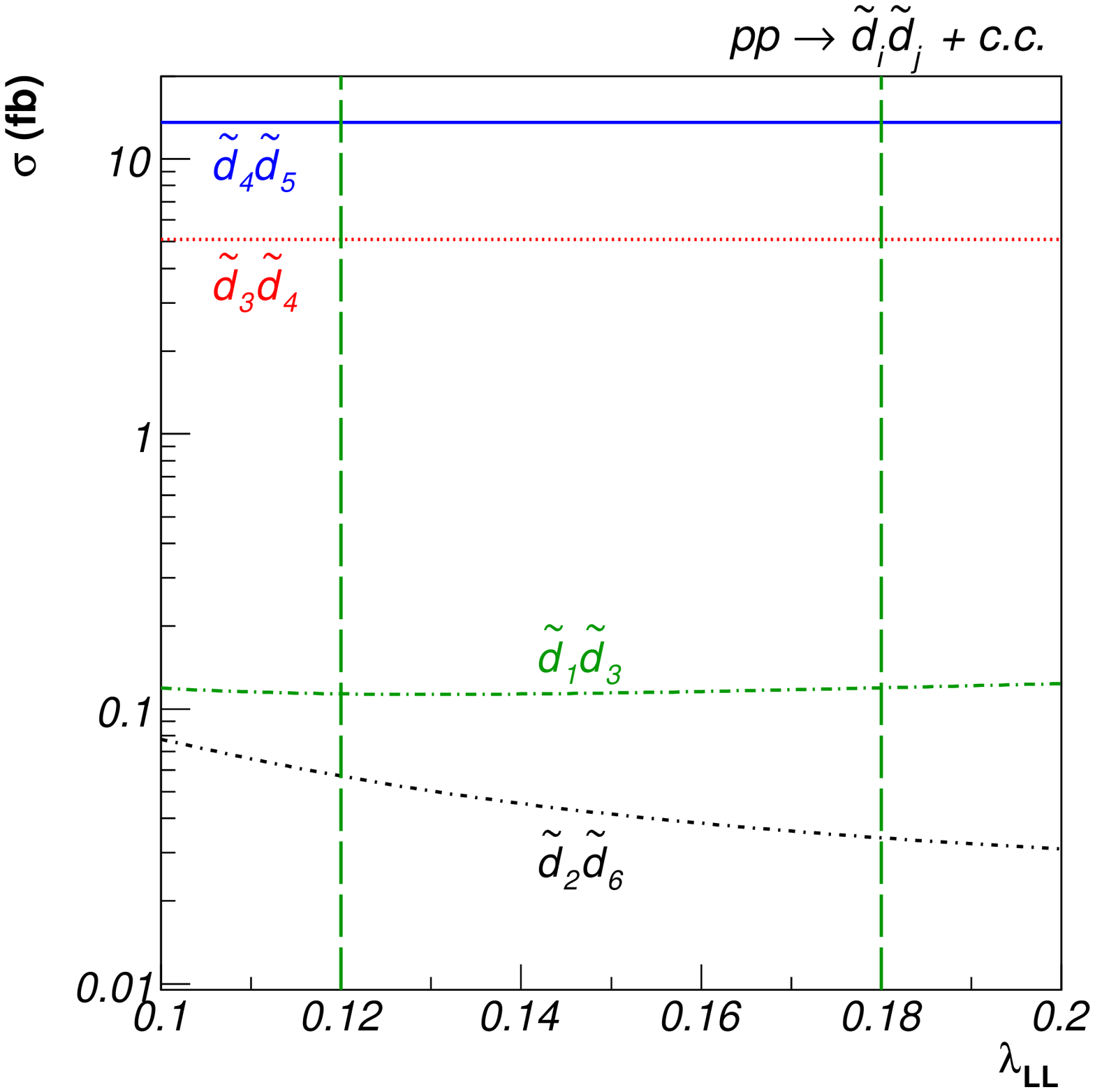} 
    \includegraphics[scale=0.28]{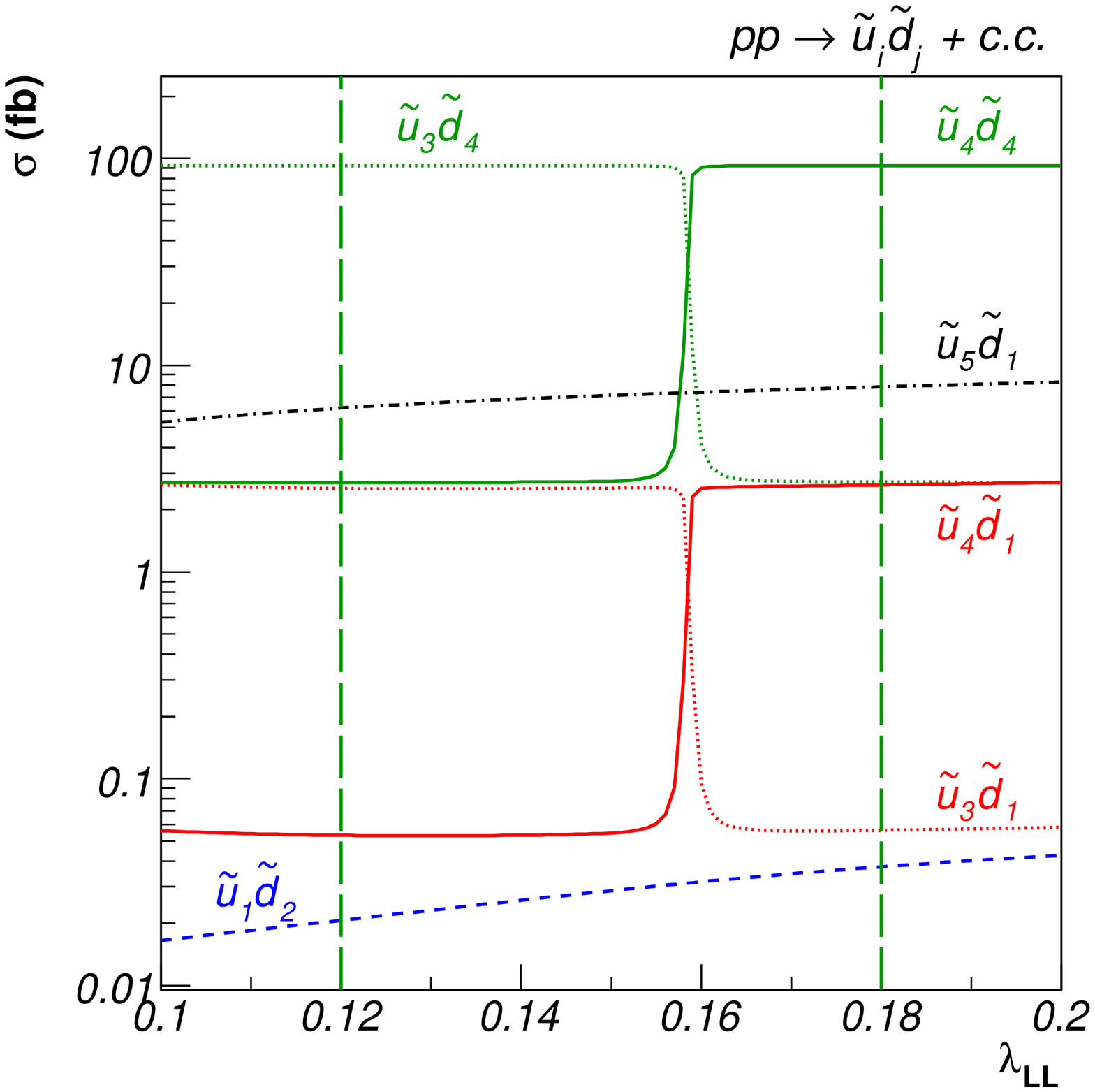} 
    \includegraphics[scale=0.28]{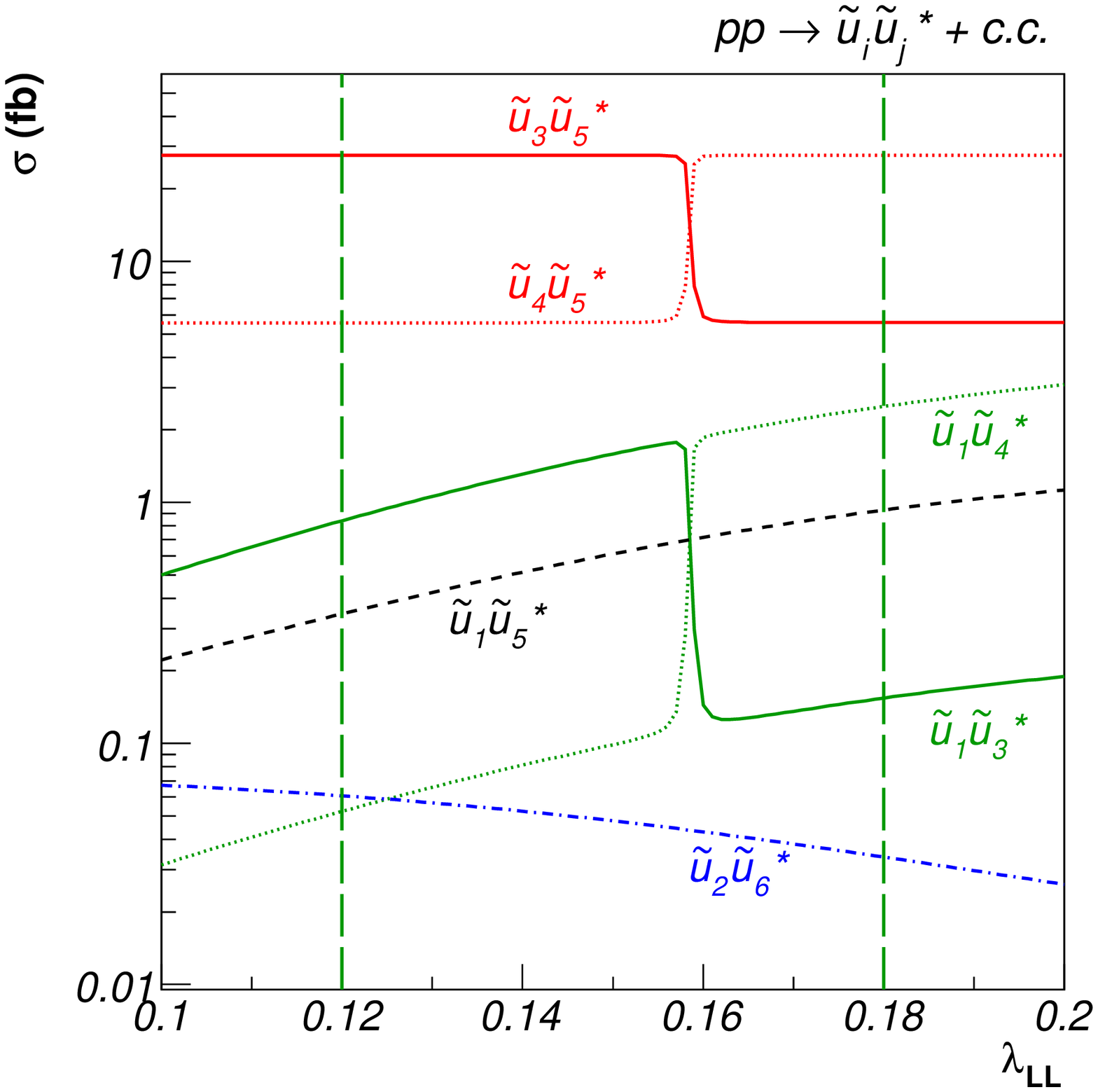} 
    \includegraphics[scale=0.28]{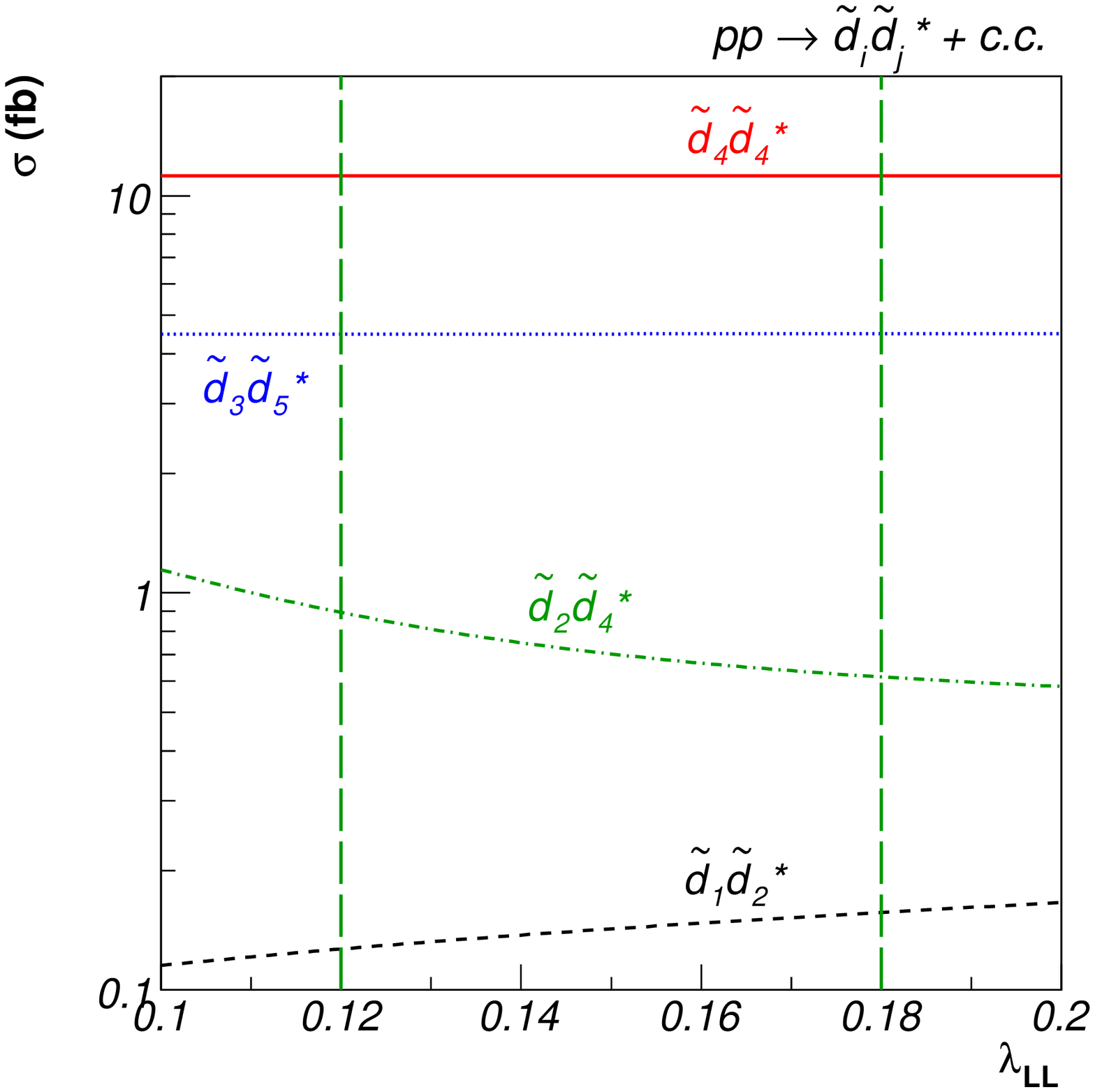} 
    \includegraphics[scale=0.28]{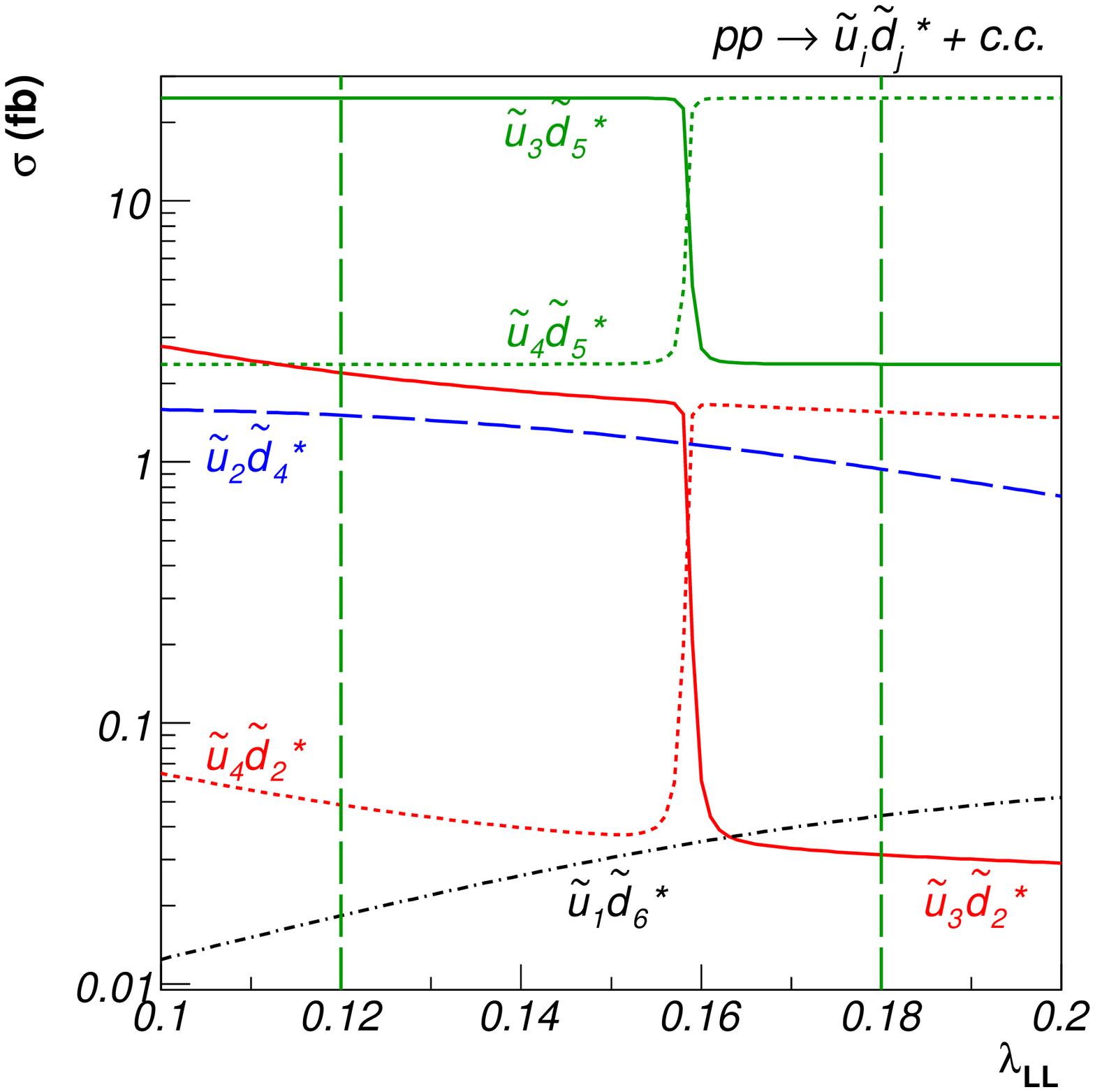} 
    \includegraphics[scale=0.28]{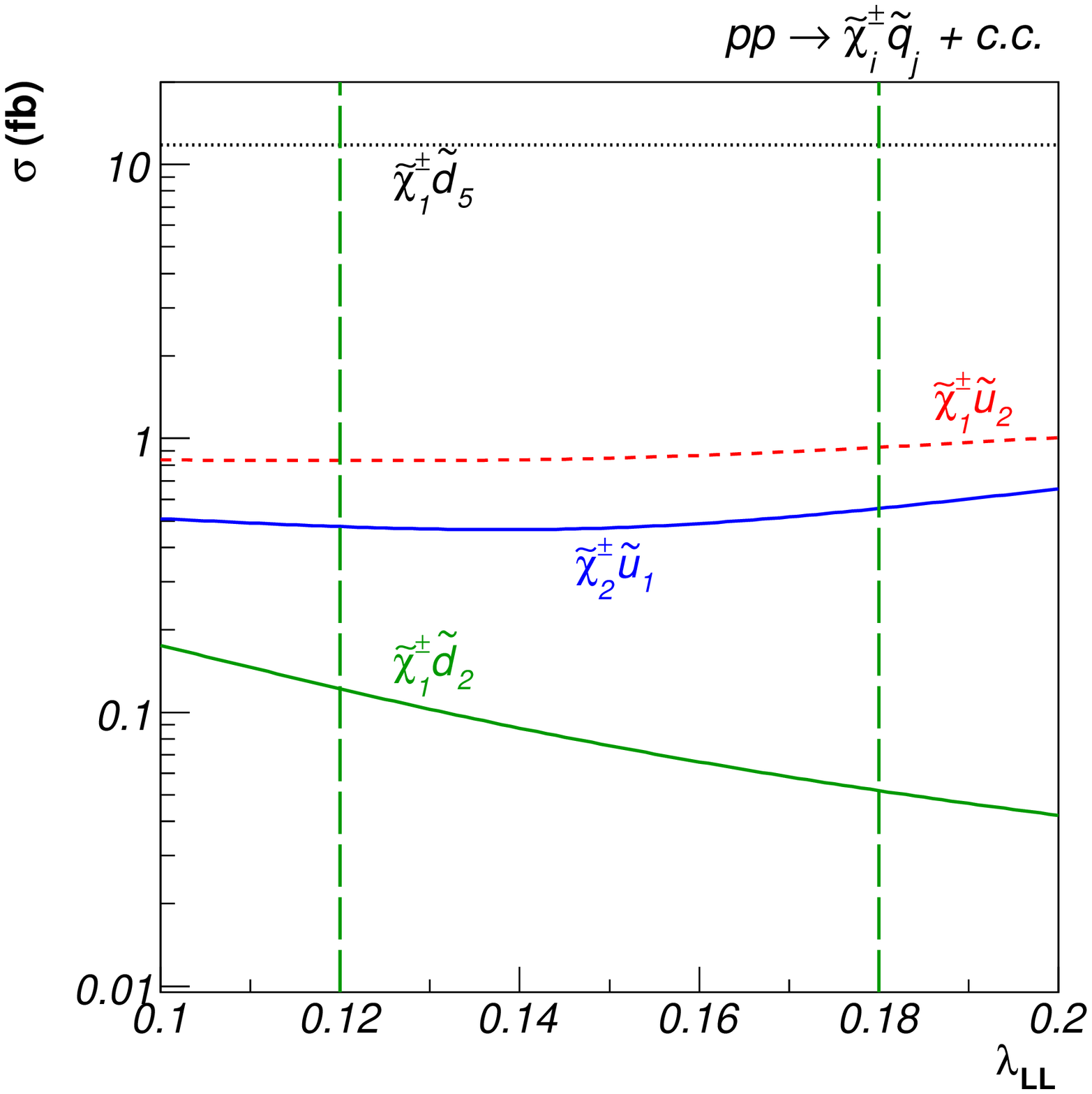} 
    \includegraphics[scale=0.28]{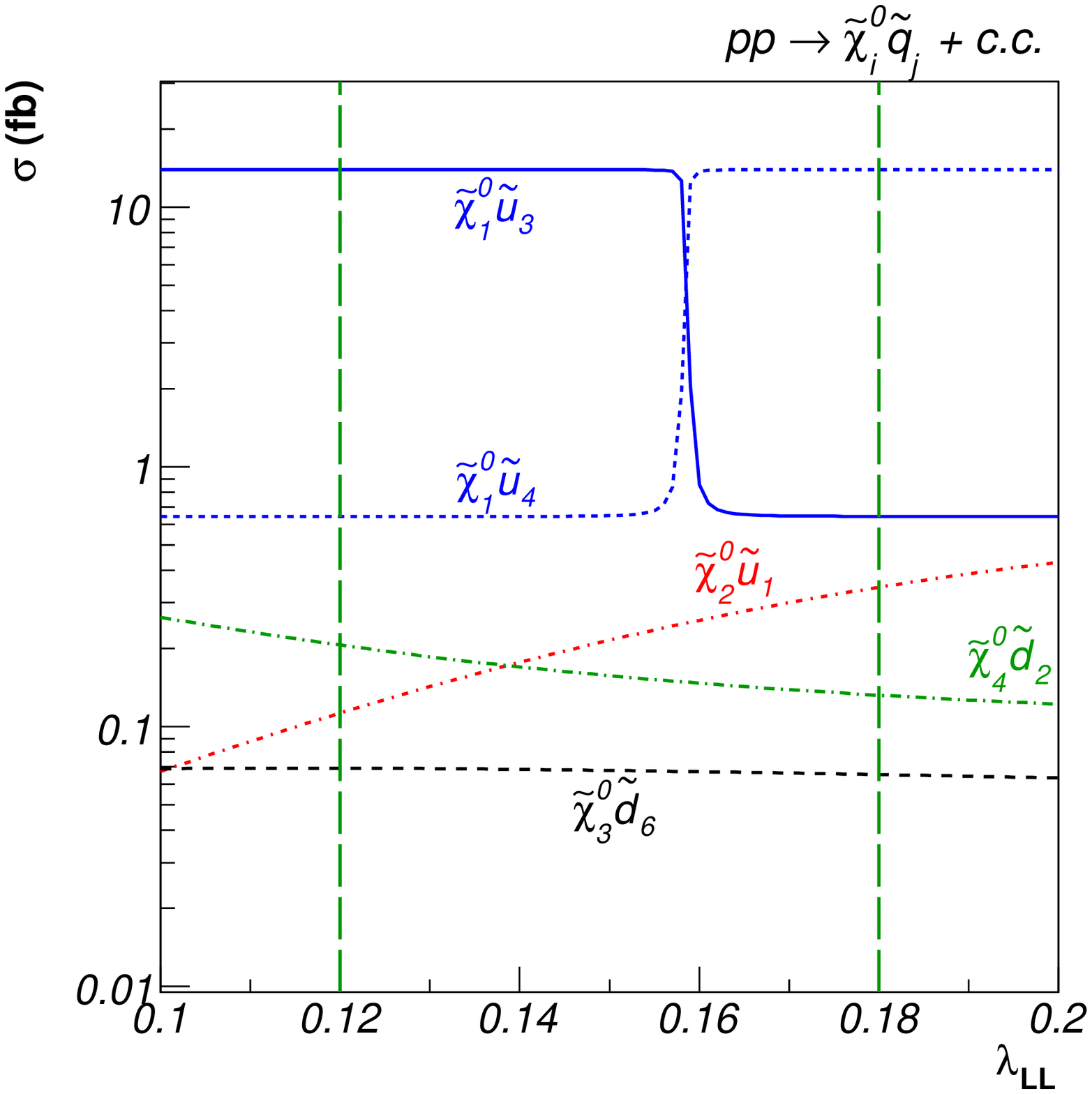} 
    \includegraphics[scale=0.28]{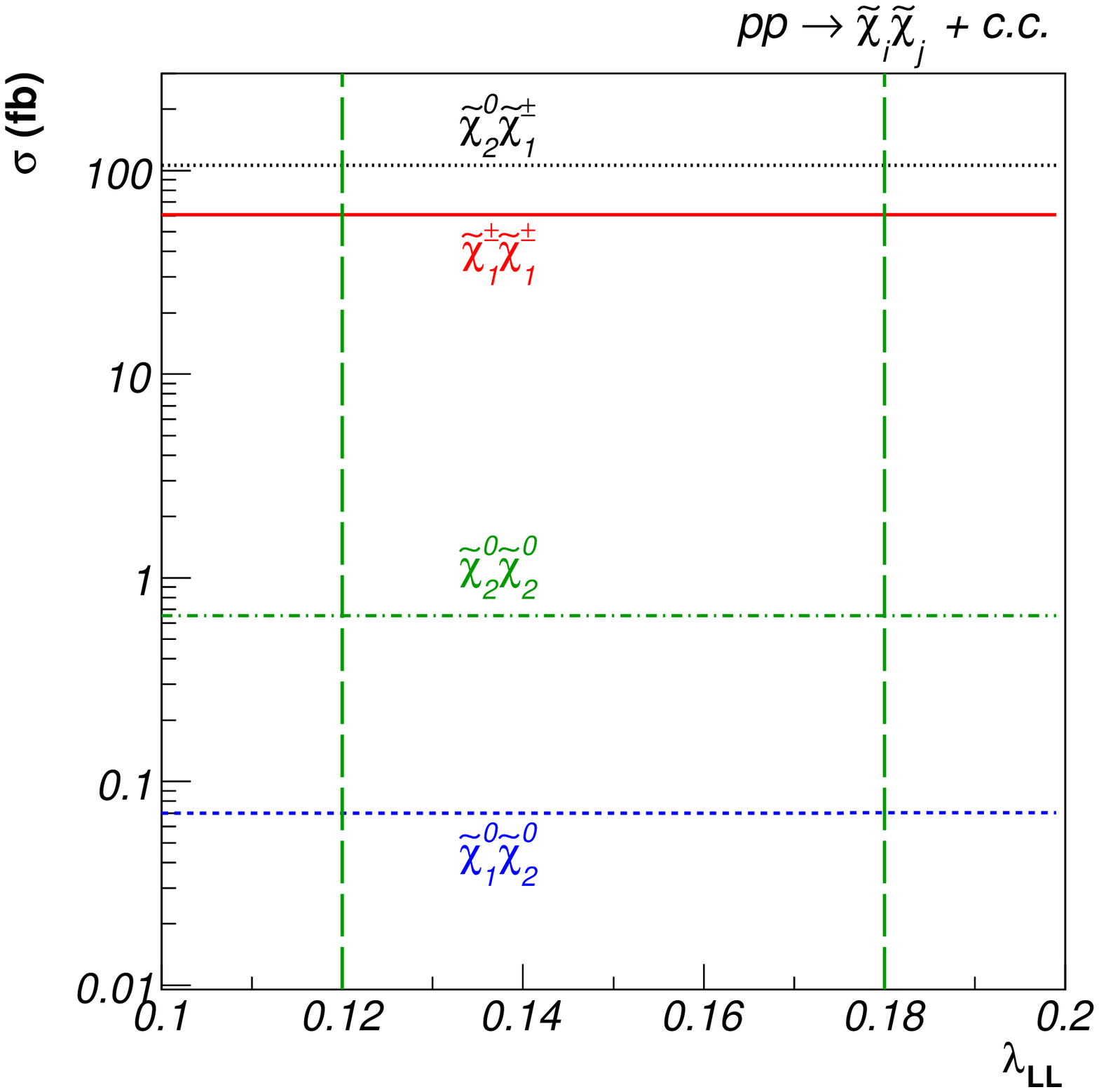} 
  \end{center}
  \vspace*{-5mm}
\caption{Same as Fig.\ \ref{fig15} for our benchmark scenario H.}
\label{fig21}
\end{figure}

\begin{figure}
  \begin{center}
    \includegraphics[scale=0.28]{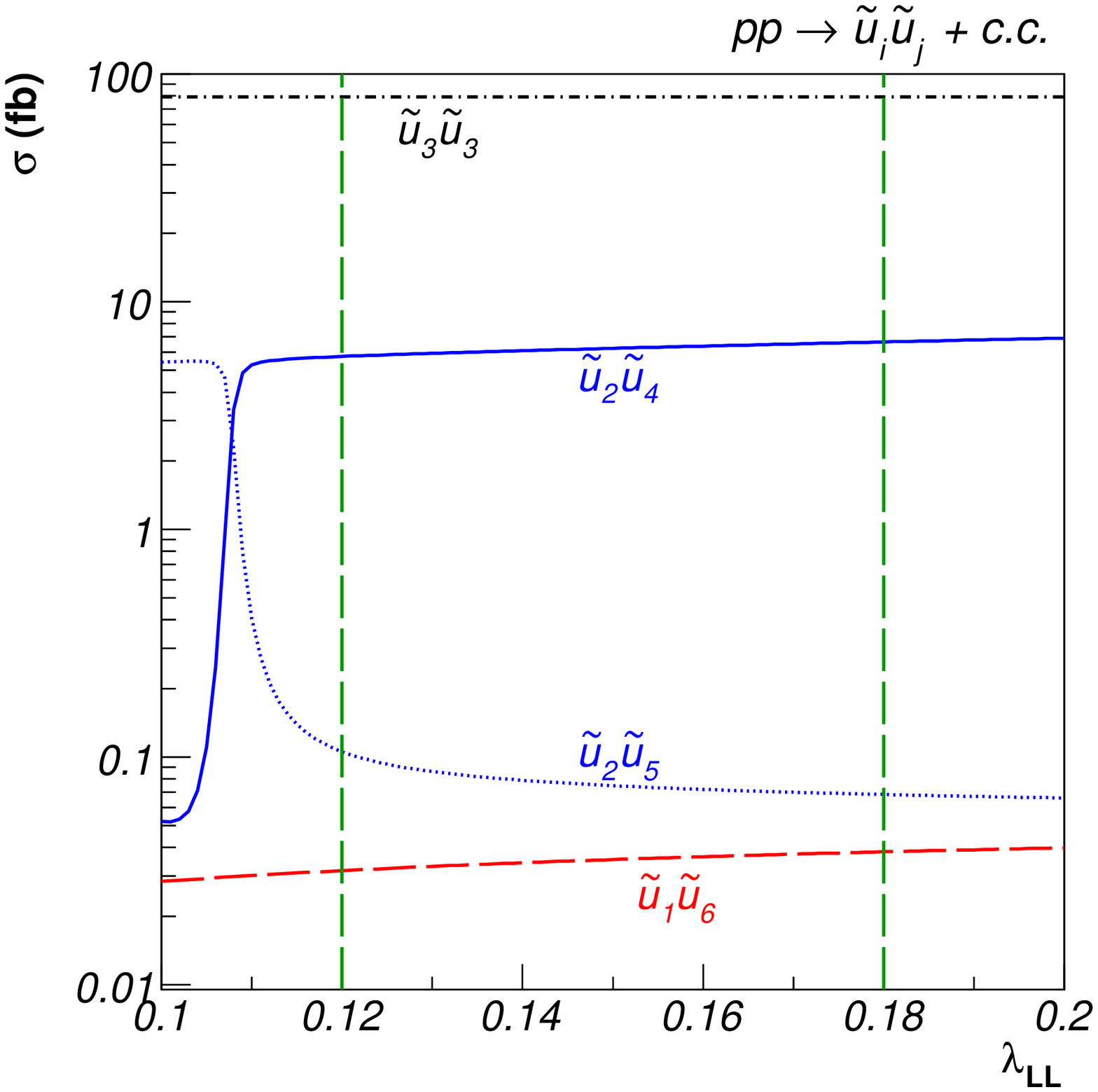} 
    \includegraphics[scale=0.28]{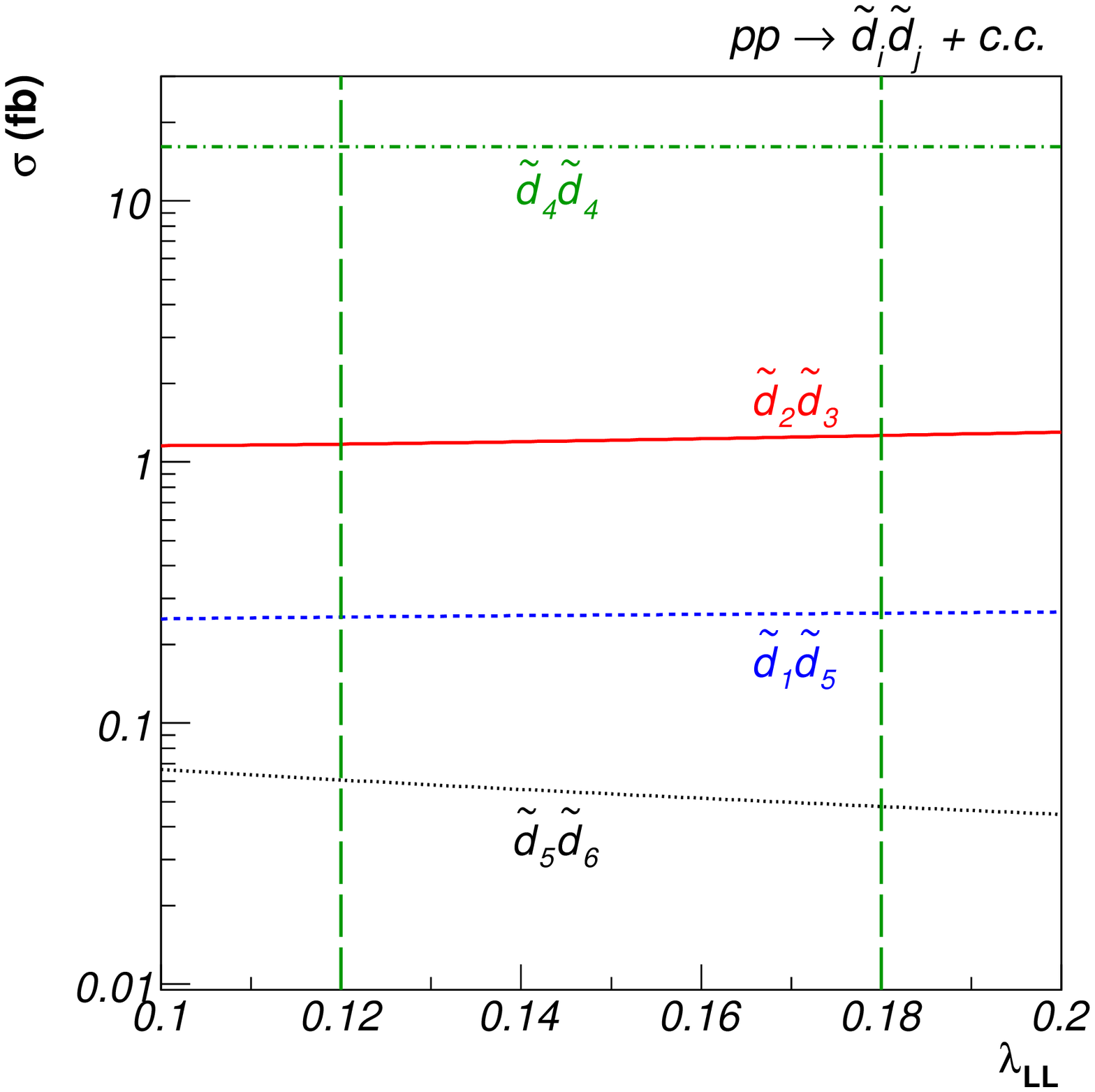} 
    \includegraphics[scale=0.28]{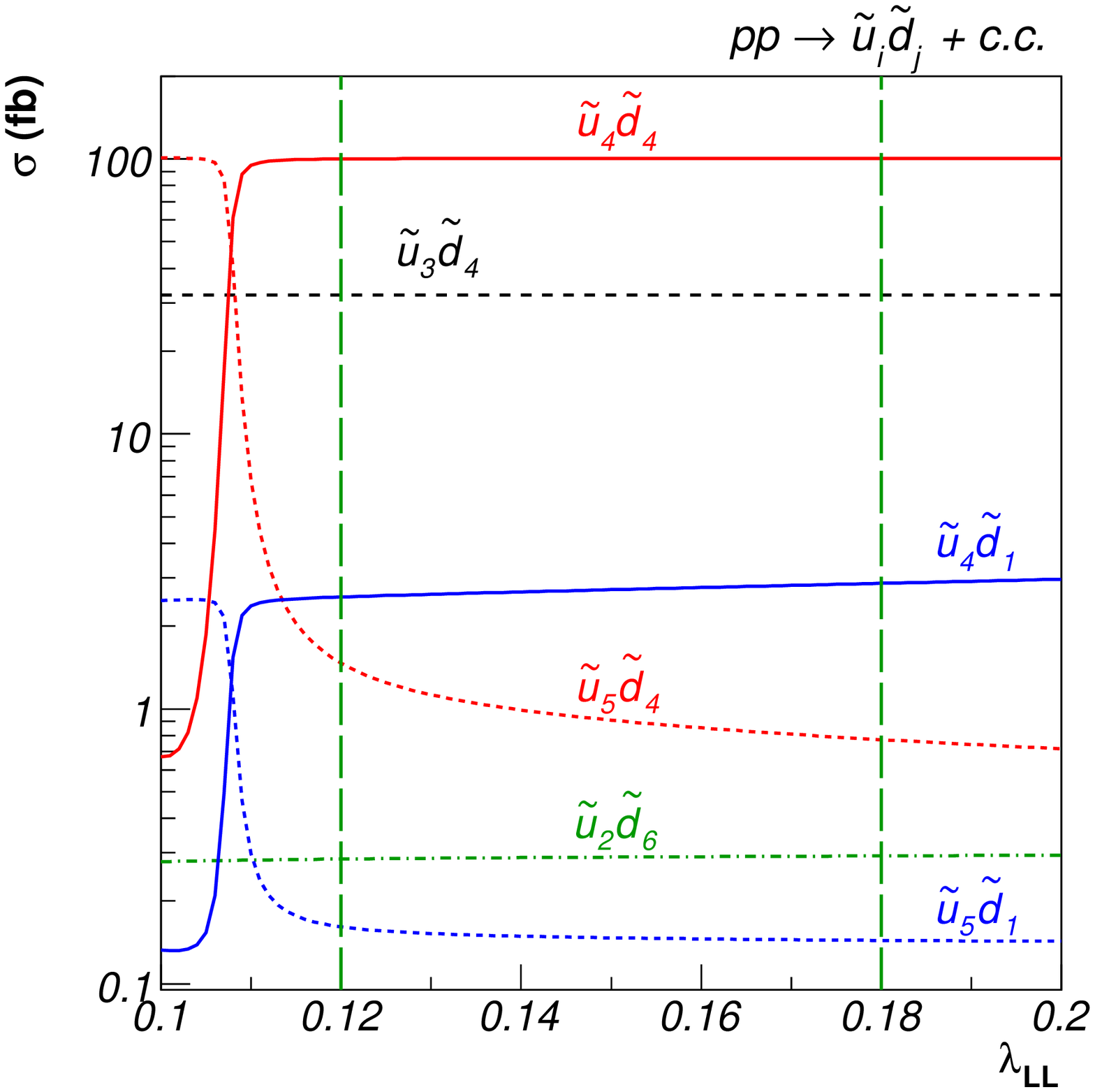} 
    \includegraphics[scale=0.28]{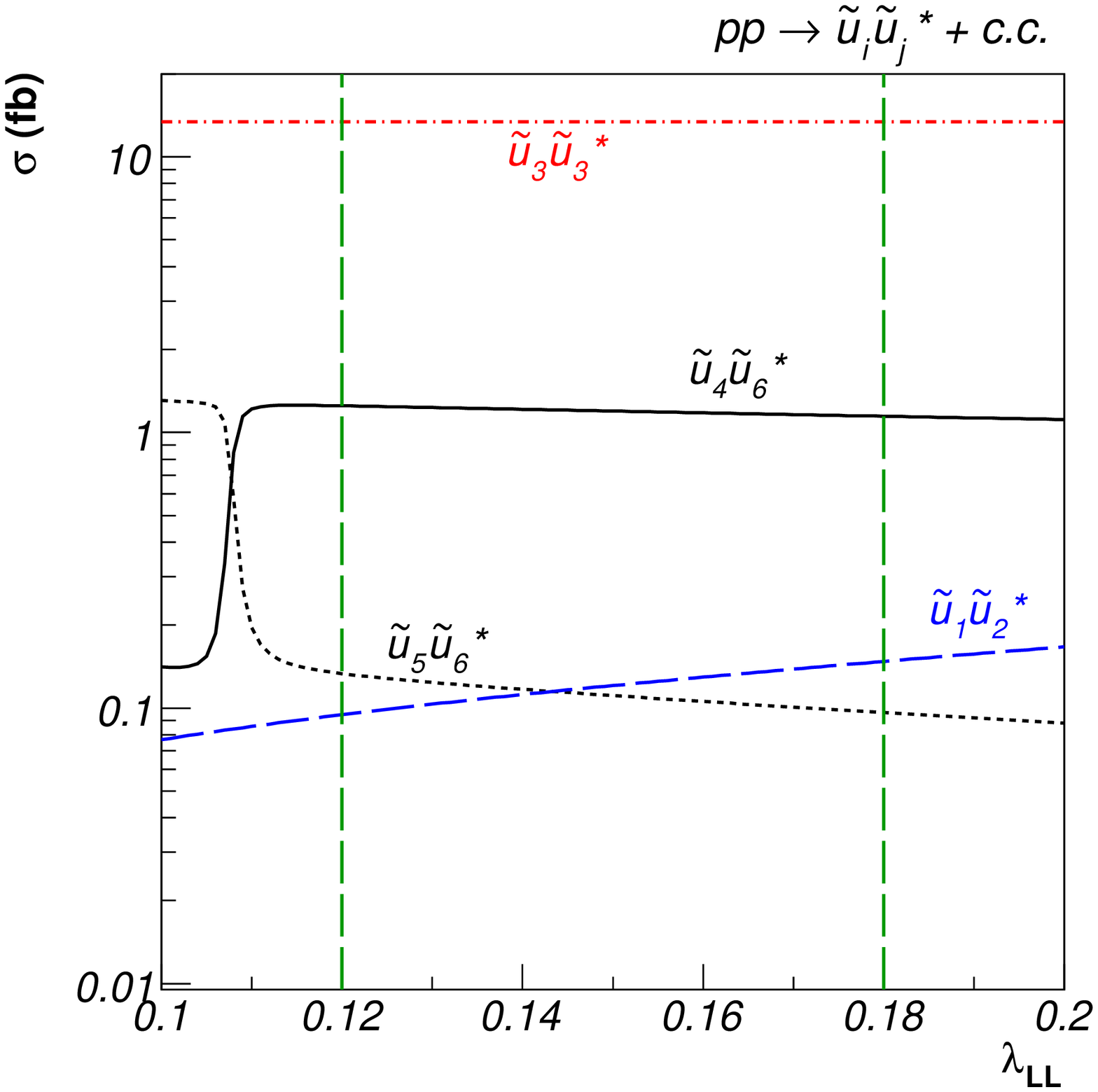} 
    \includegraphics[scale=0.28]{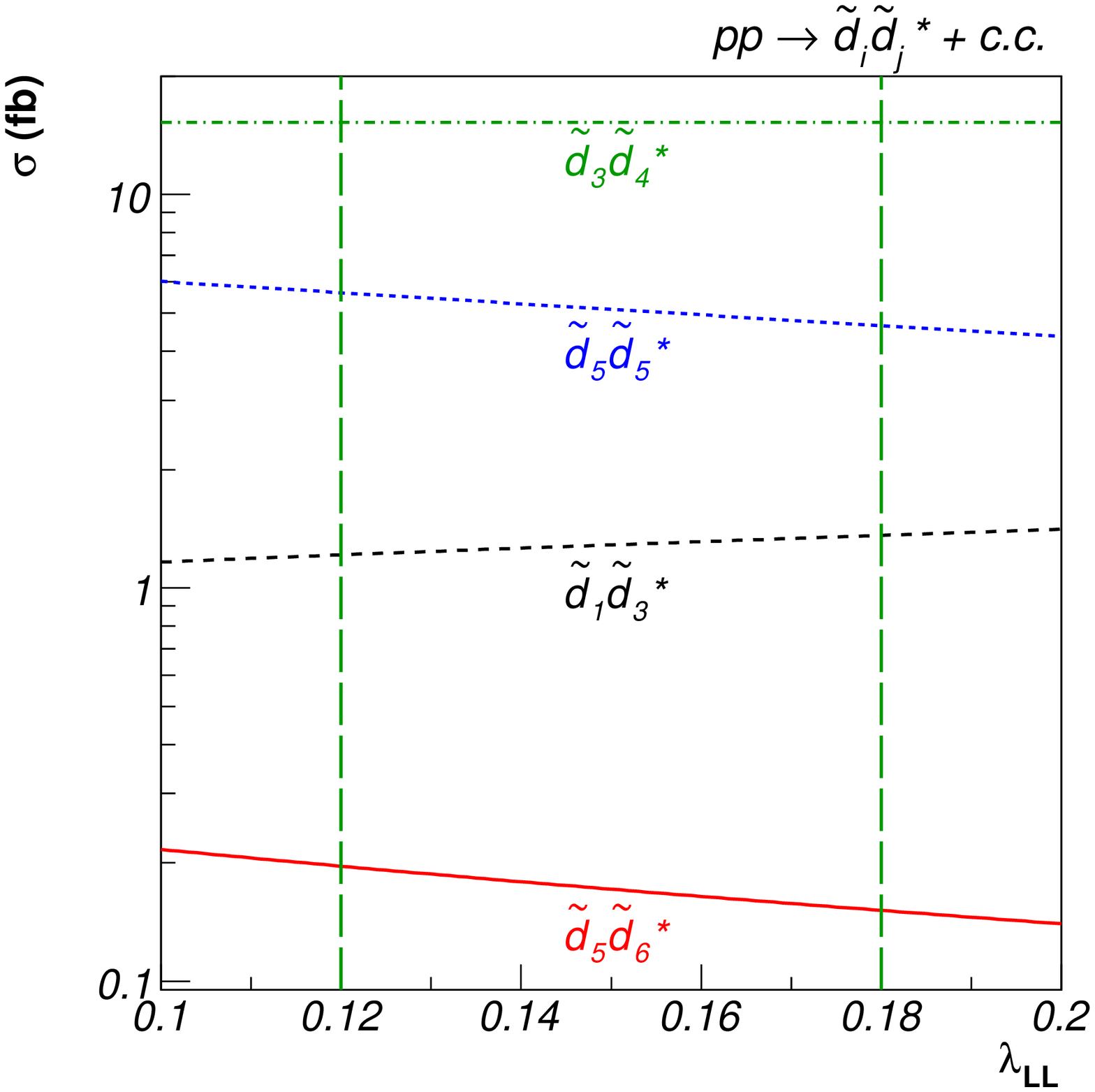} 
    \includegraphics[scale=0.28]{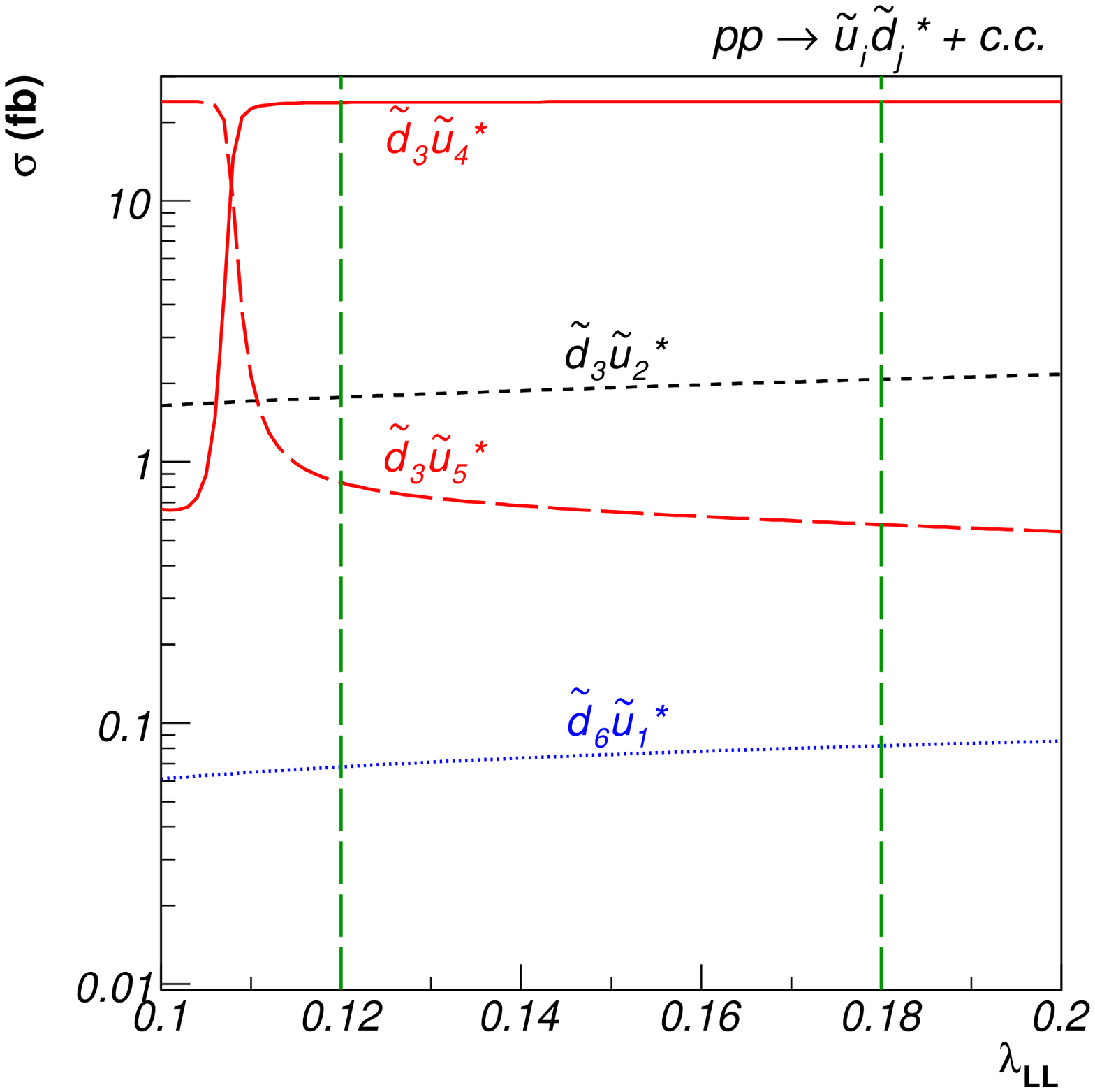} 
    \includegraphics[scale=0.28]{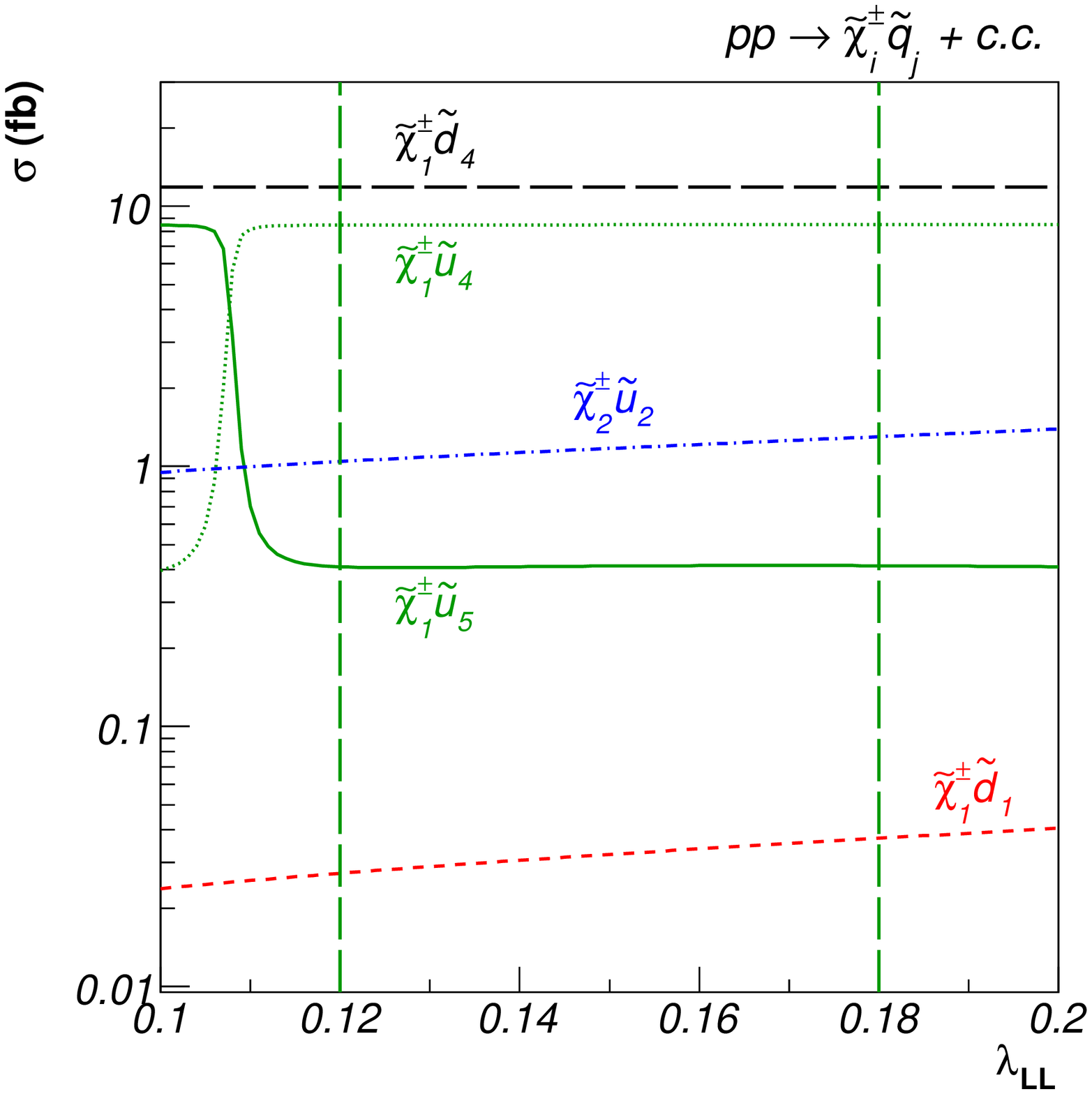} 
    \includegraphics[scale=0.28]{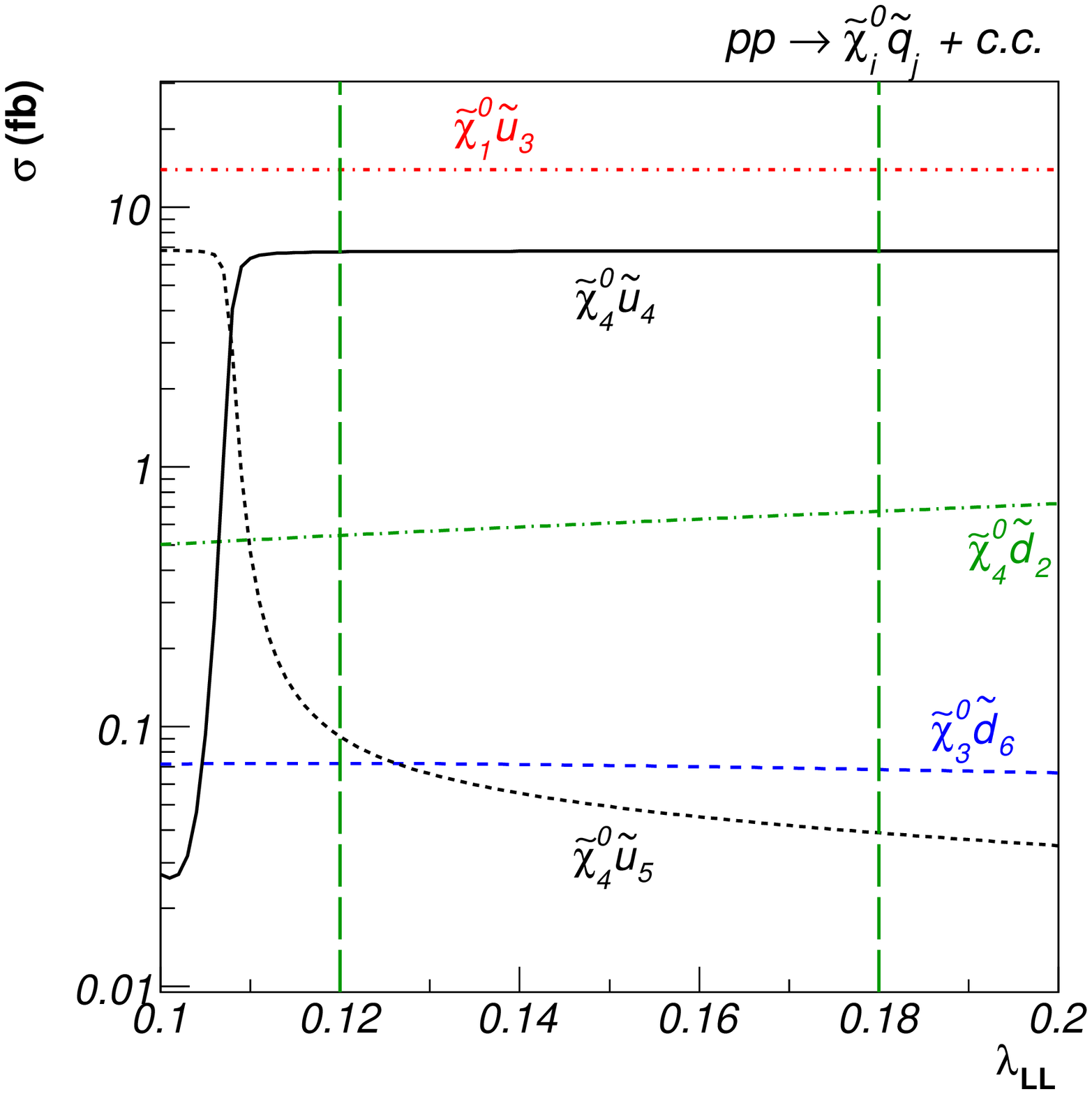} 
    \includegraphics[scale=0.28]{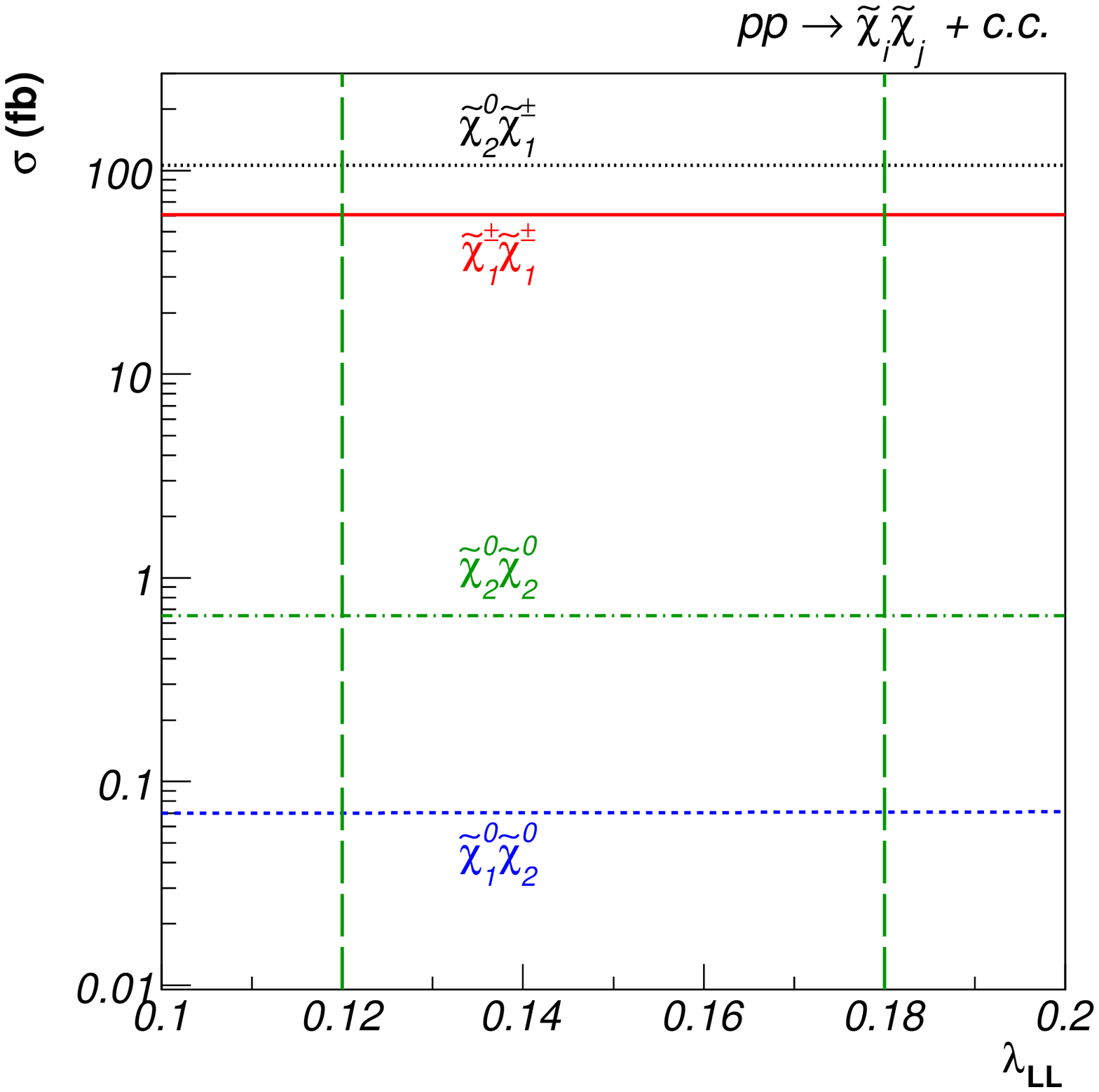} 
  \end{center}
  \vspace*{-5mm}
\caption{Same as Fig.\ \ref{fig16} for our benchmark scenario H.}
\label{fig22}
\end{figure}

\begin{figure}
  \begin{center}
    \includegraphics[scale=0.28]{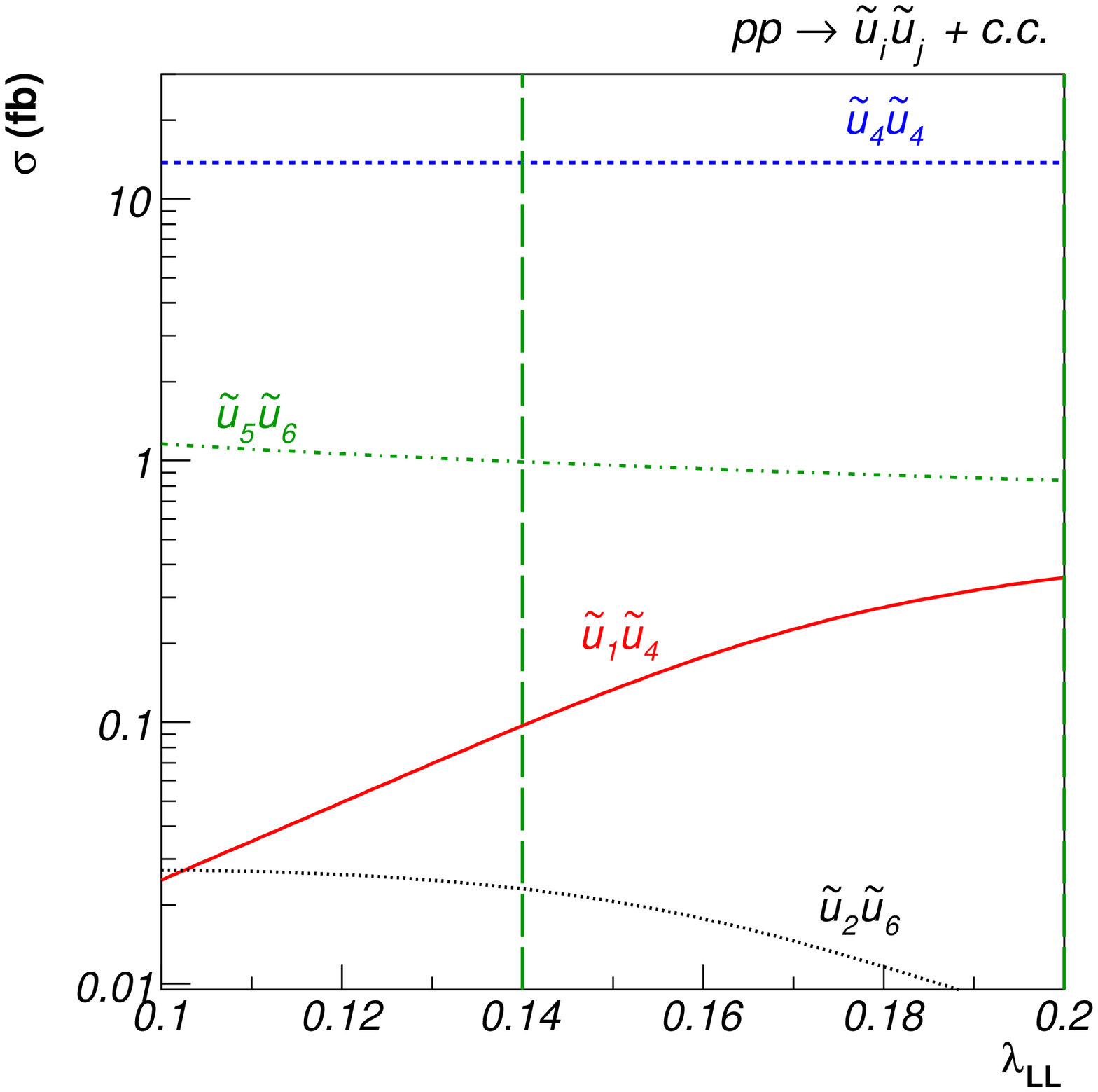} 
    \includegraphics[scale=0.28]{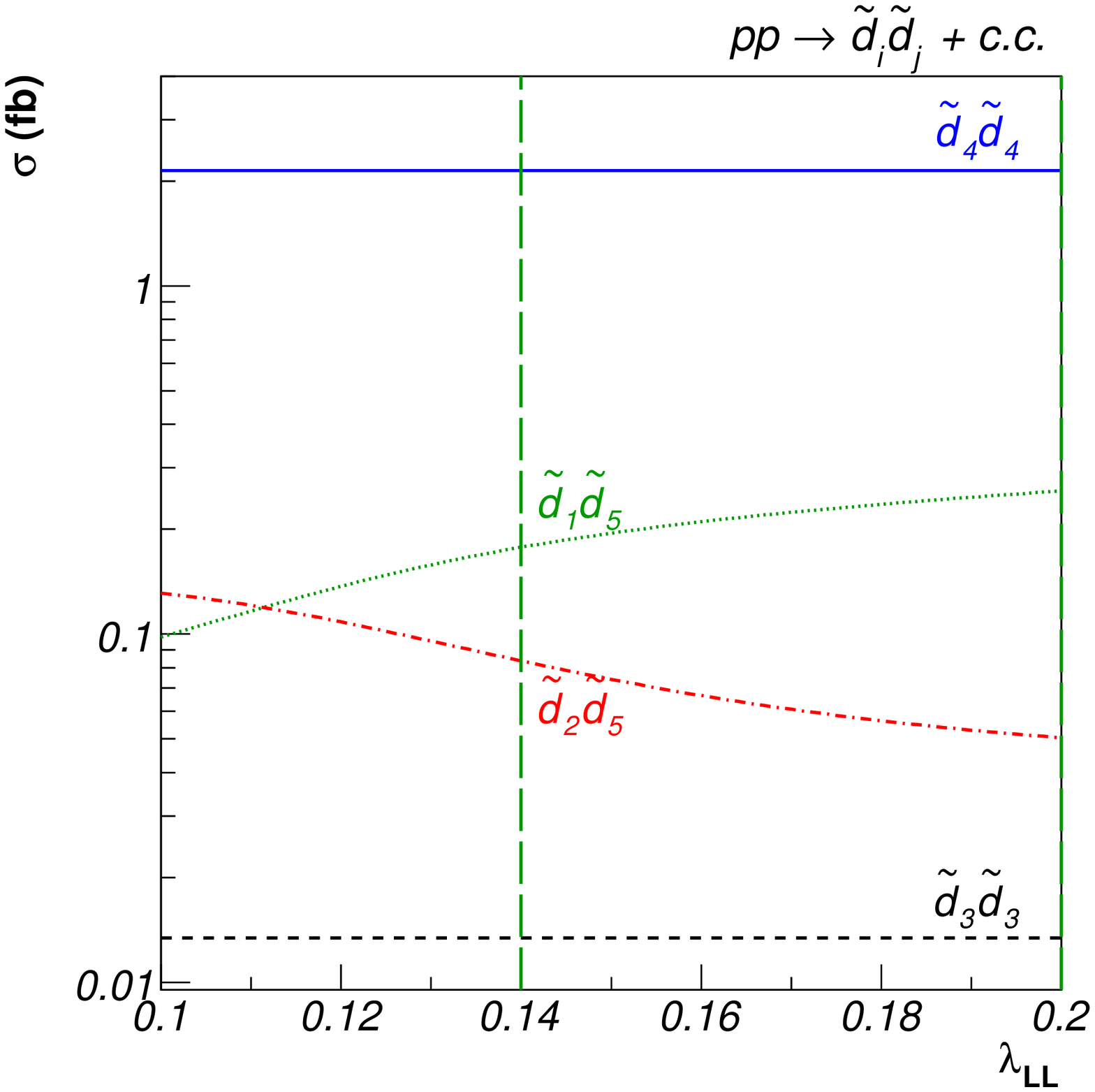} 
    \includegraphics[scale=0.28]{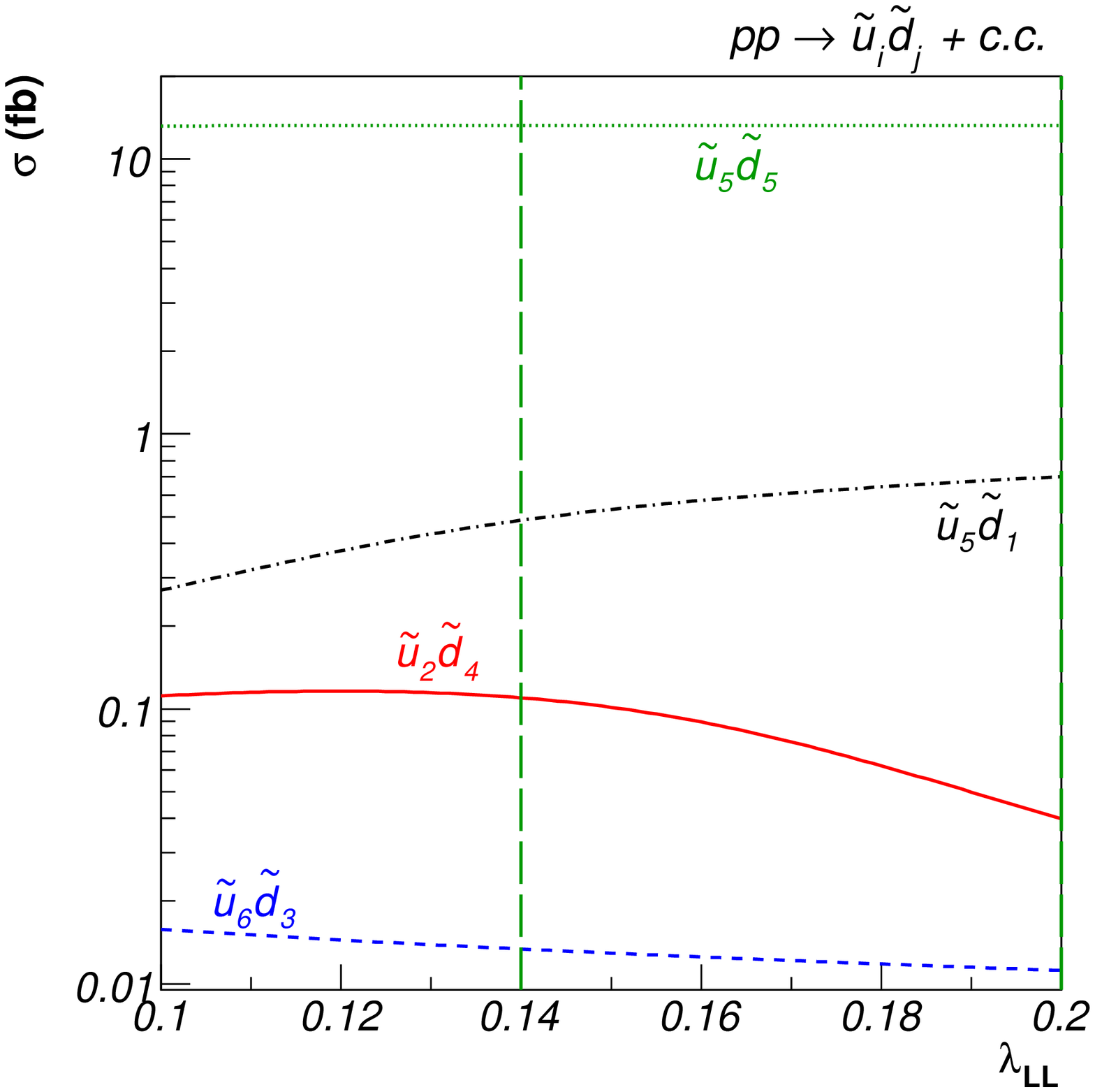} 
    \includegraphics[scale=0.28]{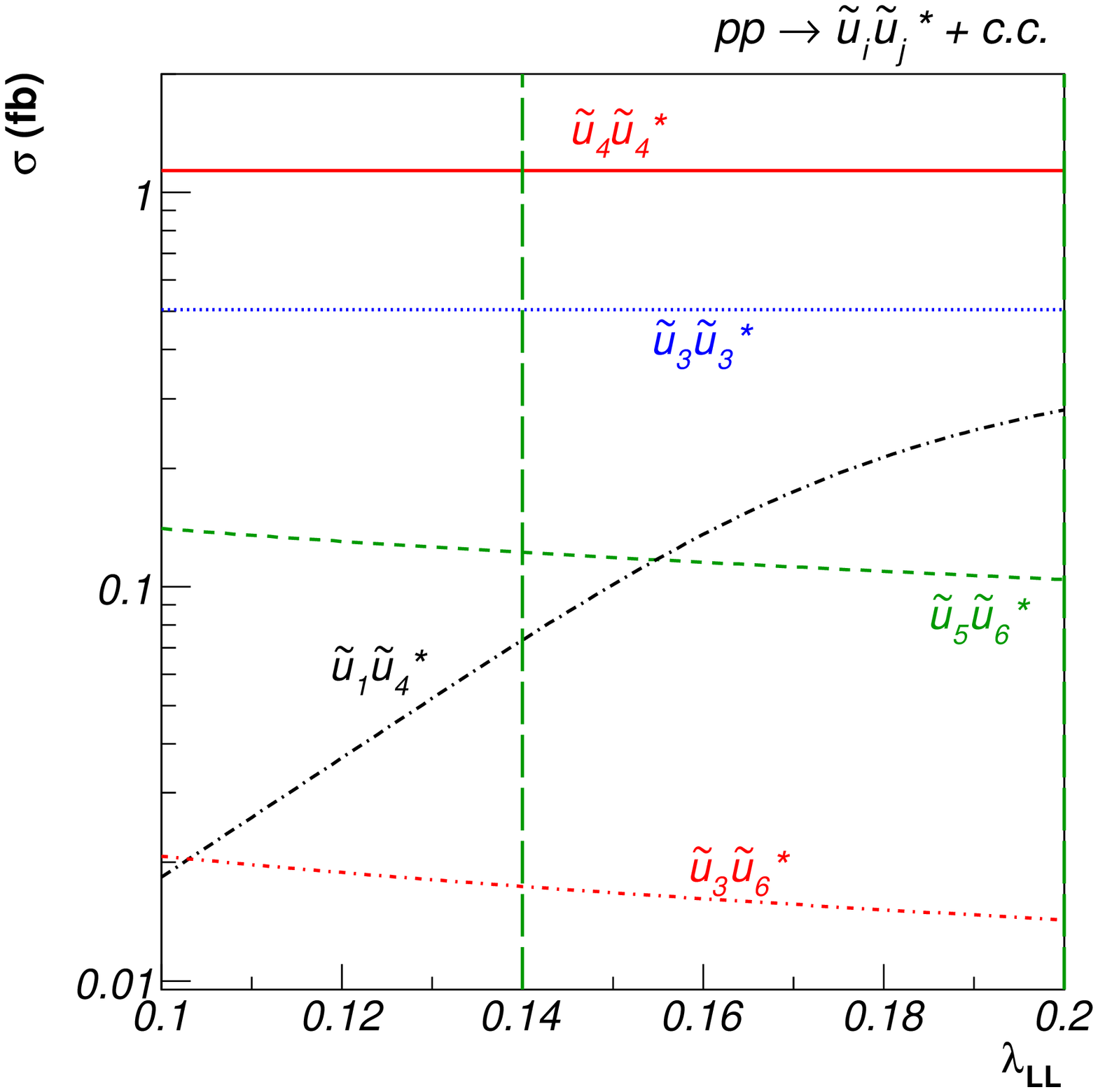} 
    \includegraphics[scale=0.28]{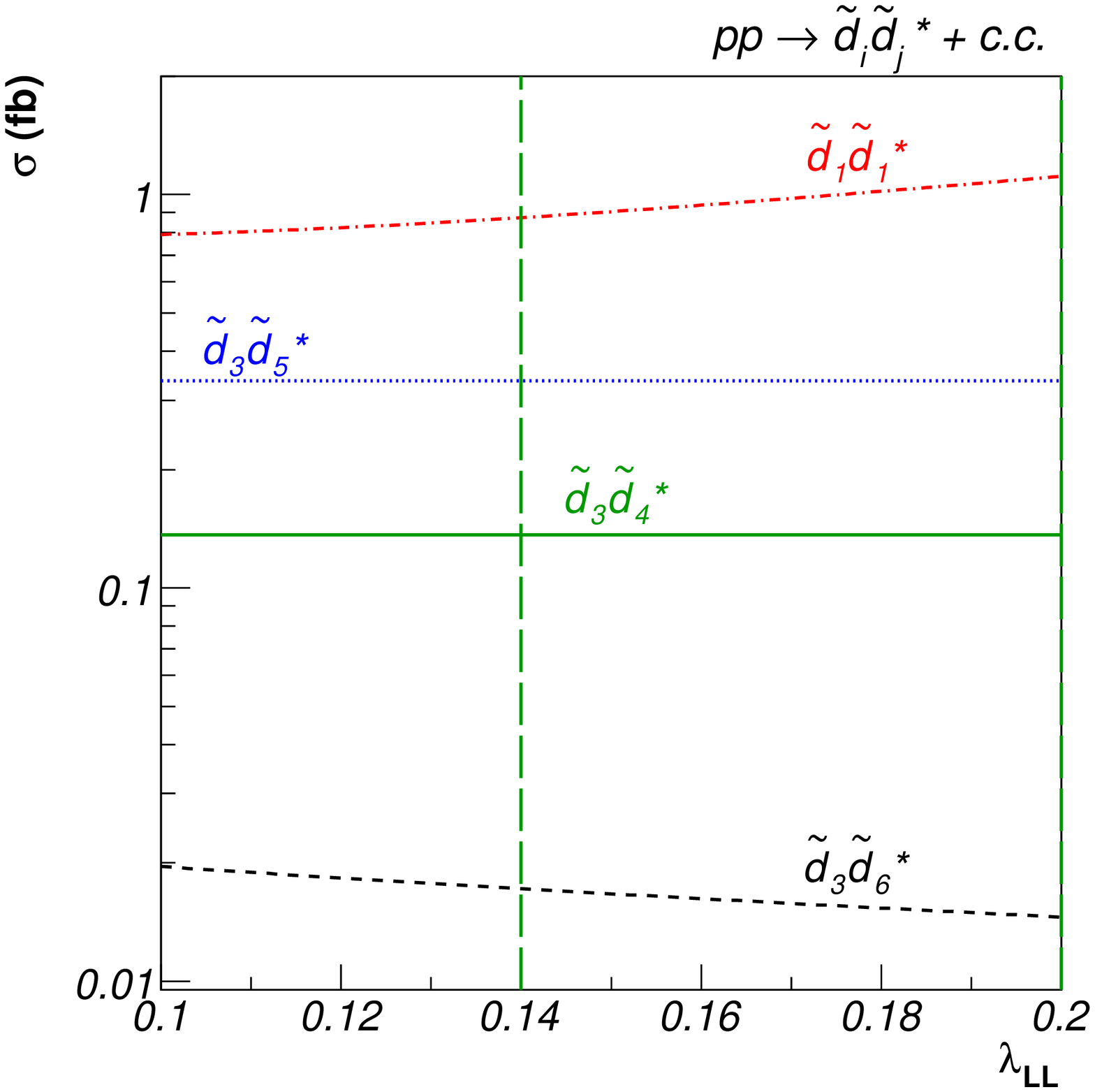} 
    \includegraphics[scale=0.28]{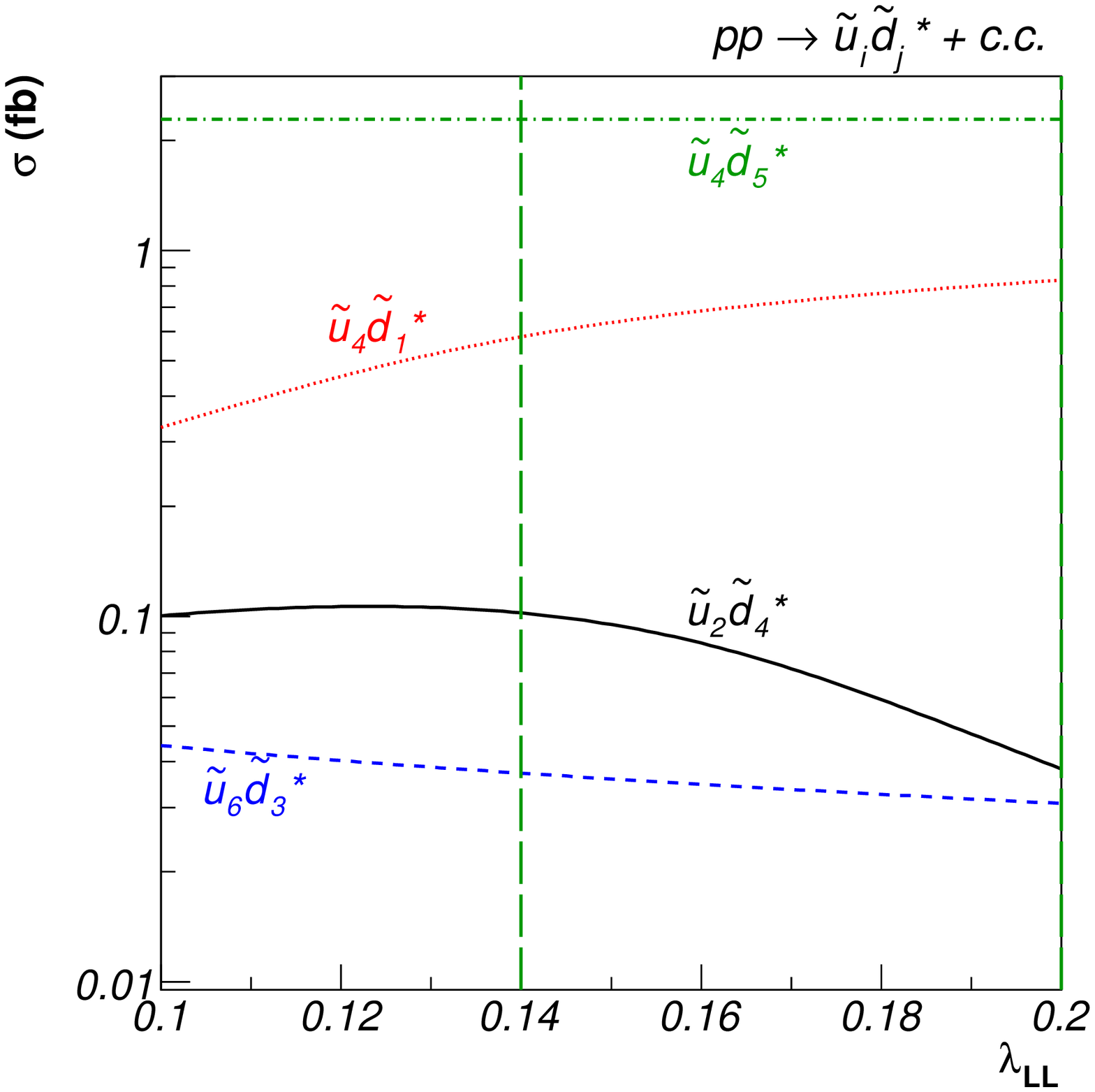} 
    \includegraphics[scale=0.28]{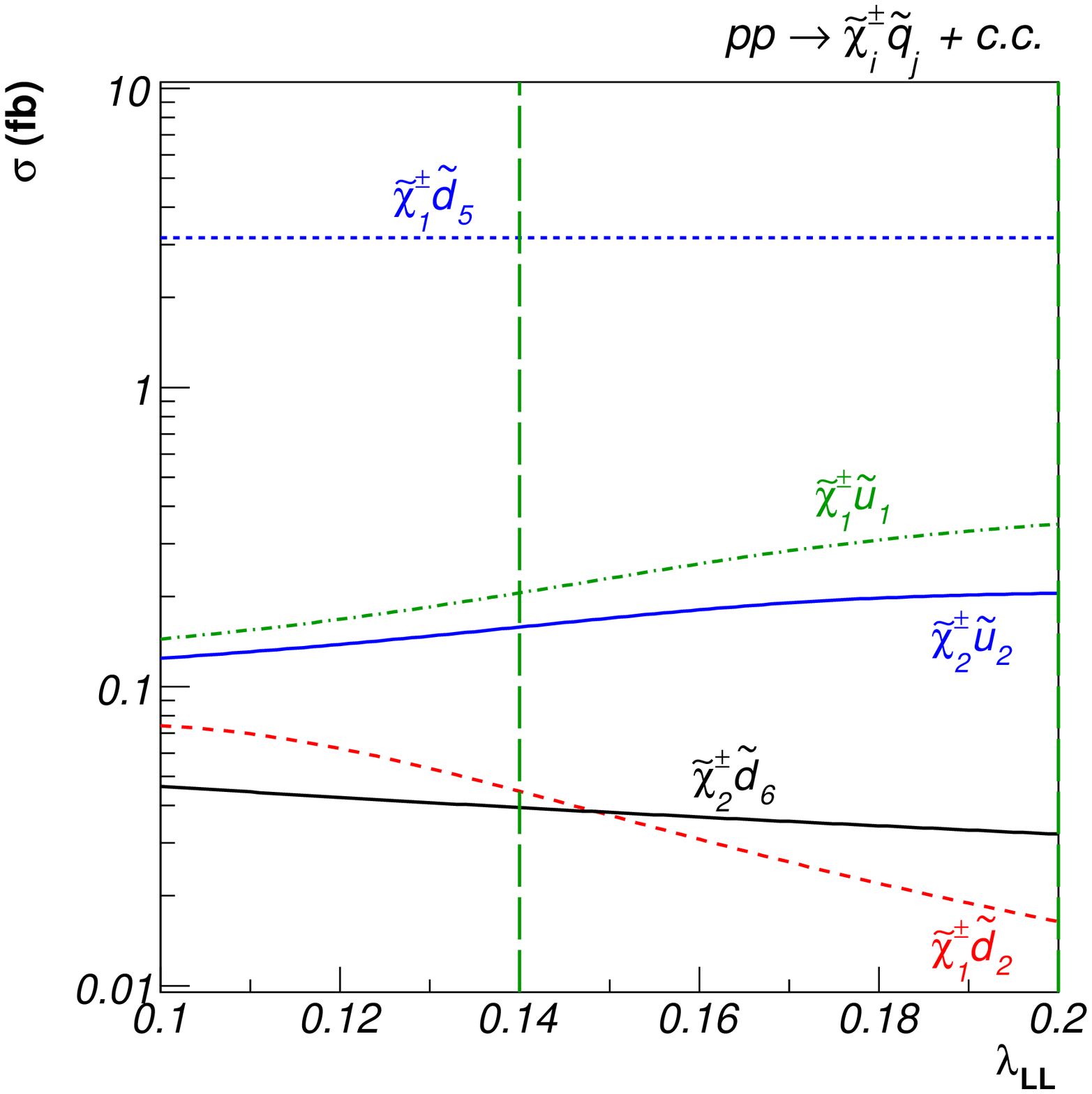} 
    \includegraphics[scale=0.28]{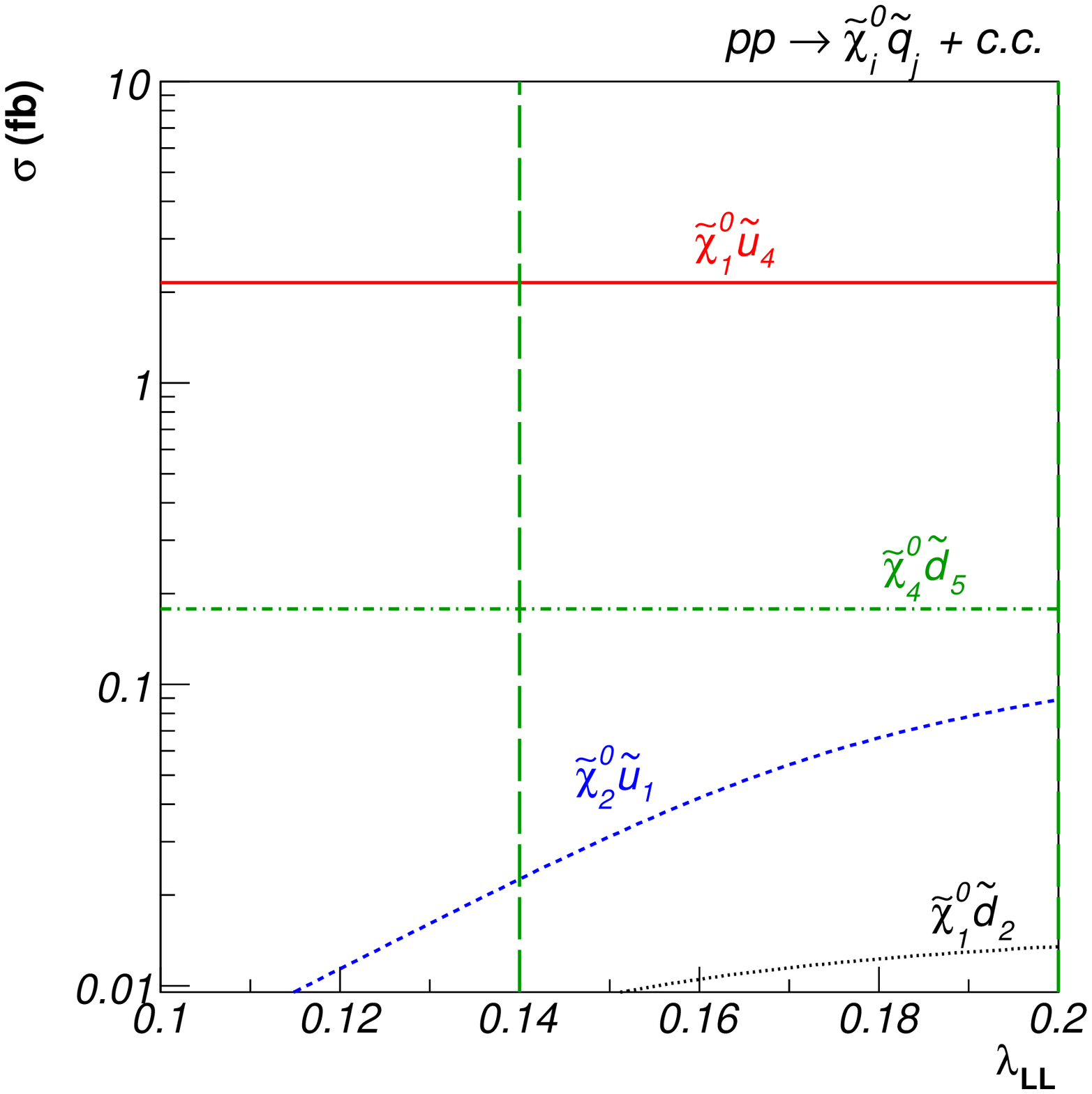} 
    \includegraphics[scale=0.28]{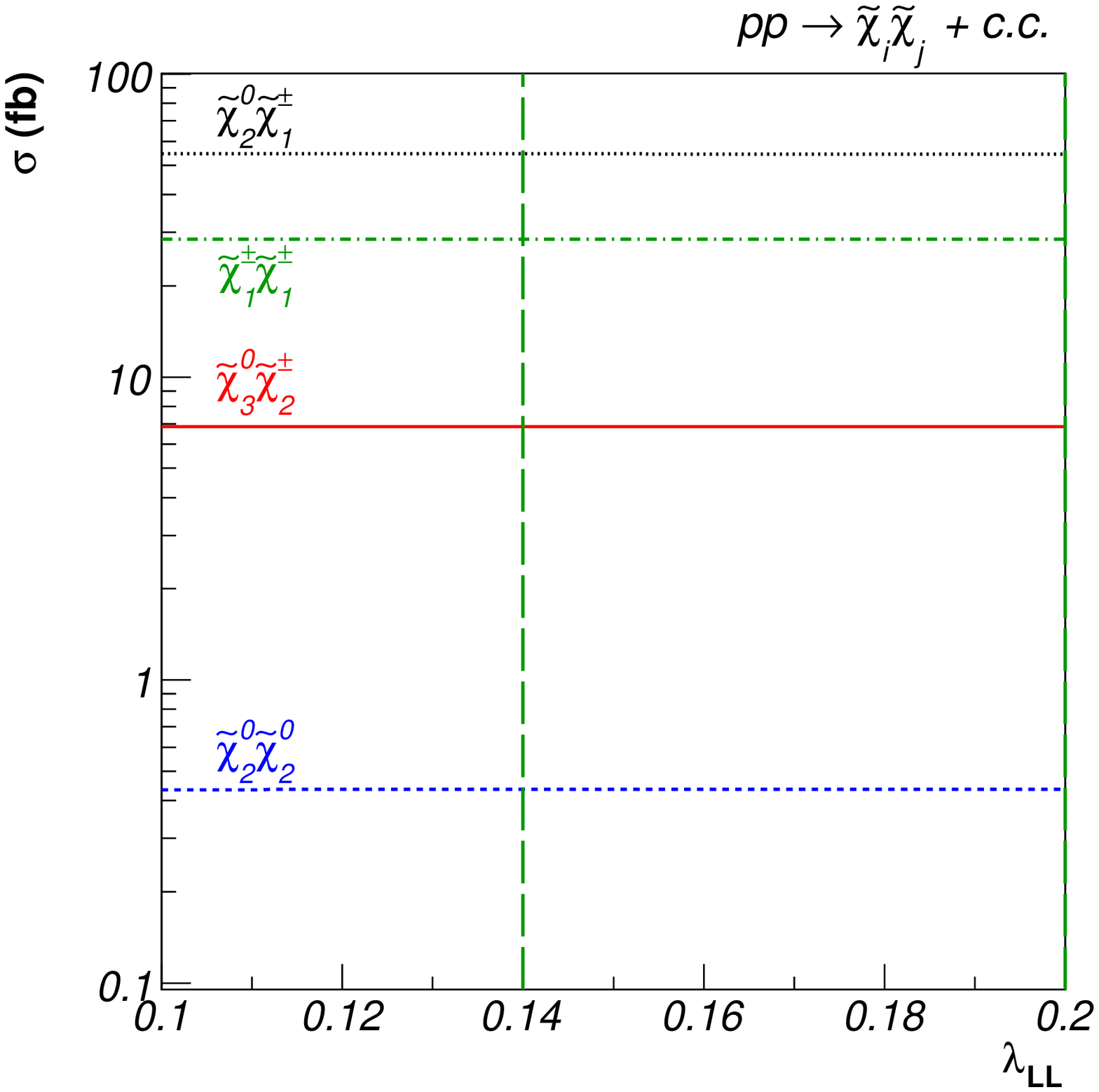} 
  \end{center}
  \vspace*{-5mm}
\caption{Same as Fig.\ \ref{fig15} for our benchmark scenario I.}
\label{fig23}
\end{figure}

\begin{figure}
  \begin{center}
    \includegraphics[scale=0.28]{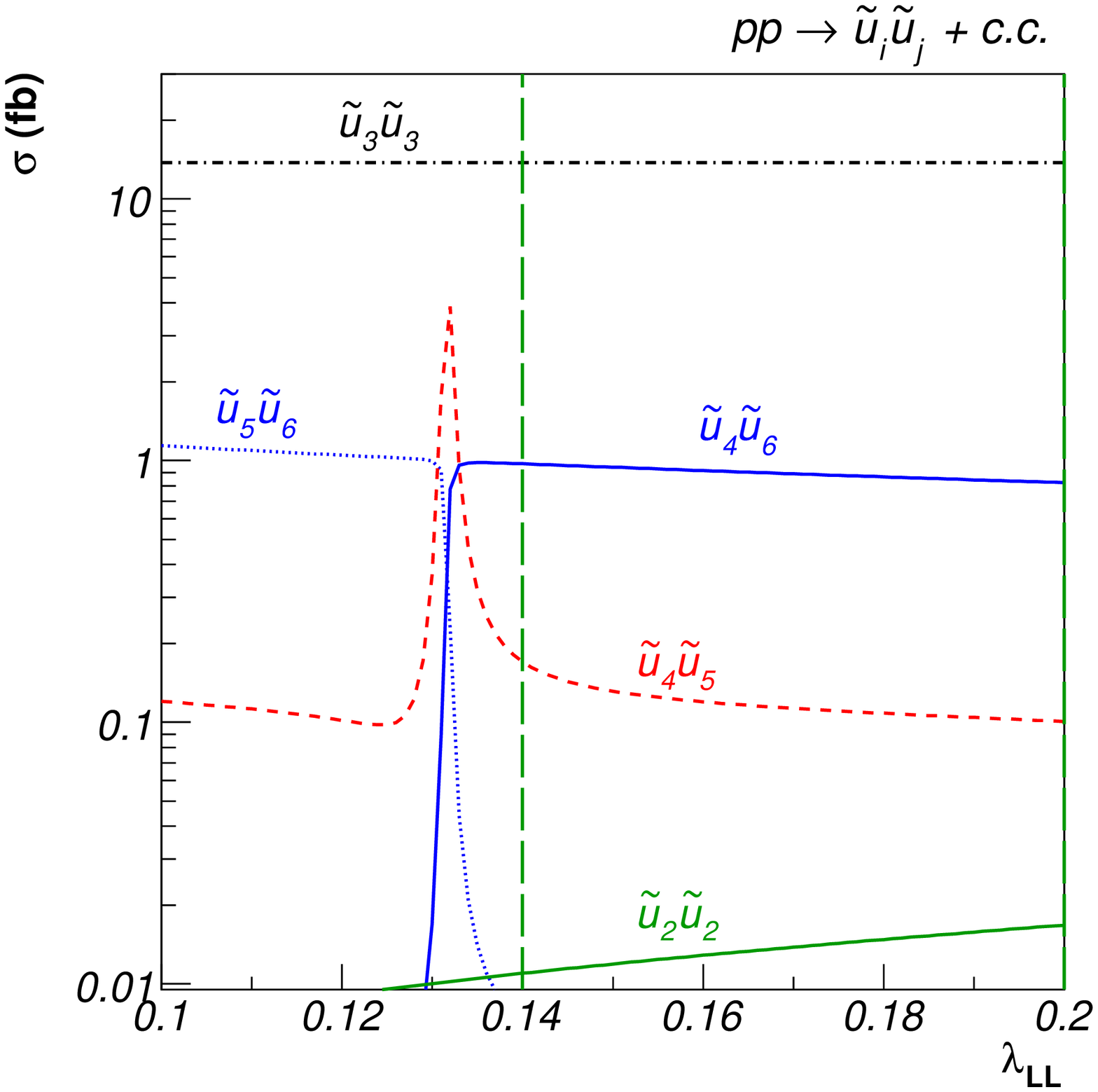} 
    \includegraphics[scale=0.28]{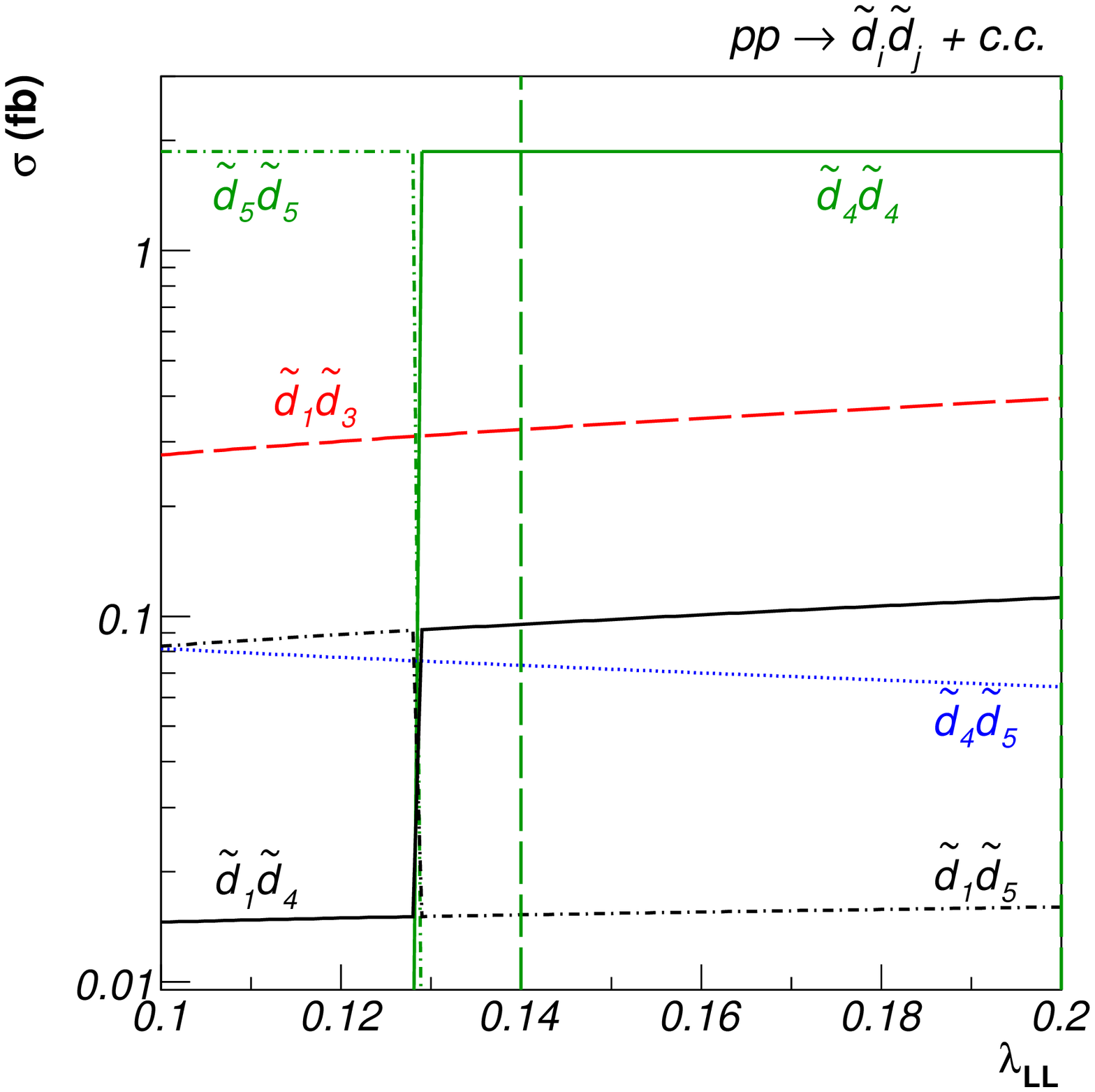} 
    \includegraphics[scale=0.28]{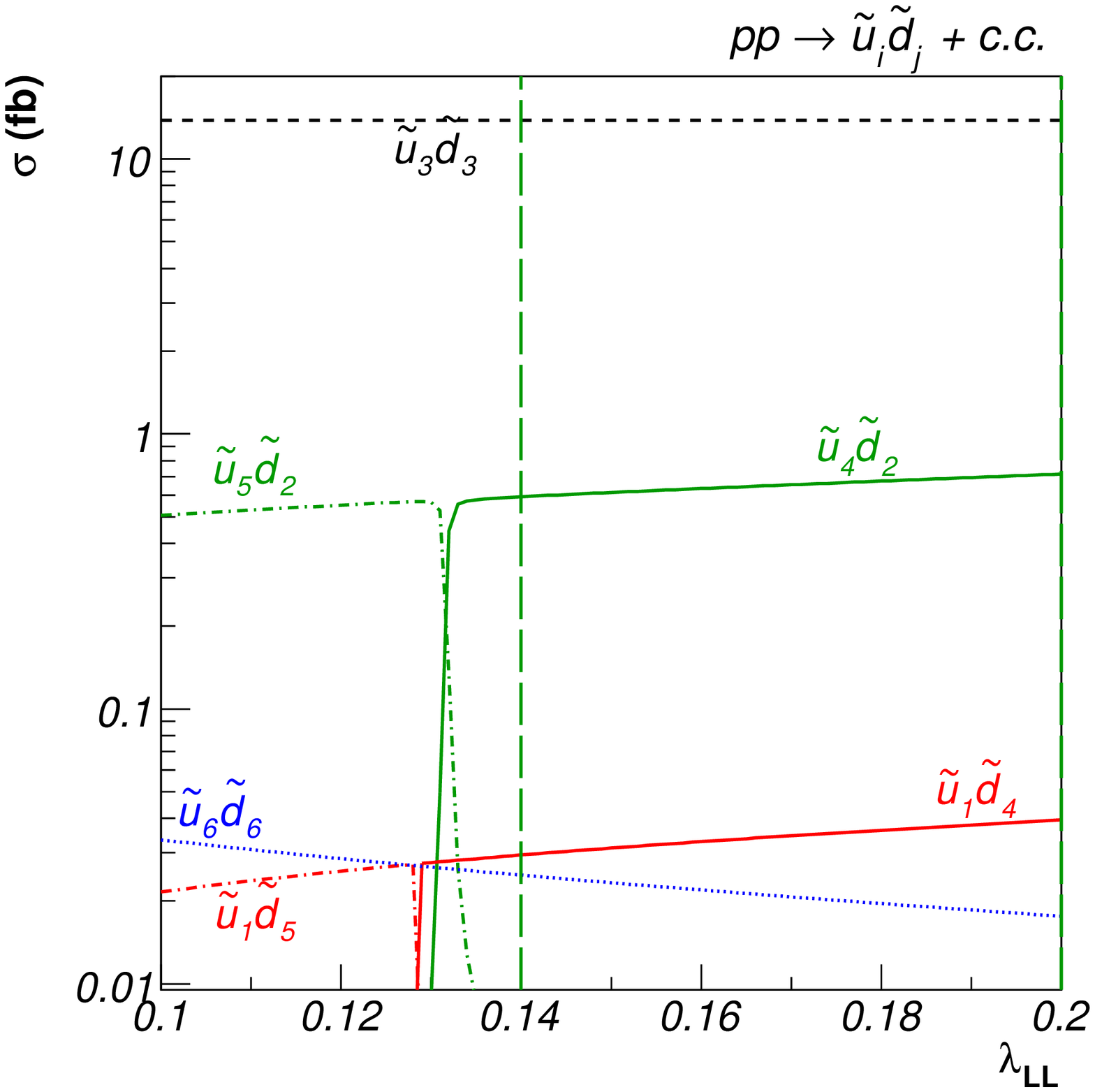} 
    \includegraphics[scale=0.28]{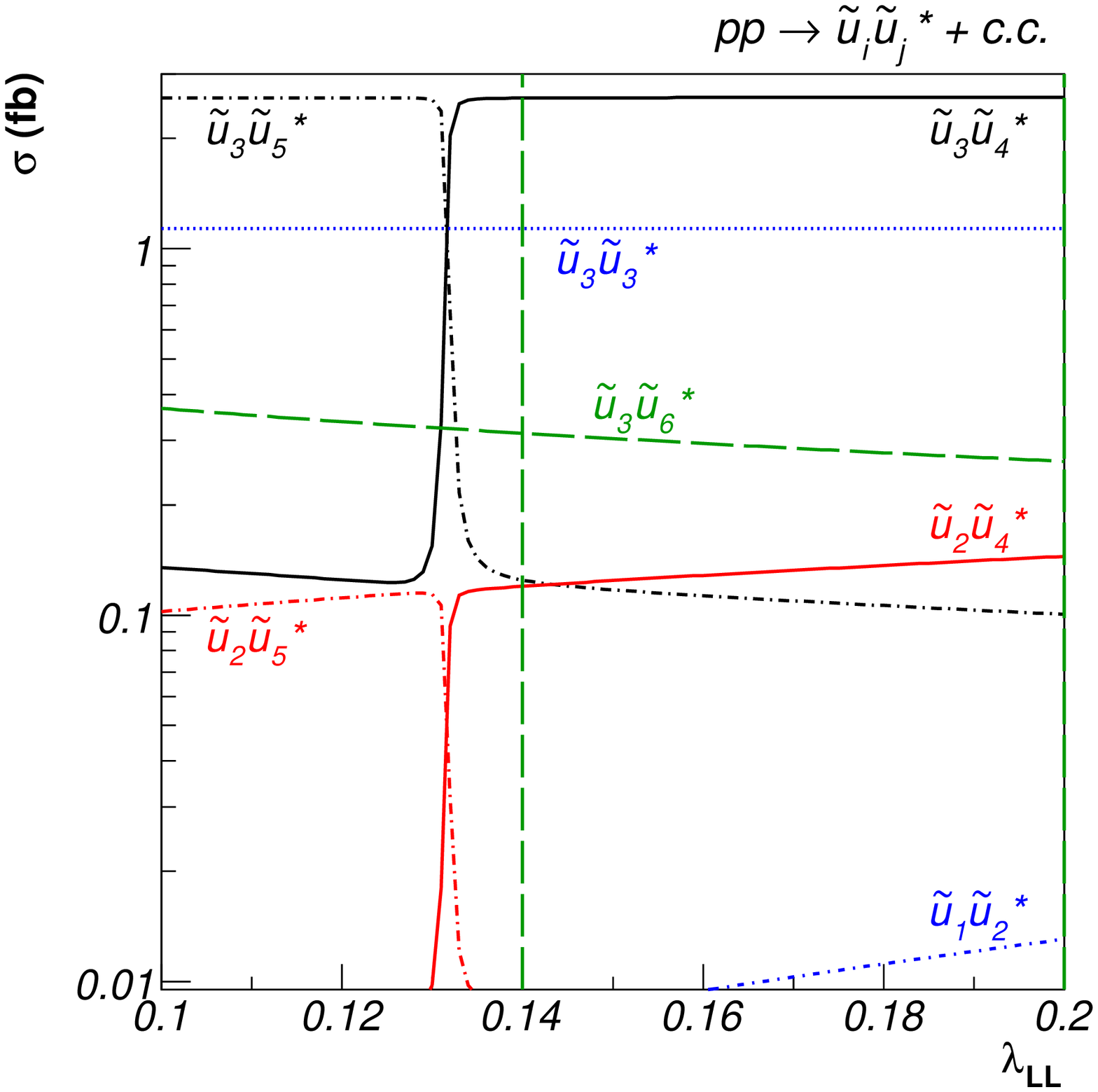} 
    \includegraphics[scale=0.28]{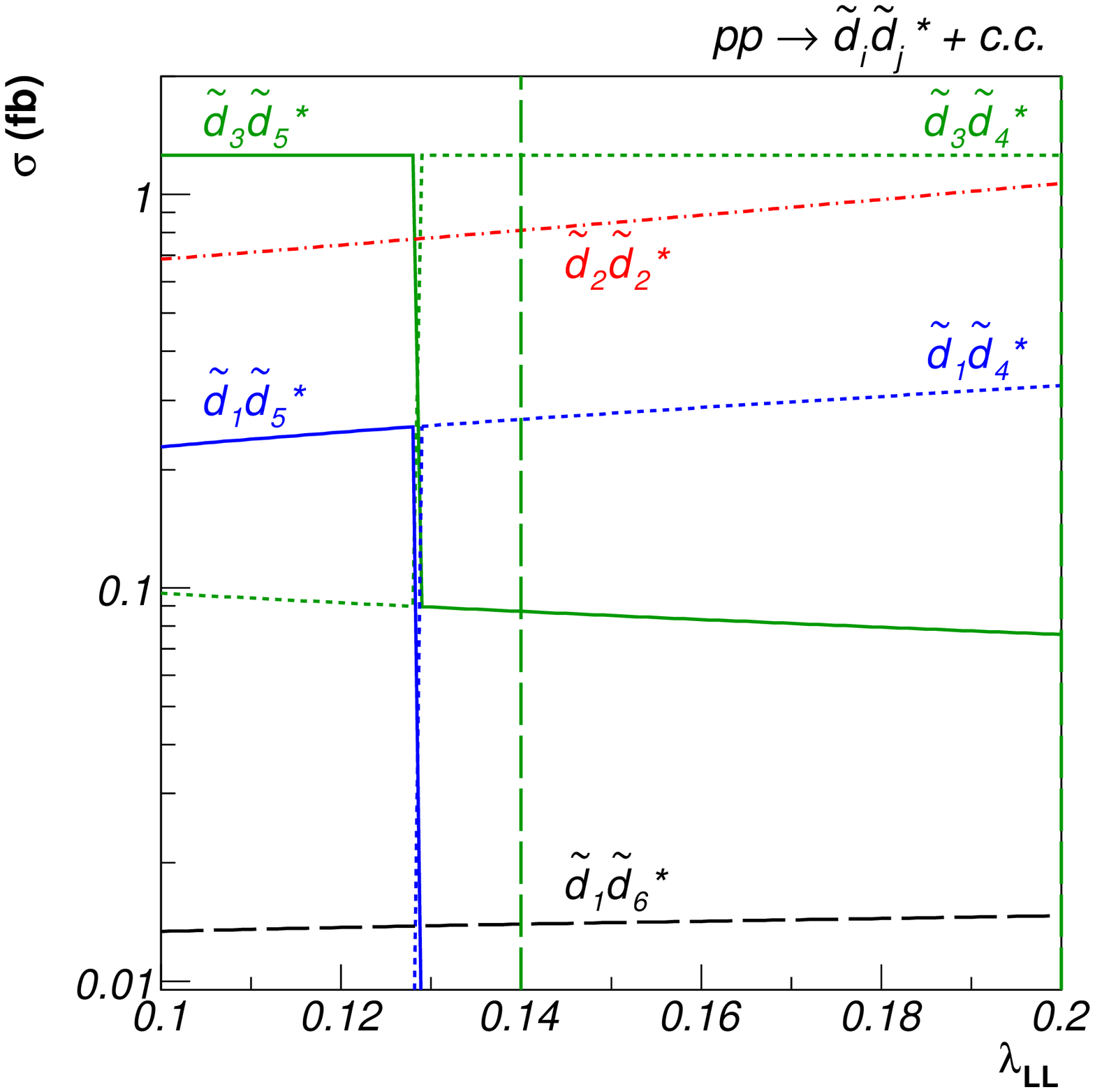} 
    \includegraphics[scale=0.28]{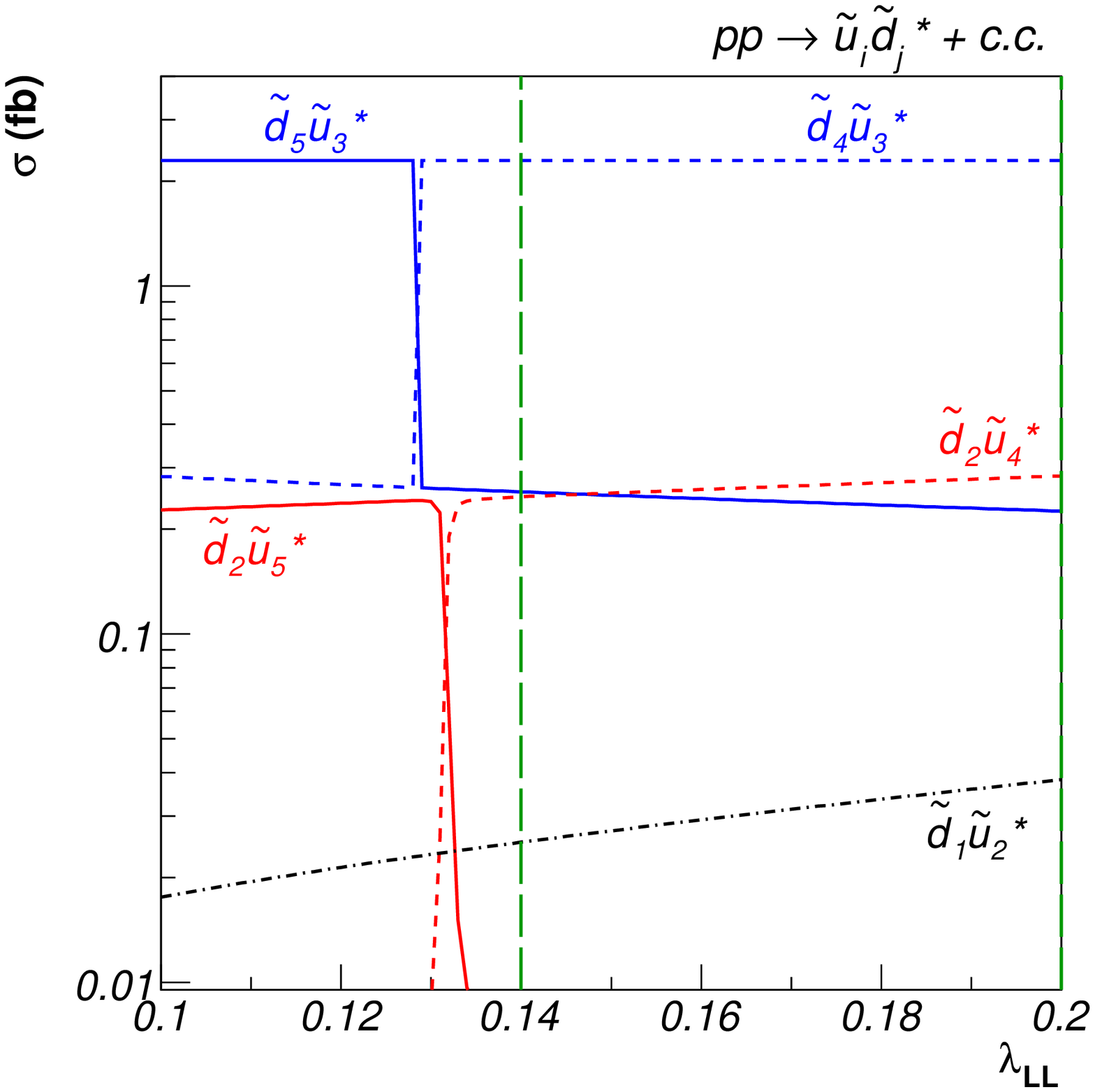} 
    \includegraphics[scale=0.28]{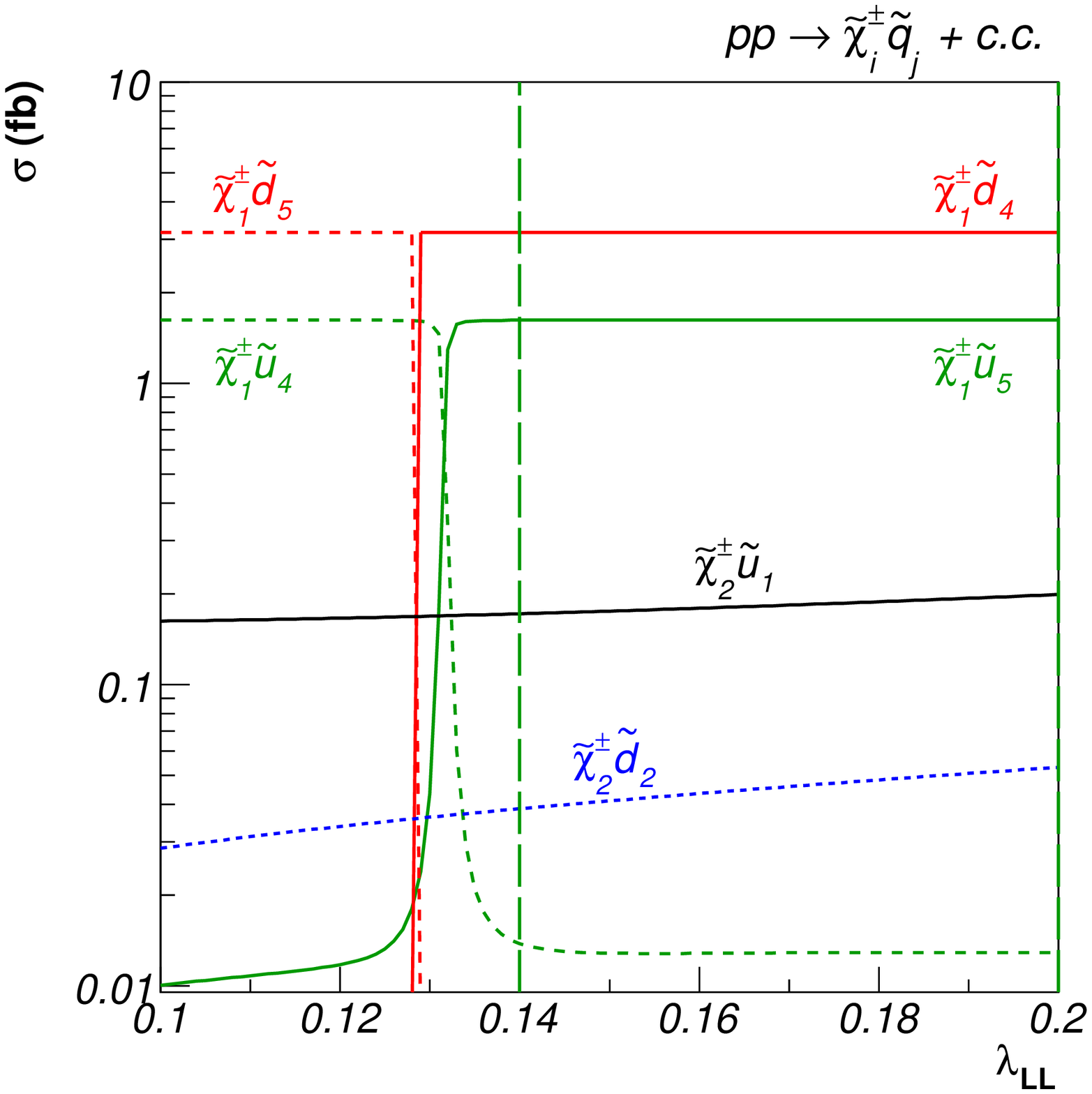} 
    \includegraphics[scale=0.28]{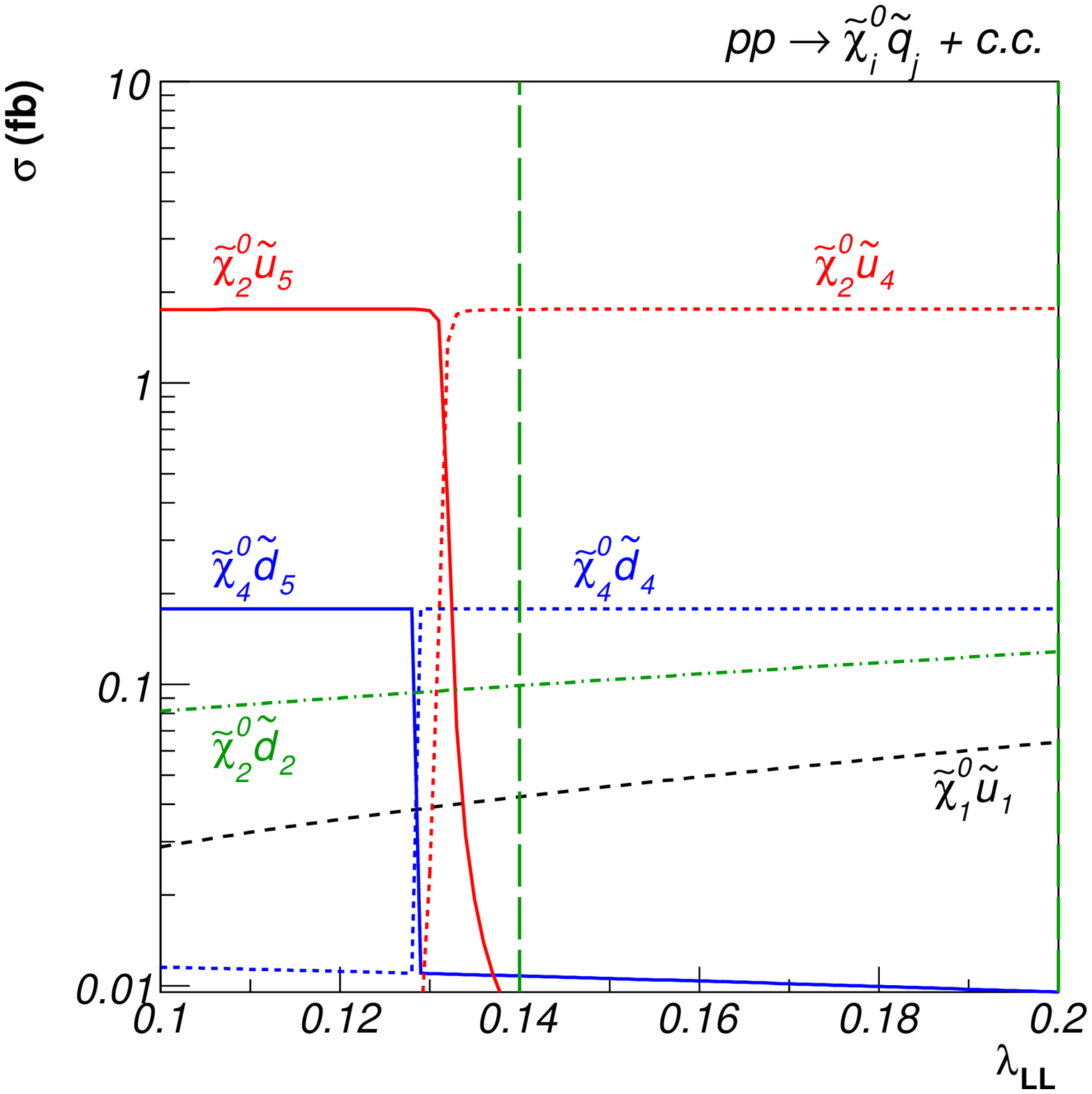} 
    \includegraphics[scale=0.28]{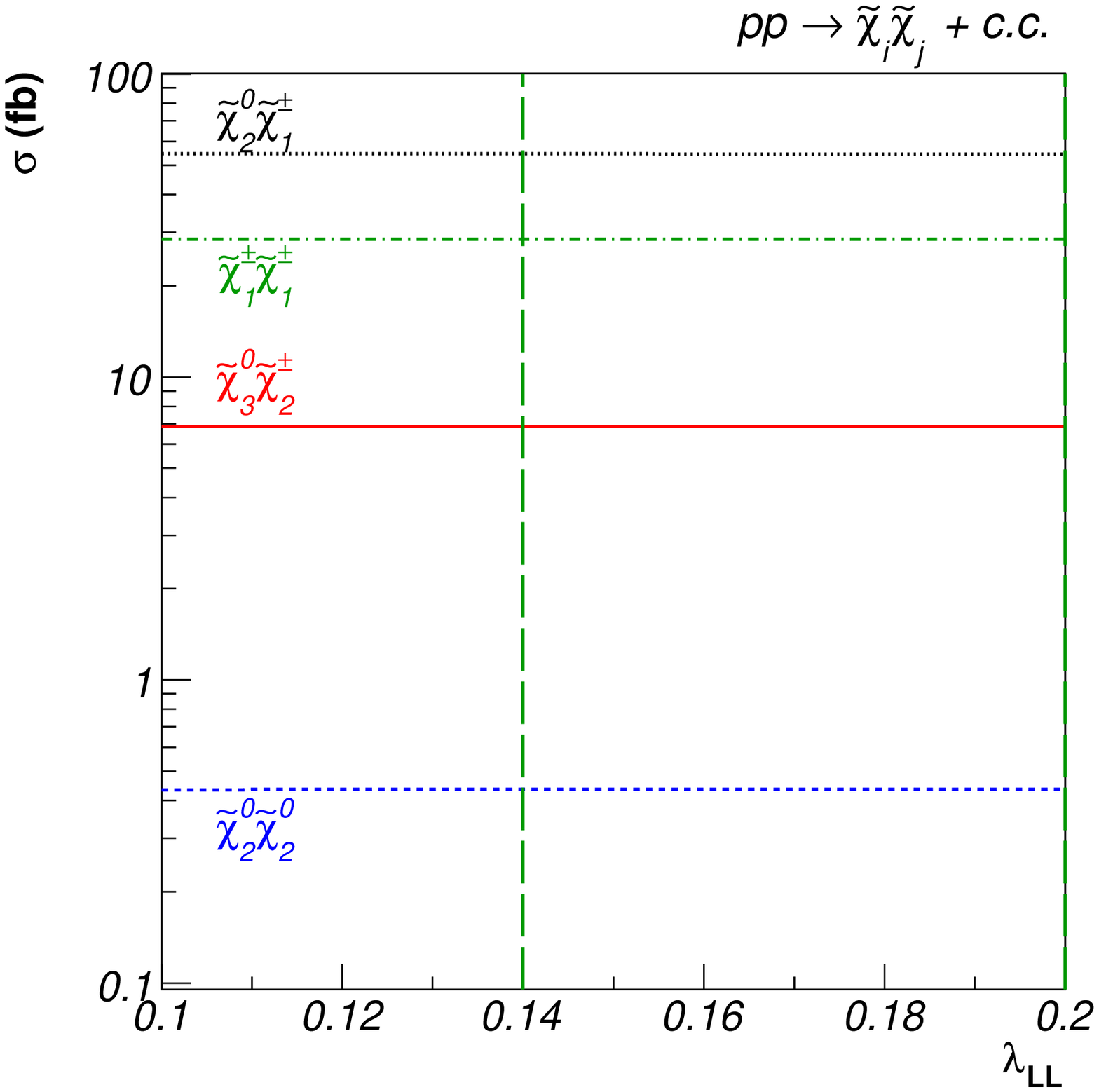} 
  \end{center}
  \vspace*{-5mm}
\caption{Same as Fig.\ \ref{fig16} for our benchmark scenario I.}
\label{fig24}
\end{figure}

\begin{figure}
  \begin{center}
    \includegraphics[scale=0.28]{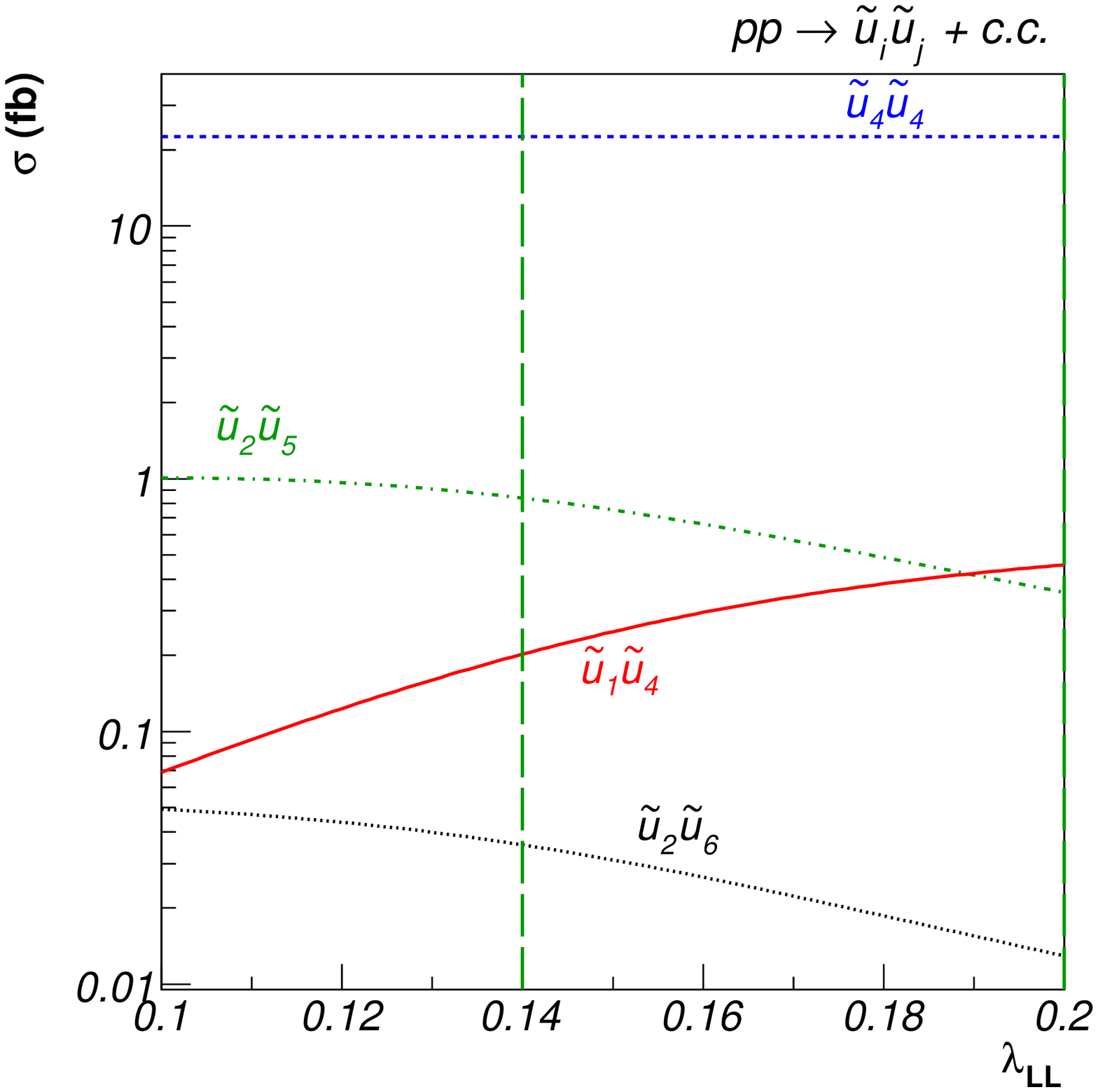} 
    \includegraphics[scale=0.28]{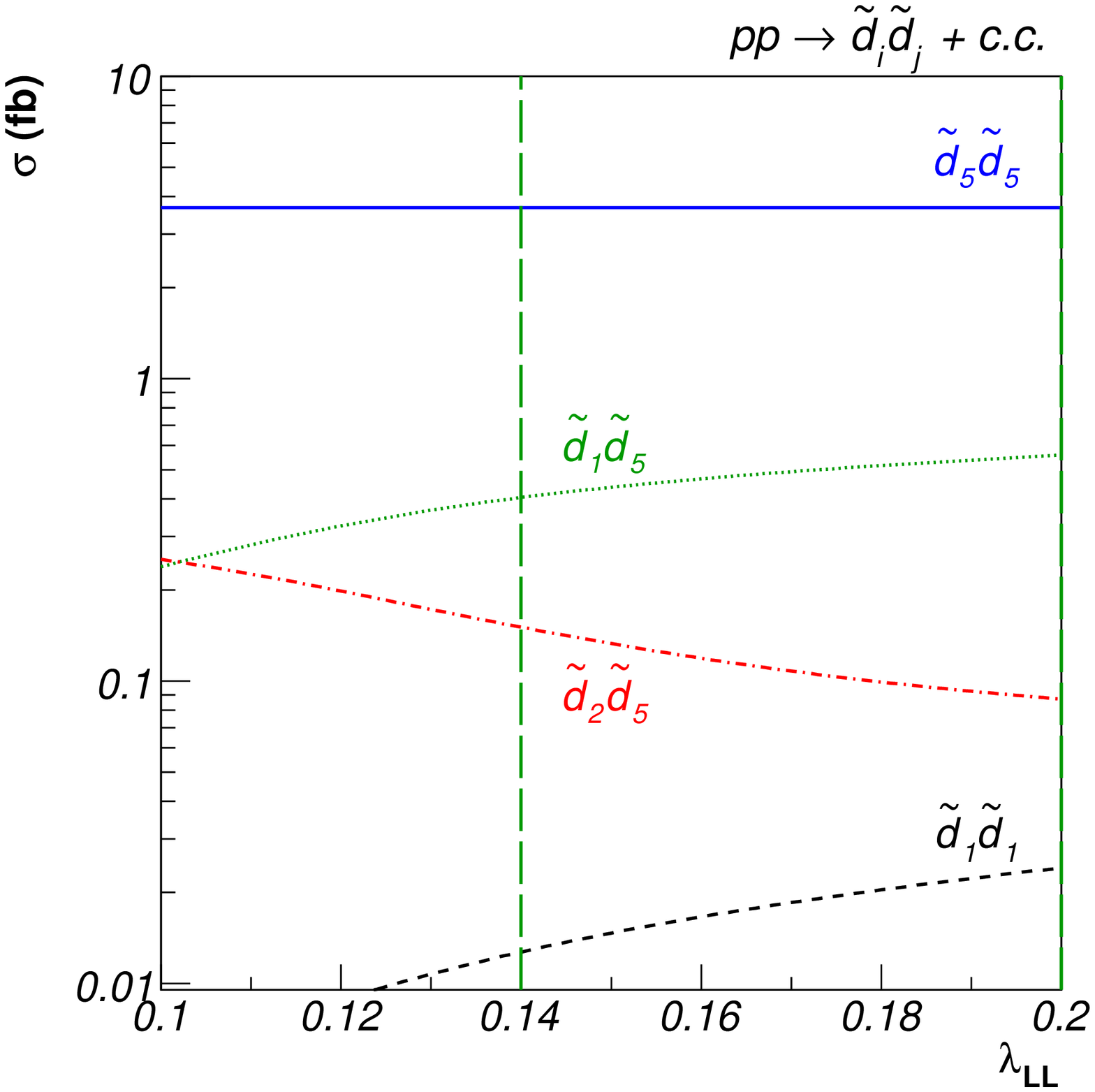} 
    \includegraphics[scale=0.28]{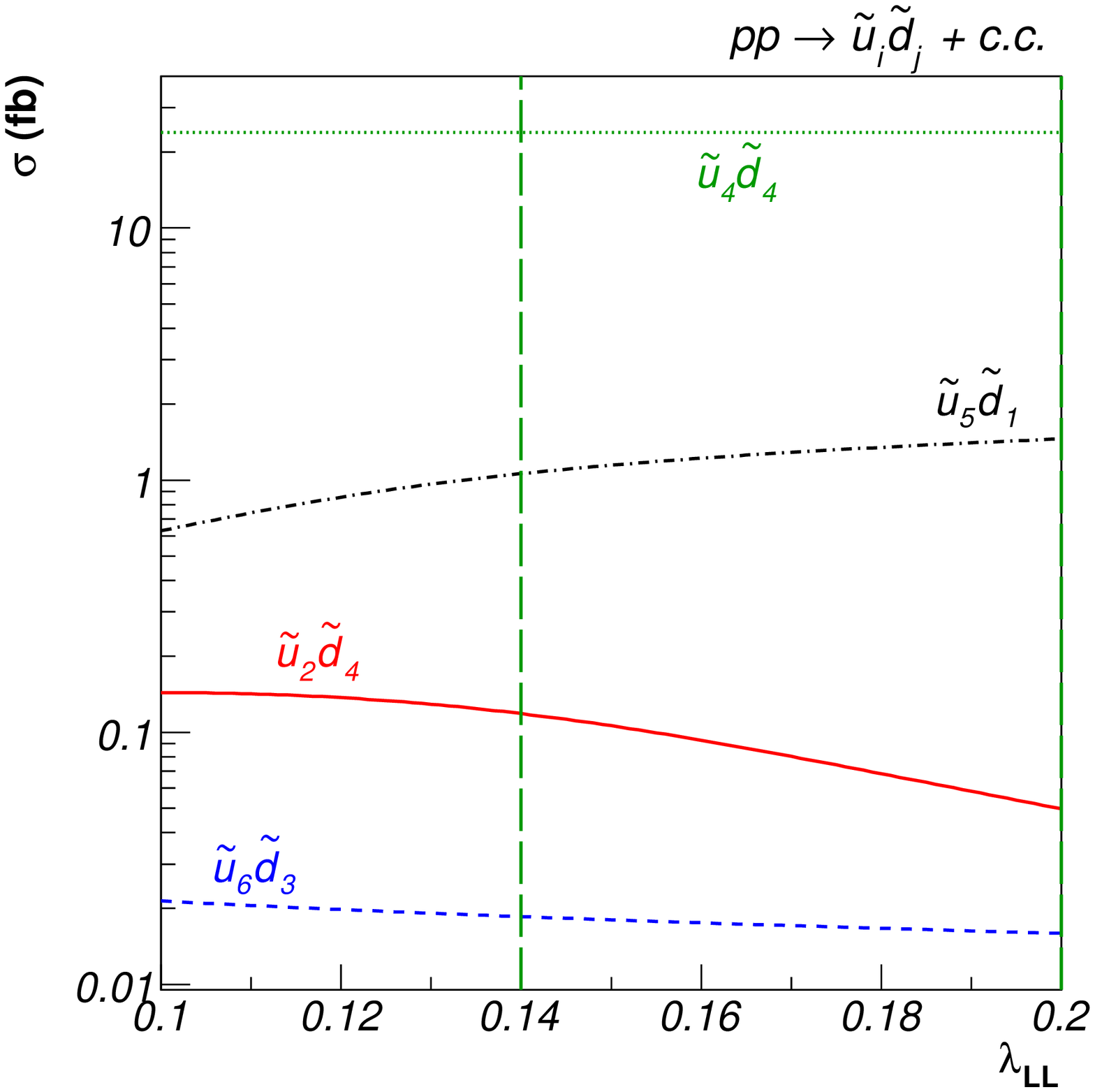} 
    \includegraphics[scale=0.28]{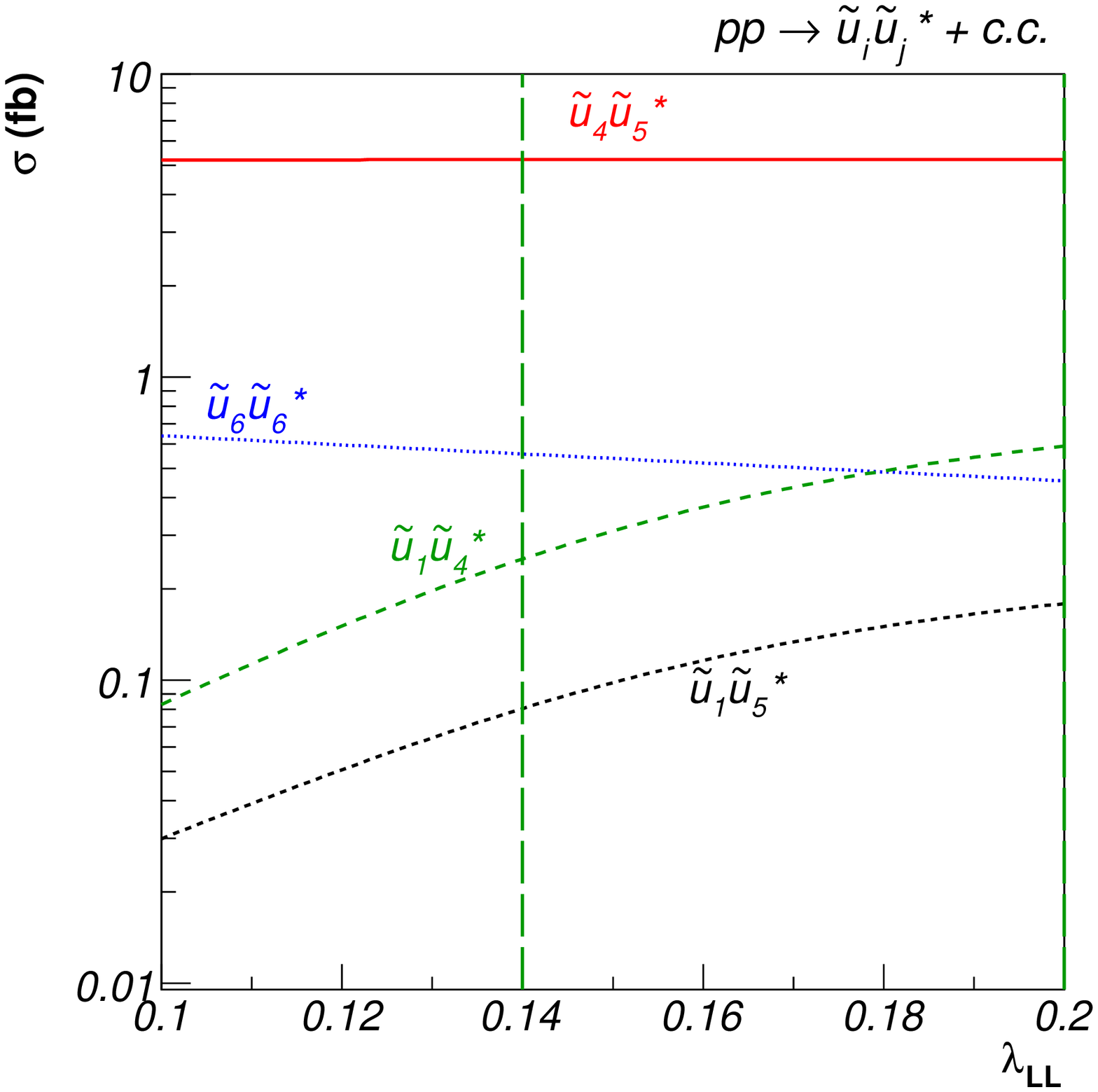} 
    \includegraphics[scale=0.28]{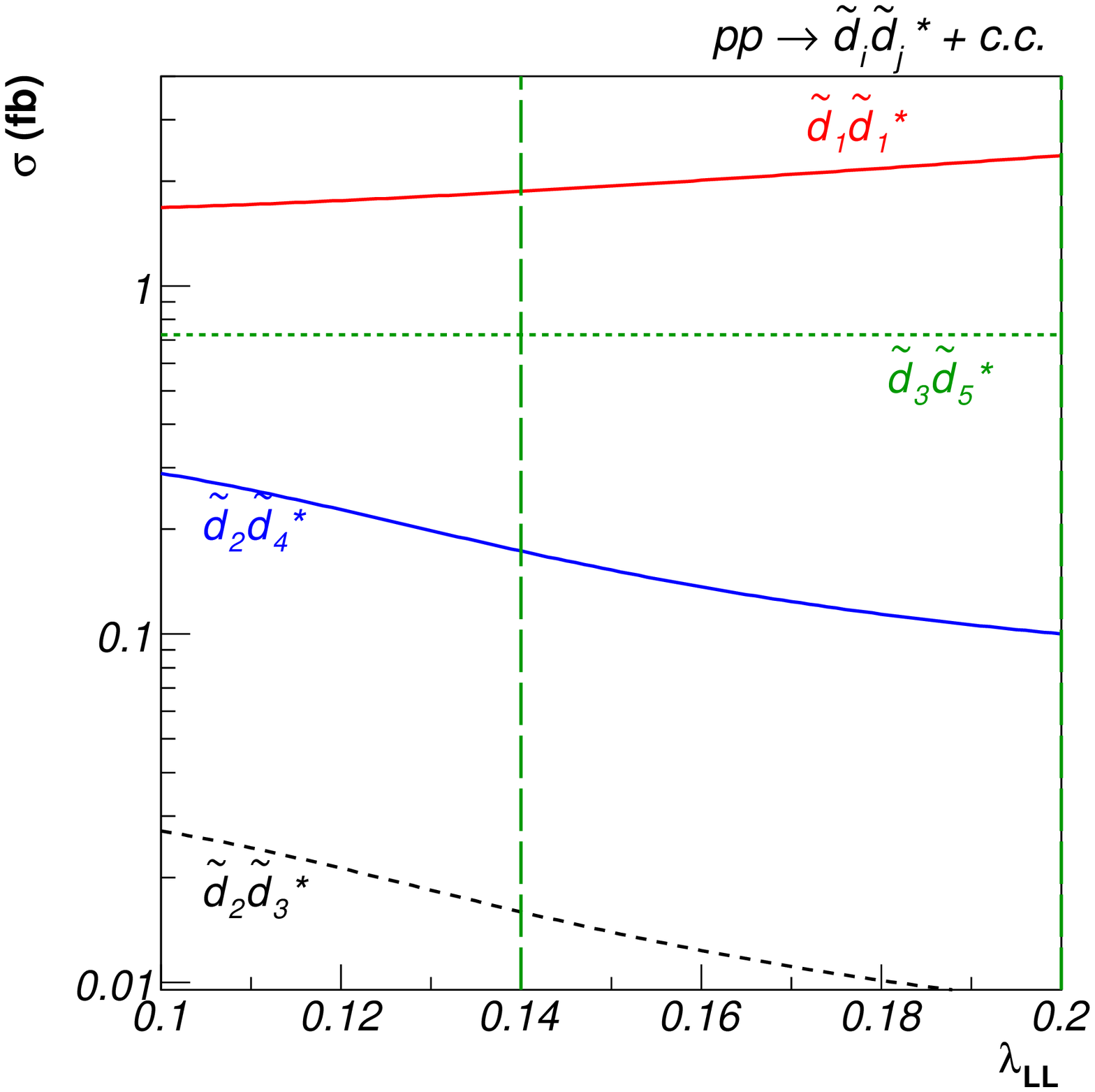} 
    \includegraphics[scale=0.28]{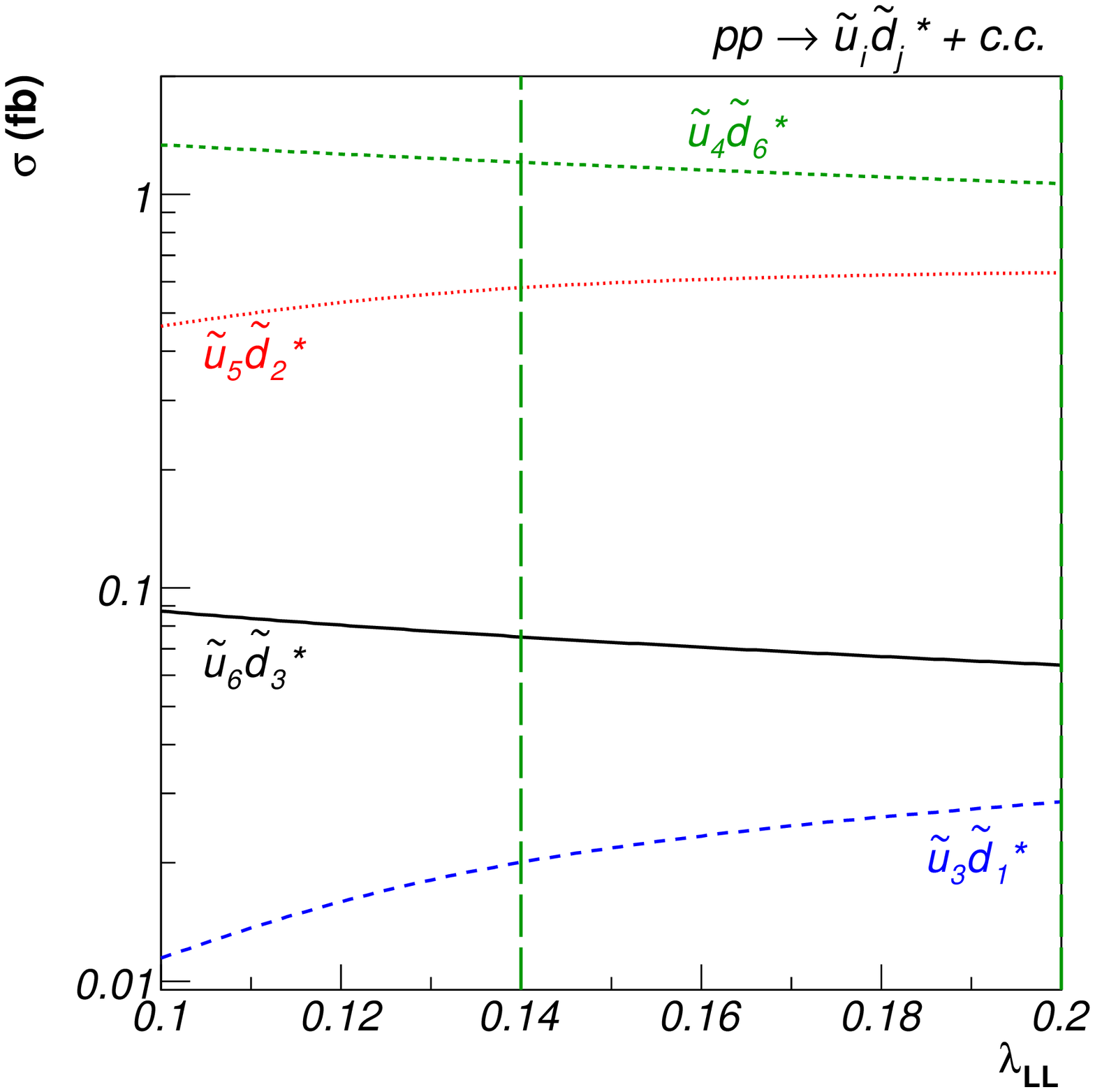} 
    \includegraphics[scale=0.28]{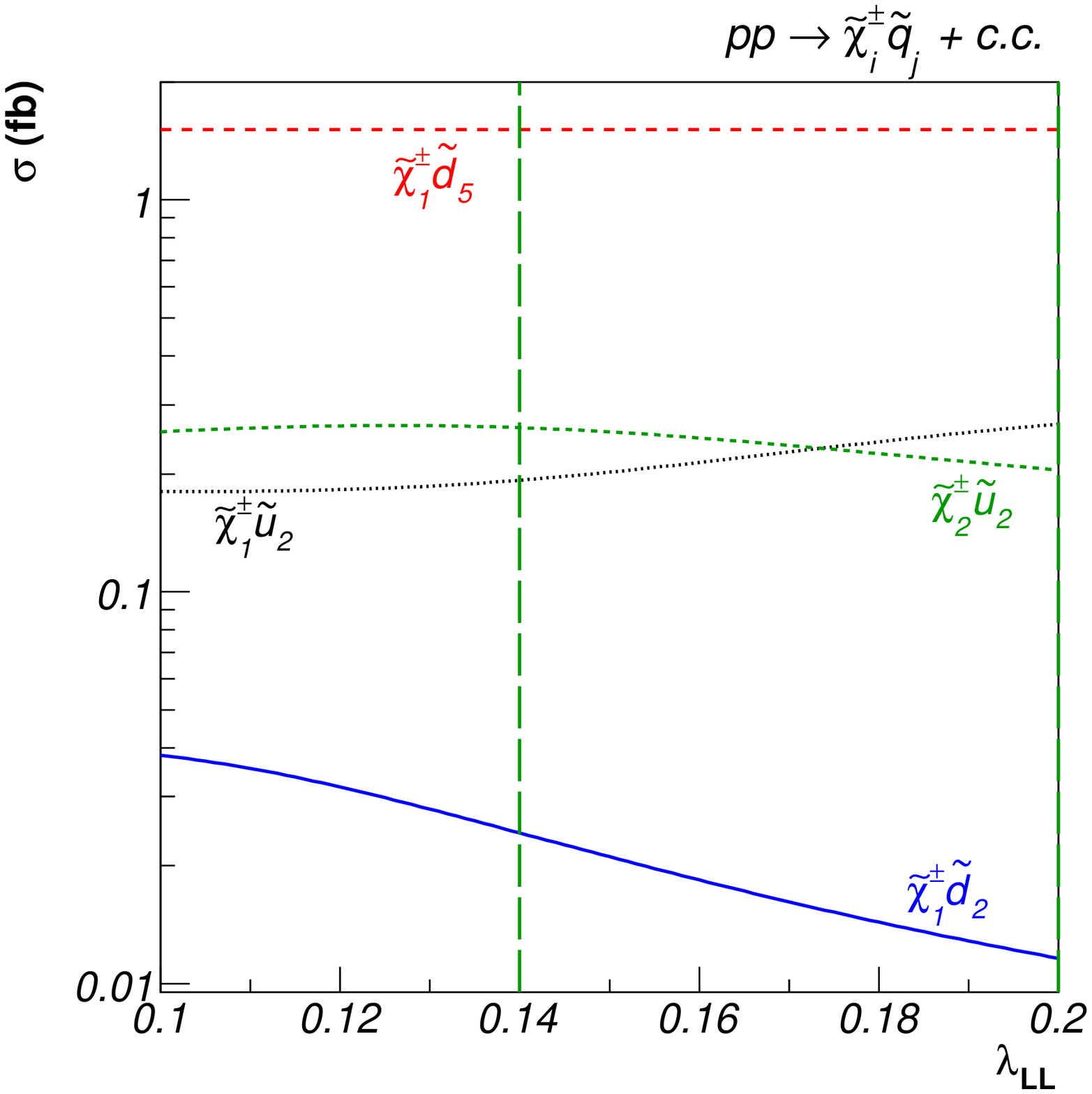} 
    \includegraphics[scale=0.28]{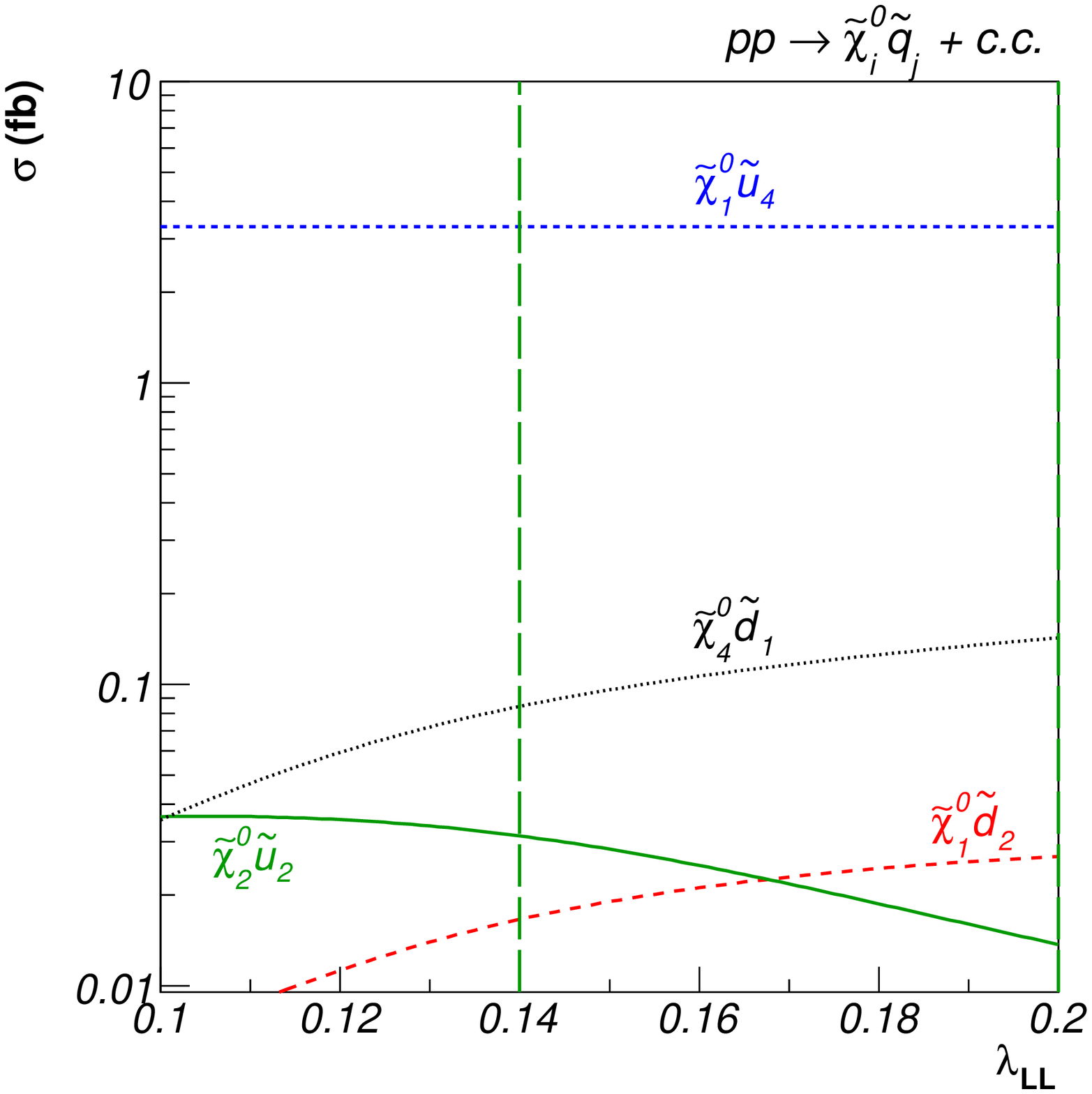} 
    \includegraphics[scale=0.28]{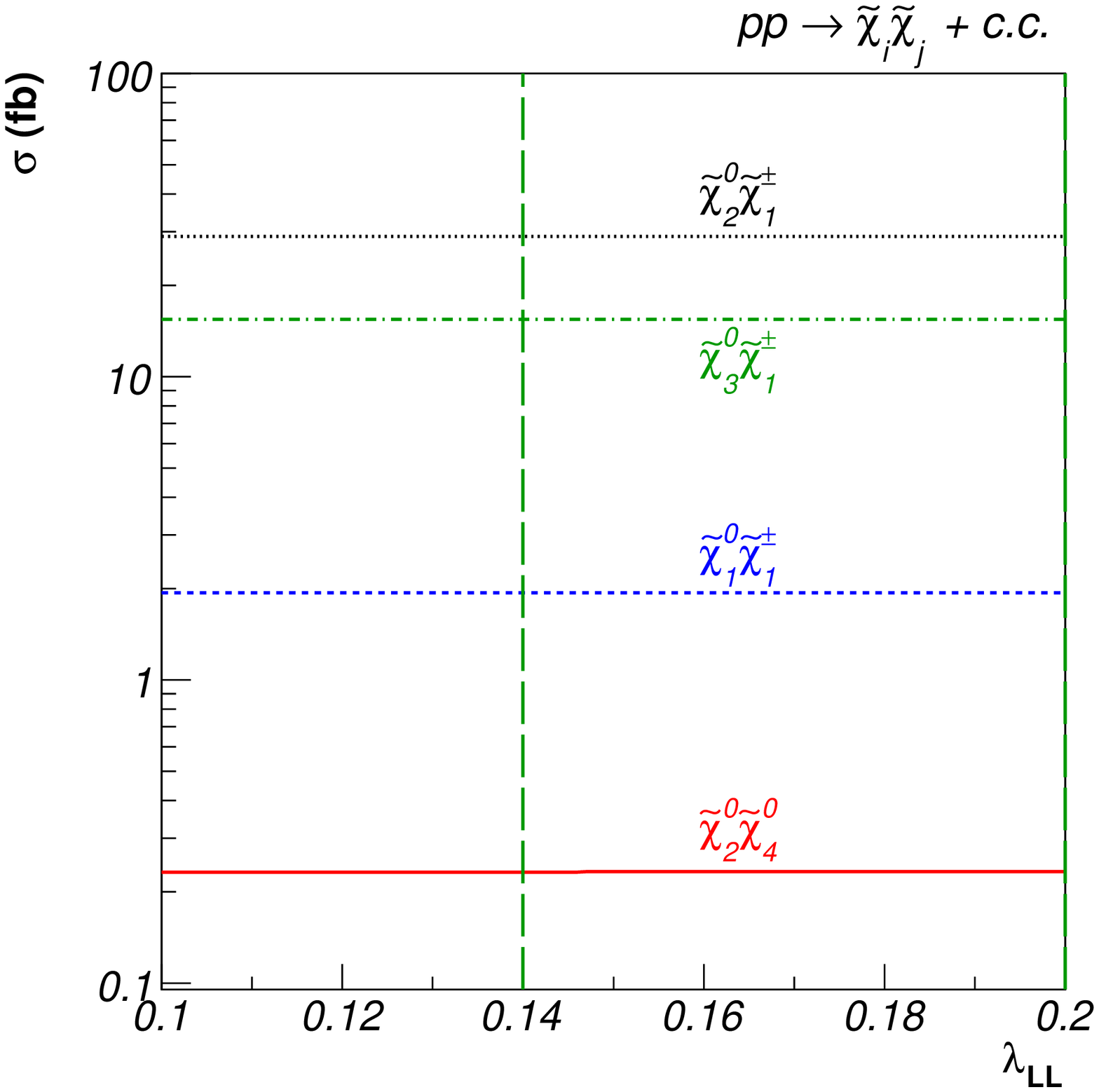} 
  \end{center}
  \vspace*{-5mm}
\caption{Same as Fig.\ \ref{fig15} for our benchmark scenario J.}
\label{fig25}
\end{figure}

\begin{figure}
  \begin{center}
    \includegraphics[scale=0.28]{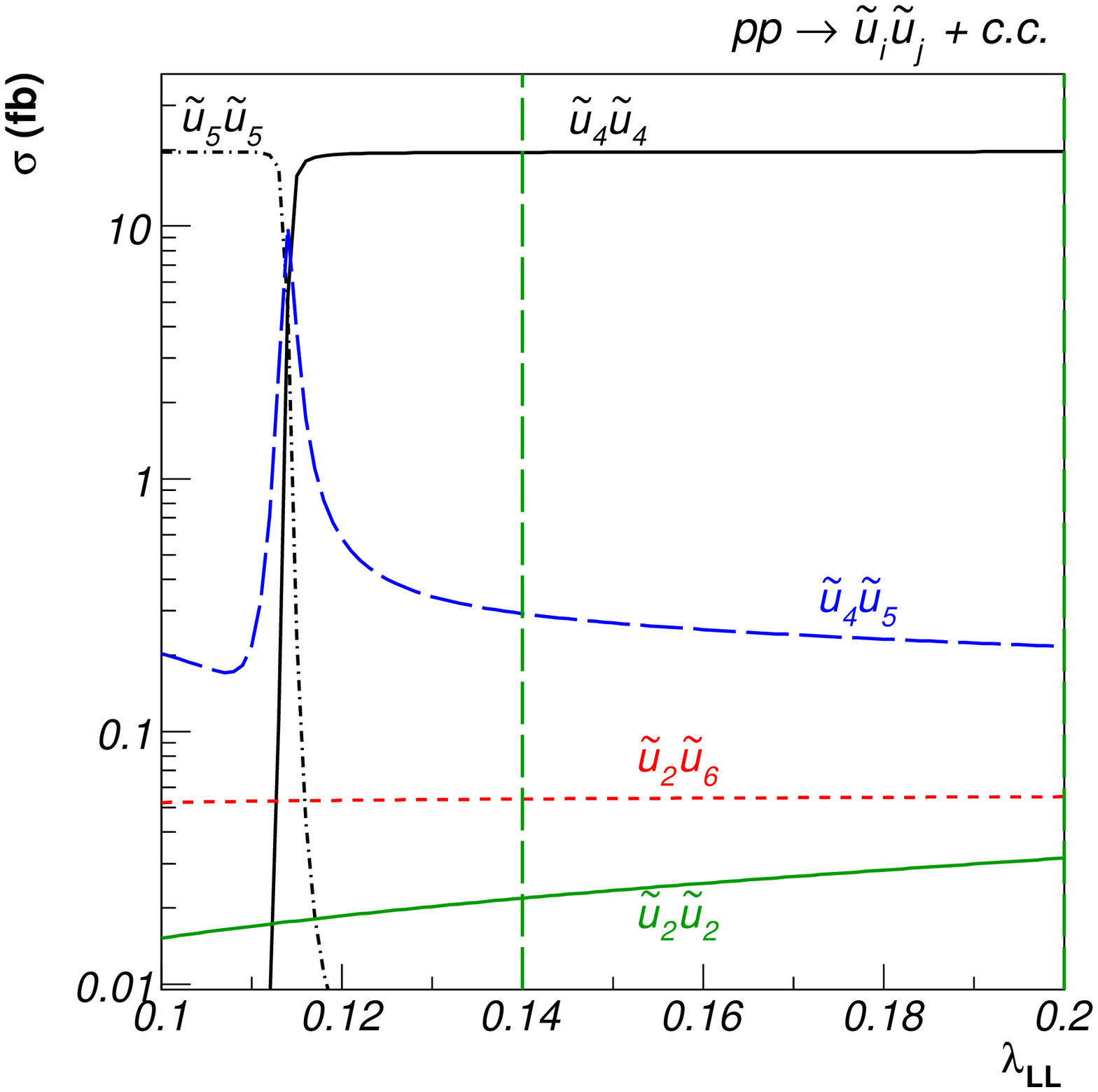} 
    \includegraphics[scale=0.28]{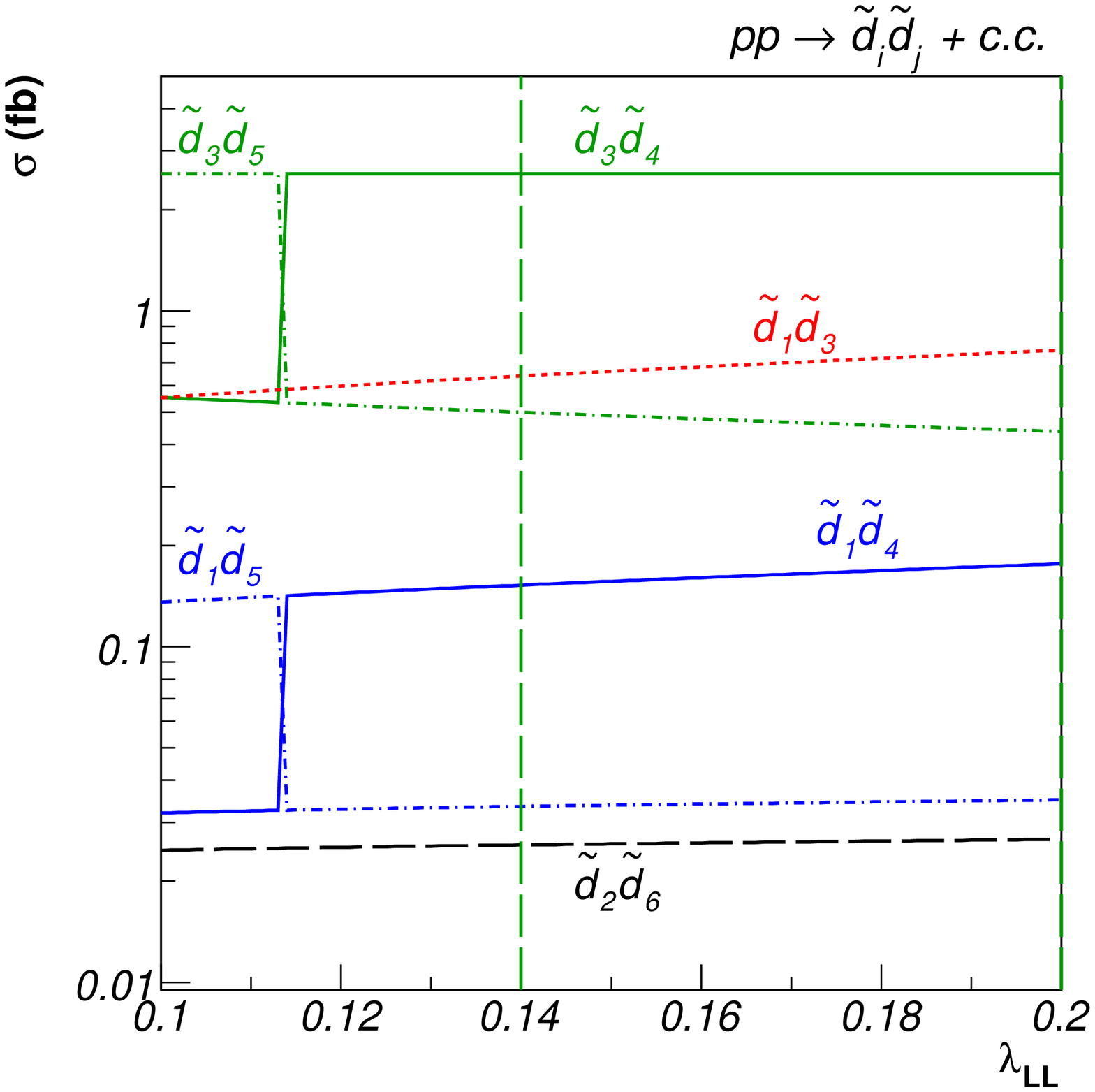} 
    \includegraphics[scale=0.28]{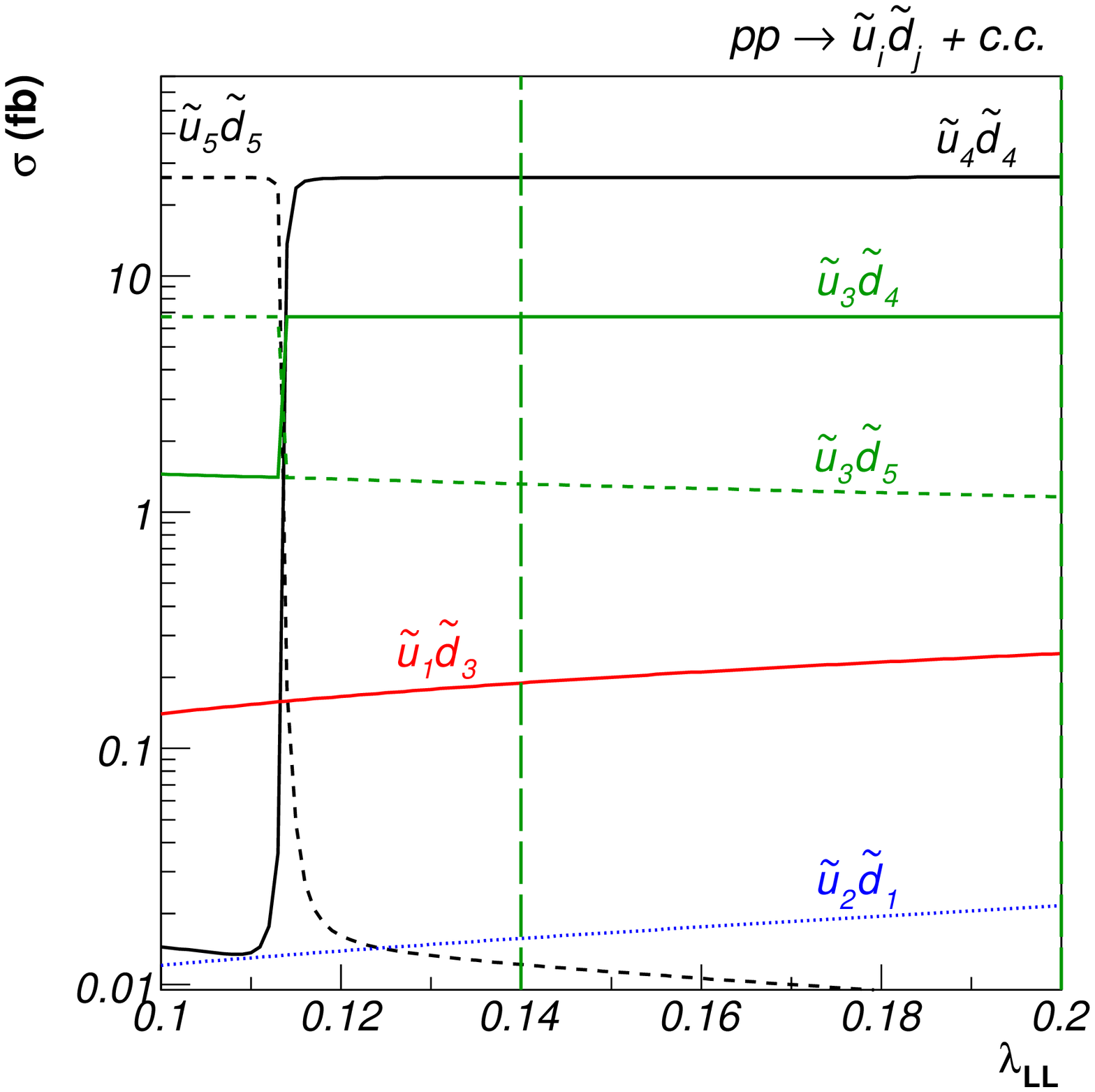} 
    \includegraphics[scale=0.28]{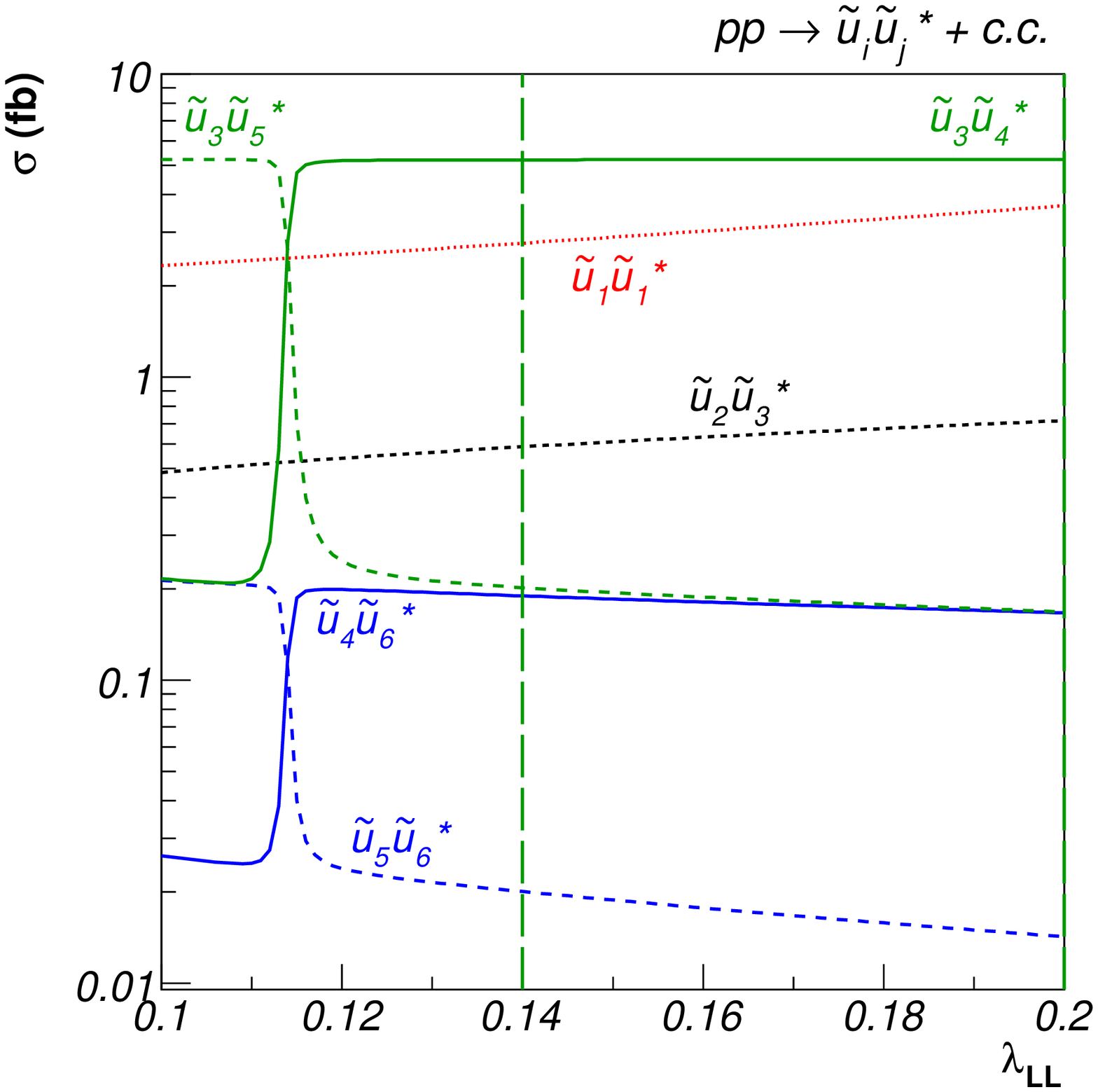} 
    \includegraphics[scale=0.28]{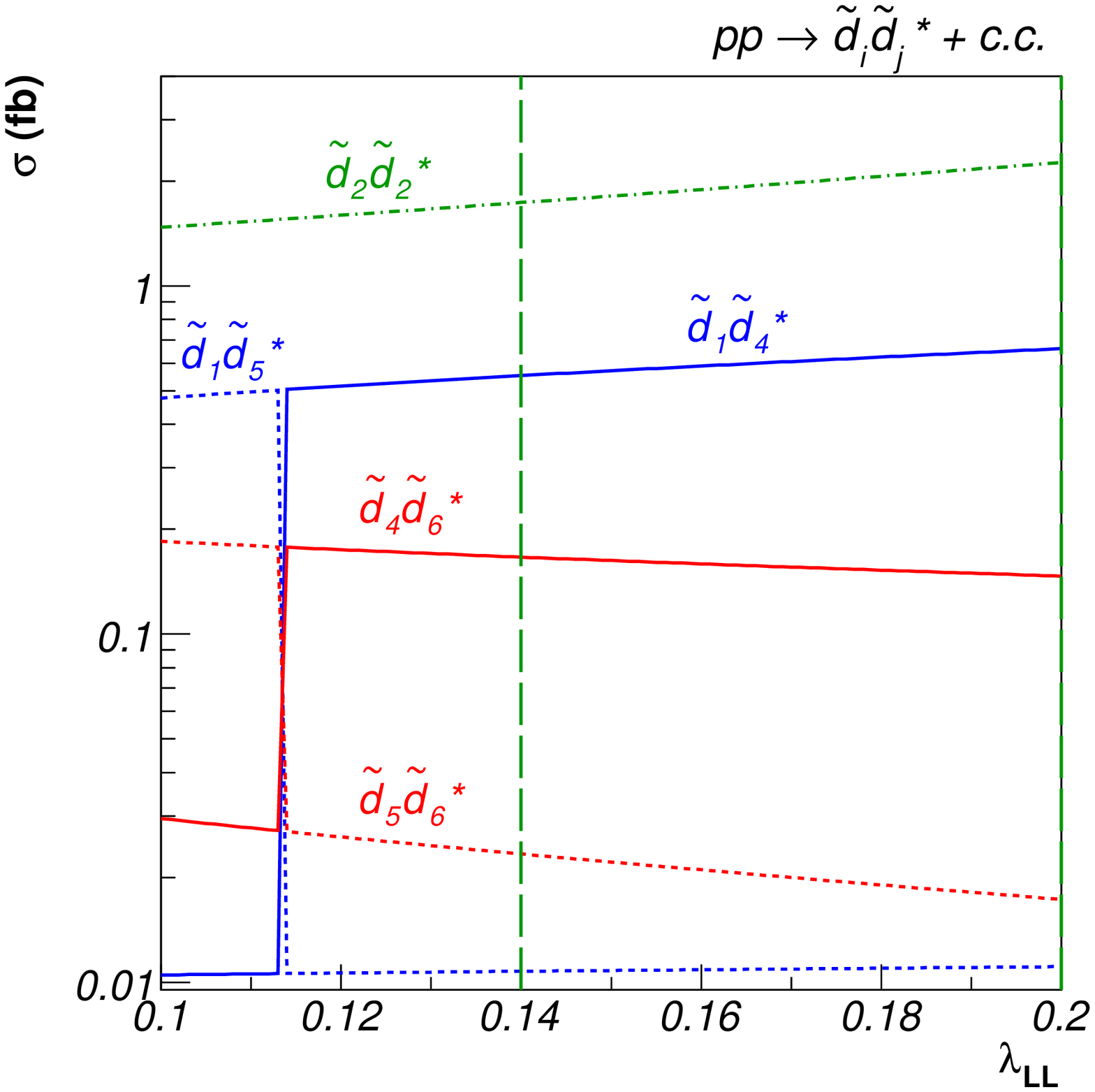} 
    \includegraphics[scale=0.28]{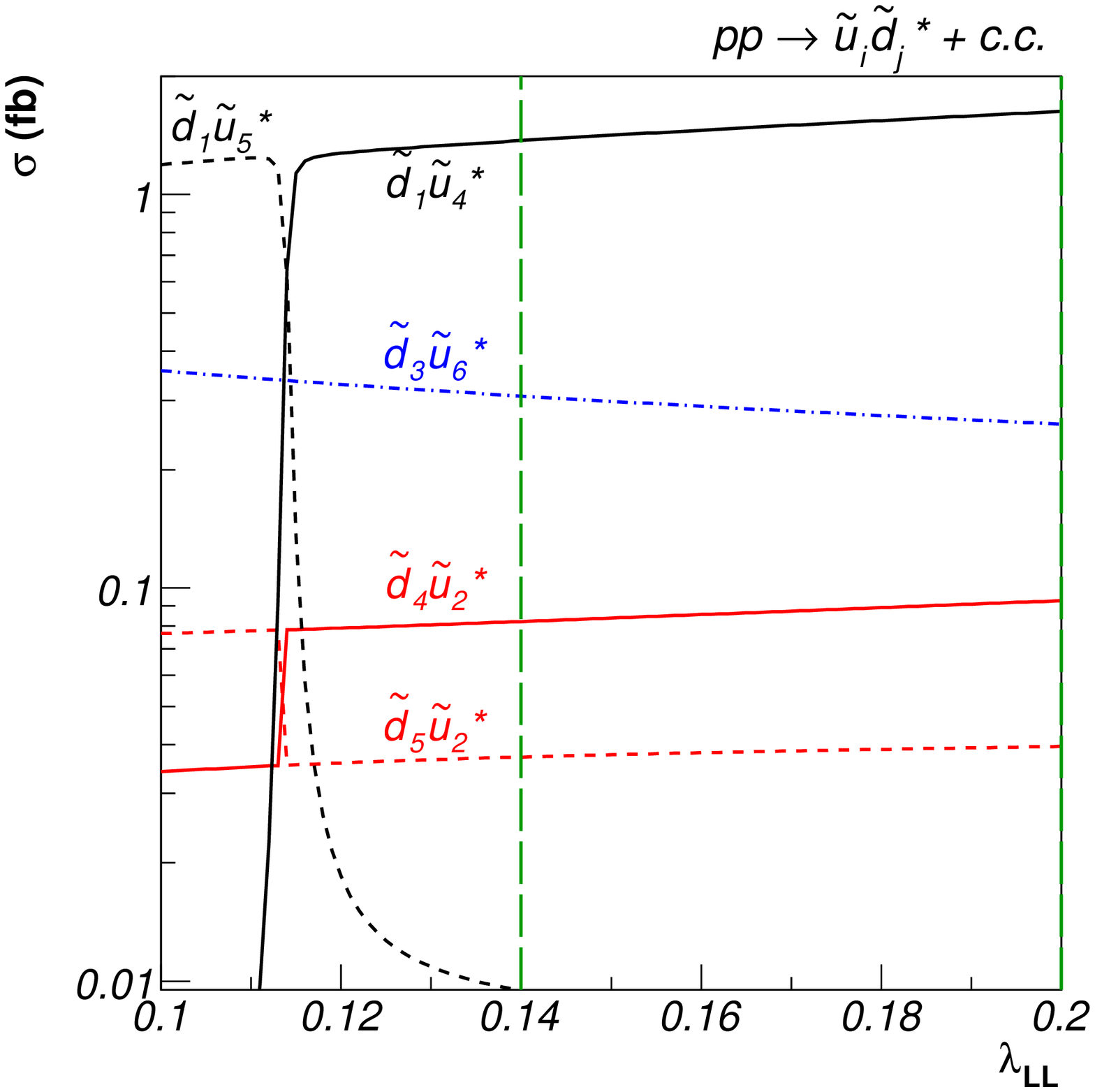} 
    \includegraphics[scale=0.28]{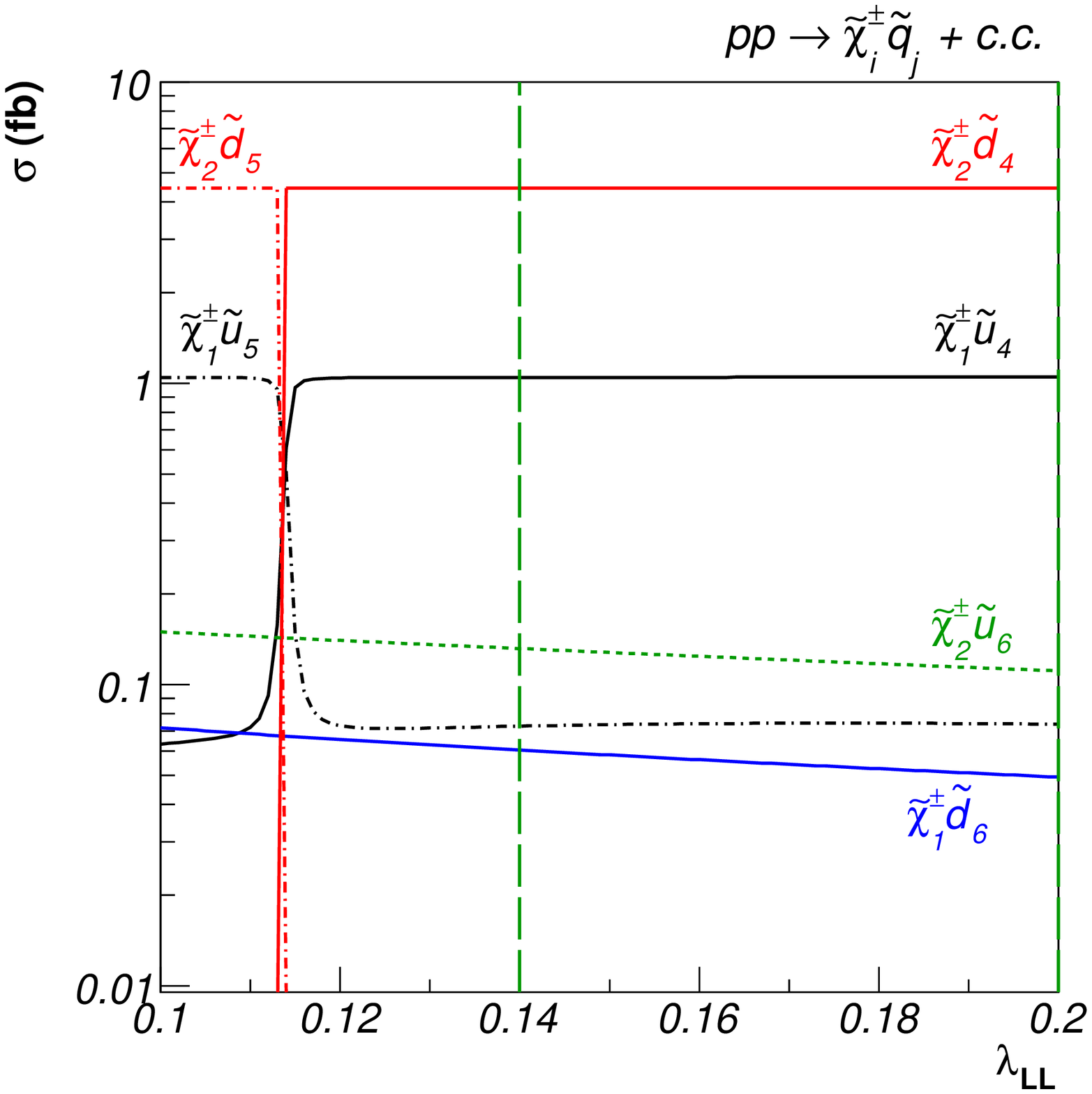} 
    \includegraphics[scale=0.28]{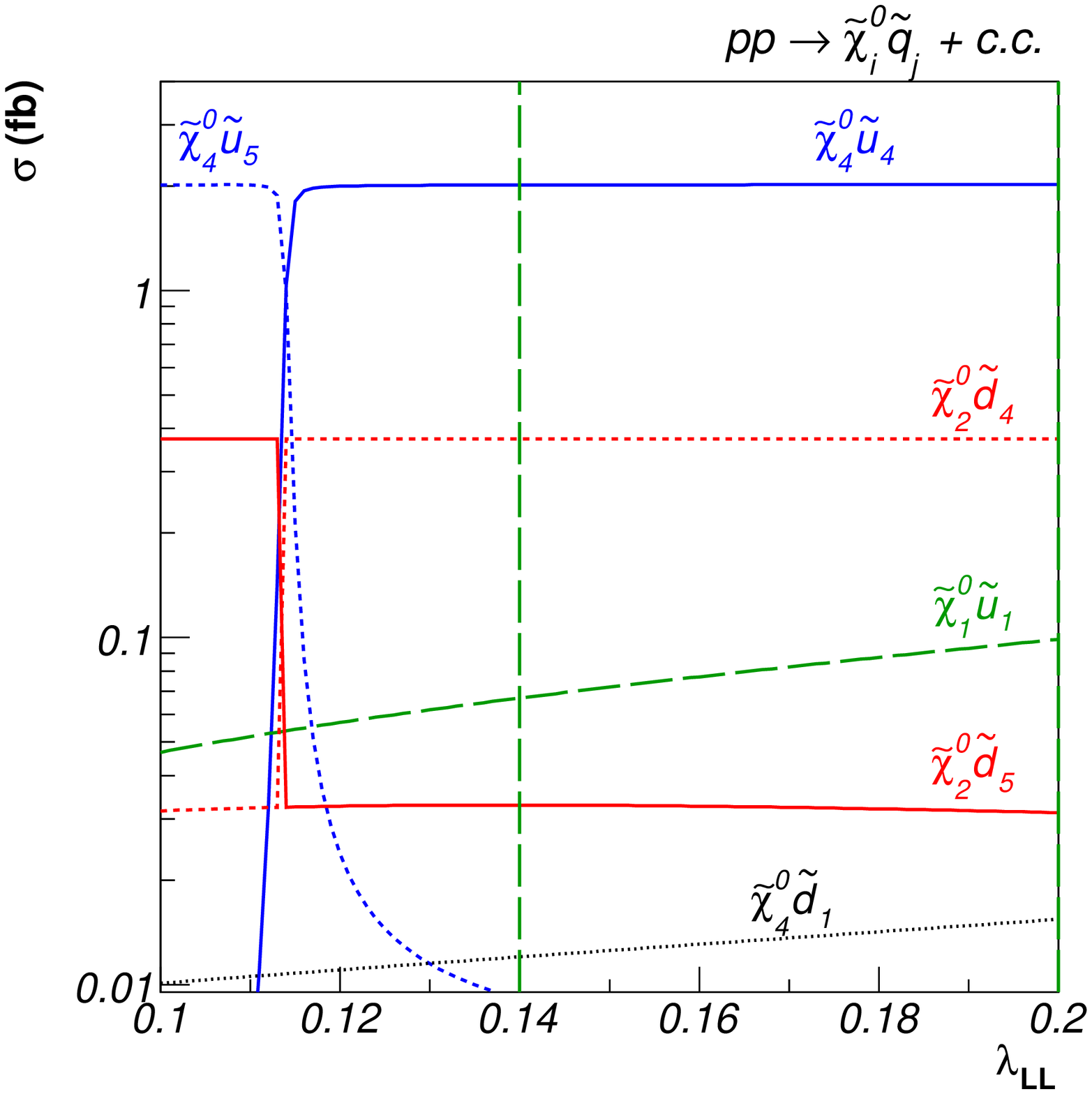} 
    \includegraphics[scale=0.28]{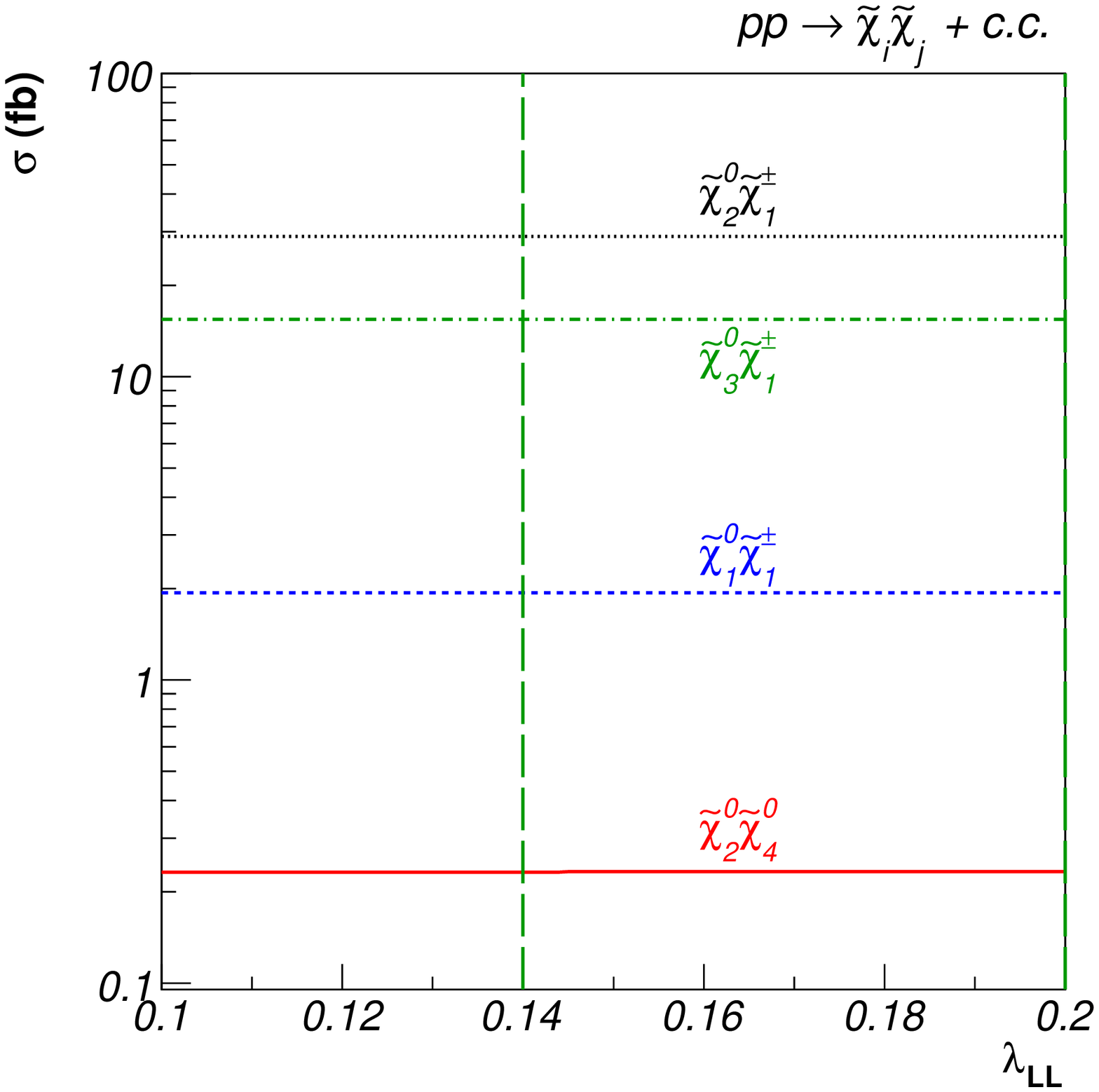} 
  \end{center}
  \vspace*{-5mm}
\caption{Same as Fig.\ \ref{fig16} for our benchmark scenario J.}
\label{fig26}
\end{figure}

Analytic expressions for the relevant partonic cross sections can be found in
Ref.\ \cite{Bozzi:2007me}. In particular, neutral squark-antisquark pair
production can proceed from a neutral quark-antiquark pair in the
initial state. At the tree-level, there are electroweak ($s$-channel photon- or
$Z$- and $t$-channel neutralino- or chargino-exchange) and strong
($s$-channel gluon- and $t$-channel gluino-exchange) contributions. The
contributing Feynman diagrams are shown in Fig.\ \ref{fig27}, where the 
chargino contribution in the third diagram is missing in the expressions given
in Ref.\ \cite{Bozzi:2007me}. The chargino exchanges, which are numerically
unimportant for the dominating channels, have to be taken into account
for up-type (down-type) quarks in the initial state and down-type (up-type)
squarks in the final state. The corrected expression for the quark-initiated
differential partonic cross section (Eq.\ (28) in Ref.\ \cite{Bozzi:2007me}) reads
\bea
 \frac{{\rm d} \hat{\sigma}^{q\bar{q}'}_{h_a, h_b}}{\d t} &=& 
   \Big(1 - h_a\Big) \Big(1 + h_b\Big) 
   \Bigg[
     \frac{\mathcal{Y}}{s^2} + 
     \frac{\mathcal{Z}_1}{s_z^2} +
     \frac{\mathcal{G}}{s^2} + 
     \frac{\widetilde{\mathcal{G}}_{11}}{t_{\tilde{g}}^2} + 
     \frac{\mathcal{[YZ]}_1}{s\, s_z} +
     \frac{[\widetilde{\mathcal{G}}\mathcal{Y}]_1}{t_{\tilde{g}}\, s} +
     \frac{[\widetilde{\mathcal{G}}\mathcal{Z}]_1}{t_{\tilde{g}}\, s_z} + 
     \frac{[\widetilde{\mathcal{G}}\mathcal{G}]_1}{t_{\tilde{g}}\, s} +
    \sum_{k,l=1,...,4}
    \bigg(
      \frac{\mathcal{N}^{kl}_{11}}{t_{\tilde{\chi}^0_k}\, t_{\tilde{\chi}^0_l}}
    \bigg) 
    \nn\\
 && + 
    \sum_{k=1,...,4} \bigg(
      \frac{\mathcal{[NY]}^k_1}{t_{\tilde{\chi}^0_k}\, s} +
      \frac{\mathcal{[NZ]}^k_1}{t_{\tilde{\chi}^0_k}\, s_z} +
      \frac{\mathcal{[NG]}^k_1}{t_{\tilde{\chi}^0_k}\, s} 
    \bigg) + 
    \sum_{k,l=1,2} \bigg(
      \frac{\mathcal{C}^{kl}_{11}}{t_{\tilde{\chi}^\pm_k}\, t_{\tilde{\chi}^\pm_l}}
    \bigg) + 
    \sum_{k=1,2} \bigg(
      \frac{\mathcal{[CY]}^k_1}{t_{\tilde{\chi}^\pm_k}\, s} +
      \frac{\mathcal{[CZ]}^k_1}{t_{\tilde{\chi}^\pm_k}\, s_z} +
      \frac{\mathcal{[CG]}^k_1}{t_{\tilde{\chi}^\pm_k}\, s} 
    \bigg) 
  \Bigg] \nn
\eea
\bea
 &+& \Big(1 + h_a\Big) \Big(1 - h_b\Big) 
   \Bigg[
     \frac{\mathcal{Y}}{s^2} + 
     \frac{\mathcal{Z}_2}{s_z^2} +
     \frac{\mathcal{G}}{s^2} + 
     \frac{\widetilde{\mathcal{G}}_{22}}{t_{\tilde{g}}^2} + 
     \frac{\mathcal{[YZ]}_2}{s\, s_z} +
     \frac{[\widetilde{\mathcal{G}}\mathcal{Y}]_2}{t_{\tilde{g}}\, s} +
     \frac{[\widetilde{\mathcal{G}}\mathcal{Z}]_2}{t_{\tilde{g}}\, s_z} + 
     \frac{[\widetilde{\mathcal{G}}\mathcal{G}]_2}{t_{\tilde{g}}\, s} +
    \sum_{k,l=1,...,4}
    \bigg(
      \frac{\mathcal{N}^{kl}_{22}}{t_{\tilde{\chi}^0_k}\, t_{\tilde{\chi}^0_l}}
    \bigg) 
    \nn\\
 && + 
    \sum_{k=1,...,4} \bigg(
      \frac{\mathcal{[NY]}^k_2}{t_{\tilde{\chi}^0_k}\, s} +
      \frac{\mathcal{[NZ]}^k_2}{t_{\tilde{\chi}^0_k}\, s_z} +
      \frac{\mathcal{[NG]}^k_2}{t_{\tilde{\chi}^0_k}\, s} 
    \bigg) + 
    \sum_{k,l=1,2} \bigg(
      \frac{\mathcal{C}^{kl}_{22}}{t_{\tilde{\chi}^\pm_k}\, t_{\tilde{\chi}^\pm_l}}
    \bigg) + 
    \sum_{k=1,2} \bigg(
      \frac{\mathcal{[CY]}^k_2}{t_{\tilde{\chi}^\pm_k}\, s} +
      \frac{\mathcal{[CZ]}^k_2}{t_{\tilde{\chi}^\pm_k}\, s_z} +
      \frac{\mathcal{[CG]}^k_2}{t_{\tilde{\chi}^\pm_k}\, s} 
    \bigg) 
  \Bigg] \nn\\
&+& \Big(1 - h_a\Big) \Big(1 - h_b\Big) 
   \Bigg[
     \frac{\widetilde{\mathcal{G}}_{12}}{t_{\tilde{g}}^2} + 
    \sum_{k,l=1,...,4}
    \bigg(
      \frac{\mathcal{N}^{kl}_{12}}{t_{\tilde{\chi}^0_k}\, t_{\tilde{\chi}^0_l}}
    \bigg) +
    \sum_{k,l=1,2} \bigg(
      \frac{\mathcal{C}^{kl}_{12}}{t_{\tilde{\chi}^\pm_k}\, t_{\tilde{\chi}^\pm_l}}
    \bigg) 
  \Bigg] \nn \\  
&+& \Big(1 + h_a\Big) \Big(1 + h_b\Big) 
   \Bigg[
     \frac{\widetilde{\mathcal{G}}_{21}}{t_{\tilde{g}}^2} + 
    \sum_{k,l=1,...,4}
    \bigg(
      \frac{\mathcal{N}^{kl}_{21}}{t_{\tilde{\chi}^0_k}\, t_{\tilde{\chi}^0_l}}
    \bigg) +
    \sum_{k,l=1,2} \bigg(
      \frac{\mathcal{C}^{kl}_{21}}{t_{\tilde{\chi}^\pm_k}\, t_{\tilde{\chi}^\pm_l}}
    \bigg) 
  \Bigg],
\eea 
where the propagators appear as mass-subtracted Mandelstam variables,
\beq
   s_z ~=~ s - m_Z^2, \quad t_{\tilde{\chi}^0} ~=~ t - m_{\tilde{\chi}^0}^2,
   \quad t_{\tilde{\chi}^{\pm}} ~=~ t - m_{\tilde{\chi}^{\pm}}^2,
   \quad t_{\tilde{g   }      } ~=~ t - m_{\tilde{g   }  }^2.
\eeq
Rather compact expressions for the appearing form factors $\mathcal{Y}$,
$\mathcal{Z}_m$, $\mathcal{G}$, $\big[\mathcal{YZ}\big]_m$,
$\big[\mathcal{\widetilde{G}Y}\big]_m$, $\big[\mathcal{\widetilde{G}Z}\big]_m$,
$\big[\mathcal{\widetilde{G}G}\big]_m$, $\big[\mathcal{NY}\big]^k_m$,
$\big[\mathcal{NZ}\big]^k_m$, $\big[\mathcal{NG}\big]^k_m$,
$\widetilde{\mathcal{G}}_{nm}$, and
$\mathcal{N}^{kl}_{nm}$
($n,m=1,2$ and $k,l=1,\dots,4$) and the unchanged gluon-initiated contribution
to the cross section are given in Ref.\ \cite{Bozzi:2007me}. The additional form
factors $\big[\mathcal{CY}\big]^k_m$, $\big[\mathcal{CZ}\big]^k_m$,
$\big[\mathcal{CG}\big]^k_m$, and $\mathcal{C}^{kl}_{nm}$ related to the chargino
exchanges read
\bea
  \mathcal{C}_{mn}^{kl} &=& \frac{\pi\,\alpha^2}{4\, x_W^2\, s^2}\,
    \mathcal{C}^{m\ast}_{\tilde{q}_i q \tilde{\chi}_k^\pm}\, 
    \mathcal{C}^m_{\tilde{q}_i q \tilde{\chi}_l^\pm}\,
    \mathcal{C}^n_{\tilde{q}_j q^\prime \tilde{\chi}_k^\pm}
    \mathcal{C}^{n\ast}_{\tilde{q}_j q^\prime \tilde{\chi}_l^\pm} 
    \Bigg[
      \left( u\, t - m^2_{\tilde{q}_i}\, m^2_{\tilde{q}_j}\right)\,
        \delta_{mn} + 
      \left( m_{\tilde{\chi}^\pm_k}\, m_{\tilde{\chi}^\pm_l}\, s \right)\, 
        \left( 1-\delta_{mn} \right) \Bigg] ,~ \nonumber \\ 
  \mathcal{[CY]}_m^k &=& 
    \frac{\pi\, \alpha^2\, e_q\, e_{\tilde{q}}\, \delta_{ij}\, \delta_{qq^\prime}}
         {3\, x_W\, s^2}\, {\rm Re } \left[
	    \mathcal{C}^m_{\tilde{q}_i q \tilde{\chi}_k^\pm}\, 
            \mathcal{C}^{m\ast}_{\tilde{q}_j q^\prime \tilde{\chi}_k^\pm} 
         \right] 
         \left( u\, t - m^2_{\tilde{q}_i}\, m^2_{\tilde{q}_j}\right),~ \nn
\eea
\bea
  \mathcal{[CZ]}_m^k &=& \frac{\pi\, \alpha^2}{12\, x_W^2\, (1 - x_W) s^2}\, 
    {\rm Re} \left[
       \mathcal{C}^m_{\tilde{q}_i q \tilde{\chi}_k^\pm}\,
       \mathcal{C}^{m\ast}_{\tilde{q}_j q^\prime \tilde{\chi}_k^\pm} 
         \left( 
           L_{\tilde{q}_i \tilde{q}_j Z} + R_{\tilde{q}_i \tilde{q}_j Z} 
         \right)
       \right]\, 
       \mathcal{C}^m_{q q^\prime Z}\, 
       \left( u\, t -  m^2_{\tilde{q}_i}\, m^2_{\tilde{q}_j}\right),~ \nn \\
  \mathcal{[CG]}_m^k &=& 
    \frac{4\, \pi\, \alpha\, \alpha_s\, \delta_{ij}\, \delta_{qq^\prime}}
         {9\, x_W\, s^2}\, {\rm Re} \left[
	   \mathcal{C}^m_{\tilde{q}_i q \tilde{\chi}_k^\pm}\, 
	   \mathcal{C}^{m\ast}_{\tilde{q}_j q^\prime \tilde{\chi}_k^\pm}
         \right] 
	 \left( u\, t - m^2_{\tilde{q}_i}\, m^2_{\tilde{q}_j}\right)
\eea
with all other variables defined as in Ref.\ \cite{Bozzi:2007me}.
We take the opportunity to also correct a few minor typographical errors
in some of the form
factors for the pair production of two up- or down-type squarks (Eq.\ (36) in
Ref.\ \cite{Bozzi:2007me})
\bea 
  \large[\mathcal{NTU}\large]_{mn}^{kl} &=& 
   \frac{2 \,\pi\,\alpha^2}{3\, x_W^2\, \big(1-x_W\big)^2 s^2} {\rm Re} 
     \left[\mathcal{C}^{m\ast}_{\tilde{q}_i q \tilde{\chi}_k^0}\, 
           \mathcal{C}^{n\ast}_{\tilde{q}_j q^\prime \tilde{\chi}_k^0}\,
	   \mathcal{C}^n_{\tilde{q}_i q^\prime \tilde{\chi}_l^0}\,
	   \mathcal{C}^m_{\tilde{q}_j q \tilde{\chi}_l^0}
     \right]\, 
     \Bigg[
        \left( u\, t - m^2_{\tilde{q}_i}\, m^2_{\tilde{q}_j}\right) 
        \left( \delta_{mn} - 1 \right) 
      + m_{\tilde{\chi}^0_k}\, m_{\tilde{\chi}^0_l}\, s\, \delta_{mn} 
     \Bigg],~ \nonumber\\
  \large[\mathcal{GU}\large]_{mn} &=& \frac{2\, \pi\, \alpha_s^2 }{9\, s^2} 
    \left| \mathcal{C}^n_{\tilde{q}_i q^\prime \tilde{g}}\, 
           \mathcal{C}^m_{\tilde{q}_j q \tilde{g}}  
    \right|^2
    \Bigg[ 
       \left( u\, t - m^2_{\tilde{q}_i}\, m^2_{\tilde{q}_j}\right)
       \left( 1-\delta_{mn} \right) 
     + m_{\tilde{g}}^2\, s\,\delta_{mn} 
    \Bigg],~ \nonumber
\eea
\bea
  \large[\mathcal{GTU}\large]_{mn} &=&
   \frac{-4\, \pi\, \alpha_s^2 }{27\, s^2} {\rm Re} 
     \left[\mathcal{C}^m_{\tilde{q}_i q \tilde{g}}\,
           \mathcal{C}^n_{\tilde{q}_j q^\prime \tilde{g}}\, 
           \mathcal{C}^{n \ast}_{\tilde{q}_i q^\prime \tilde{g}}\, 
           \mathcal{C}^{m \ast}_{\tilde{q}_j q \tilde{g}} 
     \right] 
     \Bigg[ 
        \left( u\, t - m^2_{\tilde{q}_i}\, m^2_{\tilde{q}_j}\right) 
	\left( \delta_{mn} - 1 \right)  
      + m_{\tilde{g}}^2\, s\,\delta_{mn}  
     \Bigg],~ \nonumber \\
 \large[\mathcal{NGA}\large]_{mn}^k &=& 
  \frac{8\, \pi\, \alpha \alpha_s}{9\, s^2\, x_W \big(1 - x_W\big)} {\rm Re} 
    \left[\mathcal{C}^{n\ast}_{\tilde{q}_j q^\prime \tilde{\chi}_k^0}\,
          \mathcal{C}^{m\ast}_{\tilde{q}_i q \tilde{\chi}_k^0}\,
	  \mathcal{C}^{n \ast}_{\tilde{q}_i q^\prime \tilde{g}}\,
	  \mathcal{C}^{m \ast}_{\tilde{q}_j q \tilde{g}} 
    \right] 
    \Bigg[
       \left( u\, t - m^2_{\tilde{q}_i}\, m^2_{\tilde{q}_j}\right) 
       \left( \delta_{mn} - 1 \right) 
      + m_{\tilde{\chi}^0_k}\, m_{\tilde{g}}\, s\,\delta_{mn}  
    \Bigg],~ \nonumber \\
 \large[\mathcal{NGB}\large]_{mn}^k &=& 
  \frac{8\, \pi\, \alpha \alpha_s}{9\, s^2\, x_W \big(1 - x_W\big)} {\rm Re} 
    \left[\mathcal{C}^{n\ast}_{\tilde{q}_i q^\prime \tilde{\chi}_k^0}\,
          \mathcal{C}^{m\ast}_{\tilde{q}_j q \tilde{\chi}_k^0}\,
	  \mathcal{C}^{n \ast}_{\tilde{q}_j q^\prime \tilde{g}}\, 
	  \mathcal{C}^{m\ast}_{\tilde{q}_i q \tilde{g}} 
    \right]
    \Bigg[ 
       \left( u\, t - m^2_{\tilde{q}_i}\, m^2_{\tilde{q}_j}\right)
       \left( \delta_{mn} - 1 \right) 
     + m_{\tilde{\chi}^0_k}\,  m_{\tilde{g}}\, s\,\delta_{mn} 
    \Bigg],
\eea
and in the differential cross section for the production of gaugino pairs (Eq.\
(40) in Ref.\ \cite{Bozzi:2007me})
\bea
 \frac{\d \hat{\sigma}^{q\bar{q}'}_{h_a, h_b}}{\d t} &=&
 \frac{\pi \alpha^2}{3 s^2}\Bigg[
    (1-h_a) (1+h_b) \Big[ 
      \left| Q^u_{LL} \right|^2 u_{\tilde{\chi}_i} u_{\tilde{\chi}_j} + 
      \left| Q_{LL}^t \right|^2 t_{\tilde{\chi}_i} t_{\tilde{\chi}_j} + 
      2 {\rm Re} [Q_{LL}^{u\ast} Q_{LL}^t] m_{\tilde{\chi}_{i}} m_{\tilde{\chi}_{j}} s 
    \Big] \nonumber
\eea
\bea
 &+& (1+h_a) (1-h_b) \Big[ 
      \left|Q_{RR}^u \right|^2  u_{\tilde{\chi}_i} u_{\tilde{\chi}_j} + 
      \left| Q_{RR}^t \right|^2 t_{\tilde{\chi}_i} t_{\tilde{\chi}_j} + 
      2 {\rm Re} [Q_{RR}^{u\ast} Q_{RR}^t] m_{\tilde{\chi}_{i}} m_{\tilde{\chi}_{j}} s
    \Big]\nonumber \\
 &+& (1+h_a) (1+h_b) \Big[ 
      \left| Q_{RL}^u \right|^2 u_{\tilde{\chi}_i} u_{\tilde{\chi}_j} + 
      \left| Q_{RL}^t \right|^2 t_{\tilde{\chi}_i} t_{\tilde{\chi}_j} + 
      2 {\rm Re} [Q_{RL}^{u\ast} Q_{RL}^t] (u t - m^2_{\tilde{\chi}_{i}} m^2_{\tilde{\chi}_{j}}) 
    \Big]\nonumber \\ 
 &+& (1-h_a) (1-h_b) \Big[ 
      \left| Q_{LR}^u \right|^2 u_{\tilde{\chi}_i} u_{\tilde{\chi}_j} + 
      \left| Q_{LR}^t \right|^2 t_{\tilde{\chi}_i} t_{\tilde{\chi}_j} +
      2 {\rm Re} [Q_{LR}^{u\ast} Q_{LR}^t] (u t - m^2_{\tilde{\chi}_{i}} m^2_{\tilde{\chi}_{j}}) 
    \Big]\Bigg].
\eea

In Figs.\ \ref{fig15} -- \ref{fig26}, we show examples of the obtained
numerical cross sections for charged squark-squark pair production,
neutral and charged squark-antisquark pair production, associated production of
squarks with charginos and neutralinos, and gaugino-pair production at the
LHC for our benchmark points E, F, G, H, I, and J and for
both of the two considered implementations of non-minimal flavour violation in
the GMSB model discussed in Sec.\ \ref{sec2}. We recall that the first is
based on mixing between matter and fundamental messengers, leading to flavour
mixing only in the left-left chiral squark sector and implemented at the
electroweak scale through the parameter $\lambda_{\rm LL}$, while $\lambda_{\rm
RR}$ is set to zero. The second scenario involves mixing with antisymmetric
messengers, giving rise to flavour violation in both the left-left and
right-right chiral squark sectors governed by the parameter $\lambda_{\rm
LL} = \lambda_{\rm RR}$. For the sake of better readability, we show only the
numerically most important curves as well as a selection of those that involve
visible flavour-violating effects. 

The magnitudes of the cross sections vary from the barely visible level of
$10^{-2}$
fb for weak production of heavy final states over the semi-strong production of
average squarks and gauginos and quark-gluon initial states to large cross
sections of $10^2$ to $10^3$ fb for the strong production of diagonal
squark-squark and squark-antisquark pairs or weak production of very light
gaugino pairs. Unfortunately, the processes whose cross sections are largest are
mostly insensitive to the parameter $\lambda_{\rm LL}$ in both flavour violation
scenarios, as the strong gauge interaction is insensitive to quark flavours and
gaugino pair production cross sections are summed over exchanged squark
flavours.  

Some of the subleading, non-diagonal cross sections show, however, sharp
transitions in particular squark production channels. These transitions are
directly related to the ``avoided crossings'' of the mass eigenvalues discussed
in Sec.\ \ref{sec3}. At the point, where two levels should cross, the
involved squarks change character and are subject to an exchange of their
flavour contents. Rather than the mass dependence on $\lambda_{\rm LL}$, these
exchanges then lead, together with the different parton densities in the proton,
to more or less sharp transitions in the production cross sections, where the
corresponding squarks are involved. This phenomenon is analogously observed in
the case of squark and gaugino hadroproduction in minimal supergravity
\cite{Bozzi:2007me}.  

As an example, let us discuss in detail the production of squarks and gauginos
for our benchmark point E. The cross sections in our flavour violation scenario
based on fundamental messengers are shown in Fig.\ \ref{fig15}. ``Avoided
crossings'' of mass eigenvalues occur here, e.g., for down-type squarks at a
value of $\lambda_{\rm LL} \approx 0.145$ between the squarks $\tilde{d}_3$ and
$\tilde{d}_4$, see also Fig.\ \ref{fig8}. Before this point, $\tilde{d}_3$ is
characterized by a dominant sdown content, while $\tilde{d}_4$ has first a
dominant sbottom and then sstrange content. For $\lambda_{\rm LL} \gtrsim 0.145$,
these contents are exchanged, i.e.\ $\tilde{d}_3$ is then a strange-squark and
$\tilde{d}_4$ becomes sdown-like. As a consequence, the cross sections involving
the two mass eigenstates exchange their values, since the production of first
generation squarks is preferred due to the more important parton density of up-
and down-type quarks in the proton. This can be seen in our example for the
production of down-type squark-squark and squark-antisquark pairs, mixed up- and
down-type squark-squark and squark-antisquark pair production, as well as for
the associated production of down-type squarks and charginos or neutralinos. For
up-type squarks, the level-reordering phenomenon occurs at values of
$\lambda_{\rm LL}\simeq0.09$ in the range excluded by BR($b\to s\gamma$) (left of the
vertical dashed/green line) and is therefore not shown here. However, another
effect becomes visible in the case of production cross sections that involve final
states with up-type squarks. Some of the mass eigenstates do not present sharp
transitions, but rather a continuous change in their flavour content. This is,
e.g., the case for the lightest mass eigenstate $\tilde{u}_1$. The corresponding
production cross sections increase smoothly with the flavour violation parameter
$\lambda_{\rm LL}$, which is explained by the fact that for lower values of
$\lambda_{\rm LL}$ the lightest up-type squark $\tilde{u}_1$ is mostly stop-like,
but receives sizable contributions of the light flavours for higher
$\lambda_{\rm LL}$. Together with the more important parton densities,
this results in an increase of the corresponding production cross sections. In
the same way, we also observe cross sections that decrease with $\lambda_{\rm
LL}$, due to a decrease of their light flavour content. 

The same phenomena are observed in the case of our second flavour violation
scenario with antisymmetric messengers, see Fig.\ \ref{fig16} for the benchmark
point E. Note that here also ``avoided crossings'' between up-type squark mass
eigenstates are observed, e.g.\ between $\tilde{u}_4$ and $\tilde{u}_5$ at
$\lambda_{\rm LL} \approx 0.11$, which lies, however, already in the
range excluded by BR($b\to s\gamma$). In this example, the $\tilde{u}_5$ loses its
important up-squark content to the scharm-dominated $\tilde{u}_4$. The latter
becomes then purely sup-like, enhancing its production cross section due to
the parton density in the proton, while the cross sections involving
$\tilde{u}_5$ become less important. 

For the benchmark points H with fundamental (Fig.\ \ref{fig21}) and I (Fig.\
\ref{fig24}) and J (Fig.\ \ref{fig26}) with antisymmetric messengers, we observe a
third effect at $\lambda_{\rm LL}=0.158$, 0.132 and 0.114, respectively. Here, the
pair production of up-type squark pairs ($\tilde{u}_{3}\tilde{u}_{4}$ for
fundamental and $\tilde{u}_{4}\tilde{u}_{5}$ for antisymmetric messengers)
exhibits an interesting resonance-like behaviour. It is generated by the fact that
these squark mass eigenstates exchange their up and charm flavour contents
(and also their chiralities in the case of antisymmetric messengers) at the
critical $\lambda_{\rm LL}$-values in a rather smooth way, so that both
squark mass eigenstates receive significant up- (valence-) quark
contributions to their production cross sections in the vicinity.

We remind the reader that in the case of flavour mixing only
in the left-left chiral sector, ``avoided crossings'' occur among
the $\tilde{q}_{1,2}$, $\tilde{q}_{3,4}$, and $\tilde{q}_{5,6}$ mass eigenstates,
whereas in the case of flavour mixing in both the left-left and right-right chiral
squark sectors, we rather observe the mass flips among the $\tilde{q}_{2,3}$ and
$\tilde{q}_{4,5}$ mass eigenstates, respectively.
Note also that the
difference between the two flavour violation scenarios is invisible for the
gaugino pair production in the bottom right panels of Figs.\ \ref{fig15} --
\ref{fig26}, respectively, that are practically insensitive to flavour
violation in the squark sector. 

Concerning the production of gravitinos, the cross sections achieve sizable
orders of magnitude only in the case of a rather light gravitino, see e.g.\ Ref.\
\cite{Klasen:2006kb}. If the latter is too heavy, its couplings are
too small to yield discoverable cross sections, since they are proportional to
the inverse of the gravitino mass squared. In particular, this is the case for
our scenarios with gravitino cold dark matter, where we have found a value of
the order of $m_{\tilde{G}} \sim 10^{-1}$ GeV derived from the different
cosmological constraints. Note that in order to have a very light gravitino and
consequently sizable production cross sections, one could consider a GMSB
scenario with gravitino hot dark matter ($m_{\tilde{G}} \lesssim 1$ keV) and
additional cold dark matter from stable messenger particles \cite{Falk:1994es,
Dimopoulos:1996gy, Hooper:2004fh, Ibe:2005xc}. In scenarios with a mixing
between messenger and matter fields, however, the stability of the lightest
messenger might be lost. We therefore do not consider the production of light
gravitinos in our GMSB scenarios with additional flavour violation in the squark
sector.

\section{Conclusions \label{sec6}}

While SUSY-breaking mediated by gauge interactions may be in principle attractive,
since the gauge interactions do not induce flavour violation as do the
gravitational
interactions, the parameter space of minimal GMSB models is today severely
constrained by low-energy, electroweak precision, and cosmological constraints,
in particular from the flavour-changing neutral-current decay $b\to s\gamma$.

We have discussed several possibilities how flavour violation may still be
induced in GMSB models. Focusing on messenger-matter mixing scenarios with
fundamental or antisymmetric messengers and flavour violation in the left-left
only or left-left and right-right chiral squark sectors, we have established
collider-friendly regions of parameter space that are at the same time
cosmologically viable and allow for the definition of benchmark points with
neutralino or stau NLSPs.

Depending on the strength of the flavour-violating parameter $\lambda_{\rm LL}$,
we showed that splittings and avoided crossings appear in the mass spectra,
inducing at the same time smooth or sharp transitions in the squark flavour
contents. This induces interesting phenomenological consequences, in particular
for squark and gaugino production cross sections at the LHC induced by valence
and sea quark parton densities that differ largely in magnitude.

We had hoped to also be able to investigate gravitino production and decay with
flavour violation at the LHC. This was, however, not possible, since cosmological
constraints require either a relatively heavy gravitino as a CDM candidate, with
masses of 10$^{-4}-10^{-1}$ GeV and consequently unobservably small cross
sections, or additional messenger dark matter, which would have been incompatible
with our messenger-matter mixing scenario.

\acknowledgments
The authors would like to thank L.\ Covi, G.\ Hiller, W.\ Hollik, S.\ Kraml and
F.D.\ Steffen for useful discussions. This work was supported by a Ph.D.\
fellowship of the French ministry for education and research and by the
{\em Theory LHC France} initiative of CNRS-IN2P3.


\end{document}